%% file: paper.tex
\documentclass[conference]{IEEEtran}
\usepackage[table,xcdraw]{xcolor}
\usepackage{cite}
\usepackage{amsmath,amssymb,amsfonts}
\usepackage[linesnumbered,ruled,vlined]{algorithm2e}
\usepackage{textcomp}
\usepackage{tabularx}
\usepackage{lscape}

\usepackage{mathtools}
\usepackage{xspace}
\usepackage{xargs}
\usepackage{ifthen}
\usepackage{xifthen}
\usepackage{comment}
\usepackage{underscore}
\usepackage{caption}
\usepackage{subcaption}
\usepackage{listings}
\usepackage{float}
\usepackage{placeins}
\usepackage{afterpage}

\usepackage{etoolbox}
\usepackage[textsize=scriptsize,textwidth=1.1cm]{todonotes}
\usepackage[hidelinks]{hyperref}
\usepackage{multirow} 

\def\BibTeX{{\rm B\kern-.05em{\sc i\kern-.025em b}\kern-.08em
    T\kern-.1667em\lower.7ex\hbox{E}\kern-.125emX}}

\usepackage[switch]{lineno}

\begin{document}

\title{Arm DynamIQ Shared Unit and Real-Time:\\An Empirical Evaluation%
}%

\iftrue
\author{%
	\IEEEauthorblockN{%
		Ashutosh Pradhan,
		Daniele Ottaviano,
            Yi Jiang,
            Haozheng Huang,
		Alexander Zuepke,
		Andrea Bastoni,
		Marco Caccamo
	}
	\IEEEauthorblockA{%
		Technical University of Munich
	}%
    \IEEEauthorblockA{Email: \{ashutosh.pradhan, daniele.ottaviano, yi9.jiang, haozheng.huang, alex.zuepke, andrea.bastoni, mcaccamo\}@tum.de}%
}%
\else
\author{%
}%
\fi

\maketitle


\input{0-notation}
\input{0-abstract}
\input{1-introduction}
\input{2-related}
\input{3-background}
\input{5-implementation}
\input{6-eval}
\input{7-conclusion}

\iftrue
\section*{Acknowledgments}
Marco Caccamo was supported by an Alexander von Humboldt Professorship
endowed by the German Federal Ministry of Education and Research.
\fi

\IEEEtriggeratref{49}
\bibliographystyle{IEEEtran}
\bibliography{paper}

\clearpage
\onecolumn
\appendix

\input{8-appendix}

\end{document}

%% file: 0-notation.tex
\definecolorseries{test}{rgb}{grad}[rgb]{.95,.55,.55}{11,11,17}
\resetcolorseries[10]{test}
\newcommand{\addtodoeditor}[1]{%
    \colorlet{#1}{test!!+!50}
    \expandafter\newcommand\csname#1\endcsname [1]{%
        \todo[color=#1,size=\tiny]{\sffamily\textbf{\uppercase{#1}:}
    ##1}\xspace%
    }
    \expandafter\newcommand\csname#1i\endcsname [1]{%
        \todo[inline, color=#1]{\sffamily\textbf{\uppercase{#1}:} ##1}\xspace%
    }
}


\addtodoeditor{az}
\addtodoeditor{ab}
\addtodoeditor{ot}
\addtodoeditor{ap}

\newcommand{\code}[1]{\mbox{\small\tt #1}\xspace}%
\newcommand{\bigo}[1]{\mbox{$\mathcal{O}(#1$)}\xspace}%
\newcommand{\bigolog}[1]{\bigo{\log{}#1}}%
\newcommand{\etal}{\mbox{\emph{et~al.\@}}\xspace}%
\newcommand{\wrt}{\mbox{\emph{w.r.t.\@}}\xspace}%
\newcommand{\eg}{\mbox{\emph{e.g.,\@}\@}\xspace}%
\newcommand{\ie}{\mbox{\emph{i.e.,\@}\@}\xspace}%
\newcommand{\cf}{\mbox{\emph{c.f.\@}}\xspace}%
\newcommand{\etc}{\mbox{\emph{etc.\@}}\xspace}%




\DeclarePairedDelimiter\ceil{\lceil}{\rceil}
\DeclarePairedDelimiter\floor{\lfloor}{\rfloor}

\newcommand{\optarg}[1][]{%
  \ifthenelse{\isempty{#1}}%
             {}
             {_{#1}}
}

\newcommand{\optpar}[1][]{%
  \ifthenelse{\isempty{#1}}%
             {}
             {(#1)}
}

\newtheorem{observation}{Obs.\@\xspace}

\newcommand{\FIXME}[1]{  \textbf{(FIXME: #1)}}
\newcommand{\memguard}{\emph{MemGuard}\xspace}
\newcommand{\mempol}{\emph{MemPol}\xspace}
\newcommand{\jailjouse}{\emph{Jailhouse}\xspace}
\newcommand{\sdvbshort}{\mbox{\emph{SD-VBS}}\xspace}
\newcommand{\sdvblong}{\emph{San Diego Vision Benchmark Suite}\xspace}

\newcommand{\scupftchhit}{\code{SCU\_PFTCH\_CPU\_HIT}}
\newcommand{\scupftchmiss}{\code{SCU\_PFTCH\_CPU\_MISS}}
\newcommand{\scupftchacc}{\code{SCU\_PFTCH\_CPU\_ACCESS}}
\newcommand{\llcdcache}{\code{L3D\_CACHE}}
\newcommand{\llcdaccess}{\code{L3D\_ACCESS}}
\newcommand{\busaccessdsu}{\code{BUS\_ACCESS}}
\newcommand{\loneinner}{\code{L1D\_CACHE\_REFILL\_INNER}}
\newcommand{\loneouter}{\code{L1D\_CACHE\_REFILL\_OUTER}}
\newcommand{\cpuectrl}{\code{CPUECTRL}}
\newcommand{\lthreeews}{\code{L3D\_WS\_MODE}}
\newcommand{\lthreecachealloc}{\code{L3D\_CACHE\_ALLOCATE}}


\newcommand{\readInterf}{\textit{read}\xspace}
\newcommand{\writeInterf}{\textit{write}\xspace}
\newcommand{\modifyInterf}{\textit{modify}\xspace}
\newcommand{\prefetchInterf}{\textit{prefetch}\xspace}

\newcommand{\localization}{\textit{localization}\xspace}
\newcommand{\sift}{\textit{sift}\xspace}
\newcommand{\tracking}{\textit{tracking}\xspace}
\newcommand{\mser}{\textit{mser}\xspace}
\newcommand{\disparity}{\textit{disparity}\xspace}

\newcommand{\vga}{\textit{vga}\xspace}
\newcommand{\cif}{\textit{cif}\xspace}

%% file: 0-abstract.tex
\begin{abstract}
The increasing complexity of embedded hardware platforms poses significant
challenges for real-time workloads.
Architectural features such as Intel RDT, Arm QoS, and Arm MPAM
are either unavailable on commercial embedded platforms or
designed primarily for server environments optimized for average-case performance
and might fail to deliver the expected real-time guarantees.
Arm DynamIQ Shared Unit (DSU) includes isolation features---among others,
hardware per-way cache partitioning---that can improve the real-time
guarantees of complex embedded multicore systems and facilitate
real-time analysis.
However, the DSU also targets average cases, and its real-time
capabilities have not yet been evaluated.

This paper presents the first comprehensive analysis of three
real-world deployments of the Arm DSU on Rockchip RK3568, Rockchip RK3588,
and NVIDIA Orin platforms.
We integrate support for the DSU at the operating system
and
hypervisor level
and conduct a large-scale evaluation using both synthetic and real-world
benchmarks with varying types and intensities of interference.
Our results make extensive use of performance counters and indicate that,
although effective, the quality of partitioning and isolation provided by the DSU depends on the
type and the intensity of the interfering workloads.
In addition, we uncover and analyze in detail the correlation between
benchmarks and different types and intensities of interference.
\end{abstract}

\begin{IEEEkeywords}
real-time system,
multi-core,
cache partitioning
\end{IEEEkeywords}

%% file: 1-introduction.tex
\section{Introduction}
\label{sec:intro}

The complexity of embedded real-time systems is growing exponentially.
Powerful AI-driven algorithms are increasingly being deployed at the edge,
driven by the growing data demands of advanced sensors. 
Embedded hardware platforms are evolving to meet these trends by
delivering enhanced computational capabilities while minimizing energy
consumption.

These platforms not only integrate dedicated accelerator units
(\eg TPUs or DSPs) but also
combine multiple types of cores with different computational
and energy profiles~\cite{arm-big-little, intel-pe-cores}.
Notably, to facilitate communication and optimize data flows, cores
and accelerators share the same memory hierarchy.

From a real-time perspective, the increasing complexity of such hardware
platforms introduce significant challenges to system analysis.
Over the years, the real-time systems community has proposed numerous techniques
at both the software and hardware levels to address these challenges, aiming to
reduce complexity and enhance the overall predictability of embedded multicore
systems.
The shared memory hierarchy has received particular attention, with prior
research extensively focusing on cache, memory, and interconnect
subsystems (\eg~\cite{MDBCCP:13, YYPCS:13, YMWP:14, KWCFAS:16}).

On the architectural side, technologies like Intel RDT~\cite{intel-rdt},
Arm QoS~\cite{arm-qos-overview}, and Arm MPAM~\cite{arm-mpam}
enable different levels of quality of service for software layers.
The real-time characteristics of RDT and Arm QoS have been exploited in
previous studies (\eg~\cite{XPCLLLL:19, STDM:20, STPMBZC:21}).
However, these techniques generally target
average-case scenarios, and corner cases may undermine the
predictability characteristics they can offer to real-time
applications~\cite{SBMYK:22}.
Furthermore, the lack of
processors implementing MPAM prevents its practical evaluation~\cite{ZiniCB:23}.

As an evolution of Arm's big.LITTLE architecture, DynamIQ technology combines
performance and energy-efficient cores into fully integrated clusters. The
DynamIQ Shared Unit (DSU)~\cite{arm-dynamiciq} provides a per-cluster last-level
cache (L3) and includes logic for interconnecting multiple DynamIQ clusters.
%
The DSU is adopted in most of the new Arm v8 and v9 designs.
For instance,
the recent
Rockchip RK3568 and RK3588, as well as
the NVIDIA Orin AGX~\cite{rockchip-rk3568, rockchip-rk3588, nvidia-jetson-orin},
incorporate up to three DSUs to connect single or multiple clusters of
Cortex-A55, A76, or A78 cores.
For real-time applications, the DSU is particularly interesting as it introduces,
among other features, 
the ability to partition the L3 cache into up to four way-based
partitions.

Similar to Intel RDT~\cite{XPCLLLL:19}, it is foreseeable that the
DSU's partitioning capabilities will become common in real-time
systems community.
However, the real-time capabilities of the DSU require close scrutiny.
For example, is its hardware-based way partitioning to be preferred to
software-based set partitioning (\emph{cache coloring})? Or,
what are the corner cases and limits of the DSU to support real-time
requirements?

This paper provides the first comprehensive evaluation of Arm DSU capabilities
for real-time workloads. Specifically, we present a large-scale evaluation
of three different deployments of the DSU on
Rockchip RK3568, 
RK3588, and NVIDIA Orin, 
as well as a comparison with the well-studied AMD Zynq UltraScale+
ZCU102~\cite{zcu-102} (Sec.~\ref{sec:background} shows details and
differences among these boards).
Using both synthetic and real-world benchmarks from the \sdvblong (\sdvbshort)~\cite{venkata2009sd}, we investigate the partitioning capabilities of the DSU
when applications are subject to different types and intensities of
interference.
Furthermore, using the Jailhouse hypervisor (HV)~\cite{jailhouse},
which has proven effective in prior
studies~\cite{KSMCVB:19, STDM:20, STPMBZC:21},
we evaluate and compare way and set
partitioning strategies on DSU-based boards in such environments.
On the ZCU102, we further underline differences with color-based partitioning,
which is the only viable cache-isolation mechanism on previous generations of Arm chips.
In all evaluations, we use carefully selected performance counters
to provide detailed insights.

Our results show a complex picture.
Although DSU-based partitioning is effective in protecting real-time workloads
from interference, the quality of isolation strongly depends on
the interfering workload and its intensity.
Our findings also indicate that platform-specific optimizations, such as
\emph{prefetching} or \emph{write-streaming}, can considerably impact the
results.
Additionally, to the best of our knowledge, this paper is the first to
provide in-depth details on the correlations between different \sdvbshort benchmarks
and both types 
and intensity of the interference.

In summary, the paper makes the following contributions:
\begin{itemize}
	\item Details how to support DSU features on three different boards for
both Linux- and hypervisor-based setups.
	\item Uses both synthetic and real-world benchmarks
to evaluate the real-time capabilities of DSU for different types and intensities
of interference.
	\item Compares way and set partitioning techniques and highlights
differences 
to setups for previous core generations.
	\item Exploits the rich performance counter architecture to provide
in-depth insights and unveil 
corner case behaviors on different platforms.
\end{itemize}

%

In the following: Sec.~\ref{sec:related-work} discusses previous works.
Sec.~\ref{sec:background} presents background on the memory subsystem for
Arm embedded-boards. Sec.~\ref{sec:setup} and Sec.~\ref{sec:eval} detail
the experimental setup, experiments, and observations.
Sec.~\ref{sec:conclusion} concludes.

%

%% file: 2-related.tex
\section{Related Work}
\label{sec:related-work}

The issue of shared resource management in multicore architectures has been well-studied in the context of real-time systems. Uncontrolled interference from co-running cores introduces non-determinism in the computation of response time and worst-case execution time (WCET) analysis. Previously, attempts have been made to define a single-core equivalent of a multicore commercial off-the-shelf (COTS) architecture~\cite{Mancuso-WCET-Estimation, Sha-RT-Multicore}. 
To define such a single-core equivalent, one must identify and account for the major sources of interference on COTS multicore architectures---namely, the shared caches, the shared interconnect, and the shared main memory~\cite{Lugo-Survey-Interference}. 

To improve the predictability in accessing shared caches, diverse cache management techniques have been proposed, which can be mostly aggregated into \emph{cache-partitioning techniques} and \emph{cache-locking techniques}~\cite{Lugo-Survey-Interference, Gracioli-Survey-Cache-Management}. 
Cache partitioning divides the cache into smaller chunks that can be assigned to specific tasks or cores. In a set-associative cache architecture, the cache can be divided into specific ways (way partitioning) or 
specific sets 
(set partitioning).
Generally,
the methods proposed for way partitioning require modifications of the cache implementation in hardware~\cite{Qureshi-Utility-Cache, PiPP, CCPiPP, Survey-Way-Part}.
The DSU implements way partitioning at the hardware level.
Set partitioning, on the other hand, has been implemented leveraging hardware features~\cite{Sri-Hardware-CP, Suh-Hardware-CP, Iyer-Hardware-CP} or purely in software~\cite{Wolfe-Software-CP, Ward-Software-CP, Kim-Software-CP, COLORIS, KSMCVB:19}. A popular method for implementing set partitioning in software is \emph{page coloring}, which can be implemented by changing the page table management at the operating system (OS) level~\cite{MDBCCP:13, COLORIS}, or at the hypervisor level~\cite{KSMCVB:19, xilinx-xen-cache-color}.
Contrary to cache partitioning, cache locking
pins certain cachelines to make the cache content and WCET more predictable~\cite{Gracioli-Survey-Cache-Management, Mittal-Survey-Locking}.
However, it often requires hardware support which is not readily available on current-generation COTS platforms~\cite{Mittal-Survey-Locking}.
For example, earlier 32-bit Arm v7-A architectures with at most L2 caches supported \emph{cache lockdown} and \emph{Lockdown-by-Master} (in the L2C-310 controller~\cite{arm-l2c-310}).
This feature was phased out in the transition from Cortex-A9 to Cortex-A15.
The numerous differences between Arm v7-A CPUs that supported cache lockdown and those with DSU make systematic evaluations and direct comparisons with our work impractical.
A combination of locking and partitioning can be used in synchronization to obtain higher levels of predictability~\cite{MDBCCP:13}.



Apart from shared caches, interference from competing cores in the shared interconnect and main memory can affect real-time performance. Most COTS platforms employ Dynamic Random Access Memory (DRAM) as the main memory, which is ill-suited to provide real-time guarantees~\cite{hassan2019reduced}. Memory access requests from non-real-time tasks can complicate the ability to provide a tight WCET for real-time tasks. To improve the predictability of memory accesses, various works propose changes to the existing hardware ~\cite{hassan2019reduced, BRU:20, MEDUSA:15}. Notably, the DRAM controller has been redesigned to provide an analytical bound on the maximum latency for each memory request~\cite{Paolieri-Mem-Controller,Reineke-PRET-DRAM,DRAMbulism:20}. In contrast to the methods that require hardware changes, several works rely on making OS-level software modifications to allow deterministic main memory access. For example, in~\cite{YMWP:14}, the authors propose PALLOC, a memory allocator in OS that reduces inter-core memory interference by partitioning the DRAM banks. Approaches like MemGuard~\cite{YYPCS:13} and MemPol~\cite{MemPol} propose a per-core memory bandwidth regulation system using performance counters. Along similar lines, Sohal \etal~\cite{STDM:20} propose a framework for profiling the temporal workload on both CPU and accelerators using a similar approach to memory bandwidth regulation. 

In recent years, new features have been introduced in COTS platforms that can aid real-time execution of applications. In 2015, Intel introduced the Resource Director Technology (RDT) to monitor shared platform resources~\cite{intel-rdt}. Similarly, QoS features exist at the interconnect level in Arm and AMD~\cite{arm-qos-overview,amd-qos}. Arm MPAM~\cite{arm-mpam} holds promise from the real-time perspective, but it is yet to be implemented in COTS platforms. Xu \etal~\cite{XPCLLLL:19} use the Cache Allocation Technology (CAT) feature of Intel RDT to provide an algorithm for partitioned scheduling of tasks in a multicore environment. However, 
the real-time features in RDT, such as cache partitioning,
are not always effective in practice~\cite{SBMYK:22}.
The recently introduced Arm DSU offers cache-partitioning of the shared cache in a cluster. However, a detailed study of the feature and its real-time applications has yet to be done.

%% file: 3-background.tex
\section{Background}
\label{sec:background}

\input{figures/cache_partitioning}

\subsection{General}
\label{cache-general}

We present a brief overview of the cache architecture on current 64-bit Arm platforms.
The Arm cores\footnote{In the paper, we use the terms \emph{core} and \emph{CPU} interchangeably.}
discussed in this paper
use a cacheline size of 64~B
at all levels
of their memory hierarchy.
%
Therefore,
the lowest 6 bits of an address
\code{addr[5:0]}
denote a specific \emph{byte offset} in a cacheline.
The next higher $\log_2{n_{sets}}$ bits of an address,
\eg \code{addr[15:6]} for 1024 sets in the L2 cache on the ZCU102,
define the \emph{set index} into a specific cache
(see Fig.~\ref{fig:cache-coloring-set}).
Additionally, the set index can include information of further address bits, as in the case of the XOR permutation on the Cortex-A78; see Sec.~\ref{cortex-a78}.

\subsection{Cache Set Partitioning}
\label{set-partitioning}

In a cache,
if the address range covered by set index and offset
is larger than the page size,
\emph{page coloring} can be used,
as Fig.~\ref{fig:cache-coloring-set} shows:
The physical address bits beyond the page size defines the specific page color,
\eg 16~colors (4~bits) for a page size of 4~KB (\code{addr[11:0]}) in the L2 cache on the ZCU102,
while the virtual address bits ensure 
a contiguous mapping of the physical memory to the application, set by the OS~\cite{MDBCCP:13} or the hypervisor~\cite{xilinx-xen-cache-color}.

\subsection{Cache Way Partitioning}
\label{way-partitioning}

In contrast,
way partitioning reserves a specific number of ways for a partition.
%
In the Arm DSU, a group of four ways can be assigned to one or more \emph{DSU scheme IDs} (up to eight). Each core can then be assigned a scheme ID
as Fig.~\ref{fig:cache-coloring-way} shows.
%
The L3 cache in the DSU 
has 16 ways,
so effectively, four partitions are available
to the 
eight cores supported by a DSU~\cite{arm-dynamiciq}.
%
On Arm, 
way partitioning effectively selects cache ways in the L3 cache 
eligible to \emph{allocate} data to 
when data is \emph{evicted} from the higher level L1 or L2 caches.
Therefore,
applications can always access 
all cachelines
in the cache,
unlike in the MMU-based set partitioning.
%
As way groups can be assigned to multiple partitions, arbitrary combinations 
of data sharing are possible.

\subsection{Memory Accesses and Architectural Optimizations}  
\label{memory-accesses}  
Modern architectures optimize memory accesses through various mechanisms that influence cache behavior. Memory operations generally fall into three categories: \textit{reads}, which load data from memory into registers; \textit{writes}, which store data from registers back to memory; and \textit{modifications}, which involve read-modify-write sequences such as atomic operations.  

To improve performance,
recent
Arm CPUs employ \emph{hardware prefetching}, which anticipates future memory accesses and loads data into caches before it is explicitly requested. While this reduces memory latency, it introduces
additional
memory operations. Additionally, to optimize memory bandwidth for large data transfer, the \emph{write-streaming} mechanism enables writes to bypass the L3 cache entirely. This can introduce latency inconsistencies depending on memory bandwidth contention. Prefetching and write-streaming are unavoidable and activated automatically in Arm systems today, and even simple operations like \texttt{memcpy} or \texttt{memset} can trigger them.
Understanding their impact on cache partitioning is necessary for accurately assessing system behavior and ensuring performance isolation.

\subsection{Cortex-A53}
\label{cortex-a53}

Introduced in 2012,
the Cortex-A53 is a power-efficient in-order design
in the first generation of 64-bit Arm cores~\cite{arm-a53trm}.
The Cortex-A53 supports a private 2-way set-associative L1 instruction cache
and a private 4-way set-associative L1 data cache, 
both with configurable sizes between 8~KB to 64~KB.
%
According to the documentation~\cite{arm-a53trm}, both L1 caches use a \emph{pseudo-random cache replacement policy}.
The Cortex-A53 predates the DSU,
and thus cannot be used together with it,
however,
cores can be instantiated in clusters of up to four cores
that share the L2 memory subsystem.
For more cores,
multiple clusters
of the same or different core types
must be instantiated.
%
%
All cores in a cluster share the same optional 16-way set-associative L2 cache
with sizes from 128~KB to 2~MB.
%
The shared L2 cache is used as a \emph{victim cache}
and is \emph{exclusive} to the L1 caches,
\ie data is not allocated in the L2 cache on misses in L1.
Cachelines get allocated in the L2 cache only when they are evicted from the cores' L1 caches.
%
%
Arm documents the internal format of the tag RAMs of TLBs and L1 caches,
but not of the L2 cache~\cite{arm-a53trm}.


\subsection{Cortex-A55}
\label{cortex-a55}

The Cortex-A55 is the power-efficient successor of the Cortex-A53 for DSU-based designs~\cite{arm-a55trm}.
%
Like its predecessor,
it uses an in-order pipeline design.
It supports 4-way set-associative instruction and data L1 caches
with sizes between 16~KB and 64~KB.
%
%
The \emph{age} bits in the documentation of the tag RAMs
indicate LRU or pseudo-LRU (PLRU) as replacement policy for the L1 data cache.
The optional 16-way set-associative L2 cache is private to the core
with sizes between 64~KB and 256~KB.
Like on the Cortex-A53, the L2 cache is a \emph{victim cache}
and is \emph{exclusive} to the L1 data cache.


\subsection{Cortex-A76}
\label{cortex-a76}

The Cortex-A76 is the second generation of out-of-order performance cores for DSU-based designs~\cite{arm-a76trm}
and typically paired with Cortex-A55 cores.
The Cortex-A76 has 4-way set-associative L1 data and instruction caches of 64~KB size.
Arm documents a PLRU cache replacement policy for both.
%
%
%
The 8-way set-associative L2 cache is of 128~KB to 512~KB size.
%
The L2 cache is mandatory and \emph{inclusive} to the L1 data cache,
\ie any data allocated in the L1 cache also takes space in the L2 cache.
The documentation mentions a \emph{dynamic biased replacement policy}
with seven \emph{PLRU} bits in the L2 victim RAM~\cite{arm-a76trm}.
The Cortex-A76 is also available in a Cortex-A76AE variant
for safety-critical applications~\cite{arm-a76aetrm}.


\subsection{Cortex-A78}
\label{cortex-a78}

The Cortex-A78 is the fourth generation of out-of-order performance cores for DSU-based designs~\cite{arm-a78trm}
and can also be paired with power-efficient Cortex-A55 cores.
%
%
The Cortex-A78 uses 4-way set-associative L1 data and instruction caches of 32~KB or 64~KB size.
Arm documents a PLRU cache replacement policy for both L1 caches.
The 8-way set-associative L2 cache is 256~KB or 512~KB size.
As in the Cortex-A76, the L2 cache is inclusive with the L1 cache.
Arm lists 24 bits in the L2 victim RAM for the replacement policy,
but differently from the other cores,
the L2 cache uses a permutation function to derive the set index,
\eg \code{addr[22:15] XOR addr[14:7]}
for the 256~KB configuration~\cite{arm-a78trm},
as shown in Fig.~\ref{fig:cache-coloring-way}.
%
%
A Cortex-A78AE variant is also available~\cite{arm-a78aetrm}.


\subsection{DSU}
\label{dsu}

Announced in 2017 with the Cortex-A55 and A75 cores,
the \emph{DynamIQ Shared Unit} (DSU)
interfaces the cores
and provides the L3 memory system in an SoC~\cite{arm-dynamiciq}.
The DSU supports up to eight cores
in up to three internal clusters of \emph{big}, \emph{little} and \emph{other} cores
of the same microarchitecture.
To support more than eight cores,
multiple instances of the DSU must be instantiated.
The DSU-AE variant additionally implements lock-step functionality
for 
pairs of two
AE cores~\cite{arm-dynamiciq-ae}.
The DSU provides an optional 16-way set-associative shared L3 cache
with sizes from 256~KB to 4096~KB.
The L3 cache can be implemented physically in one or two slices (memory blocks).
The address bit to select the slice is configurable.
Power management allows \emph{all}, $\frac{3}{4}$, $\frac{1}{2}$, $\frac{1}{4}$ or \emph{none}
of the cache to be powered down.
%
%
For cache sizes of 1536~KB and 3072~KB,
the number of available ways in the cache is reduced to 12,
and $\frac{1}{4}$ of the cache is permanently powered down.
The DSU
implements \emph{cache-way partitioning}
in the L3 cache
with a granularity of groups of four ways,
as described 
in Sec.~\ref{way-partitioning}.
%
%
%
%
When cores like the Cortex-A55 are configured without L2 caches,
the shared L3 cache appears to them as L2 cache.
However,
adhering to Arm documentation,
we indicate the DSU cache as L3 cache and denote the L2 cache as missing.
%
The DSU provides its own Performance Monitoring Unit (PMU).
The \code{CLUSTERPM} PMU registers are shared between the cores
and monitor cache and bus activity of all cores.

%% file: figures/cache_partitioning.tex
\begin{figure*}[!t]%
\centering%
    \begin{subfigure}[b]{0.99\columnwidth}%
    	\includegraphics[width=\columnwidth]{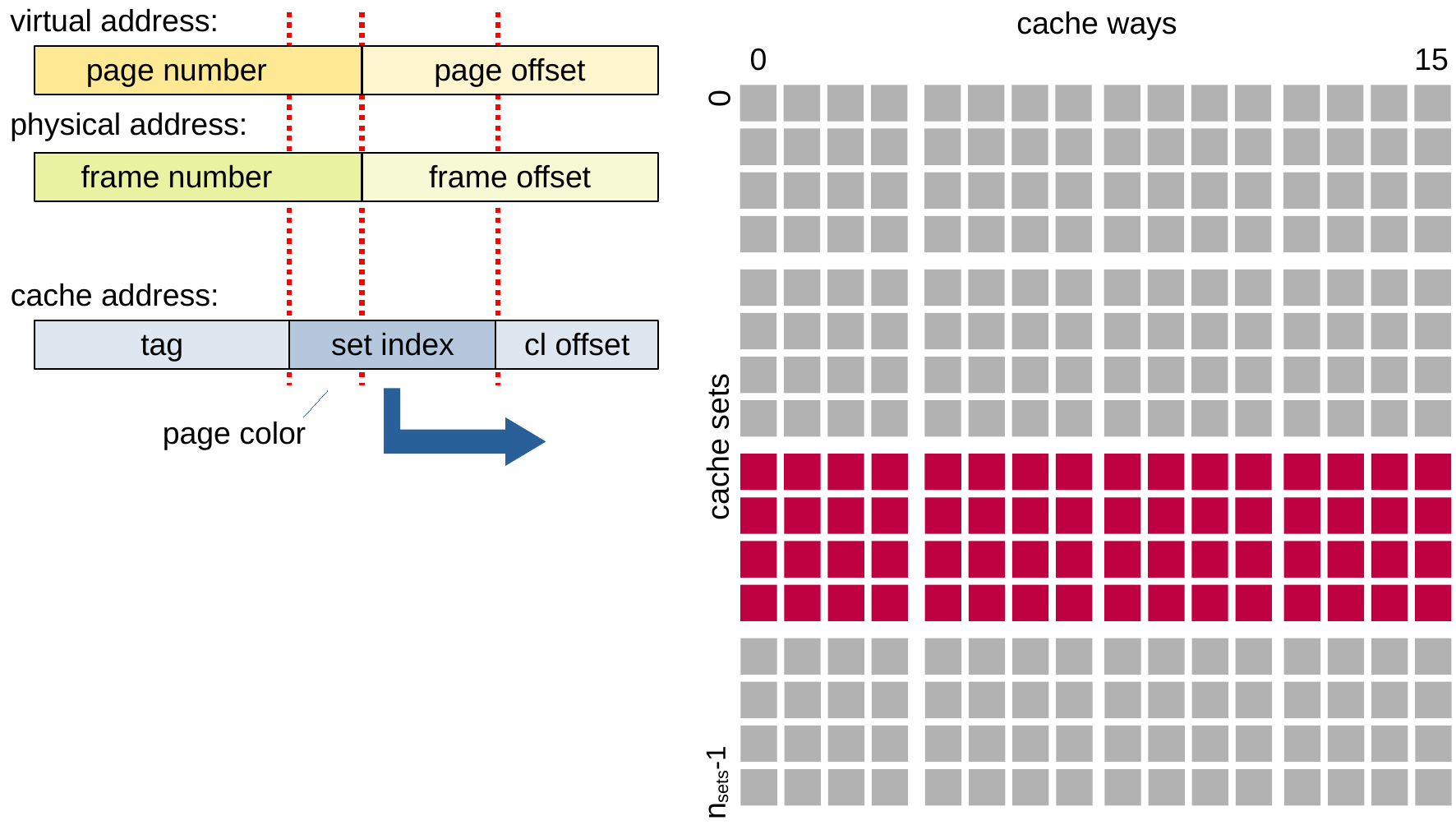}%
        \caption{Cache set partitioning using page coloring.}%
        \label{fig:cache-coloring-set}%
    \end{subfigure}%
    \hspace{\columnsep}%
    \begin{subfigure}[b]{0.99\columnwidth}%
     	\includegraphics[width=\columnwidth]{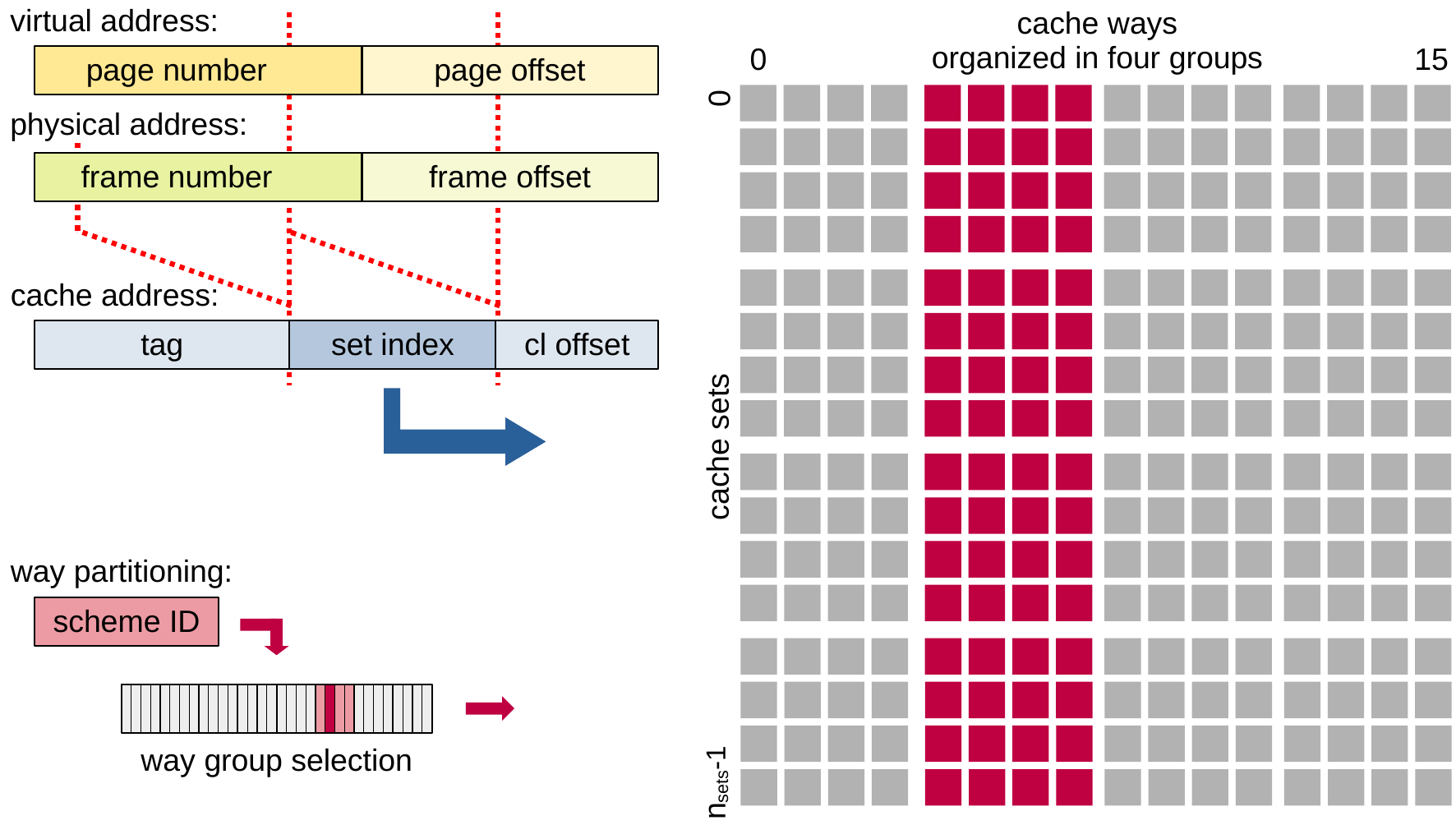}%
        \caption{Cache way partitioning in Arm DSU.}%
        \label{fig:cache-coloring-way}%
    \end{subfigure}%
\caption{%
    Cache tag and set index are derived from the physical address.
    The set index selects a set (row) in the cache.
    For a cache hit, the tag matches a corresponding tag in one of the ways (column).
    %
    (a) Set partitioning:
    bits beyond the page size in the set index select the cache color,
    \ie a specific range of sets.
    The set index requires a linear dependency on the address.
    (b) Way partitioning:
    the ID register selects one of eight schemes
    in the group selection register.
    The four bits indicate the group of ways private to the scheme.
    The set index can be a non-linear function,
    \eg permutated from upper and lower address bits.
}%
\label{fig:cache-coloring}%
\vspace{-1em}
\end{figure*}%

%% file: 5-implementation.tex
\section{Setup}
\label{sec:setup}

Our in-depth, exhaustive 
experiments
focus
on isolation and real-time properties. To this end, 
we compare and analyze
two implementations of cache-partitioning: set partitioning, 
enabled through Jailhouse's cache coloring~\cite{minerva-jailhouse},
and way partitioning, 
achieved through a custom driver controlling the Arm DSU cache 
partitioning feature.

\subsection{SoCs, Boards, and Configuration}
We refer to the specific boards by the name of their SoCs,
as the ability to use cache partitioning
is a property of the SoC and the firmware.
Boards using the same SoC might have different configurations,
\eg use DDR4 vs LP-DDR4,
but this does not change the general observations.

\subsubsection*{ZCU102}
The ZCU102 is a reference board
for the AMD Zynq UltraScale+ SoC family,
which supports four Cortex-A53 cores
with 32~KB private L1 caches
and a shared L2 cache of 1~MB size~\cite{zcu-102}.
We use Linux kernel 6.1\_LTS of the Linux repository from Xilinx~\cite{xilinx-linux}. 
We note that this board lacks the DSU and is primarily included as a baseline platform 
for comparison, given its frequent use in prior research on cache 
partitioning and specifically on 
set partitioning~\cite{RM:20,CCOSMMTBP:23}.

\subsubsection*{RK3568}
The Rockchip RK3568 SoC supports
four Cortex-A55 cores with 32~KB private L1 caches
and a shared LLC of 512~KB size in the DSU~\cite{rockchip-rk3568}.
This SoC lacks private L2 caches for the cores
and, as previously noted,
we refer to the DSU cache as L3 cache
and denote an L2 cache size of zero
for consistency with the other SoCs.
We evaluate the SoC on a Youyeetoo YY3568 board
using Debian 12 and a mainline Linux kernel 6.1.

\subsubsection*{RK3588}
The Rockchip RK3588 SoC supports
four Cortex-A55 cores with 32~KB private L1 caches and 128~KB private L2 caches,
four Cortex-A76 cores with 64~KB private L1 caches and 512~KB private L2 caches,
and a shared L3 cache of 3~MB size in the DSU~\cite{rockchip-rk3588}.
%
%
We performed our experiments on both
the Orange Pi 5 Plus
and Raxda ROCK 5B boards.
The results are similar, and we consider only the 
Orange Pi 5 Plus
for the rest of this paper.
We again use Debian 12 and a mainline Linux kernel 6.11-rc4.\footnote{The vendor kernel 5.15 is tuned for aggressive power saving and Android use cases and produces unstable benchmark results.}

\subsubsection*{Orin}
We evaluate the NVIDIA Orin SoC
on the NVIDIA Orin AGX Devkit. 
In this configuration,
the Orin provides three DSU-AE clusters
of four Cortex-A78AE each.
The cores are configured with 64~KB private L1 caches,
256~KB private L2 caches,
2~MB L3 caches per DSU,
and a 16-way set-associative L4 cache of 4~MB size.
We use 
Jetson Linux 36.4 (based on Ubuntu 22.04) with Linux kernel 5.15
and the default power mode (30W, mode ID 2).
We note that extending Jailhouse support to the Orin is a non-trivial effort that is still ongoing.
Therefore, we defer the analysis of set partitioning for the NVIDIA Orin
to future work.



\subsection{Evaluation Methodology}
\label{sec:eval-setup}
For each hardware setup, we evaluate the real-time performance of 
the two cache-partitioning implementations by assessing the temporal 
behavior of \sdvbshort~\cite{venkata2009sd, rt-bench}
running alongside a memory-intensive 
synthetic interference application on separate cores, both described 
in detail in Sec.~\ref{subsec:bench} and Sec.~\ref{subsec:interference}.
The evaluation is conducted by varying two factors: the cache 
allocation between the benchmark and the interference, and the 
work-set size of the interference task.

Specifically, each benchmark is executed first without any cache 
partition. Then, the cache is partitioned using both way 
and set techniques (as applicable) by allocating four cache 
partitions between the benchmark and interference applications in 
three ratios: 
\begin{itemize}
    \item \textbf{1/3}: $\frac{1}{4}$ to the benchmark and $\frac{3}{4}$ to the interference
    \item \textbf{2/2}: $\frac{1}{2}$ to the benchmark and $\frac{1}{2}$ to the interference
    \item \textbf{3/1}: $\frac{3}{4}$ to the benchmark and $\frac{1}{4}$ to the interference
\end{itemize}
All configurations are tested by varying the working-set size
(\emph{``intensity''})
of the interference from 8~KB to 8~MB. We evaluate the 
 real-time performance of partitioning when the interference impacts 
different layers of the memory hierarchy, from the L1 cache to the 
DRAM. The minimum working set size is smaller than the L1 
cache and the maximum is larger than two times the size of the 
LLC for each platform.

\subsubsection{Set Partitioning Setup}
\label{subsec:set-setup}
We rely on the cache-coloring implementation in the Minerva fork of
Jailhouse~\cite{minerva-jailhouse} for the \emph{ZCU102}, as it provides fine-grained control over cache set partitioning (\eg~\cite{KSMCVB:19}).
We extended Jailhouse to support our Rockchip platforms (both Cortex-A55 and A76 cores), integrating support for cache coloring.
The work has been upstreamed~\cite{minerva-jailhouse}.
To have a fair comparison between set and way partitioning,
we avoid choosing
configurations that would color the L1 cache of a core, and we carefully 
select configurations that---like on way partitioning---only partition the LLC.

\subsubsection{Way Partitioning Setup}
\label{subsec:way-setup}
We develop an ad-hoc \emph{driver}
(specifically, an OS driver and a HV-extension)
to control 
the DSU registers at both Linux and Jailhouse levels. 
Using the driver, we (I) initially assign a unique ID $0..n-1$ to each 
core's \code{CLUSTERTHREADSID} registers, and then we (II) assign the way groups in the \code{CLUSTERPARTCR} register.

However, by default, access to the DSU PMU and cluster configuration registers 
is disabled at the hypervisor and operating system level.
Therefore, we also modify the Arm Trusted Firmware (ATF)
and unconditionally enable the
\code{SMEN},    
\code{TSIDEN},  
and \code{CLUSTERPMUEN} bits
in both \code{ACTLR_EL3} and \code{ACTLR_EL2} registers
on all available cores.
This allows us to configure DSU cache partitioning
at Linux kernel level for our experiments.
%
Note that hypervisors should not configure \code{ACTLR_EL2}
to disable access to the cache configuration registers at OS level in a VM.

\subsection{Performance Counters in Cortex-A55, A76, A78 and DSU}
\label{subsec:perfcounters}
Nearly one hundred PMU events are available per Arm core, and over forty within the DSU. Still, only a subset can be counted at any given time due to the limited number of Performance Monitoring Counters (PMCs) available on both the cores and DSU.
We use the following per-core and DSU events to analyze cache and memory interference. 

\subsubsection{Arm Cores Events}
\label{subsec:core-events}
\loneinner \code{(0x44)} and \loneouter \code{(0x45)} measure L1 data cache refills, distinguishing between refills sourced from within the cluster (like the L3 cache or from another core) versus those from external sources. This distinction aids in assessing the locality of memory access and the pressure placed on the L1 cache by external memory transactions. 
The \lthreeews \code{(0xc7)} event (only available on Cortex-A55) records cycles in which the core operates in \emph{write-streaming} mode, avoiding L3 allocations. 
This allows for assessing the impact of sustained write operations on cache utilization and memory coherence.

\subsubsection{DSU Events}
\label{subsec:DSU-events}
\scupftchacc \code{(0x500)} tracks L3 prefetch transactions initiated by the CPU, giving insight into the frequency and distribution of L3 cache prefetching. 
Complementing this, \scupftchhit \code{(0x502)} and \scupftchmiss \code{(0x501)} count prefetch hits and misses, respectively, helping to measure prefetching effectiveness in reducing memory latency. 
Additionally, \llcdaccess \code{(0x2b)} is a comprehensive counter that captures all read and write transactions directed to the DSU, providing an aggregated view of L3 activity. 
\busaccessdsu\code{(0x19)} monitors 
transactions
between the DSU and the interconnect, indicating the volume of external memory traffic used to understand memory transaction demands. 
Finally, \lthreecachealloc \code{(0x29)} counts L3 cache allocations that do not involve refills. We use it to observe how cache space is utilized without triggering additional memory latency.

\subsection{Dataset – RT-Bench Vision}
\label{subsec:bench}
This setup uses the RT-Bench framework~\cite{rt-bench}, which 
wraps the \sdvbshort suite to set up real-time behavior: RT-Bench enables 
periodic task execution, the use of real-time task priorities, 
and precise temporal measurements such as latency and execution time.%
\footnote{We use \texttt{dev/unstable} RT-Bench (\url{https://gitlab.com/rt-bench/rt-bench/}). It was cross-compiled on x86 Ubuntu 22.04 using aarch64-linux-gnu-gcc version 11.4.0}

The \sdvblong used for the assessment includes the 
following applications:
\begin{itemize}
    \item \disparity computes depth information by identifying 
    pixel correspondences between stereo images. 
    \item \sift (Scale-Invariant Feature Transform) detects and 
    matches distinctive features in images, providing robustness across scale and orientation changes.
    \item \localization assesses spatial positioning within an 
    environment by leveraging visual landmarks. 
    \item \mser (Maximally Stable Extremal Regions) identifies 
    stable regions across varying thresholds. 
    \item \tracking monitors the position of objects over time 
    in video sequences, enabling applications in surveillance 
    and interactive environments.
\end{itemize}
To improve the quality of the evaluation, we decided to use 
two datasets with different image resolutions, namely \cif (352×288) 
and \vga (640×480). The different datasets help understand the impact of
computational-intensive workloads 
on real-time performances while partitioning the cache.

\subsection{Interference-Bench}
\label{subsec:interference}
To obtain deep insights from experiments on \sdvbshort,
we focus on analyzing the effects of interfering 
workloads with varying working set sizes under different 
cache partitioning configurations. To achieve this, we 
developed a synthetic interference program (henceforth 
called interference-bench) which enables the generation 
of four types of interference workloads---\readInterf, \writeInterf, \modifyInterf,
and \prefetchInterf---each with configurable working set sizes.
\begin{itemize}
    \item \readInterf refers to reading of data from memory 
    using a \emph{load} instruction of a single byte. This requires the core to 
    load the full cacheline to the L1 cache.
    \item \writeInterf refers to writing of a cacheline 
    to  memory using \emph{store} instructions to the whole cacheline.
    \item \modifyInterf changes a single byte of a cacheline using a \emph{store} instruction. 
    The core thus needs to load the full cacheline and then 
    apply the modification to it.
    \item \prefetchInterf refers to prefetching a cacheline for 
    future use. In our experiments, we refer to prefetching to 
    the L3 cache in RK3568, RK3588 and the Orin boards as 
    prefetch.
\end{itemize}

\input{figures/fig-plru}

We chose not to use other types of realistic applications as 
interference sources, as this synthetic interference benchmarks 
provides the most extreme memory usage patterns, revealing 
worst-case behaviors. This approach allows us to expose 
platform-specific corner cases that could arise in real-world 
scenarios under maximum stress conditions.
Both the benchmarks and interference are implemented on Linux and
the OS itself can be considered part of the working set.

Given the extensive number of experiments conducted, this 
paper presents only the most significant results that contribute 
directly to our observations in Sec.~\ref{sec:eval}.
\footnote{%
The extended version of the paper with all results is available in~\cite{rtas2025-extended}.}

\subsection{Bit Assignment for Set Partitioning}
\label{sec:set-hash-scrambling}


Set partitioning using \emph{page coloring} techniques relies on the fact that the set index can be computed by masking specific (coloring)  bits of a given physical address (see Sec.~\ref{sec:background}).
The presence of undocumented bit-scrambling and hash functions can result in losing control over the coloring bits and, consequently, in unexpected and erroneous behavior of the set partitioning implementation.
Arm's DSU~\cite{arm-dynamiciq} does not document the presence or absence of bit-scrambling in set-index computation. To verify the correct behavior of set partitioning on the boards under test, we experimentally validate the following hypothesis: If we load a page of a given color as computed from the set index, it will replace the previous pages of the same color loaded into the cache. The experiment involves the following steps:
\begin{itemize}
    \item Similar to~\cite{PM:15}, we use 2~MB Linux huge pages to target specific set-index in the cache. Since we need 21 bits to address a location inside a page of size 2~MB (the lower 21 bits of the virtual address), a memory location inside the page will be equal to the physical address.
    \item Using the huge page and software prefetch command, we sequentially load $n$ 4~KB memory chunks of the same color to the L3 cache.
    \item We prefetch $n$ 4~KB memory chunks of the same color to the L3 cache again and count the number of prefetch hits and misses using the PMCs (see Sec.~\ref{subsec:perfcounters}). 
\end{itemize}
We executed the experiment on both RK3568 and RK3588 using one Cortex-A55 core.
Hardware prefetching was turned off by writing to the \code{L3PCTL} bit field of the \code{CPUECTRL} register to avoid interference from the hardware prefetcher. 
Following a similar calculation as in Sec.~\ref{set-partitioning}, we obtain 8 colors on the RK3568 and 64 on the RK3588. Since the RK3568 has an L3 cache size of 512~KB, each color has a size of 64~KB. Similarly, each color on the RK3588 has a size of 48~KB.

In Fig.~\ref{fig:prefetch-plru}, we plot the miss ratio (number of misses observed divided by the maximum number of misses possible) for a given access size. We observe that the number of misses rises linearly as we cross the size of one color, indicating cache thrashing as content is loaded to the same previously occupied locations. This result, combined with the results from
Obs.~\ref{obs:way-conflict} and Obs.~\ref{obs:way-set-same}, strongly indicate
that the DSU does not implement bit-scrambling in the color bits of the set index computation.

%% file: figures/fig-plru.tex
\begin{figure}[!t]%
\centering%
\includegraphics[width=\columnwidth, trim=10 10 10 10, clip]{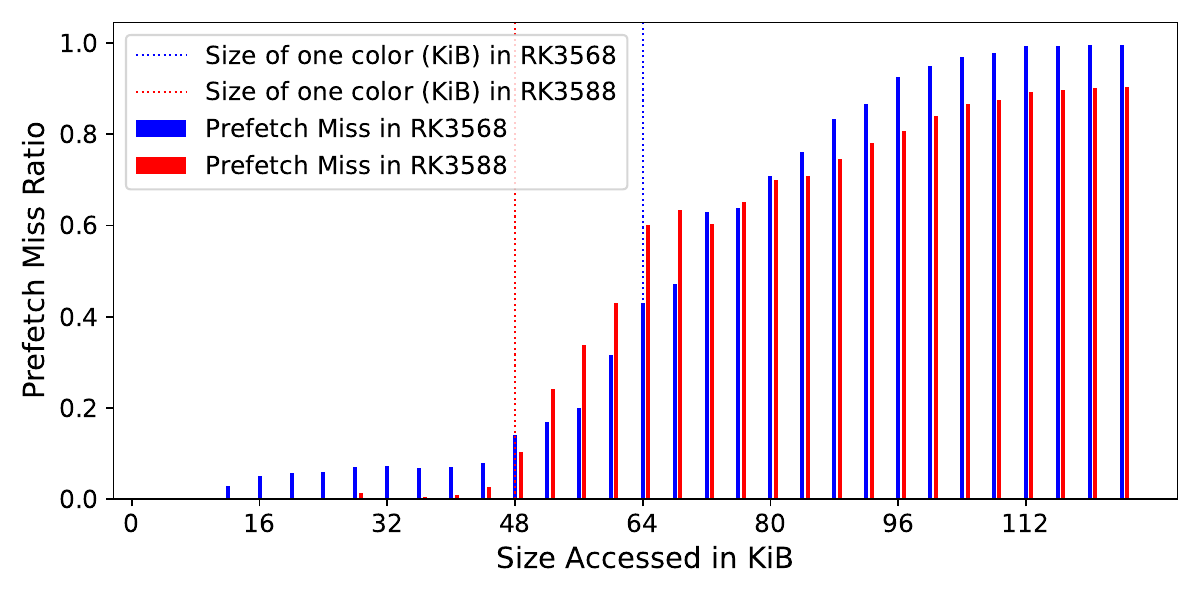}%
\caption{
	Prefetch Miss Ratio vs the Size of memory accessed in the same color on the RK3568 and RK3588 platforms
}%
\label{fig:prefetch-plru}%
\vspace{-1em}
\end{figure}%

%% file: 6-eval.tex
\section{Experiments and Discussion}
\label{sec:eval}
\input{tables}

\begin{observation}
\label{obs:way-conflict}
\textbf{Way partitioning can increase conflict misses}
\label{obs:synthetic-way-part}
\end{observation}
Way partitioning effectively reduces the associativity of the cache.
Therefore, access to memory that heavily relies on set associativity to achieve optimal performance can be penalized under way partitioning.
For example, for workloads with strided access pattern (\ie that continuously write to the same sets), way-partitioned cache can have a higher execution time than a set-partitioned cache.
To test this, we designed a synthetic benchmark for the RK3568 that simulates a workload that makes even use of the caches.
The benchmark operates at three levels,
using 1/4, 1/2, and 3/4 of the
total sets available in the RK3568's L3 cache.
Since the RK3568 has 512 sets, the levels correspond to
128, 256, and 384 sets, respectively.
For each level, the benchmark linearly accesses the sets and modifies one cacheline $s \times w$ times, where $s$ is the number of sets, and $w$ is the number of ways---16 on the RK3568. 
We record the execution time of the benchmark
under set and way partitioning and without partitioning.
As shown in Fig.~\ref{fig:synthetic-plot}, the execution time for way partitioning is much higher than set partitioning for the same configuration due to the higher rate of 
conflict misses in the way-partitioned cache.
Further, we observe that for 384 sets, 
the execution time under way-partitioning is slightly lower than for 256 sets.
This hints at a recency bias in the replacement policy.

Despite the non-negligible effect that way partitioning can have on workloads that evenly access caches, in real environments, workloads are unlikely to present the strided access pattern used in our synthetic benchmark.
%

\begin{observation}
\label{obs:way-set-same}
\textbf{Way and set partitioning are effective in reducing interference}
\end{observation}
Following the methodology described in Sec.~\ref{sec:eval-setup},
we evaluate and visualize the behavior of cache-partitioning.
Fig.~\ref{fig:set-way-comparison-mser-vga} shows a subset of our results
for both set and way partitioning using the \mser/\vga benchmark
on the RK3568 (a--d) and RK3588 (e--h).\footnote{%
Other samples of the effects of cache-partitioning are shown in
Figs.~\ref{fig:way-compare-orin}, ~\ref{fig:set-compare-write-streaming}, and~\ref{fig:set-compare-l1-modify}.
The full set of graphs is available in~\cite{rtas2025-extended}.
}
The figures report the slowdown ratio of the execution time of the benchmarks
compared to the no-interference case for increasing size of the interference.
Insets (a), (b), (e), (f) present results for \modifyInterf interference, while (c), (d), (g), (h) for \prefetchInterf interference.
%

We observe that both way-partitioning and set partitioning significantly reduce the slowdown ratio compared to the case without partitioning.
The effect of cache partitioning is more pronounced in the RK3588 boards than in the RK3568.
In contrast to the synthetic example (see Obs.~\ref{obs:synthetic-way-part}), the slowdown ratio in way partitioning is comparable to the slowdown ratio from set partitioning.
In Fig.~\ref{fig:rk3568-set-mser-modify} and \ref{fig:rk3568-way-mser-modify}, we can see that beyond the L3 cache size, set partitioning
performs slightly better than way partitioning.
On the other hand, in Fig.~\ref{fig:rk3588-set-mser-prefetch} and \ref{fig:rk3588-way-mser-prefetch}, way partitioning performs slightly better than set partitioning.

\input{figures/fig-synthetic}

\input{figures/fig-set-vs-way}

\tableTracking
\begin{observation}
\label{obs:no-partitioning-dominate}
\textbf{No partitioning method dominates the other}
\end{observation}
Motivated by the results from Fig.~\ref{fig:set-way-comparison-mser-vga}, we
carefully analyzed whether one of the two partitioning methods would
consistently perform better than the other across several benchmarks and
type/intensity of interference.
Table~\ref{table-tracking} presents an example of our results
for the maximum slowdown execution time ratio (relative to the no-interference case)
for the \tracking benchmark on the RK3568 and RK3588 platforms.
For computing the execution times, the interference-bench is run on one of the A55 cores, and the benchmark is run on a different A55 core. We take the maximum of the slowdown ratio as the size of interference in the interference-bench varies from 8~KB to 8192~KB.
Table~\ref{table-tracking} shows that the maximum slowdown ratio is similar between way and set partitioning across different interference patterns and datasets.
In particular, for some scenarios, set partitioning performs better than way partitioning, while in other scenarios, the opposite holds true.
We observed similar behaviors across the full set of benchmarks/interference types.

In summary, our results on practical benchmarks subject to interference indicate that no significant differences exist between the partitioning methods.
Given the complexity of implementing cache coloring in software, way partitioning might be a more suitable choice for most of the real-world scenarios where no pathological access patterns (see Obs.~\ref{obs:synthetic-way-part}) exist.

\input{figures/fig-no-partition}

\begin{observation}
\label{obs:private-cache}
\textbf{Cache Partitioning affects performance only when the working-set size exceeds the private cache size}
\end{observation}
If the working set of the benchmark fits in the private L1 and L2 caches, there is no interference from the other cores, and partitioning is not required.
However, if the benchmark uses the shared LLC, we observe an expected slowdown depending on the interaction between the competing accesses to the LLC.
Fig.~\ref{fig:rk3588-benchmarks-no-partition}, reports the execution time slowdown (\wrt the no-interference case) of
\localization, \sift, \tracking, \mser, and \disparity
(\vga dataset) with interference from \modifyInterf workload.
The benchmarks and interference run on distinct A55 on the RK3588. 
Slowdown is negligible for \localization and \sift, but is very noticeable for \disparity and \mser datasets.

To further investigate this behavior,
in Fig.~\ref{fig:rk3588-perf-no-partition}, we plot the value of performance counter \llcdcache
and \busaccessdsu.
The values provide information on the L3 cache accesses and main-memory accesses from the L3 cache (Sec.~\ref{subsec:perfcounters}). 
The performance counters are normalized by multiplying for the size of the cacheline and dividing by the execution time of the benchmark (\ie converting to MB/s).
We observe that the L3 access in \localization is negligible, while \disparity and \mser have a high volume of transactions in the L3 cache and main memory.
Therefore, when no longer executing on private caches, the effect of interference is high for \disparity and \mser, while it is negligible for \localization.
The rapid growth of \mser slowdown in
Fig.~\ref{fig:rk3588-benchmarks-no-partition}
is related to both the lower number of transactions in LLC and to the absolute execution time of \mser and \disparity (baselines: 0.19s and 0.71s, respectively). In fact, the absolute increase in execution time peaks at 0.10s in \disparity, while it is less than 0.08s for \mser.
%

\begin{observation}
\label{obs:mem-controller}
\textbf{Accesses to the memory controller impact partitioning}
\end{observation}
Fig.~\ref{fig:set-way-comparison-mser-vga} shows that, despite limiting interference, both set and way partitioning still suffer from slowdowns.
The slowdown starts increasing when the size of interference is less than the L3 cache
and stabilizes beyond the L3 cache size.
%
As visible in Fig.~\ref{fig:64-modify-rk3568},
such a slowdown is directly connected to an increase in the contention at the memory controller.
Fig.~\ref{fig:64-modify-rk3568} shows the \busaccessdsu
PMC for \modifyInterf interference (used here as benchmark) on the RK3568 (units are normalized to MB/s as in Obs.~\ref{obs:private-cache}). This reflects the pressure on the memory controller.
Fig.~\ref{fig:64-modify-rk3568} underlines that, without partitioning, the memory controller traffic increases around the size of the L3 cache.
In fact,
beyond the L3 cache size, \modifyInterf starts accessing data that can no longer fit in the caches. Memory controller traffic therefore increases till it saturates to the maximum bandwidth.

\input{figures/fig-bus-access}

However, under way partitioning, memory controller traffic increases much below the L3 cache size. This is because fewer LLCs is available 
to \modifyInterf,
hence early increasing the accesses to main memory.
This point corresponds to the increase in slowdown for way partitioning
as observed in Fig.~\ref{fig:rk3568-way-mser-modify} for each \textbf{1/3}, \textbf{2/2}, and \textbf{3/1} ratios.
When the working set of the benchmarks does not fit in the L3 cache,
the increased memory controller traffic from interference adversely affects their execution times.

\begin{observation}
\label{obs:spike}
\textbf{Reduced interference pressure causes a sharp speedup at the LLC size}
\end{observation}
Most insets in Fig.~\ref{fig:set-way-comparison-mser-vga}, and \eg Fig.~\ref{fig:rk3588-benchmarks-no-partition} and Fig.~\ref{fig:orin-way-disparity-read} present a slowdown ``spike'' around the L3 cache size for the no-partition case.
Counter-intuitively, the execution time of many benchmarks decreases when the working-set size of the interference exceeds the size of the L3.
%
We observe similar spikes on all tested boards.
On the Nvidia Orin, the spike is observed even when interference
and benchmarks execute on different clusters, \ie when only the system L4 cache is shared.
However, such spikes are not observed in either set or way partitioned scenarios.
To further investigate this effect, we analyzed the Cortex-A55 PMCs \loneinner and \loneouter (Sec.~\ref{subsec:perfcounters}). 
\input{figures/fig-l1-refill-counter}
Fig.~\ref{fig:44-45-mser-modify} shows the two PMCs for \mser/\vga benchmark.
\loneinner dips at the L3 cache size and rises again, indicating that more L3 is allocated to \mser as the interference size increases. In contrast, the PMC values are nearly constant for a partitioned cache, indicating a constant usage of the L3 cache. 
%
The spikes correlate, therefore, with the interaction between interference and benchmarks.
Below the L3 cache size, the fast loops of the interference workload are almost always cache-hot in L1, L2, and L3 caches.
Cache lines used by the benchmark are quickly evicted by the interference.
As noted in Obs.~\ref{obs:mem-controller},
memory accesses increase sharply after the L3 cache size, leading to overall higher latencies and slower execution of the requests that must be served from memory.
Interference-related memory accesses are also served slower, leading to less frequent evictions for the benchmarks.
In turn, this yields faster execution time for the benchmarks and the decrease of the slowdown ratio.
Instead, benchmarks and interference don't share common cache lines in the partitioned cache scenarios, leading to a constant execution time as the memory controller bandwidth saturates to the highest value.

\input{figures/fig-orin-way-disparity-read-cif}
\begin{observation}
\label{obs:way-orin}
\textbf{Way partitioning is less effective on Orin}
\end{observation}
Fig.~\ref{fig:way-compare-orin} shows
the slowdown ratio of \disparity/\cif with \readInterf interference for RK3568, RK3588, and Orin. On Orin, we run the vision-benchmark on one core and the interference-bench on another core in the \emph{same} cluster.
As visible in Fig.~\ref{fig:orin-way-disparity-read}, the benefits of way partitioning on Orin are lower than on RK3568 (Fig.~\ref{fig:rk3568-way-disparity-read}) and RK3588 (Fig.~\ref{fig:rk3588-way-disparity-read}).
In particular, on Orin, the \textbf{1/3} and \textbf{2/2} cases produce a higher slowdown than without partitioning.
Such slowdown increases considerably as we cross the L3 cache size.
The lower effectiveness of way partitioning can be attributed to the presence of the non-partitioned system cache (L4) in the Orin. Here inter-core interference is unregulated.

\input{figures/fig-dsu-clusters}

\begin{observation}
\label{obs:multi-DSU}
\textbf{Interference within DSU clusters is significantly higher than across clusters}
\end{observation}
On the NVIDIA Orin, we measure the slowdown when the interference workload and the benchmark application run on different DSU clusters.
The Orin is the only tested platform that features multiple DSU clusters and an additional system-level L4 cache.
Fig.~\ref{fig:dsu-clusters} presents the slowdown of \mser/\vga under \modifyInterf interference. As expected, in the different cluster configuration, there is no measurable slowdown when the interference workload fits entirely within the L3 cache.
However, even when the interference workload uses the L4 cache, the slowdown of this configuration is minimal compared to the same cluster one.
We attribute this behavior to the different pressure on the L4 cache in the two configurations. Due to the exclusive availability of the L3 cache in the different-cluster configuration, the pressure on the L4 cache is lower than in the same-cluster one.

\input{figures/fig-write-streaming}
\input{figures/fig-nowritestream}
\begin{observation}
\label{obs:write-streaming}
\textbf{Partitioning does not protect against streaming \writeInterf interference}
\end{observation}
The A55 cores support a \emph{write-streaming} mode, where continuous store operations bypass the cache. This minimizes cache pollution by not allocating data that is unlikely to be accessed soon.
\emph{Write-streaming} is controlled from the \cpuectrl register.
By default, the A55 cores apply \emph{write-streaming} mode to the 4th, 128th, and 1024th consecutive streaming write access on the L1, L2, and L3 caches, respectively~\cite{arm-a55trm}.
With a cache line size of 64~B, after a continuous write of 64~KB, data does not allocate to the L3 cache.
On RK3568 and RK3588, in \emph{write-streaming} mode, cache partitioning techniques do not improve the slowdown ratio compared to a case without partitioning.

Fig.~\ref{fig:set-compare-write-streaming} illustrates the slowdown (w.r.t. no-interference) of \disparity/\cif on RK3568, RK3588, and ZCU102 when the interference workload is \writeInterf.
As can be observed in Fig.~\ref{fig:rk3588-set-disparity-write} for the RK3588 platform (and in Table~\ref{table-tracking} on the RK3568),
under \writeInterf interference,
partitioning techniques can produce higher slowdowns.
In contrast, as shown in Fig.~\ref{fig:zcu102-set-disparity-write}, on the ZCU102---which does not have \emph{write-streaming}---we observe that partitioning improves the slowdown ratio. 
In fact, on the RK3568 and RK3588
when caches are not partitioned, the maximum interference
in \emph{write-streaming} mode is limited to 64~KB.
However, in cases \textbf{1/3}, \textbf{2/2} and \textbf{3/1}, we force a reduction of
the cache available to the benchmark by a size much larger than a 64~KB reduction.
The remaining source of interference comes from contention in memory accesses, which saturates to the same value in both partitioned and non-partitioned (see Fig.~\ref{fig:64-modify-rk3568}).

Cache partitioning is therefore ineffective in \emph{write-streaming} mode since caches and memory controller interference are anyway less polluted.
To evaluate further, we repeated the experiment by disabling the \emph{write-streaming} mode on the RK3568 platform. Fig.~\ref{fig:rk3568-way-mser-write-stream} and Fig.~\ref{fig:rk3568-way-mser-write-nostream} show the effect of \writeInterf on \mser/\vga benchmark, when \emph{write-streaming} mode, is enabled and disabled respectively. From Fig.~\ref{fig:rk3568-way-mser-write-nostream}, we observe that disabling \emph{write-streaming} mode makes cache partitioning effective in reducing slowdowns. 
\input{figures/fig-l1-increase}

\begin{observation}
\label{obs:platform-specific}
\textbf{Platform Specific Observations}
\end{observation}

\textit{The slowdown ratio increases around the L1 cache size in RK3568 and ZCU102 platforms irrespective of cache partitioning.}
When the size of the interference benchmark exceeds the size of the L1 cache, benchmark's execution times on RK3568 and ZCU102 increase.
This is clearly visible in Fig.~\ref{fig:set-compare-l1-modify} where
the execution time jumps by 6\% and 10\% on RK3568 and ZCU102, respectively.
This effect is also present in the \tracking and \mser datasets for \readInterf and \modifyInterf cases, but not for \prefetchInterf in RK3568. This slowdown is not observed on the RK3588 or Orin platforms, likely due to the presence of a private L2 cache mitigating the effect. However, there is no conclusive evidence from the performance counters on any platform to fully explain this behavior. 

\textit{Slowdown from \writeInterf interference is much higher on RK3568.}
From a cache point of view, write operations are expensive compared to read since they necessitate invalidating the cacheline, maintaining coherency across memory hierarchies, and writing back to the slower memory. \writeInterf produces much worse slowdowns compared to \readInterf/\modifyInterf/\prefetchInterf on RK3568. Instead, on RK3588, performance is comparable (Table~\ref{table-tracking} and Fig.~\ref{fig:rk3568-set-disparity-write}~\ref{fig:rk3588-set-disparity-write}).
We suspect the higher slowdowns on RK3568 to be a side effect of the \emph{write-streaming} mode. When \emph{write-streaming} mode is disabled, the slowdown decreases on the RK3568 platform. The maximum slowdown reduced from 6.2x to 2.2x in a non-partitioned cache for the \mser/\vga dataset, and for a \textbf{3/1} way-partitioned cache with the same dataset, the slowdown reduced from 6.1x to 1.72x (Fig.~\ref{fig:rk3568-write-streaming-mode}).

\textit{In \emph{write-streaming} mode on RK3568 and RK3588 platforms, performance counters indicate allocation into the L3 cache.}
Using the \lthreeews A55 performance counter (Sec.~\ref{subsec:perfcounters}), one can monitor the transition
into \emph{write-streaming} mode in the L3 cache.
If we make continuous write accesses to a size greater than 64~KB, we observe a high value in this counter, confirming that the core enters a write-streaming mode.
However, \lthreecachealloc performance counter in the DSU shows comparable values when \emph{write-streaming} mode is enabled or disabled (by modifying the \cpuectrl register).
This counters the expectation that \lthreecachealloc does not count in write-streaming mode as the data accessed are not being allocated to the cache.
We are not sure if this is an error in the behavior of \emph{write-streaming} mode or counting implementation of \lthreecachealloc counter.

\textit{The effect of \prefetchInterf is stronger on the A55 cores than on the A76 or A78 cores.}
We compare the bandwidth of \prefetchInterf accesses obtained from the DSU \scupftchacc counter and from the execution time of the program. 
As seen in Fig.~\ref{fig:prefetch-compare}, on the A55 cores, the execution time is higher, but the bandwidth obtained from it matches the bandwidth obtained from performance counters. On the A76 and the A78 cores, the execution time is much lower, but the bandwidth obtained from the execution time is much higher than that obtained from the performance counter. This might imply that the A76 and A78 cores take the prefetch instructions as more of a hint, while the A55 cores execute it and load the contents to the cache as specified in the software prefetch command. 

\input{figures/fig-prefetch}

%



\smallskip
\paragraph*{\textbf{Summary}}
Our findings indicate that both way and set partitioning techniques effectively reduce interference between cores (Obs.~\ref{obs:way-set-same}), but neither consistently outperforms the other across the tested benchmarks (Obs.~\ref{obs:no-partitioning-dominate}). Way-partitioning is simpler to implement and remains unaffected by cache scrambling mechanisms that could hinder set-partitioning (Sec.~\ref{sec:set-hash-scrambling}); however, it can increase conflict misses under specific
access patterns (Obs.~\ref{obs:synthetic-way-part}). 
Regardless of the method used, cache partitioning increases pressure on the memory controller, even when the workload is smaller than the L3 cache, potentially degrading overall performance (Obs.~\ref{obs:mem-controller}).
Furthermore, architectural optimizations, such as write-streaming, can diminish the effectiveness of cache partitioning, with their impact varying across different SoCs due to implementation differences (Obs.~\ref{obs:write-streaming}).

The tests also show that increasing the interference workload size without cache partitioning leads to execution time spikes (Obs.~\ref{obs:spike}). This behavior originates from the memory controller becoming overloaded, reducing the capability of the interference-bench to evict cache lines of the tested benchmarks. The effect disappears with cache partitioning since benchmarks and interference do not share cache lines.

Our results on the Orin---the only platform tested with multiple DSU clusters---highlight that, despite way-partitioning at the L3 cache, interference is significant due to the shared, unpartitioned L4 cache across DSU clusters (Obs.~\ref{obs:way-orin}).
However, interference between cores in different DSU clusters is much lower than within the same cluster, emphasizing the benefits of cluster-level isolation (Obs.~\ref{obs:multi-DSU}).

%% file: tables.tex
\newcommand{\tableTracking}{
\begin{table*}[]
\centering
\caption{Maximum execution time slowdown \wrt the no-interference case for the \tracking benchmark on the RK3568 and RK3588 platforms for way/set partitioning and all interference types.}
\resizebox{\textwidth}{!}{%
\begin{tabular}{|llllllllllllllllll|}
\hline
\rowcolor[HTML]{FFFFFF} 
\multicolumn{18}{|c|}{\cellcolor[HTML]{FFFFFF}RK3568}                                                                                                                                                                                                                                                                                                                                                                                                                                                                                                                                                                                                                                                                                                                                                                                                                                                                                                                                                                                                                                                                                                                                                                                                                                                                                                                                                                                                                                                                                                                                                            \\ \hline
\rowcolor[HTML]{FFFFFF} 
\multicolumn{2}{|l|}{\cellcolor[HTML]{FFFFFF}}                                                                                                                                                                            & \multicolumn{16}{c|}{\cellcolor[HTML]{FFFFFF}\tracking}                                                                                                                                                                                                                                                                                                                                                                                                                                                                                                                                                                                                                                                                                                                                                                                                                                                                                                                                                                                                                                                                                                                                                                                                                                                                                                                               \\
\rowcolor[HTML]{FFFFFF} 
\multicolumn{2}{|l|}{\multirow{-2}{*}{\cellcolor[HTML]{FFFFFF}}}                                                                                                                                                          & \multicolumn{4}{c|}{\cellcolor[HTML]{FFFFFF}\readInterf}                                                                                                                                                                                                                                                                                                      & \multicolumn{4}{c|}{\cellcolor[HTML]{FFFFFF}\writeInterf}                                                                                                                                                                                                                                                                                                     & \multicolumn{4}{c|}{\cellcolor[HTML]{FFFFFF}\modifyInterf}                                                                                                                                                                                                                                                                                                    & \multicolumn{4}{c|}{\cellcolor[HTML]{FFFFFF}\prefetchInterf}                                                                                                                                                                                                                                                     \\ \cline{2-18} 
\rowcolor[HTML]{FFFFFF} 
\multicolumn{1}{|l|}{\cellcolor[HTML]{FFFFFF}}                                             & \multicolumn{1}{l|}{\cellcolor[HTML]{FFFFFF}\begin{tabular}[c]{@{}l@{}}Partition\\ Size\end{tabular}}                        & \multicolumn{1}{l|}{\cellcolor[HTML]{FFFFFF}No}                                            & \multicolumn{1}{l|}{\cellcolor[HTML]{FFFFFF}1/3}                                  & \multicolumn{1}{l|}{\cellcolor[HTML]{FFFFFF}2/2}                                  & \multicolumn{1}{l|}{\cellcolor[HTML]{FFFFFF}3/1}                                  & \multicolumn{1}{l|}{\cellcolor[HTML]{FFFFFF}No}                                            & \multicolumn{1}{l|}{\cellcolor[HTML]{FFFFFF}1/3}                                  & \multicolumn{1}{l|}{\cellcolor[HTML]{FFFFFF}2/2}                                  & \multicolumn{1}{l|}{\cellcolor[HTML]{FFFFFF}3/1}                                  & \multicolumn{1}{l|}{\cellcolor[HTML]{FFFFFF}No}                                            & \multicolumn{1}{l|}{\cellcolor[HTML]{FFFFFF}1/3}                                  & \multicolumn{1}{l|}{\cellcolor[HTML]{FFFFFF}2/2}                                  & \multicolumn{1}{l|}{\cellcolor[HTML]{FFFFFF}3/1}                                  & \multicolumn{1}{l|}{\cellcolor[HTML]{FFFFFF}No}                                            & \multicolumn{1}{l|}{\cellcolor[HTML]{FFFFFF}1/3}                                  & \multicolumn{1}{l|}{\cellcolor[HTML]{FFFFFF}2/2}                                  & 3/1                                  \\ \hline
\rowcolor[HTML]{FFFFFF} 
\multicolumn{1}{|l|}{\cellcolor[HTML]{FFFFFF}}                                             & \multicolumn{1}{l|}{\cellcolor[HTML]{FFFFFF}way}                                                                             & \multicolumn{1}{l|}{\cellcolor[HTML]{FFFFFF}}                                              & \multicolumn{1}{l|}{\cellcolor[HTML]{FFFFFF}\textbf{1.03}}                        & \multicolumn{1}{l|}{\cellcolor[HTML]{FFFFFF}1.03}                                 & \multicolumn{1}{l|}{\cellcolor[HTML]{FFFFFF}1.03}                                 & \multicolumn{1}{l|}{\cellcolor[HTML]{FFFFFF}}                                              & \multicolumn{1}{l|}{\cellcolor[HTML]{FFFFFF}\textbf{1.38}}                        & \multicolumn{1}{l|}{\cellcolor[HTML]{FFFFFF}\textbf{1.36}}                        & \multicolumn{1}{l|}{\cellcolor[HTML]{FFFFFF}1.35}                                 & \multicolumn{1}{l|}{\cellcolor[HTML]{FFFFFF}}                                              & \multicolumn{1}{l|}{\cellcolor[HTML]{FFFFFF}1.02}                                 & \multicolumn{1}{l|}{\cellcolor[HTML]{FFFFFF}1.02}                                 & \multicolumn{1}{l|}{\cellcolor[HTML]{FFFFFF}1.02}                                 & \multicolumn{1}{l|}{\cellcolor[HTML]{FFFFFF}}                                              & \multicolumn{1}{l|}{\cellcolor[HTML]{FFFFFF}\textbf{1.13}}                        & \multicolumn{1}{l|}{\cellcolor[HTML]{FFFFFF}\textbf{1.12}}                        & \textbf{1.12}                        \\ \cline{2-2} \cline{4-6} \cline{8-10} \cline{12-14} \cline{16-18} 
\rowcolor[HTML]{FFFFFF} 
\multicolumn{1}{|l|}{\multirow{-2}{*}{\cellcolor[HTML]{FFFFFF}\cif}}                        & \multicolumn{1}{l|}{\cellcolor[HTML]{FFFFFF}set}                                                                             & \multicolumn{1}{l|}{\multirow{-2}{*}{\cellcolor[HTML]{FFFFFF}1.05}}                        & \multicolumn{1}{l|}{\cellcolor[HTML]{FFFFFF}1.04}                                 & \multicolumn{1}{l|}{\cellcolor[HTML]{FFFFFF}\textbf{1.03}}                        & \multicolumn{1}{l|}{\cellcolor[HTML]{FFFFFF}\textbf{1.03}}                        & \multicolumn{1}{l|}{\multirow{-2}{*}{\cellcolor[HTML]{FFFFFF}1.33}}                        & \multicolumn{1}{l|}{\cellcolor[HTML]{FFFFFF}1.46}                                 & \multicolumn{1}{l|}{\cellcolor[HTML]{FFFFFF}1.38}                                 & \multicolumn{1}{l|}{\cellcolor[HTML]{FFFFFF}\textbf{1.33}}                        & \multicolumn{1}{l|}{\multirow{-2}{*}{\cellcolor[HTML]{FFFFFF}1.03}}                        & \multicolumn{1}{l|}{\cellcolor[HTML]{FFFFFF}\textbf{1.02}}                        & \multicolumn{1}{l|}{\cellcolor[HTML]{FFFFFF}\textbf{1.02}}                        & \multicolumn{1}{l|}{\cellcolor[HTML]{FFFFFF}\textbf{1.01}}                        & \multicolumn{1}{l|}{\multirow{-2}{*}{\cellcolor[HTML]{FFFFFF}1.13}}                        & \multicolumn{1}{l|}{\cellcolor[HTML]{FFFFFF}1.13}                                 & \multicolumn{1}{l|}{\cellcolor[HTML]{FFFFFF}1.12}                                 & 1.12                                 \\ \hline
\rowcolor[HTML]{FFFFFF} 
\multicolumn{1}{|l|}{\cellcolor[HTML]{FFFFFF}}                                             & \multicolumn{1}{l|}{\cellcolor[HTML]{FFFFFF}way}                                                                             & \multicolumn{1}{l|}{\cellcolor[HTML]{FFFFFF}}                                              & \multicolumn{1}{l|}{\cellcolor[HTML]{FFFFFF}1.17}                                 & \multicolumn{1}{l|}{\cellcolor[HTML]{FFFFFF}\textbf{1.08}}                        & \multicolumn{1}{l|}{\cellcolor[HTML]{FFFFFF}1.09}                                 & \multicolumn{1}{l|}{\cellcolor[HTML]{FFFFFF}}                                              & \multicolumn{1}{l|}{\cellcolor[HTML]{FFFFFF}2.8}                                  & \multicolumn{1}{l|}{\cellcolor[HTML]{FFFFFF}\textbf{1.63}}                        & \multicolumn{1}{l|}{\cellcolor[HTML]{FFFFFF}\textbf{1.64}}                        & \multicolumn{1}{l|}{\cellcolor[HTML]{FFFFFF}}                                              & \multicolumn{1}{l|}{\cellcolor[HTML]{FFFFFF}1.09}                                 & \multicolumn{1}{l|}{\cellcolor[HTML]{FFFFFF}\textbf{1.07}}                        & \multicolumn{1}{l|}{\cellcolor[HTML]{FFFFFF}\textbf{1.07}}                        & \multicolumn{1}{l|}{\cellcolor[HTML]{FFFFFF}}                                              & \multicolumn{1}{l|}{\cellcolor[HTML]{FFFFFF}1.55}                                 & \multicolumn{1}{l|}{\cellcolor[HTML]{FFFFFF}1.34}                                 & 1.34                                 \\ \cline{2-2} \cline{4-6} \cline{8-10} \cline{12-14} \cline{16-18} 
\rowcolor[HTML]{FFFFFF} 
\multicolumn{1}{|l|}{\multirow{-2}{*}{\cellcolor[HTML]{FFFFFF}\vga}}                        & \multicolumn{1}{l|}{\cellcolor[HTML]{FFFFFF}set}                                                                             & \multicolumn{1}{l|}{\multirow{-2}{*}{\cellcolor[HTML]{FFFFFF}1.28}}                        & \multicolumn{1}{l|}{\cellcolor[HTML]{FFFFFF}\textbf{1.15}}                        & \multicolumn{1}{l|}{\cellcolor[HTML]{FFFFFF}1.09}                                 & \multicolumn{1}{l|}{\cellcolor[HTML]{FFFFFF}\textbf{1.07}}                        & \multicolumn{1}{l|}{\multirow{-2}{*}{\cellcolor[HTML]{FFFFFF}1.59}}                        & \multicolumn{1}{l|}{\cellcolor[HTML]{FFFFFF}\textbf{2.74}}                        & \multicolumn{1}{l|}{\cellcolor[HTML]{FFFFFF}1.68}                                 & \multicolumn{1}{l|}{\cellcolor[HTML]{FFFFFF}1.70}                                 & \multicolumn{1}{l|}{\multirow{-2}{*}{\cellcolor[HTML]{FFFFFF}1.14}}                        & \multicolumn{1}{l|}{\cellcolor[HTML]{FFFFFF}\textbf{1.07}}                        & \multicolumn{1}{l|}{\cellcolor[HTML]{FFFFFF}1.07}                                 & \multicolumn{1}{l|}{\cellcolor[HTML]{FFFFFF}1.07}                                 & \multicolumn{1}{l|}{\multirow{-2}{*}{\cellcolor[HTML]{FFFFFF}1.44}}                        & \multicolumn{1}{l|}{\cellcolor[HTML]{FFFFFF}\textbf{1.48}}                        & \multicolumn{1}{l|}{\cellcolor[HTML]{FFFFFF}\textbf{1.34}}                        & \textbf{1.32}                        \\ \hline
\rowcolor[HTML]{FFFFFF} 
\multicolumn{18}{|c|}{\cellcolor[HTML]{FFFFFF}{\color[HTML]{333333} RK3588}}                                                                                                                                                                                                                                                                                                                                                                                                                                                                                                                                                                                                                                                                                                                                                                                                                                                                                                                                                                                                                                                                                                                                                                                                                                                                                                                                                                                                                                                                                                                                     \\ \hline
\rowcolor[HTML]{FFFFFF} 
\multicolumn{2}{|l|}{\cellcolor[HTML]{FFFFFF}{\color[HTML]{333333} }}                                                                                                                                                     & \multicolumn{16}{c|}{\cellcolor[HTML]{FFFFFF}{\color[HTML]{333333} \tracking}}                                                                                                                                                                                                                                                                                                                                                                                                                                                                                                                                                                                                                                                                                                                                                                                                                                                                                                                                                                                                                                                                                                                                                                                                                                                                                                        \\
\rowcolor[HTML]{FFFFFF} 
\multicolumn{2}{|l|}{\multirow{-2}{*}{\cellcolor[HTML]{FFFFFF}{\color[HTML]{333333} }}}                                                                                                                                   & \multicolumn{4}{c|}{\cellcolor[HTML]{FFFFFF}{\color[HTML]{333333} \readInterf}}                                                                                                                                                                                                                                                                               & \multicolumn{4}{c|}{\cellcolor[HTML]{FFFFFF}{\color[HTML]{333333} \writeInterf}}                                                                                                                                                                                                                                                                              & \multicolumn{4}{c|}{\cellcolor[HTML]{FFFFFF}{\color[HTML]{333333} \modifyInterf}}                                                                                                                                                                                                                                                                             & \multicolumn{4}{c|}{\cellcolor[HTML]{FFFFFF}{\color[HTML]{333333} \prefetchInterf}}                                                                                                                                                                                                                              \\ \cline{2-18} 
\rowcolor[HTML]{FFFFFF} 
\multicolumn{1}{|l|}{\cellcolor[HTML]{FFFFFF}{\color[HTML]{333333} }}                      & \multicolumn{1}{l|}{\cellcolor[HTML]{FFFFFF}{\color[HTML]{333333} \begin{tabular}[c]{@{}l@{}}Partition\\ Size\end{tabular}}} & \multicolumn{1}{l|}{\cellcolor[HTML]{FFFFFF}{\color[HTML]{333333} No}}                     & \multicolumn{1}{l|}{\cellcolor[HTML]{FFFFFF}{\color[HTML]{333333} 1/3}}           & \multicolumn{1}{l|}{\cellcolor[HTML]{FFFFFF}{\color[HTML]{333333} 2/2}}           & \multicolumn{1}{l|}{\cellcolor[HTML]{FFFFFF}{\color[HTML]{333333} 3/1}}           & \multicolumn{1}{l|}{\cellcolor[HTML]{FFFFFF}{\color[HTML]{333333} No}}                     & \multicolumn{1}{l|}{\cellcolor[HTML]{FFFFFF}{\color[HTML]{333333} 1/3}}           & \multicolumn{1}{l|}{\cellcolor[HTML]{FFFFFF}{\color[HTML]{333333} 2/2}}           & \multicolumn{1}{l|}{\cellcolor[HTML]{FFFFFF}{\color[HTML]{333333} 3/1}}           & \multicolumn{1}{l|}{\cellcolor[HTML]{FFFFFF}{\color[HTML]{333333} No}}                     & \multicolumn{1}{l|}{\cellcolor[HTML]{FFFFFF}{\color[HTML]{333333} 1/3}}           & \multicolumn{1}{l|}{\cellcolor[HTML]{FFFFFF}{\color[HTML]{333333} 2/2}}           & \multicolumn{1}{l|}{\cellcolor[HTML]{FFFFFF}{\color[HTML]{333333} 3/1}}           & \multicolumn{1}{l|}{\cellcolor[HTML]{FFFFFF}{\color[HTML]{333333} No}}                     & \multicolumn{1}{l|}{\cellcolor[HTML]{FFFFFF}{\color[HTML]{333333} 1/3}}           & \multicolumn{1}{l|}{\cellcolor[HTML]{FFFFFF}{\color[HTML]{333333} 2/2}}           & {\color[HTML]{333333} 3/1}           \\ \hline
\rowcolor[HTML]{FFFFFF} 
\multicolumn{1}{|l|}{\cellcolor[HTML]{FFFFFF}{\color[HTML]{333333} }}                      & \multicolumn{1}{l|}{\cellcolor[HTML]{FFFFFF}{\color[HTML]{333333} way}}                                                      & \multicolumn{1}{l|}{\cellcolor[HTML]{FFFFFF}{\color[HTML]{333333} }}                       & \multicolumn{1}{l|}{\cellcolor[HTML]{FFFFFF}{\color[HTML]{333333} \textbf{1.00}}} & \multicolumn{1}{l|}{\cellcolor[HTML]{FFFFFF}{\color[HTML]{333333} 1.01}}          & \multicolumn{1}{l|}{\cellcolor[HTML]{FFFFFF}{\color[HTML]{333333} \textbf{1.01}}} & \multicolumn{1}{l|}{\cellcolor[HTML]{FFFFFF}{\color[HTML]{333333} }}                       & \multicolumn{1}{l|}{\cellcolor[HTML]{FFFFFF}{\color[HTML]{333333} \textbf{1.03}}} & \multicolumn{1}{l|}{\cellcolor[HTML]{FFFFFF}{\color[HTML]{333333} \textbf{1.02}}} & \multicolumn{1}{l|}{\cellcolor[HTML]{FFFFFF}{\color[HTML]{333333} \textbf{1.02}}} & \multicolumn{1}{l|}{\cellcolor[HTML]{FFFFFF}{\color[HTML]{333333} }}                       & \multicolumn{1}{l|}{\cellcolor[HTML]{FFFFFF}{\color[HTML]{333333} \textbf{1.01}}} & \multicolumn{1}{l|}{\cellcolor[HTML]{FFFFFF}{\color[HTML]{333333} 1.01}}          & \multicolumn{1}{l|}{\cellcolor[HTML]{FFFFFF}{\color[HTML]{333333} \textbf{1.01}}} & \multicolumn{1}{l|}{\cellcolor[HTML]{FFFFFF}{\color[HTML]{333333} }}                       & \multicolumn{1}{l|}{\cellcolor[HTML]{FFFFFF}{\color[HTML]{333333} \textbf{1.04}}} & \multicolumn{1}{l|}{\cellcolor[HTML]{FFFFFF}{\color[HTML]{333333} \textbf{1.03}}} & {\color[HTML]{333333} \textbf{1.03}} \\ \cline{2-2} \cline{4-6} \cline{8-10} \cline{12-14} \cline{16-18} 
\rowcolor[HTML]{FFFFFF} 
\multicolumn{1}{|l|}{\multirow{-2}{*}{\cellcolor[HTML]{FFFFFF}{\color[HTML]{333333} \cif}}} & \multicolumn{1}{l|}{\cellcolor[HTML]{FFFFFF}{\color[HTML]{333333} set}}                                                      & \multicolumn{1}{l|}{\multirow{-2}{*}{\cellcolor[HTML]{FFFFFF}{\color[HTML]{333333} 1.04}}} & \multicolumn{1}{l|}{\cellcolor[HTML]{FFFFFF}{\color[HTML]{333333} 1.01}}          & \multicolumn{1}{l|}{\cellcolor[HTML]{FFFFFF}{\color[HTML]{333333} \textbf{1.01}}} & \multicolumn{1}{l|}{\cellcolor[HTML]{FFFFFF}{\color[HTML]{333333} 1.01}}          & \multicolumn{1}{l|}{\multirow{-2}{*}{\cellcolor[HTML]{FFFFFF}{\color[HTML]{333333} 1.03}}} & \multicolumn{1}{l|}{\cellcolor[HTML]{FFFFFF}{\color[HTML]{333333} 1.04}}          & \multicolumn{1}{l|}{\cellcolor[HTML]{FFFFFF}{\color[HTML]{333333} 1.02}}          & \multicolumn{1}{l|}{\cellcolor[HTML]{FFFFFF}{\color[HTML]{333333} 1.03}}          & \multicolumn{1}{l|}{\multirow{-2}{*}{\cellcolor[HTML]{FFFFFF}{\color[HTML]{333333} 1.05}}} & \multicolumn{1}{l|}{\cellcolor[HTML]{FFFFFF}{\color[HTML]{333333} 1.01}}          & \multicolumn{1}{l|}{\cellcolor[HTML]{FFFFFF}{\color[HTML]{333333} \textbf{1.01}}} & \multicolumn{1}{l|}{\cellcolor[HTML]{FFFFFF}{\color[HTML]{333333} 1.01}}          & \multicolumn{1}{l|}{\multirow{-2}{*}{\cellcolor[HTML]{FFFFFF}{\color[HTML]{333333} 1.10}}} & \multicolumn{1}{l|}{\cellcolor[HTML]{FFFFFF}{\color[HTML]{333333} 1.06}}          & \multicolumn{1}{l|}{\cellcolor[HTML]{FFFFFF}{\color[HTML]{333333} 1.04}}          & {\color[HTML]{333333} 1.04}          \\ \hline
\rowcolor[HTML]{FFFFFF} 
\multicolumn{1}{|l|}{\cellcolor[HTML]{FFFFFF}{\color[HTML]{333333} }}                      & \multicolumn{1}{l|}{\cellcolor[HTML]{FFFFFF}{\color[HTML]{333333} way}}                                                      & \multicolumn{1}{l|}{\cellcolor[HTML]{FFFFFF}{\color[HTML]{333333} }}                       & \multicolumn{1}{l|}{\cellcolor[HTML]{FFFFFF}{\color[HTML]{333333} \textbf{1.01}}} & \multicolumn{1}{l|}{\cellcolor[HTML]{FFFFFF}{\color[HTML]{333333} 1.01}}          & \multicolumn{1}{l|}{\cellcolor[HTML]{FFFFFF}{\color[HTML]{333333} \textbf{1.01}}} & \multicolumn{1}{l|}{\cellcolor[HTML]{FFFFFF}{\color[HTML]{333333} }}                       & \multicolumn{1}{l|}{\cellcolor[HTML]{FFFFFF}{\color[HTML]{333333} \textbf{1.03}}} & \multicolumn{1}{l|}{\cellcolor[HTML]{FFFFFF}{\color[HTML]{333333} 1.04}}          & \multicolumn{1}{l|}{\cellcolor[HTML]{FFFFFF}{\color[HTML]{333333} 1.05}}          & \multicolumn{1}{l|}{\cellcolor[HTML]{FFFFFF}{\color[HTML]{333333} }}                       & \multicolumn{1}{l|}{\cellcolor[HTML]{FFFFFF}{\color[HTML]{333333} \textbf{1.01}}} & \multicolumn{1}{l|}{\cellcolor[HTML]{FFFFFF}{\color[HTML]{333333} 1.01}}          & \multicolumn{1}{l|}{\cellcolor[HTML]{FFFFFF}{\color[HTML]{333333} \textbf{1.01}}} & \multicolumn{1}{l|}{\cellcolor[HTML]{FFFFFF}{\color[HTML]{333333} }}                       & \multicolumn{1}{l|}{\cellcolor[HTML]{FFFFFF}{\color[HTML]{333333} \textbf{1.08}}} & \multicolumn{1}{l|}{\cellcolor[HTML]{FFFFFF}{\color[HTML]{333333} \textbf{1.08}}} & {\color[HTML]{333333} \textbf{1.07}} \\ \cline{2-2} \cline{4-6} \cline{8-10} \cline{12-14} \cline{16-18} 
\rowcolor[HTML]{FFFFFF} 
\multicolumn{1}{|l|}{\multirow{-2}{*}{\cellcolor[HTML]{FFFFFF}{\color[HTML]{333333} \vga}}} & \multicolumn{1}{l|}{\cellcolor[HTML]{FFFFFF}{\color[HTML]{333333} set}}                                                      & \multicolumn{1}{l|}{\multirow{-2}{*}{\cellcolor[HTML]{FFFFFF}{\color[HTML]{333333} 1.03}}} & \multicolumn{1}{l|}{\cellcolor[HTML]{FFFFFF}{\color[HTML]{333333} 1.01}}          & \multicolumn{1}{l|}{\cellcolor[HTML]{FFFFFF}{\color[HTML]{333333} \textbf{1.01}}} & \multicolumn{1}{l|}{\cellcolor[HTML]{FFFFFF}{\color[HTML]{333333} 1.01}}          & \multicolumn{1}{l|}{\multirow{-2}{*}{\cellcolor[HTML]{FFFFFF}{\color[HTML]{333333} 1.04}}} & \multicolumn{1}{l|}{\cellcolor[HTML]{FFFFFF}{\color[HTML]{333333} 1.04}}          & \multicolumn{1}{l|}{\cellcolor[HTML]{FFFFFF}{\color[HTML]{333333} \textbf{1.04}}} & \multicolumn{1}{l|}{\cellcolor[HTML]{FFFFFF}{\color[HTML]{333333} \textbf{1.05}}} & \multicolumn{1}{l|}{\multirow{-2}{*}{\cellcolor[HTML]{FFFFFF}{\color[HTML]{333333} 1.03}}} & \multicolumn{1}{l|}{\cellcolor[HTML]{FFFFFF}{\color[HTML]{333333} 1.01}}          & \multicolumn{1}{l|}{\cellcolor[HTML]{FFFFFF}{\color[HTML]{333333} \textbf{1.01}}} & \multicolumn{1}{l|}{\cellcolor[HTML]{FFFFFF}{\color[HTML]{333333} 1.02}}          & \multicolumn{1}{l|}{\multirow{-2}{*}{\cellcolor[HTML]{FFFFFF}{\color[HTML]{333333} 1.10}}} & \multicolumn{1}{l|}{\cellcolor[HTML]{FFFFFF}{\color[HTML]{333333} 1.09}}          & \multicolumn{1}{l|}{\cellcolor[HTML]{FFFFFF}{\color[HTML]{333333} 1.09}}          & {\color[HTML]{333333} 1.09}          \\ \hline
\end{tabular}%
}
\label{table-tracking}
\vspace{-1em}
\end{table*}
}

\newcommand{\tableRKsixeight}{\begin{table*}[h!]
\centering
\resizebox{\textwidth}{!}{%
\begin{tabular}{|
>{\columncolor[HTML]{FFFFFF}}l 
>{\columncolor[HTML]{FFFFFF}}l 
>{\columncolor[HTML]{FFFFFF}}l 
>{\columncolor[HTML]{FFFFFF}}l 
>{\columncolor[HTML]{FFFFFF}}l 
>{\columncolor[HTML]{FFFFFF}}l 
>{\columncolor[HTML]{FFFFFF}}l 
>{\columncolor[HTML]{FFFFFF}}l 
>{\columncolor[HTML]{FFFFFF}}l 
>{\columncolor[HTML]{FFFFFF}}l 
>{\columncolor[HTML]{FFFFFF}}l 
>{\columncolor[HTML]{FFFFFF}}l 
>{\columncolor[HTML]{FFFFFF}}l 
>{\columncolor[HTML]{FFFFFF}}l 
>{\columncolor[HTML]{FFFFFF}}l 
>{\columncolor[HTML]{FFFFFF}}l 
>{\columncolor[HTML]{FFFFFF}}l 
>{\columncolor[HTML]{FFFFFF}}l |}
\hline
\multicolumn{18}{|c|}{\cellcolor[HTML]{FFFFFF}RK3568}                                                                                                                                                                                                                                                                                                                                                                                                                                                                                                                                                                                                                                                                                                                                                                                                                                                                                                                                                                                                                                                                                                                                                                                                                                                                                                                                                                                                                                                                                          \\ \hline
\multicolumn{2}{|l|}{\cellcolor[HTML]{FFFFFF}}                                                                                                                                 & \multicolumn{16}{c|}{\cellcolor[HTML]{FFFFFF}\disparity}                                                                                                                                                                                                                                                                                                                                                                                                                                                                                                                                                                                                                                                                                                                                                                                                                                                                                                                                                                                                                                                                                                                                                                                                                                                                                                       \\ \cline{3-18} 
\multicolumn{2}{|l|}{\multirow{-2}{*}{\cellcolor[HTML]{FFFFFF}}}                                                                                                               & \multicolumn{4}{c|}{\cellcolor[HTML]{FFFFFF}\readInterf}                                                                                                                                                                                                                                                                               & \multicolumn{4}{c|}{\cellcolor[HTML]{FFFFFF}\writeInterf}                                                                                                                                                                                                                                                                                                     & \multicolumn{4}{c|}{\cellcolor[HTML]{FFFFFF}\modifyInterf}                                                                                                                                                                                                                                                                                                    & \multicolumn{4}{c|}{\cellcolor[HTML]{FFFFFF}\prefetchInterf}                                                                                                                                                                                                                                                     \\ \cline{2-18} 
\multicolumn{1}{|l|}{\cellcolor[HTML]{FFFFFF}{\color[HTML]{333333} }} & \multicolumn{1}{l|}{\cellcolor[HTML]{FFFFFF}\begin{tabular}[c]{@{}l@{}}Partition \\ Size\end{tabular}} & \multicolumn{1}{l|}{\cellcolor[HTML]{FFFFFF}No}                     & \multicolumn{1}{l|}{\cellcolor[HTML]{FFFFFF}1/3}                                  & \multicolumn{1}{l|}{\cellcolor[HTML]{FFFFFF}2/2}                                  & \multicolumn{1}{l|}{\cellcolor[HTML]{FFFFFF}3/1}                                  & \multicolumn{1}{l|}{\cellcolor[HTML]{FFFFFF}No}                                            & \multicolumn{1}{l|}{\cellcolor[HTML]{FFFFFF}1/3}                                  & \multicolumn{1}{l|}{\cellcolor[HTML]{FFFFFF}2/2}                                  & \multicolumn{1}{l|}{\cellcolor[HTML]{FFFFFF}3/1}                                  & \multicolumn{1}{l|}{\cellcolor[HTML]{FFFFFF}No}                                            & \multicolumn{1}{l|}{\cellcolor[HTML]{FFFFFF}1/3}                                  & \multicolumn{1}{l|}{\cellcolor[HTML]{FFFFFF}2/2}                                  & \multicolumn{1}{l|}{\cellcolor[HTML]{FFFFFF}3/1}                                  & \multicolumn{1}{l|}{\cellcolor[HTML]{FFFFFF}No}                                            & \multicolumn{1}{l|}{\cellcolor[HTML]{FFFFFF}1/3}                                  & \multicolumn{1}{l|}{\cellcolor[HTML]{FFFFFF}2/2}                                  & 3/1                                  \\ \hline
\multicolumn{1}{|c|}{\cellcolor[HTML]{FFFFFF}}                        & \multicolumn{1}{l|}{\cellcolor[HTML]{FFFFFF}way}                                                       & \multicolumn{1}{l|}{\cellcolor[HTML]{FFFFFF}}                       & \multicolumn{1}{l|}{\cellcolor[HTML]{FFFFFF}1.21}                                 & \multicolumn{1}{l|}{\cellcolor[HTML]{FFFFFF}1.21}                                 & \multicolumn{1}{l|}{\cellcolor[HTML]{FFFFFF}1.19}                                 & \multicolumn{1}{l|}{\cellcolor[HTML]{FFFFFF}}                                              & \multicolumn{1}{l|}{\cellcolor[HTML]{FFFFFF}2.90}                                 & \multicolumn{1}{l|}{\cellcolor[HTML]{FFFFFF}2.68}                                 & \multicolumn{1}{l|}{\cellcolor[HTML]{FFFFFF}2.54}                                 & \multicolumn{1}{l|}{\cellcolor[HTML]{FFFFFF}}                                              & \multicolumn{1}{l|}{\cellcolor[HTML]{FFFFFF}1.10}                                 & \multicolumn{1}{l|}{\cellcolor[HTML]{FFFFFF}1.10}                                 & \multicolumn{1}{l|}{\cellcolor[HTML]{FFFFFF}1.09}                                 & \multicolumn{1}{l|}{\cellcolor[HTML]{FFFFFF}}                                              & \multicolumn{1}{l|}{\cellcolor[HTML]{FFFFFF}1.69}                                 & \multicolumn{1}{l|}{\cellcolor[HTML]{FFFFFF}1.68}                                 & 1.64                                 \\ \cline{2-2} \cline{4-6} \cline{8-10} \cline{12-14} \cline{16-18} 
\multicolumn{1}{|c|}{\multirow{-2}{*}{\cellcolor[HTML]{FFFFFF}\cif}}   & \multicolumn{1}{l|}{\cellcolor[HTML]{FFFFFF}set}                                                       & \multicolumn{1}{l|}{\multirow{-2}{*}{\cellcolor[HTML]{FFFFFF}1.25}} & \multicolumn{1}{l|}{\cellcolor[HTML]{FFFFFF}{\color[HTML]{333333} \textbf{1.16}}} & \multicolumn{1}{l|}{\cellcolor[HTML]{FFFFFF}\textbf{1.16}}                        & \multicolumn{1}{l|}{\cellcolor[HTML]{FFFFFF}\textbf{1.15}}                        & \multicolumn{1}{l|}{\multirow{-2}{*}{\cellcolor[HTML]{FFFFFF}2.57}}                        & \multicolumn{1}{l|}{\cellcolor[HTML]{FFFFFF}{\color[HTML]{333333} \textbf{2.63}}} & \multicolumn{1}{l|}{\cellcolor[HTML]{FFFFFF}\textbf{2.47}}                        & \multicolumn{1}{l|}{\cellcolor[HTML]{FFFFFF}\textbf{2.29}}                        & \multicolumn{1}{l|}{\multirow{-2}{*}{\cellcolor[HTML]{FFFFFF}1.15}}                        & \multicolumn{1}{l|}{\cellcolor[HTML]{FFFFFF}{\color[HTML]{333333} \textbf{1.08}}} & \multicolumn{1}{l|}{\cellcolor[HTML]{FFFFFF}\textbf{1.08}}                        & \multicolumn{1}{l|}{\cellcolor[HTML]{FFFFFF}\textbf{1.07}}                        & \multicolumn{1}{l|}{\multirow{-2}{*}{\cellcolor[HTML]{FFFFFF}1.66}}                        & \multicolumn{1}{l|}{\cellcolor[HTML]{FFFFFF}{\color[HTML]{333333} \textbf{1.60}}} & \multicolumn{1}{l|}{\cellcolor[HTML]{FFFFFF}\textbf{1.60}}                        & \textbf{1.58}                        \\ \hline
\multicolumn{1}{|c|}{\cellcolor[HTML]{FFFFFF}}                        & \multicolumn{1}{l|}{\cellcolor[HTML]{FFFFFF}way}                                                       & \multicolumn{1}{l|}{\cellcolor[HTML]{FFFFFF}}                       & \multicolumn{1}{l|}{\cellcolor[HTML]{FFFFFF}1.17}                                 & \multicolumn{1}{l|}{\cellcolor[HTML]{FFFFFF}1.15}                                 & \multicolumn{1}{l|}{\cellcolor[HTML]{FFFFFF}1.15}                                 & \multicolumn{1}{l|}{\cellcolor[HTML]{FFFFFF}}                                              & \multicolumn{1}{l|}{\cellcolor[HTML]{FFFFFF}\textbf{2.59}}                        & \multicolumn{1}{l|}{\cellcolor[HTML]{FFFFFF}2.56}                                 & \multicolumn{1}{l|}{\cellcolor[HTML]{FFFFFF}2.56}                                 & \multicolumn{1}{l|}{\cellcolor[HTML]{FFFFFF}}                                              & \multicolumn{1}{l|}{\cellcolor[HTML]{FFFFFF}1.11}                                 & \multicolumn{1}{l|}{\cellcolor[HTML]{FFFFFF}1.10}                                 & \multicolumn{1}{l|}{\cellcolor[HTML]{FFFFFF}1.10}                                 & \multicolumn{1}{l|}{\cellcolor[HTML]{FFFFFF}}                                              & \multicolumn{1}{l|}{\cellcolor[HTML]{FFFFFF}1.70}                                 & \multicolumn{1}{l|}{\cellcolor[HTML]{FFFFFF}1.66}                                 & 1.66                                 \\ \cline{2-2} \cline{4-6} \cline{8-10} \cline{12-14} \cline{16-18} 
\multicolumn{1}{|c|}{\multirow{-2}{*}{\cellcolor[HTML]{FFFFFF}\vga}}   & \multicolumn{1}{l|}{\cellcolor[HTML]{FFFFFF}set}                                                       & \multicolumn{1}{l|}{\multirow{-2}{*}{\cellcolor[HTML]{FFFFFF}1.17}} & \multicolumn{1}{l|}{\cellcolor[HTML]{FFFFFF}\textbf{1.14}}                        & \multicolumn{1}{l|}{\cellcolor[HTML]{FFFFFF}\textbf{1.14}}                        & \multicolumn{1}{l|}{\cellcolor[HTML]{FFFFFF}\textbf{1.13}}                        & \multicolumn{1}{l|}{\multirow{-2}{*}{\cellcolor[HTML]{FFFFFF}2.58}}                        & \multicolumn{1}{l|}{\cellcolor[HTML]{FFFFFF}2.76}                                 & \multicolumn{1}{l|}{\cellcolor[HTML]{FFFFFF}\textbf{2.54}}                        & \multicolumn{1}{l|}{\cellcolor[HTML]{FFFFFF}\textbf{2.49}}                        & \multicolumn{1}{l|}{\multirow{-2}{*}{\cellcolor[HTML]{FFFFFF}1.12}}                        & \multicolumn{1}{l|}{\cellcolor[HTML]{FFFFFF}\textbf{1.08}}                        & \multicolumn{1}{l|}{\cellcolor[HTML]{FFFFFF}\textbf{1.08}}                        & \multicolumn{1}{l|}{\cellcolor[HTML]{FFFFFF}\textbf{1.08}}                        & \multicolumn{1}{l|}{\multirow{-2}{*}{\cellcolor[HTML]{FFFFFF}1.69}}                        & \multicolumn{1}{l|}{\cellcolor[HTML]{FFFFFF}\textbf{1.63}}                        & \multicolumn{1}{l|}{\cellcolor[HTML]{FFFFFF}\textbf{1.63}}                        & \textbf{1.63}                        \\ \hline
\multicolumn{2}{|l|}{\cellcolor[HTML]{FFFFFF}}                                                                                                                                 & \multicolumn{16}{c|}{\cellcolor[HTML]{FFFFFF}\mser}                                                                                                                                                                                                                                                                                                                                                                                                                                                                                                                                                                                                                                                                                                                                                                                                                                                                                                                                                                                                                                                                                                                                                                                                                                                                                                            \\ \cline{3-18} 
\multicolumn{2}{|l|}{\multirow{-2}{*}{\cellcolor[HTML]{FFFFFF}}}                                                                                                               & \multicolumn{4}{c|}{\cellcolor[HTML]{FFFFFF}\readInterf}                                                                                                                                                                                                                                                                               & \multicolumn{4}{c|}{\cellcolor[HTML]{FFFFFF}\writeInterf}                                                                                                                                                                                                                                                                                                     & \multicolumn{4}{c|}{\cellcolor[HTML]{FFFFFF}\modifyInterf}                                                                                                                                                                                                                                                                                                    & \multicolumn{4}{c|}{\cellcolor[HTML]{FFFFFF}\prefetchInterf}                                                                                                                                                                                                                                                     \\ \cline{2-18} 
\multicolumn{1}{|l|}{\cellcolor[HTML]{FFFFFF}}                        & \multicolumn{1}{l|}{\cellcolor[HTML]{FFFFFF}\begin{tabular}[c]{@{}l@{}}Partition \\ Size\end{tabular}} & \multicolumn{1}{l|}{\cellcolor[HTML]{FFFFFF}No}                     & \multicolumn{1}{l|}{\cellcolor[HTML]{FFFFFF}1/3}                                  & \multicolumn{1}{l|}{\cellcolor[HTML]{FFFFFF}2/2}                                  & \multicolumn{1}{l|}{\cellcolor[HTML]{FFFFFF}3/1}                                  & \multicolumn{1}{l|}{\cellcolor[HTML]{FFFFFF}No}                                            & \multicolumn{1}{l|}{\cellcolor[HTML]{FFFFFF}1/3}                                  & \multicolumn{1}{l|}{\cellcolor[HTML]{FFFFFF}2/2}                                  & \multicolumn{1}{l|}{\cellcolor[HTML]{FFFFFF}3/1}                                  & \multicolumn{1}{l|}{\cellcolor[HTML]{FFFFFF}No}                                            & \multicolumn{1}{l|}{\cellcolor[HTML]{FFFFFF}1/3}                                  & \multicolumn{1}{l|}{\cellcolor[HTML]{FFFFFF}2/2}                                  & \multicolumn{1}{l|}{\cellcolor[HTML]{FFFFFF}3/1}                                  & \multicolumn{1}{l|}{\cellcolor[HTML]{FFFFFF}No}                                            & \multicolumn{1}{l|}{\cellcolor[HTML]{FFFFFF}1/3}                                  & \multicolumn{1}{l|}{\cellcolor[HTML]{FFFFFF}2/2}                                  & 3/1                                  \\ \hline
\multicolumn{1}{|c|}{\cellcolor[HTML]{FFFFFF}}                        & \multicolumn{1}{l|}{\cellcolor[HTML]{FFFFFF}way}                                                       & \multicolumn{1}{l|}{\cellcolor[HTML]{FFFFFF}}                       & \multicolumn{1}{l|}{\cellcolor[HTML]{FFFFFF}{\color[HTML]{333333} 1.46}}          & \multicolumn{1}{l|}{\cellcolor[HTML]{FFFFFF}{\color[HTML]{333333} 1.40}}          & \multicolumn{1}{l|}{\cellcolor[HTML]{FFFFFF}{\color[HTML]{333333} 1.35}}          & \multicolumn{1}{l|}{\cellcolor[HTML]{FFFFFF}{\color[HTML]{333333} }}                       & \multicolumn{1}{l|}{\cellcolor[HTML]{FFFFFF}{\color[HTML]{333333} \textbf{6.20}}} & \multicolumn{1}{l|}{\cellcolor[HTML]{FFFFFF}{\color[HTML]{333333} 5.79}}          & \multicolumn{1}{l|}{\cellcolor[HTML]{FFFFFF}{\color[HTML]{333333} 5.57}}          & \multicolumn{1}{l|}{\cellcolor[HTML]{FFFFFF}{\color[HTML]{333333} }}                       & \multicolumn{1}{l|}{\cellcolor[HTML]{FFFFFF}{\color[HTML]{333333} 1.29}}          & \multicolumn{1}{l|}{\cellcolor[HTML]{FFFFFF}{\color[HTML]{333333} 1.24}}          & \multicolumn{1}{l|}{\cellcolor[HTML]{FFFFFF}{\color[HTML]{333333} 1.21}}          & \multicolumn{1}{l|}{\cellcolor[HTML]{FFFFFF}{\color[HTML]{333333} }}                       & \multicolumn{1}{l|}{\cellcolor[HTML]{FFFFFF}{\color[HTML]{333333} 2.12}}          & \multicolumn{1}{l|}{\cellcolor[HTML]{FFFFFF}{\color[HTML]{333333} 2.0}}           & {\color[HTML]{333333} 1.93}          \\ \cline{2-2} \cline{4-6} \cline{8-10} \cline{12-14} \cline{16-18} 
\multicolumn{1}{|c|}{\multirow{-2}{*}{\cellcolor[HTML]{FFFFFF}\cif}}   & \multicolumn{1}{l|}{\cellcolor[HTML]{FFFFFF}set}                                                       & \multicolumn{1}{l|}{\multirow{-2}{*}{\cellcolor[HTML]{FFFFFF}1.82}} & \multicolumn{1}{l|}{\cellcolor[HTML]{FFFFFF}{\color[HTML]{333333} \textbf{1.33}}} & \multicolumn{1}{l|}{\cellcolor[HTML]{FFFFFF}{\color[HTML]{333333} \textbf{1.31}}} & \multicolumn{1}{l|}{\cellcolor[HTML]{FFFFFF}{\color[HTML]{333333} \textbf{1.28}}} & \multicolumn{1}{l|}{\multirow{-2}{*}{\cellcolor[HTML]{FFFFFF}{\color[HTML]{333333} 5.39}}} & \multicolumn{1}{l|}{\cellcolor[HTML]{FFFFFF}{\color[HTML]{333333} 6.25}}          & \multicolumn{1}{l|}{\cellcolor[HTML]{FFFFFF}{\color[HTML]{333333} \textbf{5.15}}} & \multicolumn{1}{l|}{\cellcolor[HTML]{FFFFFF}{\color[HTML]{333333} \textbf{4.47}}} & \multicolumn{1}{l|}{\multirow{-2}{*}{\cellcolor[HTML]{FFFFFF}{\color[HTML]{333333} 1.49}}} & \multicolumn{1}{l|}{\cellcolor[HTML]{FFFFFF}{\color[HTML]{333333} \textbf{1.20}}} & \multicolumn{1}{l|}{\cellcolor[HTML]{FFFFFF}{\color[HTML]{333333} \textbf{1.19}}} & \multicolumn{1}{l|}{\cellcolor[HTML]{FFFFFF}{\color[HTML]{333333} \textbf{1.16}}} & \multicolumn{1}{l|}{\multirow{-2}{*}{\cellcolor[HTML]{FFFFFF}{\color[HTML]{333333} 2.45}}} & \multicolumn{1}{l|}{\cellcolor[HTML]{FFFFFF}{\color[HTML]{333333} \textbf{1.91}}} & \multicolumn{1}{l|}{\cellcolor[HTML]{FFFFFF}{\color[HTML]{333333} \textbf{1.85}}} & {\color[HTML]{333333} \textbf{1.8}}  \\ \hline
\multicolumn{1}{|c|}{\cellcolor[HTML]{FFFFFF}}                        & \multicolumn{1}{l|}{\cellcolor[HTML]{FFFFFF}way}                                                       & \multicolumn{1}{l|}{\cellcolor[HTML]{FFFFFF}}                       & \multicolumn{1}{l|}{\cellcolor[HTML]{FFFFFF}{\color[HTML]{333333} 1.47}}          & \multicolumn{1}{l|}{\cellcolor[HTML]{FFFFFF}{\color[HTML]{333333} 1.45}}          & \multicolumn{1}{l|}{\cellcolor[HTML]{FFFFFF}{\color[HTML]{333333} 1.42}}          & \multicolumn{1}{l|}{\cellcolor[HTML]{FFFFFF}{\color[HTML]{333333} }}                       & \multicolumn{1}{l|}{\cellcolor[HTML]{FFFFFF}{\color[HTML]{333333} \textbf{6.00}}} & \multicolumn{1}{l|}{\cellcolor[HTML]{FFFFFF}{\color[HTML]{333333} 6.16}}          & \multicolumn{1}{l|}{\cellcolor[HTML]{FFFFFF}{\color[HTML]{333333} 6.12}}          & \multicolumn{1}{l|}{\cellcolor[HTML]{FFFFFF}{\color[HTML]{333333} }}                       & \multicolumn{1}{l|}{\cellcolor[HTML]{FFFFFF}{\color[HTML]{333333} 1.30}}          & \multicolumn{1}{l|}{\cellcolor[HTML]{FFFFFF}{\color[HTML]{333333} 1.28}}          & \multicolumn{1}{l|}{\cellcolor[HTML]{FFFFFF}{\color[HTML]{333333} 1.26}}          & \multicolumn{1}{l|}{\cellcolor[HTML]{FFFFFF}{\color[HTML]{333333} }}                       & \multicolumn{1}{l|}{\cellcolor[HTML]{FFFFFF}{\color[HTML]{333333} 2.16}}          & \multicolumn{1}{l|}{\cellcolor[HTML]{FFFFFF}{\color[HTML]{333333} 2.11}}          & {\color[HTML]{333333} 2.05}          \\ \cline{2-2} \cline{4-6} \cline{8-10} \cline{12-14} \cline{16-18} 
\multicolumn{1}{|c|}{\multirow{-2}{*}{\cellcolor[HTML]{FFFFFF}\vga}}   & \multicolumn{1}{l|}{\cellcolor[HTML]{FFFFFF}set}                                                       & \multicolumn{1}{l|}{\multirow{-2}{*}{\cellcolor[HTML]{FFFFFF}1.69}} & \multicolumn{1}{l|}{\cellcolor[HTML]{FFFFFF}{\color[HTML]{333333} \textbf{1.37}}} & \multicolumn{1}{l|}{\cellcolor[HTML]{FFFFFF}{\color[HTML]{333333} \textbf{1.40}}} & \multicolumn{1}{l|}{\cellcolor[HTML]{FFFFFF}{\color[HTML]{333333} \textbf{1.40}}} & \multicolumn{1}{l|}{\multirow{-2}{*}{\cellcolor[HTML]{FFFFFF}{\color[HTML]{333333} 6.17}}} & \multicolumn{1}{l|}{\cellcolor[HTML]{FFFFFF}{\color[HTML]{333333} 6.45}}          & \multicolumn{1}{l|}{\cellcolor[HTML]{FFFFFF}{\color[HTML]{333333} \textbf{5.44}}} & \multicolumn{1}{l|}{\cellcolor[HTML]{FFFFFF}{\color[HTML]{333333} \textbf{4.97}}} & \multicolumn{1}{l|}{\multirow{-2}{*}{\cellcolor[HTML]{FFFFFF}{\color[HTML]{333333} 1.46}}} & \multicolumn{1}{l|}{\cellcolor[HTML]{FFFFFF}{\color[HTML]{333333} \textbf{1.22}}} & \multicolumn{1}{l|}{\cellcolor[HTML]{FFFFFF}{\color[HTML]{333333} \textbf{1.21}}} & \multicolumn{1}{l|}{\cellcolor[HTML]{FFFFFF}{\color[HTML]{333333} \textbf{1.2}}}  & \multicolumn{1}{l|}{\multirow{-2}{*}{\cellcolor[HTML]{FFFFFF}{\color[HTML]{333333} 2.26}}} & \multicolumn{1}{l|}{\cellcolor[HTML]{FFFFFF}{\color[HTML]{333333} \textbf{1.91}}} & \multicolumn{1}{l|}{\cellcolor[HTML]{FFFFFF}{\color[HTML]{333333} \textbf{1.93}}} & {\color[HTML]{333333} \textbf{1.91}} \\ \hline
\multicolumn{2}{|l|}{\cellcolor[HTML]{FFFFFF}}                                                                                                                                 & \multicolumn{16}{c|}{\cellcolor[HTML]{FFFFFF}\tracking}                                                                                                                                                                                                                                                                                                                                                                                                                                                                                                                                                                                                                                                                                                                                                                                                                                                                                                                                                                                                                                                                                                                                                                                                                                                                                                        \\ \cline{3-18} 
\multicolumn{2}{|l|}{\multirow{-2}{*}{\cellcolor[HTML]{FFFFFF}}}                                                                                                               & \multicolumn{4}{c|}{\cellcolor[HTML]{FFFFFF}\readInterf}                                                                                                                                                                                                                                                                               & \multicolumn{4}{c|}{\cellcolor[HTML]{FFFFFF}{\color[HTML]{333333} \writeInterf}}                                                                                                                                                                                                                                                                              & \multicolumn{4}{c|}{\cellcolor[HTML]{FFFFFF}{\color[HTML]{333333} \modifyInterf}}                                                                                                                                                                                                                                                                             & \multicolumn{4}{c|}{\cellcolor[HTML]{FFFFFF}{\color[HTML]{333333} \prefetchInterf}}                                                                                                                                                                                                                              \\ \cline{2-18} 
\multicolumn{1}{|l|}{\cellcolor[HTML]{FFFFFF}}                        & \multicolumn{1}{l|}{\cellcolor[HTML]{FFFFFF}\begin{tabular}[c]{@{}l@{}}Partition\\ Size\end{tabular}}  & \multicolumn{1}{l|}{\cellcolor[HTML]{FFFFFF}No}                     & \multicolumn{1}{l|}{\cellcolor[HTML]{FFFFFF}{\color[HTML]{333333} 1/3}}           & \multicolumn{1}{l|}{\cellcolor[HTML]{FFFFFF}{\color[HTML]{333333} 2/2}}           & \multicolumn{1}{l|}{\cellcolor[HTML]{FFFFFF}{\color[HTML]{333333} 3/1}}           & \multicolumn{1}{l|}{\cellcolor[HTML]{FFFFFF}{\color[HTML]{333333} No}}                     & \multicolumn{1}{l|}{\cellcolor[HTML]{FFFFFF}{\color[HTML]{333333} 1/3}}           & \multicolumn{1}{l|}{\cellcolor[HTML]{FFFFFF}{\color[HTML]{333333} 2/2}}           & \multicolumn{1}{l|}{\cellcolor[HTML]{FFFFFF}{\color[HTML]{333333} 3/1}}           & \multicolumn{1}{l|}{\cellcolor[HTML]{FFFFFF}{\color[HTML]{333333} No}}                     & \multicolumn{1}{l|}{\cellcolor[HTML]{FFFFFF}{\color[HTML]{333333} 1/3}}           & \multicolumn{1}{l|}{\cellcolor[HTML]{FFFFFF}{\color[HTML]{333333} 2/2}}           & \multicolumn{1}{l|}{\cellcolor[HTML]{FFFFFF}{\color[HTML]{333333} 3/1}}           & \multicolumn{1}{l|}{\cellcolor[HTML]{FFFFFF}{\color[HTML]{333333} No}}                     & \multicolumn{1}{l|}{\cellcolor[HTML]{FFFFFF}{\color[HTML]{333333} 1/3}}           & \multicolumn{1}{l|}{\cellcolor[HTML]{FFFFFF}{\color[HTML]{333333} 2/2}}           & {\color[HTML]{333333} 3/1}           \\ \hline
\multicolumn{1}{|l|}{\cellcolor[HTML]{FFFFFF}}                        & \multicolumn{1}{l|}{\cellcolor[HTML]{FFFFFF}way}                                                       & \multicolumn{1}{l|}{\cellcolor[HTML]{FFFFFF}}                       & \multicolumn{1}{l|}{\cellcolor[HTML]{FFFFFF}{\color[HTML]{333333} \textbf{1.03}}} & \multicolumn{1}{l|}{\cellcolor[HTML]{FFFFFF}{\color[HTML]{333333} 1.03}}          & \multicolumn{1}{l|}{\cellcolor[HTML]{FFFFFF}{\color[HTML]{333333} 1.03}}          & \multicolumn{1}{l|}{\cellcolor[HTML]{FFFFFF}{\color[HTML]{333333} }}                       & \multicolumn{1}{l|}{\cellcolor[HTML]{FFFFFF}{\color[HTML]{333333} \textbf{1.38}}} & \multicolumn{1}{l|}{\cellcolor[HTML]{FFFFFF}{\color[HTML]{333333} \textbf{1.36}}} & \multicolumn{1}{l|}{\cellcolor[HTML]{FFFFFF}{\color[HTML]{333333} 1.35}}          & \multicolumn{1}{l|}{\cellcolor[HTML]{FFFFFF}{\color[HTML]{333333} }}                       & \multicolumn{1}{l|}{\cellcolor[HTML]{FFFFFF}{\color[HTML]{333333} 1.02}}          & \multicolumn{1}{l|}{\cellcolor[HTML]{FFFFFF}{\color[HTML]{333333} 1.02}}          & \multicolumn{1}{l|}{\cellcolor[HTML]{FFFFFF}{\color[HTML]{333333} 1.02}}          & \multicolumn{1}{l|}{\cellcolor[HTML]{FFFFFF}{\color[HTML]{333333} }}                       & \multicolumn{1}{l|}{\cellcolor[HTML]{FFFFFF}{\color[HTML]{333333} \textbf{1.13}}} & \multicolumn{1}{l|}{\cellcolor[HTML]{FFFFFF}{\color[HTML]{333333} \textbf{1.12}}} & {\color[HTML]{333333} \textbf{1.12}} \\ \cline{2-2} \cline{4-6} \cline{8-10} \cline{12-14} \cline{16-18} 
\multicolumn{1}{|l|}{\multirow{-2}{*}{\cellcolor[HTML]{FFFFFF}\cif}}   & \multicolumn{1}{l|}{\cellcolor[HTML]{FFFFFF}set}                                                       & \multicolumn{1}{l|}{\multirow{-2}{*}{\cellcolor[HTML]{FFFFFF}1.05}} & \multicolumn{1}{l|}{\cellcolor[HTML]{FFFFFF}{\color[HTML]{333333} 1.04}}          & \multicolumn{1}{l|}{\cellcolor[HTML]{FFFFFF}{\color[HTML]{333333} \textbf{1.03}}} & \multicolumn{1}{l|}{\cellcolor[HTML]{FFFFFF}{\color[HTML]{333333} \textbf{1.03}}} & \multicolumn{1}{l|}{\multirow{-2}{*}{\cellcolor[HTML]{FFFFFF}{\color[HTML]{333333} 1.33}}} & \multicolumn{1}{l|}{\cellcolor[HTML]{FFFFFF}{\color[HTML]{333333} 1.46}}          & \multicolumn{1}{l|}{\cellcolor[HTML]{FFFFFF}{\color[HTML]{333333} 1.38}}          & \multicolumn{1}{l|}{\cellcolor[HTML]{FFFFFF}{\color[HTML]{333333} \textbf{1.33}}} & \multicolumn{1}{l|}{\multirow{-2}{*}{\cellcolor[HTML]{FFFFFF}{\color[HTML]{333333} 1.03}}} & \multicolumn{1}{l|}{\cellcolor[HTML]{FFFFFF}{\color[HTML]{333333} \textbf{1.02}}} & \multicolumn{1}{l|}{\cellcolor[HTML]{FFFFFF}{\color[HTML]{333333} \textbf{1.02}}} & \multicolumn{1}{l|}{\cellcolor[HTML]{FFFFFF}{\color[HTML]{333333} \textbf{1.01}}} & \multicolumn{1}{l|}{\multirow{-2}{*}{\cellcolor[HTML]{FFFFFF}{\color[HTML]{333333} 1.13}}} & \multicolumn{1}{l|}{\cellcolor[HTML]{FFFFFF}{\color[HTML]{333333} 1.13}}          & \multicolumn{1}{l|}{\cellcolor[HTML]{FFFFFF}{\color[HTML]{333333} 1.13}}          & {\color[HTML]{333333} 1.12}          \\ \hline
\multicolumn{1}{|l|}{\cellcolor[HTML]{FFFFFF}}                        & \multicolumn{1}{l|}{\cellcolor[HTML]{FFFFFF}way}                                                       & \multicolumn{1}{l|}{\cellcolor[HTML]{FFFFFF}}                       & \multicolumn{1}{l|}{\cellcolor[HTML]{FFFFFF}{\color[HTML]{333333} 1.17}}          & \multicolumn{1}{l|}{\cellcolor[HTML]{FFFFFF}{\color[HTML]{333333} \textbf{1.08}}} & \multicolumn{1}{l|}{\cellcolor[HTML]{FFFFFF}{\color[HTML]{333333} 1.09}}          & \multicolumn{1}{l|}{\cellcolor[HTML]{FFFFFF}{\color[HTML]{333333} }}                       & \multicolumn{1}{l|}{\cellcolor[HTML]{FFFFFF}{\color[HTML]{333333} 2.8}}           & \multicolumn{1}{l|}{\cellcolor[HTML]{FFFFFF}{\color[HTML]{333333} \textbf{1.63}}} & \multicolumn{1}{l|}{\cellcolor[HTML]{FFFFFF}{\color[HTML]{333333} \textbf{1.64}}} & \multicolumn{1}{l|}{\cellcolor[HTML]{FFFFFF}{\color[HTML]{333333} }}                       & \multicolumn{1}{l|}{\cellcolor[HTML]{FFFFFF}{\color[HTML]{333333} 1.09}}          & \multicolumn{1}{l|}{\cellcolor[HTML]{FFFFFF}{\color[HTML]{333333} \textbf{1.07}}} & \multicolumn{1}{l|}{\cellcolor[HTML]{FFFFFF}{\color[HTML]{333333} \textbf{1.07}}} & \multicolumn{1}{l|}{\cellcolor[HTML]{FFFFFF}{\color[HTML]{333333} }}                       & \multicolumn{1}{l|}{\cellcolor[HTML]{FFFFFF}{\color[HTML]{333333} 1.55}}          & \multicolumn{1}{l|}{\cellcolor[HTML]{FFFFFF}{\color[HTML]{333333} 1.34}}          & {\color[HTML]{333333} 1.34}          \\ \cline{2-2} \cline{4-6} \cline{8-10} \cline{12-14} \cline{16-18} 
\multicolumn{1}{|l|}{\multirow{-2}{*}{\cellcolor[HTML]{FFFFFF}\vga}}   & \multicolumn{1}{l|}{\cellcolor[HTML]{FFFFFF}set}                                                       & \multicolumn{1}{l|}{\multirow{-2}{*}{\cellcolor[HTML]{FFFFFF}1.28}} & \multicolumn{1}{l|}{\cellcolor[HTML]{FFFFFF}{\color[HTML]{333333} \textbf{1.15}}} & \multicolumn{1}{l|}{\cellcolor[HTML]{FFFFFF}{\color[HTML]{333333} 1.09}}          & \multicolumn{1}{l|}{\cellcolor[HTML]{FFFFFF}{\color[HTML]{333333} \textbf{1.07}}} & \multicolumn{1}{l|}{\multirow{-2}{*}{\cellcolor[HTML]{FFFFFF}{\color[HTML]{333333} 1.59}}} & \multicolumn{1}{l|}{\cellcolor[HTML]{FFFFFF}{\color[HTML]{333333} \textbf{2.74}}} & \multicolumn{1}{l|}{\cellcolor[HTML]{FFFFFF}{\color[HTML]{333333} 1.68}}          & \multicolumn{1}{l|}{\cellcolor[HTML]{FFFFFF}{\color[HTML]{333333} 1.7}}           & \multicolumn{1}{l|}{\multirow{-2}{*}{\cellcolor[HTML]{FFFFFF}{\color[HTML]{333333} 1.14}}} & \multicolumn{1}{l|}{\cellcolor[HTML]{FFFFFF}{\color[HTML]{333333} \textbf{1.07}}} & \multicolumn{1}{l|}{\cellcolor[HTML]{FFFFFF}{\color[HTML]{333333} 1.07}}          & \multicolumn{1}{l|}{\cellcolor[HTML]{FFFFFF}{\color[HTML]{333333} 1.07}}          & \multicolumn{1}{l|}{\multirow{-2}{*}{\cellcolor[HTML]{FFFFFF}{\color[HTML]{333333} 1.44}}} & \multicolumn{1}{l|}{\cellcolor[HTML]{FFFFFF}{\color[HTML]{333333} \textbf{1.48}}} & \multicolumn{1}{l|}{\cellcolor[HTML]{FFFFFF}{\color[HTML]{333333} \textbf{1.34}}} & {\color[HTML]{333333} \textbf{1.32}} \\ \hline
\multicolumn{2}{|l|}{\cellcolor[HTML]{FFFFFF}}                                                                                                                                 & \multicolumn{16}{c|}{\cellcolor[HTML]{FFFFFF}\sift}                                                                                                                                                                                                                                                                                                                                                                                                                                                                                                                                                                                                                                                                                                                                                                                                                                                                                                                                                                                                                                                                                                                                                                                                                                                                                                            \\ \cline{3-18} 
\multicolumn{2}{|l|}{\multirow{-2}{*}{\cellcolor[HTML]{FFFFFF}}}                                                                                                               & \multicolumn{4}{c|}{\cellcolor[HTML]{FFFFFF}\readInterf}                                                                                                                                                                                                                                                                               & \multicolumn{4}{c|}{\cellcolor[HTML]{FFFFFF}{\color[HTML]{333333} \writeInterf}}                                                                                                                                                                                                                                                                              & \multicolumn{4}{c|}{\cellcolor[HTML]{FFFFFF}{\color[HTML]{333333} \modifyInterf}}                                                                                                                                                                                                                                                                             & \multicolumn{4}{c|}{\cellcolor[HTML]{FFFFFF}{\color[HTML]{333333} \prefetchInterf}}                                                                                                                                                                                                                              \\ \cline{2-18} 
\multicolumn{1}{|l|}{\cellcolor[HTML]{FFFFFF}}                        & \multicolumn{1}{l|}{\cellcolor[HTML]{FFFFFF}\begin{tabular}[c]{@{}l@{}}Partition\\ Size\end{tabular}}  & \multicolumn{1}{l|}{\cellcolor[HTML]{FFFFFF}No}                     & \multicolumn{1}{l|}{\cellcolor[HTML]{FFFFFF}{\color[HTML]{333333} 1/3}}           & \multicolumn{1}{l|}{\cellcolor[HTML]{FFFFFF}{\color[HTML]{333333} 2/2}}           & \multicolumn{1}{l|}{\cellcolor[HTML]{FFFFFF}{\color[HTML]{333333} 3/1}}           & \multicolumn{1}{l|}{\cellcolor[HTML]{FFFFFF}{\color[HTML]{333333} No}}                     & \multicolumn{1}{l|}{\cellcolor[HTML]{FFFFFF}{\color[HTML]{333333} 1/3}}           & \multicolumn{1}{l|}{\cellcolor[HTML]{FFFFFF}{\color[HTML]{333333} 2/2}}           & \multicolumn{1}{l|}{\cellcolor[HTML]{FFFFFF}{\color[HTML]{333333} 3/1}}           & \multicolumn{1}{l|}{\cellcolor[HTML]{FFFFFF}{\color[HTML]{333333} No}}                     & \multicolumn{1}{l|}{\cellcolor[HTML]{FFFFFF}{\color[HTML]{333333} 1/3}}           & \multicolumn{1}{l|}{\cellcolor[HTML]{FFFFFF}{\color[HTML]{333333} 2/2}}           & \multicolumn{1}{l|}{\cellcolor[HTML]{FFFFFF}{\color[HTML]{333333} 3/1}}           & \multicolumn{1}{l|}{\cellcolor[HTML]{FFFFFF}{\color[HTML]{333333} No}}                     & \multicolumn{1}{l|}{\cellcolor[HTML]{FFFFFF}{\color[HTML]{333333} 1/3}}           & \multicolumn{1}{l|}{\cellcolor[HTML]{FFFFFF}{\color[HTML]{333333} 2/2}}           & {\color[HTML]{333333} 3/1}           \\ \hline
\multicolumn{1}{|c|}{\cellcolor[HTML]{FFFFFF}}                        & \multicolumn{1}{l|}{\cellcolor[HTML]{FFFFFF}way}                                                       & \multicolumn{1}{l|}{\cellcolor[HTML]{FFFFFF}}                       & \multicolumn{1}{l|}{\cellcolor[HTML]{FFFFFF}{\color[HTML]{333333} 1.07}}          & \multicolumn{1}{l|}{\cellcolor[HTML]{FFFFFF}{\color[HTML]{333333} 1.04}}          & \multicolumn{1}{l|}{\cellcolor[HTML]{FFFFFF}{\color[HTML]{333333} 1.03}}          & \multicolumn{1}{l|}{\cellcolor[HTML]{FFFFFF}{\color[HTML]{333333} }}                       & \multicolumn{1}{l|}{\cellcolor[HTML]{FFFFFF}{\color[HTML]{333333} 1.99}}          & \multicolumn{1}{l|}{\cellcolor[HTML]{FFFFFF}{\color[HTML]{333333} \textbf{1.52}}} & \multicolumn{1}{l|}{\cellcolor[HTML]{FFFFFF}{\color[HTML]{333333} \textbf{1.51}}} & \multicolumn{1}{l|}{\cellcolor[HTML]{FFFFFF}{\color[HTML]{333333} }}                       & \multicolumn{1}{l|}{\cellcolor[HTML]{FFFFFF}{\color[HTML]{333333} 1.03}}          & \multicolumn{1}{l|}{\cellcolor[HTML]{FFFFFF}{\color[HTML]{333333} \textbf{1.02}}} & \multicolumn{1}{l|}{\cellcolor[HTML]{FFFFFF}{\color[HTML]{333333} \textbf{1.01}}} & \multicolumn{1}{l|}{\cellcolor[HTML]{FFFFFF}{\color[HTML]{333333} }}                       & \multicolumn{1}{l|}{\cellcolor[HTML]{FFFFFF}{\color[HTML]{333333} 1.23}}          & \multicolumn{1}{l|}{\cellcolor[HTML]{FFFFFF}{\color[HTML]{333333} 1.17}}          & {\color[HTML]{333333} 1.15}          \\ \cline{2-2} \cline{4-6} \cline{8-10} \cline{12-14} \cline{16-18} 
\multicolumn{1}{|c|}{\multirow{-2}{*}{\cellcolor[HTML]{FFFFFF}\cif}}   & \multicolumn{1}{l|}{\cellcolor[HTML]{FFFFFF}set}                                                       & \multicolumn{1}{l|}{\multirow{-2}{*}{\cellcolor[HTML]{FFFFFF}1.08}} & \multicolumn{1}{l|}{\cellcolor[HTML]{FFFFFF}{\color[HTML]{333333} \textbf{1.04}}} & \multicolumn{1}{l|}{\cellcolor[HTML]{FFFFFF}{\color[HTML]{333333} \textbf{1.03}}} & \multicolumn{1}{l|}{\cellcolor[HTML]{FFFFFF}{\color[HTML]{333333} \textbf{1.03}}} & \multicolumn{1}{l|}{\multirow{-2}{*}{\cellcolor[HTML]{FFFFFF}{\color[HTML]{333333} 1.5}}}  & \multicolumn{1}{l|}{\cellcolor[HTML]{FFFFFF}{\color[HTML]{333333} \textbf{1.82}}} & \multicolumn{1}{l|}{\cellcolor[HTML]{FFFFFF}{\color[HTML]{333333} 1.53}}          & \multicolumn{1}{l|}{\cellcolor[HTML]{FFFFFF}{\color[HTML]{333333} 1.52}}          & \multicolumn{1}{l|}{\multirow{-2}{*}{\cellcolor[HTML]{FFFFFF}{\color[HTML]{333333} 1.06}}} & \multicolumn{1}{l|}{\cellcolor[HTML]{FFFFFF}{\color[HTML]{333333} \textbf{1.02}}} & \multicolumn{1}{l|}{\cellcolor[HTML]{FFFFFF}{\color[HTML]{333333} 1.02}}          & \multicolumn{1}{l|}{\cellcolor[HTML]{FFFFFF}{\color[HTML]{333333} 1.02}}          & \multicolumn{1}{l|}{\multirow{-2}{*}{\cellcolor[HTML]{FFFFFF}{\color[HTML]{333333} 1.24}}} & \multicolumn{1}{l|}{\cellcolor[HTML]{FFFFFF}{\color[HTML]{333333} \textbf{1.17}}} & \multicolumn{1}{l|}{\cellcolor[HTML]{FFFFFF}{\color[HTML]{333333} \textbf{1.14}}} & {\color[HTML]{333333} \textbf{1.14}} \\ \hline
\multicolumn{1}{|c|}{\cellcolor[HTML]{FFFFFF}}                        & \multicolumn{1}{l|}{\cellcolor[HTML]{FFFFFF}way}                                                       & \multicolumn{1}{l|}{\cellcolor[HTML]{FFFFFF}}                       & \multicolumn{1}{l|}{\cellcolor[HTML]{FFFFFF}{\color[HTML]{333333} 1.12}}          & \multicolumn{1}{l|}{\cellcolor[HTML]{FFFFFF}{\color[HTML]{333333} \textbf{1.04}}} & \multicolumn{1}{l|}{\cellcolor[HTML]{FFFFFF}{\color[HTML]{333333} \textbf{1.03}}} & \multicolumn{1}{l|}{\cellcolor[HTML]{FFFFFF}{\color[HTML]{333333} }}                       & \multicolumn{1}{l|}{\cellcolor[HTML]{FFFFFF}{\color[HTML]{333333} \textbf{2.77}}} & \multicolumn{1}{l|}{\cellcolor[HTML]{FFFFFF}{\color[HTML]{333333} \textbf{1.63}}} & \multicolumn{1}{l|}{\cellcolor[HTML]{FFFFFF}{\color[HTML]{333333} \textbf{1.52}}} & \multicolumn{1}{l|}{\cellcolor[HTML]{FFFFFF}{\color[HTML]{333333} }}                       & \multicolumn{1}{l|}{\cellcolor[HTML]{FFFFFF}{\color[HTML]{333333} 1.06}}          & \multicolumn{1}{l|}{\cellcolor[HTML]{FFFFFF}{\color[HTML]{333333} 1.02}}          & \multicolumn{1}{l|}{\cellcolor[HTML]{FFFFFF}{\color[HTML]{333333} \textbf{1.01}}} & \multicolumn{1}{l|}{\cellcolor[HTML]{FFFFFF}{\color[HTML]{333333} }}                       & \multicolumn{1}{l|}{\cellcolor[HTML]{FFFFFF}{\color[HTML]{333333} 1.40}}          & \multicolumn{1}{l|}{\cellcolor[HTML]{FFFFFF}{\color[HTML]{333333} \textbf{1.16}}} & {\color[HTML]{333333} \textbf{1.15}} \\ \cline{2-2} \cline{4-6} \cline{8-10} \cline{12-14} \cline{16-18} 
\multicolumn{1}{|c|}{\multirow{-2}{*}{\cellcolor[HTML]{FFFFFF}\vga}}   & \multicolumn{1}{l|}{\cellcolor[HTML]{FFFFFF}set}                                                       & \multicolumn{1}{l|}{\multirow{-2}{*}{\cellcolor[HTML]{FFFFFF}1.21}} & \multicolumn{1}{l|}{\cellcolor[HTML]{FFFFFF}{\color[HTML]{333333} \textbf{1.10}}} & \multicolumn{1}{l|}{\cellcolor[HTML]{FFFFFF}{\color[HTML]{333333} 1.04}}          & \multicolumn{1}{l|}{\cellcolor[HTML]{FFFFFF}{\color[HTML]{333333} 1.09}}          & \multicolumn{1}{l|}{\multirow{-2}{*}{\cellcolor[HTML]{FFFFFF}{\color[HTML]{333333} 1.49}}} & \multicolumn{1}{l|}{\cellcolor[HTML]{FFFFFF}{\color[HTML]{333333} 2.8}}           & \multicolumn{1}{l|}{\cellcolor[HTML]{FFFFFF}{\color[HTML]{333333} 1.87}}          & \multicolumn{1}{l|}{\cellcolor[HTML]{FFFFFF}{\color[HTML]{333333} 2.23}}          & \multicolumn{1}{l|}{\multirow{-2}{*}{\cellcolor[HTML]{FFFFFF}{\color[HTML]{333333} 1.27}}} & \multicolumn{1}{l|}{\cellcolor[HTML]{FFFFFF}{\color[HTML]{333333} \textbf{1.05}}} & \multicolumn{1}{l|}{\cellcolor[HTML]{FFFFFF}{\color[HTML]{333333} \textbf{1.02}}} & \multicolumn{1}{l|}{\cellcolor[HTML]{FFFFFF}{\color[HTML]{333333} 1.07}}          & \multicolumn{1}{l|}{\multirow{-2}{*}{\cellcolor[HTML]{FFFFFF}{\color[HTML]{333333} 1.44}}} & \multicolumn{1}{l|}{\cellcolor[HTML]{FFFFFF}{\color[HTML]{333333} \textbf{1.33}}} & \multicolumn{1}{l|}{\cellcolor[HTML]{FFFFFF}{\color[HTML]{333333} 1.19}}          & {\color[HTML]{333333} 1.34}          \\ \hline
\end{tabular}%
}
\caption{Table of maximum execution time slowdown for \disparity, \mser, \tracking and \sift benchmarks on the RK3568 platform}
\label{table-rk3568}
\end{table*}
}

\newcommand{\tableRKeighteight}{\begin{table*}[h!]
\centering
\resizebox{\textwidth}{!}{%
\begin{tabular}{|
>{\columncolor[HTML]{FFFFFF}}l 
>{\columncolor[HTML]{FFFFFF}}l 
>{\columncolor[HTML]{FFFFFF}}l 
>{\columncolor[HTML]{FFFFFF}}l 
>{\columncolor[HTML]{FFFFFF}}l 
>{\columncolor[HTML]{FFFFFF}}l 
>{\columncolor[HTML]{FFFFFF}}l 
>{\columncolor[HTML]{FFFFFF}}l 
>{\columncolor[HTML]{FFFFFF}}l 
>{\columncolor[HTML]{FFFFFF}}l 
>{\columncolor[HTML]{FFFFFF}}l 
>{\columncolor[HTML]{FFFFFF}}l 
>{\columncolor[HTML]{FFFFFF}}l 
>{\columncolor[HTML]{FFFFFF}}l 
>{\columncolor[HTML]{FFFFFF}}l 
>{\columncolor[HTML]{FFFFFF}}l 
>{\columncolor[HTML]{FFFFFF}}l 
>{\columncolor[HTML]{FFFFFF}}l |}
\hline
\multicolumn{18}{|c|}{\cellcolor[HTML]{FFFFFF}{\color[HTML]{333333} RK3588}}                                                                                                                                                                                                                                                                                                                                                                                                                                                                                                                                                                                                                                                                                                                                                                                                                                                                                                                                                                                                                                                                                                                                                                                                                                                                                                                                                                                                                                                                                                                                     \\ \hline
\multicolumn{2}{|l|}{\cellcolor[HTML]{FFFFFF}{\color[HTML]{333333} }}                                                                                                                                                     & \multicolumn{16}{c|}{\cellcolor[HTML]{FFFFFF}{\color[HTML]{333333} \disparity}}                                                                                                                                                                                                                                                                                                                                                                                                                                                                                                                                                                                                                                                                                                                                                                                                                                                                                                                                                                                                                                                                                                                                                                                                                                                                                                       \\ \cline{3-18} 
\multicolumn{2}{|l|}{\multirow{-2}{*}{\cellcolor[HTML]{FFFFFF}{\color[HTML]{333333} }}}                                                                                                                                   & \multicolumn{4}{c|}{\cellcolor[HTML]{FFFFFF}{\color[HTML]{333333} \readInterf}}                                                                                                                                                                                                                                                                               & \multicolumn{4}{c|}{\cellcolor[HTML]{FFFFFF}{\color[HTML]{333333} \writeInterf}}                                                                                                                                                                                                                                                                              & \multicolumn{4}{c|}{\cellcolor[HTML]{FFFFFF}{\color[HTML]{333333} \modifyInterf}}                                                                                                                                                                                                                                                                             & \multicolumn{4}{c|}{\cellcolor[HTML]{FFFFFF}{\color[HTML]{333333} \prefetchInterf}}                                                                                                                                                                                                                              \\ \cline{2-18} 
\multicolumn{1}{|l|}{\cellcolor[HTML]{FFFFFF}{\color[HTML]{333333} }}                      & \multicolumn{1}{l|}{\cellcolor[HTML]{FFFFFF}{\color[HTML]{333333} \begin{tabular}[c]{@{}l@{}}Partition\\ Size\end{tabular}}} & \multicolumn{1}{l|}{\cellcolor[HTML]{FFFFFF}{\color[HTML]{333333} No}}                     & \multicolumn{1}{l|}{\cellcolor[HTML]{FFFFFF}{\color[HTML]{333333} 1/3}}           & \multicolumn{1}{l|}{\cellcolor[HTML]{FFFFFF}{\color[HTML]{333333} 2/2}}           & \multicolumn{1}{l|}{\cellcolor[HTML]{FFFFFF}{\color[HTML]{333333} 3/1}}           & \multicolumn{1}{c|}{\cellcolor[HTML]{FFFFFF}{\color[HTML]{333333} No}}                     & \multicolumn{1}{l|}{\cellcolor[HTML]{FFFFFF}{\color[HTML]{333333} 1/3}}           & \multicolumn{1}{l|}{\cellcolor[HTML]{FFFFFF}{\color[HTML]{333333} 2/2}}           & \multicolumn{1}{l|}{\cellcolor[HTML]{FFFFFF}{\color[HTML]{333333} 3/1}}           & \multicolumn{1}{c|}{\cellcolor[HTML]{FFFFFF}{\color[HTML]{333333} No}}                     & \multicolumn{1}{l|}{\cellcolor[HTML]{FFFFFF}{\color[HTML]{333333} 1/3}}           & \multicolumn{1}{l|}{\cellcolor[HTML]{FFFFFF}{\color[HTML]{333333} 2/2}}           & \multicolumn{1}{l|}{\cellcolor[HTML]{FFFFFF}{\color[HTML]{333333} 3/1}}           & \multicolumn{1}{c|}{\cellcolor[HTML]{FFFFFF}{\color[HTML]{333333} No}}                     & \multicolumn{1}{l|}{\cellcolor[HTML]{FFFFFF}{\color[HTML]{333333} 1/3}}           & \multicolumn{1}{l|}{\cellcolor[HTML]{FFFFFF}{\color[HTML]{333333} 2/2}}           & {\color[HTML]{333333} 3/1}           \\ \hline
\multicolumn{1}{|l|}{\cellcolor[HTML]{FFFFFF}{\color[HTML]{333333} }}                      & \multicolumn{1}{l|}{\cellcolor[HTML]{FFFFFF}{\color[HTML]{333333} way}}                                                      & \multicolumn{1}{l|}{\cellcolor[HTML]{FFFFFF}{\color[HTML]{333333} }}                       & \multicolumn{1}{l|}{\cellcolor[HTML]{FFFFFF}{\color[HTML]{333333} \textbf{1.02}}} & \multicolumn{1}{l|}{\cellcolor[HTML]{FFFFFF}{\color[HTML]{333333} \textbf{1.01}}} & \multicolumn{1}{l|}{\cellcolor[HTML]{FFFFFF}{\color[HTML]{333333} \textbf{1.01}}} & \multicolumn{1}{l|}{\cellcolor[HTML]{FFFFFF}{\color[HTML]{333333} }}                       & \multicolumn{1}{l|}{\cellcolor[HTML]{FFFFFF}{\color[HTML]{333333} \textbf{1.09}}} & \multicolumn{1}{l|}{\cellcolor[HTML]{FFFFFF}{\color[HTML]{333333} 1.10}}          & \multicolumn{1}{l|}{\cellcolor[HTML]{FFFFFF}{\color[HTML]{333333} \textbf{1.06}}} & \multicolumn{1}{l|}{\cellcolor[HTML]{FFFFFF}{\color[HTML]{333333} }}                       & \multicolumn{1}{l|}{\cellcolor[HTML]{FFFFFF}{\color[HTML]{333333} 1.02}}          & \multicolumn{1}{l|}{\cellcolor[HTML]{FFFFFF}{\color[HTML]{333333} \textbf{1.01}}} & \multicolumn{1}{l|}{\cellcolor[HTML]{FFFFFF}{\color[HTML]{333333} \textbf{1.01}}} & \multicolumn{1}{l|}{\cellcolor[HTML]{FFFFFF}{\color[HTML]{333333} }}                       & \multicolumn{1}{l|}{\cellcolor[HTML]{FFFFFF}{\color[HTML]{333333} \textbf{1.11}}} & \multicolumn{1}{l|}{\cellcolor[HTML]{FFFFFF}{\color[HTML]{333333} \textbf{1.11}}} & {\color[HTML]{333333} \textbf{1.07}} \\ \cline{2-2} \cline{4-6} \cline{8-10} \cline{12-14} \cline{16-18} 
\multicolumn{1}{|l|}{\multirow{-2}{*}{\cellcolor[HTML]{FFFFFF}{\color[HTML]{333333} \cif}}} & \multicolumn{1}{l|}{\cellcolor[HTML]{FFFFFF}{\color[HTML]{333333} set}}                                                      & \multicolumn{1}{l|}{\multirow{-2}{*}{\cellcolor[HTML]{FFFFFF}{\color[HTML]{333333} 1.15}}} & \multicolumn{1}{l|}{\cellcolor[HTML]{FFFFFF}{\color[HTML]{333333} 1.02}}          & \multicolumn{1}{l|}{\cellcolor[HTML]{FFFFFF}{\color[HTML]{333333} 1.01}}          & \multicolumn{1}{l|}{\cellcolor[HTML]{FFFFFF}{\color[HTML]{333333} 1.02}}          & \multicolumn{1}{l|}{\multirow{-2}{*}{\cellcolor[HTML]{FFFFFF}{\color[HTML]{333333} 1.02}}} & \multicolumn{1}{l|}{\cellcolor[HTML]{FFFFFF}{\color[HTML]{333333} 1.09}}          & \multicolumn{1}{l|}{\cellcolor[HTML]{FFFFFF}{\color[HTML]{333333} \textbf{1.09}}} & \multicolumn{1}{l|}{\cellcolor[HTML]{FFFFFF}{\color[HTML]{333333} 1.11}}          & \multicolumn{1}{l|}{\multirow{-2}{*}{\cellcolor[HTML]{FFFFFF}{\color[HTML]{333333} 1.09}}} & \multicolumn{1}{l|}{\cellcolor[HTML]{FFFFFF}{\color[HTML]{333333} \textbf{1.02}}} & \multicolumn{1}{l|}{\cellcolor[HTML]{FFFFFF}{\color[HTML]{333333} 1.02}}          & \multicolumn{1}{l|}{\cellcolor[HTML]{FFFFFF}{\color[HTML]{333333} 1.02}}          & \multicolumn{1}{l|}{\multirow{-2}{*}{\cellcolor[HTML]{FFFFFF}{\color[HTML]{333333} 1.48}}} & \multicolumn{1}{l|}{\cellcolor[HTML]{FFFFFF}{\color[HTML]{333333} 1.15}}          & \multicolumn{1}{l|}{\cellcolor[HTML]{FFFFFF}{\color[HTML]{333333} 1.12}}          & {\color[HTML]{333333} 1.13}          \\ \hline
\multicolumn{1}{|l|}{\cellcolor[HTML]{FFFFFF}{\color[HTML]{333333} }}                      & \multicolumn{1}{l|}{\cellcolor[HTML]{FFFFFF}{\color[HTML]{333333} way}}                                                      & \multicolumn{1}{l|}{\cellcolor[HTML]{FFFFFF}{\color[HTML]{333333} }}                       & \multicolumn{1}{l|}{\cellcolor[HTML]{FFFFFF}{\color[HTML]{333333} 1.03}}          & \multicolumn{1}{l|}{\cellcolor[HTML]{FFFFFF}{\color[HTML]{333333} \textbf{1.01}}} & \multicolumn{1}{l|}{\cellcolor[HTML]{FFFFFF}{\color[HTML]{333333} \textbf{1.01}}} & \multicolumn{1}{l|}{\cellcolor[HTML]{FFFFFF}{\color[HTML]{333333} }}                       & \multicolumn{1}{l|}{\cellcolor[HTML]{FFFFFF}{\color[HTML]{333333} 1.14}}          & \multicolumn{1}{l|}{\cellcolor[HTML]{FFFFFF}{\color[HTML]{333333} 1.08}}          & \multicolumn{1}{l|}{\cellcolor[HTML]{FFFFFF}{\color[HTML]{333333} \textbf{1.06}}} & \multicolumn{1}{l|}{\cellcolor[HTML]{FFFFFF}{\color[HTML]{333333} }}                       & \multicolumn{1}{l|}{\cellcolor[HTML]{FFFFFF}{\color[HTML]{333333} 1.05}}          & \multicolumn{1}{l|}{\cellcolor[HTML]{FFFFFF}{\color[HTML]{333333} \textbf{1.01}}} & \multicolumn{1}{l|}{\cellcolor[HTML]{FFFFFF}{\color[HTML]{333333} \textbf{1.01}}} & \multicolumn{1}{l|}{\cellcolor[HTML]{FFFFFF}{\color[HTML]{333333} }}                       & \multicolumn{1}{l|}{\cellcolor[HTML]{FFFFFF}{\color[HTML]{333333} \textbf{1.19}}} & \multicolumn{1}{l|}{\cellcolor[HTML]{FFFFFF}{\color[HTML]{333333} \textbf{1.10}}} & {\color[HTML]{333333} \textbf{1.07}} \\ \cline{2-2} \cline{4-6} \cline{8-10} \cline{12-14} \cline{16-18} 
\multicolumn{1}{|l|}{\multirow{-2}{*}{\cellcolor[HTML]{FFFFFF}{\color[HTML]{333333} \vga}}} & \multicolumn{1}{l|}{\cellcolor[HTML]{FFFFFF}{\color[HTML]{333333} set}}                                                      & \multicolumn{1}{l|}{\multirow{-2}{*}{\cellcolor[HTML]{FFFFFF}{\color[HTML]{333333} 1.17}}} & \multicolumn{1}{l|}{\cellcolor[HTML]{FFFFFF}{\color[HTML]{333333} \textbf{1.02}}} & \multicolumn{1}{l|}{\cellcolor[HTML]{FFFFFF}{\color[HTML]{333333} 1.01}}          & \multicolumn{1}{l|}{\cellcolor[HTML]{FFFFFF}{\color[HTML]{333333} 1.02}}          & \multicolumn{1}{l|}{\multirow{-2}{*}{\cellcolor[HTML]{FFFFFF}{\color[HTML]{333333} 1.06}}} & \multicolumn{1}{l|}{\cellcolor[HTML]{FFFFFF}{\color[HTML]{333333} \textbf{1.14}}} & \multicolumn{1}{l|}{\cellcolor[HTML]{FFFFFF}{\color[HTML]{333333} \textbf{1.07}}} & \multicolumn{1}{l|}{\cellcolor[HTML]{FFFFFF}{\color[HTML]{333333} 1.10}}          & \multicolumn{1}{l|}{\multirow{-2}{*}{\cellcolor[HTML]{FFFFFF}{\color[HTML]{333333} 1.15}}} & \multicolumn{1}{l|}{\cellcolor[HTML]{FFFFFF}{\color[HTML]{333333} \textbf{1.03}}} & \multicolumn{1}{l|}{\cellcolor[HTML]{FFFFFF}{\color[HTML]{333333} 1.02}}          & \multicolumn{1}{l|}{\cellcolor[HTML]{FFFFFF}{\color[HTML]{333333} 1.02}}          & \multicolumn{1}{l|}{\multirow{-2}{*}{\cellcolor[HTML]{FFFFFF}{\color[HTML]{333333} 1.44}}} & \multicolumn{1}{l|}{\cellcolor[HTML]{FFFFFF}{\color[HTML]{333333} 1.19}}          & \multicolumn{1}{l|}{\cellcolor[HTML]{FFFFFF}{\color[HTML]{333333} 1.12}}          & {\color[HTML]{333333} 1.13}          \\ \hline
\multicolumn{2}{|c|}{\cellcolor[HTML]{FFFFFF}{\color[HTML]{333333} }}                                                                                                                                                     & \multicolumn{16}{c|}{\cellcolor[HTML]{FFFFFF}{\color[HTML]{333333} \mser}}                                                                                                                                                                                                                                                                                                                                                                                                                                                                                                                                                                                                                                                                                                                                                                                                                                                                                                                                                                                                                                                                                                                                                                                                                                                                                                            \\ \cline{3-18} 
\multicolumn{2}{|c|}{\multirow{-2}{*}{\cellcolor[HTML]{FFFFFF}{\color[HTML]{333333} }}}                                                                                                                                   & \multicolumn{4}{c|}{\cellcolor[HTML]{FFFFFF}{\color[HTML]{333333} \readInterf}}                                                                                                                                                                                                                                                                               & \multicolumn{4}{c|}{\cellcolor[HTML]{FFFFFF}{\color[HTML]{333333} \writeInterf}}                                                                                                                                                                                                                                                                              & \multicolumn{4}{c|}{\cellcolor[HTML]{FFFFFF}{\color[HTML]{333333} \modifyInterf}}                                                                                                                                                                                                                                                                             & \multicolumn{4}{c|}{\cellcolor[HTML]{FFFFFF}{\color[HTML]{333333} \prefetchInterf}}                                                                                                                                                                                                                              \\ \cline{2-18} 
\multicolumn{1}{|l|}{\cellcolor[HTML]{FFFFFF}{\color[HTML]{333333} }}                      & \multicolumn{1}{l|}{\cellcolor[HTML]{FFFFFF}{\color[HTML]{333333} \begin{tabular}[c]{@{}l@{}}Partition\\ Size\end{tabular}}} & \multicolumn{1}{l|}{\cellcolor[HTML]{FFFFFF}{\color[HTML]{333333} No}}                     & \multicolumn{1}{l|}{\cellcolor[HTML]{FFFFFF}{\color[HTML]{333333} 1/3}}           & \multicolumn{1}{l|}{\cellcolor[HTML]{FFFFFF}{\color[HTML]{333333} 2/2}}           & \multicolumn{1}{l|}{\cellcolor[HTML]{FFFFFF}{\color[HTML]{333333} 3/1}}           & \multicolumn{1}{c|}{\cellcolor[HTML]{FFFFFF}{\color[HTML]{333333} No}}                     & \multicolumn{1}{l|}{\cellcolor[HTML]{FFFFFF}{\color[HTML]{333333} 1/3}}           & \multicolumn{1}{l|}{\cellcolor[HTML]{FFFFFF}{\color[HTML]{333333} 2/2}}           & \multicolumn{1}{l|}{\cellcolor[HTML]{FFFFFF}{\color[HTML]{333333} 3/1}}           & \multicolumn{1}{c|}{\cellcolor[HTML]{FFFFFF}{\color[HTML]{333333} No}}                     & \multicolumn{1}{l|}{\cellcolor[HTML]{FFFFFF}{\color[HTML]{333333} 1/3}}           & \multicolumn{1}{l|}{\cellcolor[HTML]{FFFFFF}{\color[HTML]{333333} 2/2}}           & \multicolumn{1}{l|}{\cellcolor[HTML]{FFFFFF}{\color[HTML]{333333} 3/1}}           & \multicolumn{1}{c|}{\cellcolor[HTML]{FFFFFF}{\color[HTML]{333333} No}}                     & \multicolumn{1}{l|}{\cellcolor[HTML]{FFFFFF}{\color[HTML]{333333} 1/3}}           & \multicolumn{1}{l|}{\cellcolor[HTML]{FFFFFF}{\color[HTML]{333333} 2/2}}           & {\color[HTML]{333333} 3/1}           \\ \hline
\multicolumn{1}{|l|}{\cellcolor[HTML]{FFFFFF}{\color[HTML]{333333} }}                      & \multicolumn{1}{l|}{\cellcolor[HTML]{FFFFFF}{\color[HTML]{333333} way}}                                                      & \multicolumn{1}{l|}{\cellcolor[HTML]{FFFFFF}{\color[HTML]{333333} }}                       & \multicolumn{1}{l|}{\cellcolor[HTML]{FFFFFF}{\color[HTML]{333333} \textbf{1.03}}} & \multicolumn{1}{l|}{\cellcolor[HTML]{FFFFFF}{\color[HTML]{333333} \textbf{1.03}}} & \multicolumn{1}{l|}{\cellcolor[HTML]{FFFFFF}{\color[HTML]{333333} \textbf{1.02}}} & \multicolumn{1}{l|}{\cellcolor[HTML]{FFFFFF}{\color[HTML]{333333} }}                       & \multicolumn{1}{l|}{\cellcolor[HTML]{FFFFFF}{\color[HTML]{333333} \textbf{1.38}}} & \multicolumn{1}{l|}{\cellcolor[HTML]{FFFFFF}{\color[HTML]{333333} 1.34}}          & \multicolumn{1}{l|}{\cellcolor[HTML]{FFFFFF}{\color[HTML]{333333} \textbf{1.25}}} & \multicolumn{1}{l|}{\cellcolor[HTML]{FFFFFF}{\color[HTML]{333333} }}                       & \multicolumn{1}{l|}{\cellcolor[HTML]{FFFFFF}{\color[HTML]{333333} \textbf{1.07}}} & \multicolumn{1}{l|}{\cellcolor[HTML]{FFFFFF}{\color[HTML]{333333} \textbf{1.06}}} & \multicolumn{1}{l|}{\cellcolor[HTML]{FFFFFF}{\color[HTML]{333333} \textbf{1.04}}} & \multicolumn{1}{l|}{\cellcolor[HTML]{FFFFFF}{\color[HTML]{333333} }}                       & \multicolumn{1}{l|}{\cellcolor[HTML]{FFFFFF}{\color[HTML]{333333} \textbf{1.60}}} & \multicolumn{1}{l|}{\cellcolor[HTML]{FFFFFF}{\color[HTML]{333333} \textbf{1.47}}} & {\color[HTML]{333333} \textbf{1.34}} \\ \cline{2-2} \cline{4-6} \cline{8-10} \cline{12-14} \cline{16-18} 
\multicolumn{1}{|l|}{\multirow{-2}{*}{\cellcolor[HTML]{FFFFFF}{\color[HTML]{333333} \cif}}} & \multicolumn{1}{l|}{\cellcolor[HTML]{FFFFFF}{\color[HTML]{333333} set}}                                                      & \multicolumn{1}{l|}{\multirow{-2}{*}{\cellcolor[HTML]{FFFFFF}{\color[HTML]{333333} 1.78}}} & \multicolumn{1}{l|}{\cellcolor[HTML]{FFFFFF}{\color[HTML]{333333} 1.08}}          & \multicolumn{1}{l|}{\cellcolor[HTML]{FFFFFF}{\color[HTML]{333333} 1.04}}          & \multicolumn{1}{l|}{\cellcolor[HTML]{FFFFFF}{\color[HTML]{333333} 1.06}}          & \multicolumn{1}{l|}{\multirow{-2}{*}{\cellcolor[HTML]{FFFFFF}{\color[HTML]{333333} 1.18}}} & \multicolumn{1}{l|}{\cellcolor[HTML]{FFFFFF}{\color[HTML]{333333} 1.46}}          & \multicolumn{1}{l|}{\cellcolor[HTML]{FFFFFF}{\color[HTML]{333333} \textbf{1.30}}} & \multicolumn{1}{l|}{\cellcolor[HTML]{FFFFFF}{\color[HTML]{333333} 1.33}}          & \multicolumn{1}{l|}{\multirow{-2}{*}{\cellcolor[HTML]{FFFFFF}{\color[HTML]{333333} 1.71}}} & \multicolumn{1}{l|}{\cellcolor[HTML]{FFFFFF}{\color[HTML]{333333} 1.15}}          & \multicolumn{1}{l|}{\cellcolor[HTML]{FFFFFF}{\color[HTML]{333333} 1.08}}          & \multicolumn{1}{l|}{\cellcolor[HTML]{FFFFFF}{\color[HTML]{333333} 1.07}}          & \multicolumn{1}{l|}{\multirow{-2}{*}{\cellcolor[HTML]{FFFFFF}{\color[HTML]{333333} 3.26}}} & \multicolumn{1}{l|}{\cellcolor[HTML]{FFFFFF}{\color[HTML]{333333} 1.68}}          & \multicolumn{1}{l|}{\cellcolor[HTML]{FFFFFF}{\color[HTML]{333333} 1.49}}          & {\color[HTML]{333333} 1.47}          \\ \hline
\multicolumn{1}{|l|}{\cellcolor[HTML]{FFFFFF}{\color[HTML]{333333} }}                      & \multicolumn{1}{l|}{\cellcolor[HTML]{FFFFFF}{\color[HTML]{333333} way}}                                                      & \multicolumn{1}{l|}{\cellcolor[HTML]{FFFFFF}{\color[HTML]{333333} }}                       & \multicolumn{1}{l|}{\cellcolor[HTML]{FFFFFF}{\color[HTML]{333333} \textbf{1.03}}} & \multicolumn{1}{l|}{\cellcolor[HTML]{FFFFFF}{\color[HTML]{333333} \textbf{1.03}}} & \multicolumn{1}{l|}{\cellcolor[HTML]{FFFFFF}{\color[HTML]{333333} \textbf{1.03}}} & \multicolumn{1}{l|}{\cellcolor[HTML]{FFFFFF}{\color[HTML]{333333} }}                       & \multicolumn{1}{l|}{\cellcolor[HTML]{FFFFFF}{\color[HTML]{333333} 1.39}}          & \multicolumn{1}{l|}{\cellcolor[HTML]{FFFFFF}{\color[HTML]{333333} 1.40}}          & \multicolumn{1}{l|}{\cellcolor[HTML]{FFFFFF}{\color[HTML]{333333} 1.44}}          & \multicolumn{1}{l|}{\cellcolor[HTML]{FFFFFF}{\color[HTML]{333333} }}                       & \multicolumn{1}{l|}{\cellcolor[HTML]{FFFFFF}{\color[HTML]{333333} 1.13}}          & \multicolumn{1}{l|}{\cellcolor[HTML]{FFFFFF}{\color[HTML]{333333} \textbf{1.07}}} & \multicolumn{1}{l|}{\cellcolor[HTML]{FFFFFF}{\color[HTML]{333333} \textbf{1.06}}} & \multicolumn{1}{l|}{\cellcolor[HTML]{FFFFFF}{\color[HTML]{333333} }}                       & \multicolumn{1}{l|}{\cellcolor[HTML]{FFFFFF}{\color[HTML]{333333} 1.63}}          & \multicolumn{1}{l|}{\cellcolor[HTML]{FFFFFF}{\color[HTML]{333333} \textbf{1.54}}} & {\color[HTML]{333333} \textbf{1.50}} \\ \cline{2-2} \cline{4-6} \cline{8-10} \cline{12-14} \cline{16-18} 
\multicolumn{1}{|l|}{\multirow{-2}{*}{\cellcolor[HTML]{FFFFFF}{\color[HTML]{333333} \vga}}} & \multicolumn{1}{l|}{\cellcolor[HTML]{FFFFFF}{\color[HTML]{333333} set}}                                                      & \multicolumn{1}{l|}{\multirow{-2}{*}{\cellcolor[HTML]{FFFFFF}{\color[HTML]{333333} 1.39}}} & \multicolumn{1}{l|}{\cellcolor[HTML]{FFFFFF}{\color[HTML]{333333} 1.09}}          & \multicolumn{1}{l|}{\cellcolor[HTML]{FFFFFF}{\color[HTML]{333333} 1.04}}          & \multicolumn{1}{l|}{\cellcolor[HTML]{FFFFFF}{\color[HTML]{333333} 1.04}}          & \multicolumn{1}{l|}{\multirow{-2}{*}{\cellcolor[HTML]{FFFFFF}{\color[HTML]{333333} 1.51}}} & \multicolumn{1}{l|}{\cellcolor[HTML]{FFFFFF}{\color[HTML]{333333} \textbf{1.38}}} & \multicolumn{1}{l|}{\cellcolor[HTML]{FFFFFF}{\color[HTML]{333333} \textbf{1.32}}} & \multicolumn{1}{l|}{\cellcolor[HTML]{FFFFFF}{\color[HTML]{333333} \textbf{1.41}}} & \multicolumn{1}{l|}{\multirow{-2}{*}{\cellcolor[HTML]{FFFFFF}{\color[HTML]{333333} 1.45}}} & \multicolumn{1}{l|}{\cellcolor[HTML]{FFFFFF}{\color[HTML]{333333} \textbf{1.09}}} & \multicolumn{1}{l|}{\cellcolor[HTML]{FFFFFF}{\color[HTML]{333333} 1.07}}          & \multicolumn{1}{l|}{\cellcolor[HTML]{FFFFFF}{\color[HTML]{333333} 1.07}}          & \multicolumn{1}{l|}{\multirow{-2}{*}{\cellcolor[HTML]{FFFFFF}{\color[HTML]{333333} 2.71}}} & \multicolumn{1}{l|}{\cellcolor[HTML]{FFFFFF}{\color[HTML]{333333} \textbf{1.62}}} & \multicolumn{1}{l|}{\cellcolor[HTML]{FFFFFF}{\color[HTML]{333333} 1.63}}          & {\color[HTML]{333333} 1.75}          \\ \hline
\multicolumn{2}{|l|}{\cellcolor[HTML]{FFFFFF}{\color[HTML]{333333} }}                                                                                                                                                     & \multicolumn{16}{c|}{\cellcolor[HTML]{FFFFFF}{\color[HTML]{333333} \tracking}}                                                                                                                                                                                                                                                                                                                                                                                                                                                                                                                                                                                                                                                                                                                                                                                                                                                                                                                                                                                                                                                                                                                                                                                                                                                                                                        \\ \cline{3-18} 
\multicolumn{2}{|l|}{\multirow{-2}{*}{\cellcolor[HTML]{FFFFFF}{\color[HTML]{333333} }}}                                                                                                                                   & \multicolumn{4}{c|}{\cellcolor[HTML]{FFFFFF}{\color[HTML]{333333} \readInterf}}                                                                                                                                                                                                                                                                               & \multicolumn{4}{c|}{\cellcolor[HTML]{FFFFFF}{\color[HTML]{333333} \writeInterf}}                                                                                                                                                                                                                                                                              & \multicolumn{4}{c|}{\cellcolor[HTML]{FFFFFF}{\color[HTML]{333333} \modifyInterf}}                                                                                                                                                                                                                                                                             & \multicolumn{4}{c|}{\cellcolor[HTML]{FFFFFF}{\color[HTML]{333333} \prefetchInterf}}                                                                                                                                                                                                                              \\ \cline{2-18} 
\multicolumn{1}{|l|}{\cellcolor[HTML]{FFFFFF}{\color[HTML]{333333} }}                      & \multicolumn{1}{l|}{\cellcolor[HTML]{FFFFFF}{\color[HTML]{333333} \begin{tabular}[c]{@{}l@{}}Partition\\ Size\end{tabular}}} & \multicolumn{1}{l|}{\cellcolor[HTML]{FFFFFF}{\color[HTML]{333333} No}}                     & \multicolumn{1}{l|}{\cellcolor[HTML]{FFFFFF}{\color[HTML]{333333} 1/3}}           & \multicolumn{1}{l|}{\cellcolor[HTML]{FFFFFF}{\color[HTML]{333333} 2/2}}           & \multicolumn{1}{l|}{\cellcolor[HTML]{FFFFFF}{\color[HTML]{333333} 3/1}}           & \multicolumn{1}{c|}{\cellcolor[HTML]{FFFFFF}{\color[HTML]{333333} No}}                     & \multicolumn{1}{l|}{\cellcolor[HTML]{FFFFFF}{\color[HTML]{333333} 1/3}}           & \multicolumn{1}{l|}{\cellcolor[HTML]{FFFFFF}{\color[HTML]{333333} 2/2}}           & \multicolumn{1}{l|}{\cellcolor[HTML]{FFFFFF}{\color[HTML]{333333} 3/1}}           & \multicolumn{1}{c|}{\cellcolor[HTML]{FFFFFF}{\color[HTML]{333333} No}}                     & \multicolumn{1}{l|}{\cellcolor[HTML]{FFFFFF}{\color[HTML]{333333} 1/3}}           & \multicolumn{1}{l|}{\cellcolor[HTML]{FFFFFF}{\color[HTML]{333333} 2/2}}           & \multicolumn{1}{l|}{\cellcolor[HTML]{FFFFFF}{\color[HTML]{333333} 3/1}}           & \multicolumn{1}{c|}{\cellcolor[HTML]{FFFFFF}{\color[HTML]{333333} No}}                     & \multicolumn{1}{l|}{\cellcolor[HTML]{FFFFFF}{\color[HTML]{333333} 1/3}}           & \multicolumn{1}{l|}{\cellcolor[HTML]{FFFFFF}{\color[HTML]{333333} 2/2}}           & {\color[HTML]{333333} 3/1}           \\ \hline
\multicolumn{1}{|l|}{\cellcolor[HTML]{FFFFFF}{\color[HTML]{333333} }}                      & \multicolumn{1}{l|}{\cellcolor[HTML]{FFFFFF}{\color[HTML]{333333} way}}                                                      & \multicolumn{1}{l|}{\cellcolor[HTML]{FFFFFF}{\color[HTML]{333333} }}                       & \multicolumn{1}{l|}{\cellcolor[HTML]{FFFFFF}{\color[HTML]{333333} \textbf{1.00}}} & \multicolumn{1}{l|}{\cellcolor[HTML]{FFFFFF}{\color[HTML]{333333} 1.01}}          & \multicolumn{1}{l|}{\cellcolor[HTML]{FFFFFF}{\color[HTML]{333333} \textbf{1.01}}} & \multicolumn{1}{l|}{\cellcolor[HTML]{FFFFFF}{\color[HTML]{333333} }}                       & \multicolumn{1}{l|}{\cellcolor[HTML]{FFFFFF}{\color[HTML]{333333} \textbf{1.03}}} & \multicolumn{1}{l|}{\cellcolor[HTML]{FFFFFF}{\color[HTML]{333333} \textbf{1.02}}} & \multicolumn{1}{l|}{\cellcolor[HTML]{FFFFFF}{\color[HTML]{333333} \textbf{1.02}}} & \multicolumn{1}{l|}{\cellcolor[HTML]{FFFFFF}{\color[HTML]{333333} }}                       & \multicolumn{1}{l|}{\cellcolor[HTML]{FFFFFF}{\color[HTML]{333333} \textbf{1.01}}} & \multicolumn{1}{l|}{\cellcolor[HTML]{FFFFFF}{\color[HTML]{333333} 1.01}}          & \multicolumn{1}{l|}{\cellcolor[HTML]{FFFFFF}{\color[HTML]{333333} \textbf{1.01}}} & \multicolumn{1}{l|}{\cellcolor[HTML]{FFFFFF}{\color[HTML]{333333} }}                       & \multicolumn{1}{l|}{\cellcolor[HTML]{FFFFFF}{\color[HTML]{333333} \textbf{1.04}}} & \multicolumn{1}{l|}{\cellcolor[HTML]{FFFFFF}{\color[HTML]{333333} \textbf{1.03}}} & {\color[HTML]{333333} \textbf{1.03}} \\ \cline{2-2} \cline{4-6} \cline{8-10} \cline{12-14} \cline{16-18} 
\multicolumn{1}{|l|}{\multirow{-2}{*}{\cellcolor[HTML]{FFFFFF}{\color[HTML]{333333} \cif}}} & \multicolumn{1}{l|}{\cellcolor[HTML]{FFFFFF}{\color[HTML]{333333} set}}                                                      & \multicolumn{1}{l|}{\multirow{-2}{*}{\cellcolor[HTML]{FFFFFF}{\color[HTML]{333333} 1.04}}} & \multicolumn{1}{l|}{\cellcolor[HTML]{FFFFFF}{\color[HTML]{333333} 1.01}}          & \multicolumn{1}{l|}{\cellcolor[HTML]{FFFFFF}{\color[HTML]{333333} \textbf{1.01}}} & \multicolumn{1}{l|}{\cellcolor[HTML]{FFFFFF}{\color[HTML]{333333} 1.01}}          & \multicolumn{1}{l|}{\multirow{-2}{*}{\cellcolor[HTML]{FFFFFF}{\color[HTML]{333333} 1.03}}} & \multicolumn{1}{l|}{\cellcolor[HTML]{FFFFFF}{\color[HTML]{333333} 1.04}}          & \multicolumn{1}{l|}{\cellcolor[HTML]{FFFFFF}{\color[HTML]{333333} 1.02}}          & \multicolumn{1}{l|}{\cellcolor[HTML]{FFFFFF}{\color[HTML]{333333} 1.03}}          & \multicolumn{1}{l|}{\multirow{-2}{*}{\cellcolor[HTML]{FFFFFF}{\color[HTML]{333333} 1.05}}} & \multicolumn{1}{l|}{\cellcolor[HTML]{FFFFFF}{\color[HTML]{333333} 1.01}}          & \multicolumn{1}{l|}{\cellcolor[HTML]{FFFFFF}{\color[HTML]{333333} \textbf{1.01}}} & \multicolumn{1}{l|}{\cellcolor[HTML]{FFFFFF}{\color[HTML]{333333} 1.01}}          & \multicolumn{1}{l|}{\multirow{-2}{*}{\cellcolor[HTML]{FFFFFF}{\color[HTML]{333333} 1.10}}} & \multicolumn{1}{l|}{\cellcolor[HTML]{FFFFFF}{\color[HTML]{333333} 1.06}}          & \multicolumn{1}{l|}{\cellcolor[HTML]{FFFFFF}{\color[HTML]{333333} 1.04}}          & {\color[HTML]{333333} 1.04}          \\ \hline
\multicolumn{1}{|l|}{\cellcolor[HTML]{FFFFFF}{\color[HTML]{333333} }}                      & \multicolumn{1}{l|}{\cellcolor[HTML]{FFFFFF}{\color[HTML]{333333} way}}                                                      & \multicolumn{1}{l|}{\cellcolor[HTML]{FFFFFF}{\color[HTML]{333333} }}                       & \multicolumn{1}{l|}{\cellcolor[HTML]{FFFFFF}{\color[HTML]{333333} \textbf{1.01}}} & \multicolumn{1}{l|}{\cellcolor[HTML]{FFFFFF}{\color[HTML]{333333} 1.01}}          & \multicolumn{1}{l|}{\cellcolor[HTML]{FFFFFF}{\color[HTML]{333333} \textbf{1.01}}} & \multicolumn{1}{l|}{\cellcolor[HTML]{FFFFFF}{\color[HTML]{333333} }}                       & \multicolumn{1}{l|}{\cellcolor[HTML]{FFFFFF}{\color[HTML]{333333} \textbf{1.03}}} & \multicolumn{1}{l|}{\cellcolor[HTML]{FFFFFF}{\color[HTML]{333333} 1.04}}          & \multicolumn{1}{l|}{\cellcolor[HTML]{FFFFFF}{\color[HTML]{333333} 1.05}}          & \multicolumn{1}{l|}{\cellcolor[HTML]{FFFFFF}{\color[HTML]{333333} }}                       & \multicolumn{1}{l|}{\cellcolor[HTML]{FFFFFF}{\color[HTML]{333333} \textbf{1.01}}} & \multicolumn{1}{l|}{\cellcolor[HTML]{FFFFFF}{\color[HTML]{333333} 1.01}}          & \multicolumn{1}{l|}{\cellcolor[HTML]{FFFFFF}{\color[HTML]{333333} \textbf{1.01}}} & \multicolumn{1}{l|}{\cellcolor[HTML]{FFFFFF}{\color[HTML]{333333} }}                       & \multicolumn{1}{l|}{\cellcolor[HTML]{FFFFFF}{\color[HTML]{333333} \textbf{1.08}}} & \multicolumn{1}{l|}{\cellcolor[HTML]{FFFFFF}{\color[HTML]{333333} \textbf{1.08}}} & {\color[HTML]{333333} \textbf{1.07}} \\ \cline{2-2} \cline{4-6} \cline{8-10} \cline{12-14} \cline{16-18} 
\multicolumn{1}{|l|}{\multirow{-2}{*}{\cellcolor[HTML]{FFFFFF}{\color[HTML]{333333} \vga}}} & \multicolumn{1}{l|}{\cellcolor[HTML]{FFFFFF}{\color[HTML]{333333} set}}                                                      & \multicolumn{1}{l|}{\multirow{-2}{*}{\cellcolor[HTML]{FFFFFF}{\color[HTML]{333333} 1.03}}} & \multicolumn{1}{l|}{\cellcolor[HTML]{FFFFFF}{\color[HTML]{333333} 1.01}}          & \multicolumn{1}{l|}{\cellcolor[HTML]{FFFFFF}{\color[HTML]{333333} \textbf{1.01}}} & \multicolumn{1}{l|}{\cellcolor[HTML]{FFFFFF}{\color[HTML]{333333} 1.01}}          & \multicolumn{1}{l|}{\multirow{-2}{*}{\cellcolor[HTML]{FFFFFF}{\color[HTML]{333333} 1.04}}} & \multicolumn{1}{l|}{\cellcolor[HTML]{FFFFFF}{\color[HTML]{333333} 1.04}}          & \multicolumn{1}{l|}{\cellcolor[HTML]{FFFFFF}{\color[HTML]{333333} \textbf{1.04}}} & \multicolumn{1}{l|}{\cellcolor[HTML]{FFFFFF}{\color[HTML]{333333} \textbf{1.05}}} & \multicolumn{1}{l|}{\multirow{-2}{*}{\cellcolor[HTML]{FFFFFF}{\color[HTML]{333333} 1.03}}} & \multicolumn{1}{l|}{\cellcolor[HTML]{FFFFFF}{\color[HTML]{333333} 1.01}}          & \multicolumn{1}{l|}{\cellcolor[HTML]{FFFFFF}{\color[HTML]{333333} \textbf{1.01}}} & \multicolumn{1}{l|}{\cellcolor[HTML]{FFFFFF}{\color[HTML]{333333} 1.02}}          & \multicolumn{1}{l|}{\multirow{-2}{*}{\cellcolor[HTML]{FFFFFF}{\color[HTML]{333333} 1.10}}} & \multicolumn{1}{l|}{\cellcolor[HTML]{FFFFFF}{\color[HTML]{333333} 1.09}}          & \multicolumn{1}{l|}{\cellcolor[HTML]{FFFFFF}{\color[HTML]{333333} 1.09}}          & {\color[HTML]{333333} 1.09}          \\ \hline
\multicolumn{2}{|l|}{\cellcolor[HTML]{FFFFFF}{\color[HTML]{333333} }}                                                                                                                                                     & \multicolumn{16}{c|}{\cellcolor[HTML]{FFFFFF}{\color[HTML]{333333} \sift}}                                                                                                                                                                                                                                                                                                                                                                                                                                                                                                                                                                                                                                                                                                                                                                                                                                                                                                                                                                                                                                                                                                                                                                                                                                                                                                            \\ \cline{3-18} 
\multicolumn{2}{|l|}{\multirow{-2}{*}{\cellcolor[HTML]{FFFFFF}{\color[HTML]{333333} }}}                                                                                                                                   & \multicolumn{4}{c|}{\cellcolor[HTML]{FFFFFF}{\color[HTML]{333333} \readInterf}}                                                                                                                                                                                                                                                                               & \multicolumn{4}{c|}{\cellcolor[HTML]{FFFFFF}{\color[HTML]{333333} \writeInterf}}                                                                                                                                                                                                                                                                              & \multicolumn{4}{c|}{\cellcolor[HTML]{FFFFFF}{\color[HTML]{333333} \modifyInterf}}                                                                                                                                                                                                                                                                             & \multicolumn{4}{c|}{\cellcolor[HTML]{FFFFFF}{\color[HTML]{333333} \prefetchInterf}}                                                                                                                                                                                                                              \\ \cline{2-18} 
\multicolumn{1}{|l|}{\cellcolor[HTML]{FFFFFF}{\color[HTML]{333333} }}                      & \multicolumn{1}{l|}{\cellcolor[HTML]{FFFFFF}{\color[HTML]{333333} \begin{tabular}[c]{@{}l@{}}Partition\\ Size\end{tabular}}} & \multicolumn{1}{l|}{\cellcolor[HTML]{FFFFFF}{\color[HTML]{333333} No}}                     & \multicolumn{1}{l|}{\cellcolor[HTML]{FFFFFF}{\color[HTML]{333333} 1/3}}           & \multicolumn{1}{l|}{\cellcolor[HTML]{FFFFFF}{\color[HTML]{333333} 2/2}}           & \multicolumn{1}{l|}{\cellcolor[HTML]{FFFFFF}{\color[HTML]{333333} 3/1}}           & \multicolumn{1}{c|}{\cellcolor[HTML]{FFFFFF}{\color[HTML]{333333} No}}                     & \multicolumn{1}{l|}{\cellcolor[HTML]{FFFFFF}{\color[HTML]{333333} 1/3}}           & \multicolumn{1}{l|}{\cellcolor[HTML]{FFFFFF}{\color[HTML]{333333} 2/2}}           & \multicolumn{1}{l|}{\cellcolor[HTML]{FFFFFF}{\color[HTML]{333333} 3/1}}           & \multicolumn{1}{c|}{\cellcolor[HTML]{FFFFFF}{\color[HTML]{333333} No}}                     & \multicolumn{1}{l|}{\cellcolor[HTML]{FFFFFF}{\color[HTML]{333333} 1/3}}           & \multicolumn{1}{l|}{\cellcolor[HTML]{FFFFFF}{\color[HTML]{333333} 2/2}}           & \multicolumn{1}{l|}{\cellcolor[HTML]{FFFFFF}{\color[HTML]{333333} 3/1}}           & \multicolumn{1}{c|}{\cellcolor[HTML]{FFFFFF}{\color[HTML]{333333} No}}                     & \multicolumn{1}{l|}{\cellcolor[HTML]{FFFFFF}{\color[HTML]{333333} 1/3}}           & \multicolumn{1}{l|}{\cellcolor[HTML]{FFFFFF}{\color[HTML]{333333} 2/2}}           & {\color[HTML]{333333} 3/1}           \\ \hline
\multicolumn{1}{|l|}{\cellcolor[HTML]{FFFFFF}{\color[HTML]{333333} }}                      & \multicolumn{1}{l|}{\cellcolor[HTML]{FFFFFF}{\color[HTML]{333333} way}}                                                      & \multicolumn{1}{l|}{\cellcolor[HTML]{FFFFFF}{\color[HTML]{333333} }}                       & \multicolumn{1}{l|}{\cellcolor[HTML]{FFFFFF}{\color[HTML]{333333} \textbf{1.01}}} & \multicolumn{1}{l|}{\cellcolor[HTML]{FFFFFF}{\color[HTML]{333333} 1.01}}          & \multicolumn{1}{l|}{\cellcolor[HTML]{FFFFFF}{\color[HTML]{333333} 1.01}}          & \multicolumn{1}{l|}{\cellcolor[HTML]{FFFFFF}{\color[HTML]{333333} }}                       & \multicolumn{1}{l|}{\cellcolor[HTML]{FFFFFF}{\color[HTML]{333333} \textbf{1.05}}} & \multicolumn{1}{l|}{\cellcolor[HTML]{FFFFFF}{\color[HTML]{333333} 1.05}}          & \multicolumn{1}{l|}{\cellcolor[HTML]{FFFFFF}{\color[HTML]{333333} \textbf{1.05}}} & \multicolumn{1}{l|}{\cellcolor[HTML]{FFFFFF}{\color[HTML]{333333} }}                       & \multicolumn{1}{l|}{\cellcolor[HTML]{FFFFFF}{\color[HTML]{333333} \textbf{1.01}}} & \multicolumn{1}{l|}{\cellcolor[HTML]{FFFFFF}{\color[HTML]{333333} \textbf{1.01}}} & \multicolumn{1}{l|}{\cellcolor[HTML]{FFFFFF}{\color[HTML]{333333} \textbf{1.01}}} & \multicolumn{1}{l|}{\cellcolor[HTML]{FFFFFF}{\color[HTML]{333333} }}                       & \multicolumn{1}{l|}{\cellcolor[HTML]{FFFFFF}{\color[HTML]{333333} \textbf{1.07}}} & \multicolumn{1}{l|}{\cellcolor[HTML]{FFFFFF}{\color[HTML]{333333} \textbf{1.07}}} & {\color[HTML]{333333} \textbf{1.07}} \\ \cline{2-2} \cline{4-6} \cline{8-10} \cline{12-14} \cline{16-18} 
\multicolumn{1}{|l|}{\multirow{-2}{*}{\cellcolor[HTML]{FFFFFF}{\color[HTML]{333333} \cif}}} & \multicolumn{1}{l|}{\cellcolor[HTML]{FFFFFF}{\color[HTML]{333333} set}}                                                      & \multicolumn{1}{l|}{\multirow{-2}{*}{\cellcolor[HTML]{FFFFFF}{\color[HTML]{333333} 1.04}}} & \multicolumn{1}{l|}{\cellcolor[HTML]{FFFFFF}{\color[HTML]{333333} 1.01}}          & \multicolumn{1}{l|}{\cellcolor[HTML]{FFFFFF}{\color[HTML]{333333} \textbf{1.01}}} & \multicolumn{1}{l|}{\cellcolor[HTML]{FFFFFF}{\color[HTML]{333333} \textbf{1.01}}} & \multicolumn{1}{l|}{\multirow{-2}{*}{\cellcolor[HTML]{FFFFFF}{\color[HTML]{333333} 1.06}}} & \multicolumn{1}{l|}{\cellcolor[HTML]{FFFFFF}{\color[HTML]{333333} 1.05}}          & \multicolumn{1}{l|}{\cellcolor[HTML]{FFFFFF}{\color[HTML]{333333} \textbf{1.05}}} & \multicolumn{1}{l|}{\cellcolor[HTML]{FFFFFF}{\color[HTML]{333333} 1.06}}          & \multicolumn{1}{l|}{\multirow{-2}{*}{\cellcolor[HTML]{FFFFFF}{\color[HTML]{333333} 1.04}}} & \multicolumn{1}{l|}{\cellcolor[HTML]{FFFFFF}{\color[HTML]{333333} 1.01}}          & \multicolumn{1}{l|}{\cellcolor[HTML]{FFFFFF}{\color[HTML]{333333} 1.02}}          & \multicolumn{1}{l|}{\cellcolor[HTML]{FFFFFF}{\color[HTML]{333333} 1.02}}          & \multicolumn{1}{l|}{\multirow{-2}{*}{\cellcolor[HTML]{FFFFFF}{\color[HTML]{333333} 1.12}}} & \multicolumn{1}{l|}{\cellcolor[HTML]{FFFFFF}{\color[HTML]{333333} 1.09}}          & \multicolumn{1}{l|}{\cellcolor[HTML]{FFFFFF}{\color[HTML]{333333} 1.08}}          & {\color[HTML]{333333} 1.08}          \\ \hline
\multicolumn{1}{|l|}{\cellcolor[HTML]{FFFFFF}{\color[HTML]{333333} }}                      & \multicolumn{1}{l|}{\cellcolor[HTML]{FFFFFF}{\color[HTML]{333333} way}}                                                      & \multicolumn{1}{l|}{\cellcolor[HTML]{FFFFFF}{\color[HTML]{333333} }}                       & \multicolumn{1}{l|}{\cellcolor[HTML]{FFFFFF}{\color[HTML]{333333} \textbf{1.02}}} & \multicolumn{1}{l|}{\cellcolor[HTML]{FFFFFF}{\color[HTML]{333333} \textbf{1.01}}} & \multicolumn{1}{l|}{\cellcolor[HTML]{FFFFFF}{\color[HTML]{333333} 1.02}}          & \multicolumn{1}{l|}{\cellcolor[HTML]{FFFFFF}{\color[HTML]{333333} }}                       & \multicolumn{1}{l|}{\cellcolor[HTML]{FFFFFF}{\color[HTML]{333333} \textbf{1.05}}} & \multicolumn{1}{l|}{\cellcolor[HTML]{FFFFFF}{\color[HTML]{333333} 1.06}}          & \multicolumn{1}{l|}{\cellcolor[HTML]{FFFFFF}{\color[HTML]{333333} \textbf{1.07}}} & \multicolumn{1}{l|}{\cellcolor[HTML]{FFFFFF}{\color[HTML]{333333} }}                       & \multicolumn{1}{l|}{\cellcolor[HTML]{FFFFFF}{\color[HTML]{333333} \textbf{1.02}}} & \multicolumn{1}{l|}{\cellcolor[HTML]{FFFFFF}{\color[HTML]{333333} \textbf{1.02}}} & \multicolumn{1}{l|}{\cellcolor[HTML]{FFFFFF}{\color[HTML]{333333} 1.02}}          & \multicolumn{1}{l|}{\cellcolor[HTML]{FFFFFF}{\color[HTML]{333333} }}                       & \multicolumn{1}{l|}{\cellcolor[HTML]{FFFFFF}{\color[HTML]{333333} \textbf{1.09}}} & \multicolumn{1}{l|}{\cellcolor[HTML]{FFFFFF}{\color[HTML]{333333} \textbf{1.09}}} & {\color[HTML]{333333} \textbf{1.08}} \\ \cline{2-2} \cline{4-6} \cline{8-10} \cline{12-14} \cline{16-18} 
\multicolumn{1}{|l|}{\multirow{-2}{*}{\cellcolor[HTML]{FFFFFF}{\color[HTML]{333333} \vga}}} & \multicolumn{1}{l|}{\cellcolor[HTML]{FFFFFF}{\color[HTML]{333333} set}}                                                      & \multicolumn{1}{l|}{\multirow{-2}{*}{\cellcolor[HTML]{FFFFFF}{\color[HTML]{333333} 1.02}}} & \multicolumn{1}{l|}{\cellcolor[HTML]{FFFFFF}{\color[HTML]{333333} 1.02}}          & \multicolumn{1}{l|}{\cellcolor[HTML]{FFFFFF}{\color[HTML]{333333} 1.02}}          & \multicolumn{1}{l|}{\cellcolor[HTML]{FFFFFF}{\color[HTML]{333333} \textbf{1.02}}} & \multicolumn{1}{l|}{\multirow{-2}{*}{\cellcolor[HTML]{FFFFFF}{\color[HTML]{333333} 1.08}}} & \multicolumn{1}{l|}{\cellcolor[HTML]{FFFFFF}{\color[HTML]{333333} 1.07}}          & \multicolumn{1}{l|}{\cellcolor[HTML]{FFFFFF}{\color[HTML]{333333} \textbf{1.05}}} & \multicolumn{1}{l|}{\cellcolor[HTML]{FFFFFF}{\color[HTML]{333333} 1.07}}          & \multicolumn{1}{l|}{\multirow{-2}{*}{\cellcolor[HTML]{FFFFFF}{\color[HTML]{333333} 1.03}}} & \multicolumn{1}{l|}{\cellcolor[HTML]{FFFFFF}{\color[HTML]{333333} 1.02}}          & \multicolumn{1}{l|}{\cellcolor[HTML]{FFFFFF}{\color[HTML]{333333} 1.02}}          & \multicolumn{1}{l|}{\cellcolor[HTML]{FFFFFF}{\color[HTML]{333333} \textbf{1.01}}} & \multicolumn{1}{l|}{\multirow{-2}{*}{\cellcolor[HTML]{FFFFFF}{\color[HTML]{333333} 1.10}}} & \multicolumn{1}{l|}{\cellcolor[HTML]{FFFFFF}{\color[HTML]{333333} 1.11}}          & \multicolumn{1}{l|}{\cellcolor[HTML]{FFFFFF}{\color[HTML]{333333} 1.10}}          & {\color[HTML]{333333} 1.10}          \\ \hline
\end{tabular}%
}

\caption{Table of maximum execution time slowdown for \disparity, \mser, \tracking and \sift benchmarks on the RK3588 platform}
\label{table-rk3588}
\end{table*}
}

%% file: figures/fig-synthetic.tex
\begin{figure}[!t]%
\centering%
\includegraphics[width=\columnwidth, trim=10 10 10 10, clip]{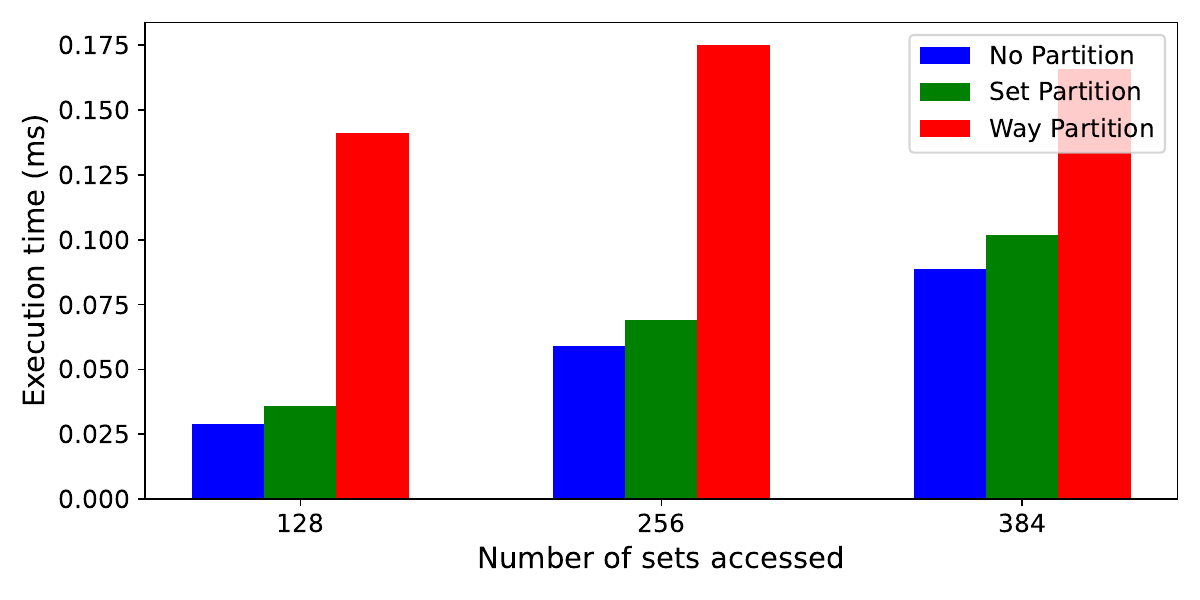}%
\caption{
	Execution time of the synthetic benchmark under no partitioning, set partitioning, and way partitioning vs the number of sets accessed fully on the RK3568 platform
}%
\label{fig:synthetic-plot}%
\vspace{-1em}
\end{figure}%

%% file: figures/fig-set-vs-way.tex
\begin{figure*}[htbp]
    \begin{subfigure}{\textwidth}
        \centering
        \includegraphics[width=0.5\textwidth]{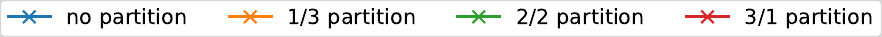}
        \label{fig:partition-legend} 
    \end{subfigure}
    \centering
    \begin{subfigure}{0.24\textwidth}
        \centering
        \includegraphics[width=\textwidth]{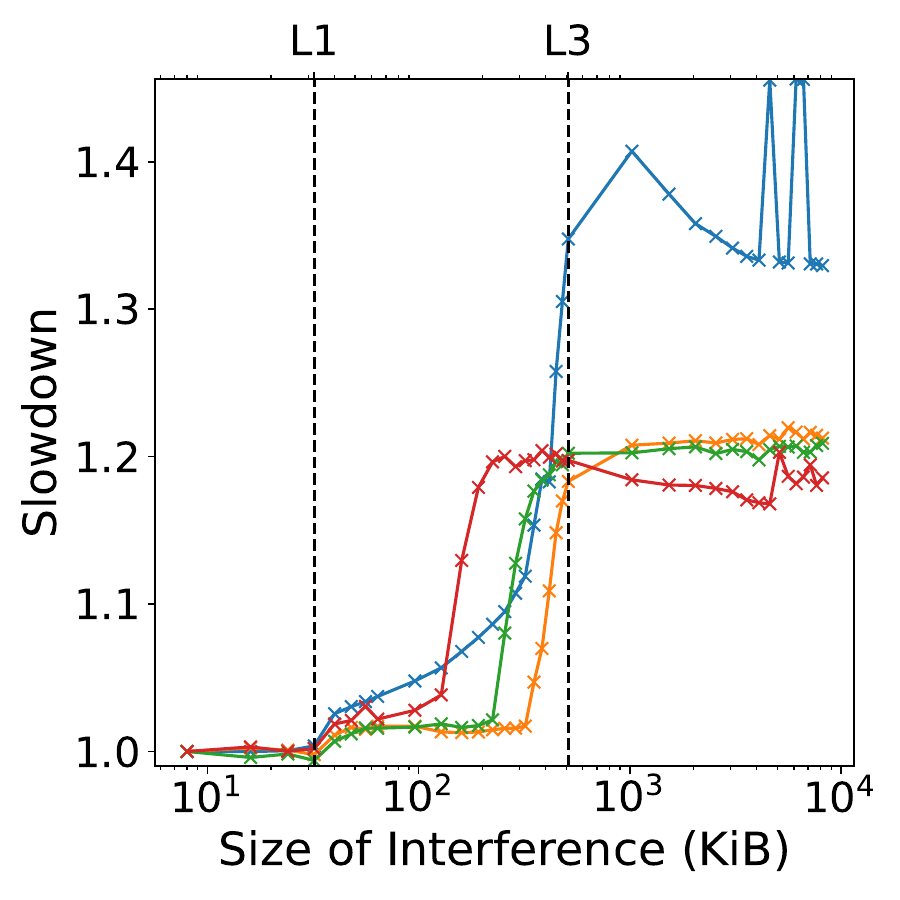}
        \captionsetup{justification=centering}
        \caption{RK3568: Set / \modifyInterf}
        \label{fig:rk3568-set-mser-modify}
    \end{subfigure}
    \begin{subfigure}{0.24\textwidth}
        \centering
        \includegraphics[width=\textwidth]{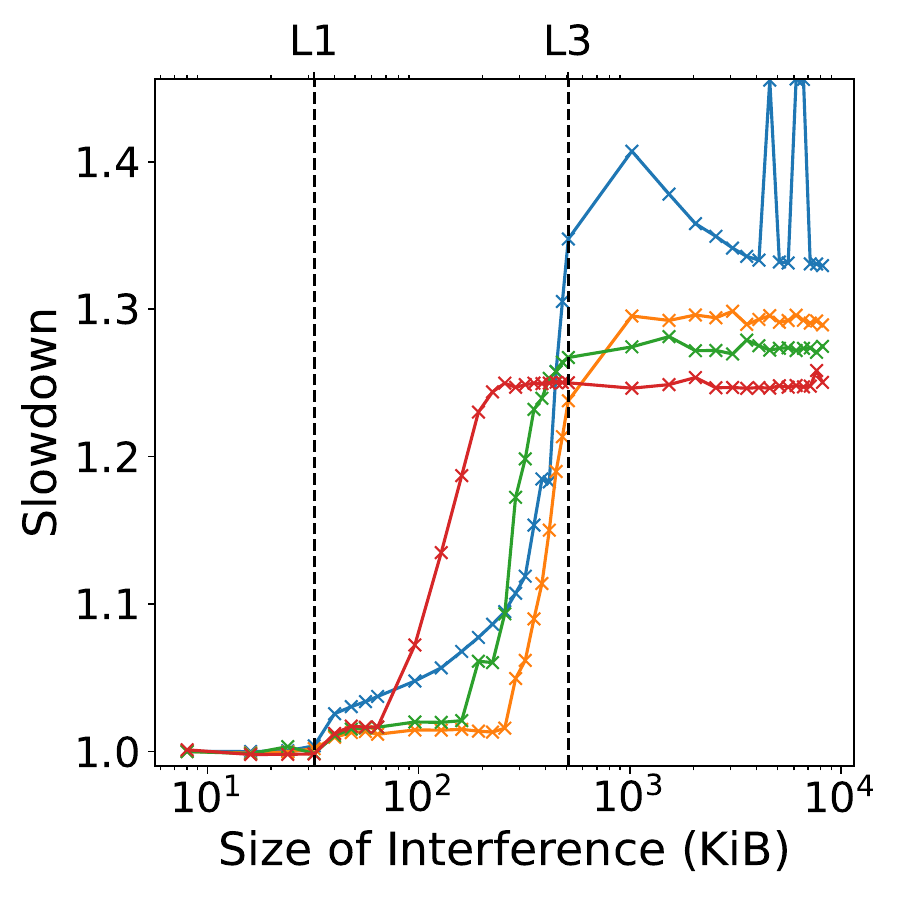}
        \captionsetup{justification=centering}
        \caption{RK3568: Way / \modifyInterf}
        \label{fig:rk3568-way-mser-modify}
    \end{subfigure}    
    \begin{subfigure}{0.24\textwidth}
        \centering
        \includegraphics[width=\textwidth]{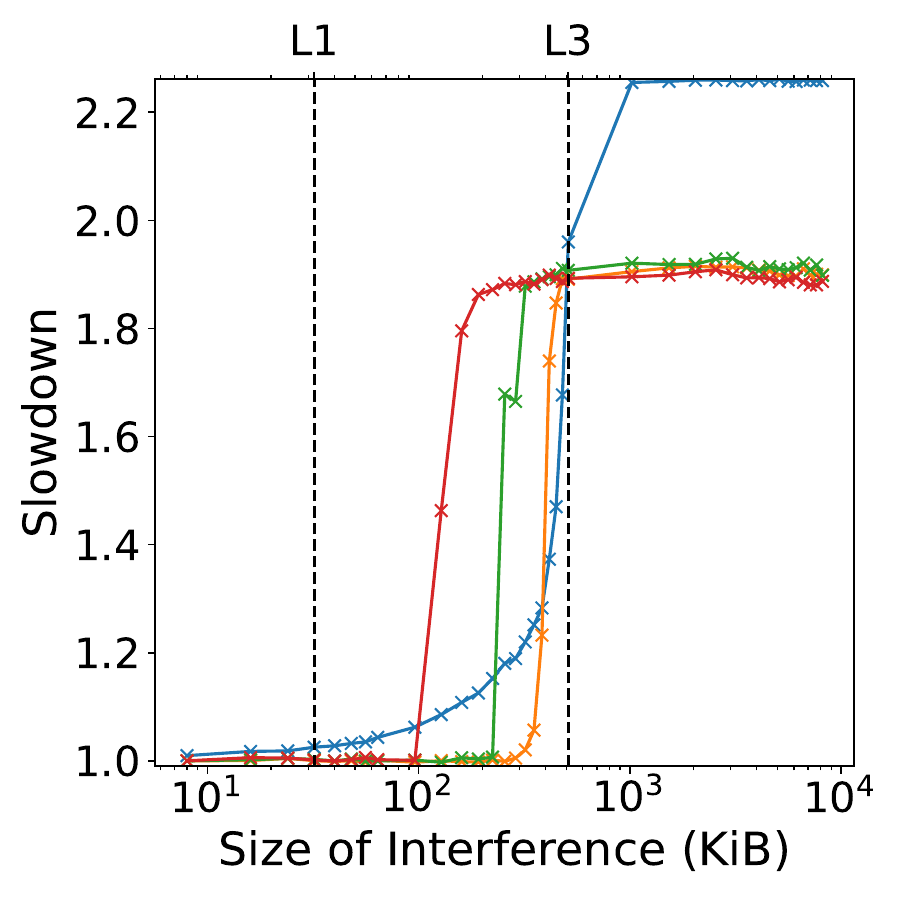}
        \captionsetup{justification=centering}
        \caption{RK3568: Set / \prefetchInterf}
        \label{fig:rk3568-set-mser-prefetch}
    \end{subfigure}
    \begin{subfigure}{0.24\textwidth}
        \centering
        \includegraphics[width=\textwidth]{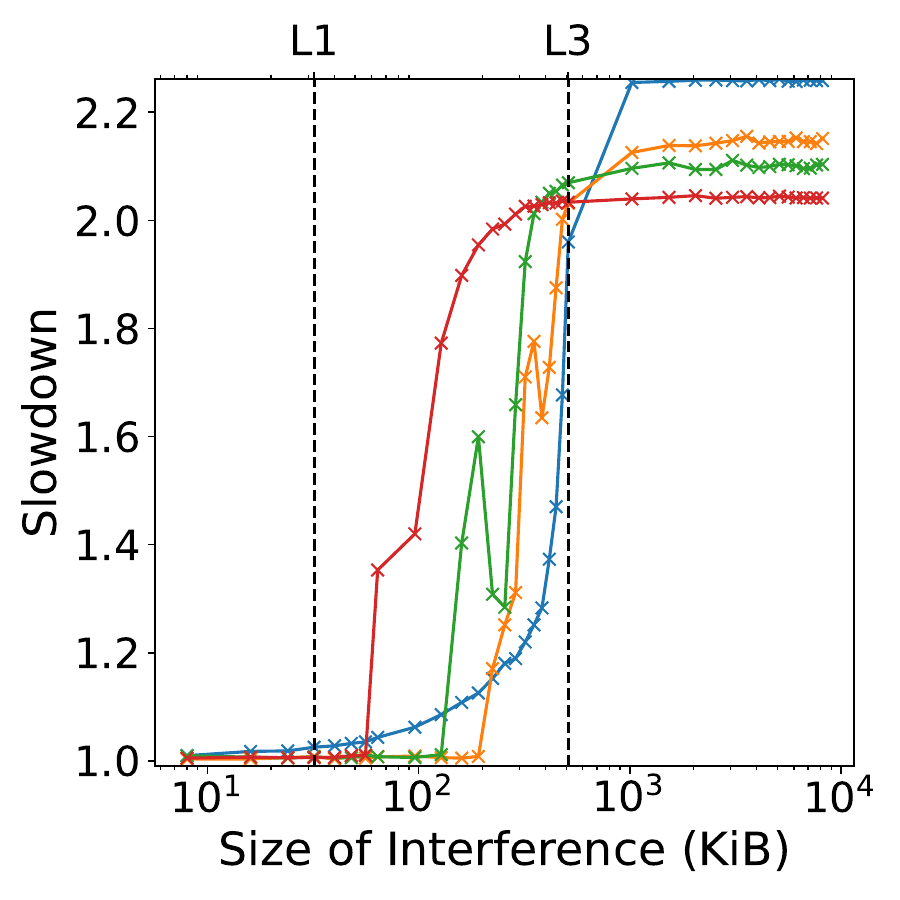}
        \captionsetup{justification=centering}
        \caption{RK3568: Way / \prefetchInterf}
        \label{fig:rk3568-way-mser-prefetch}
    \end{subfigure}

    \begin{subfigure}{0.24\textwidth}
        \centering
        \includegraphics[width=\textwidth]{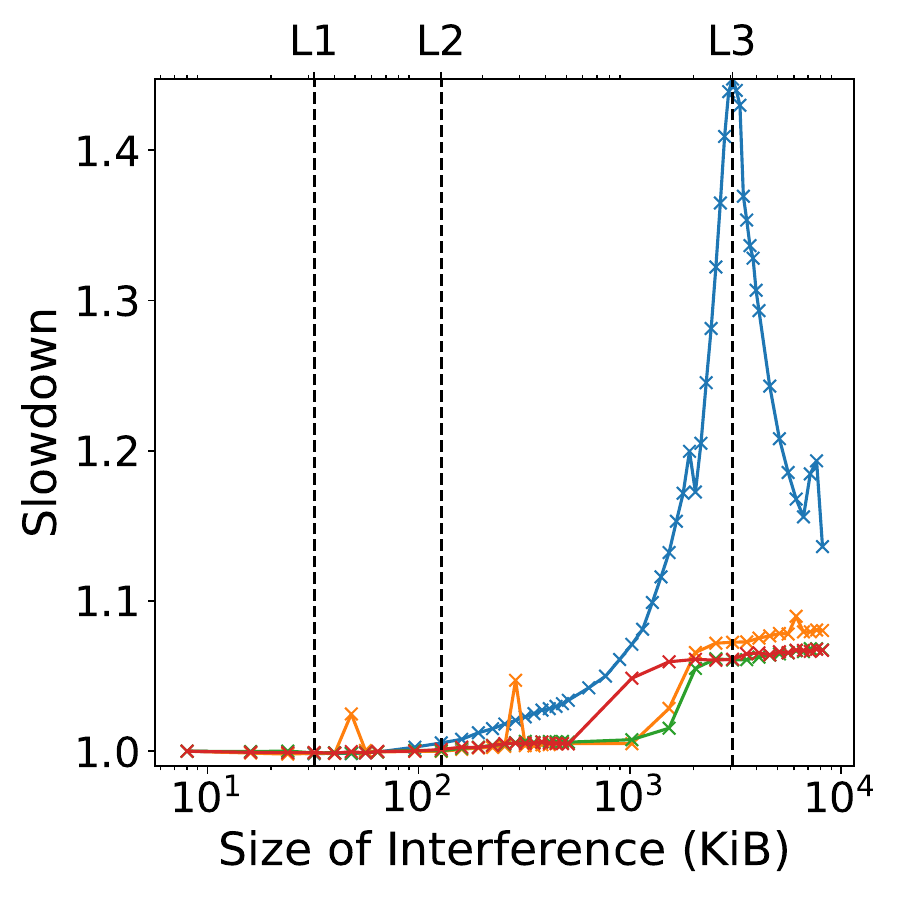}
        \captionsetup{justification=centering}
        \caption{RK3588: Set / \modifyInterf}
        \label{fig:rk3588-set-mser-modify}
    \end{subfigure}
    \begin{subfigure}{0.24\textwidth}
        \centering
        \includegraphics[width=\textwidth]{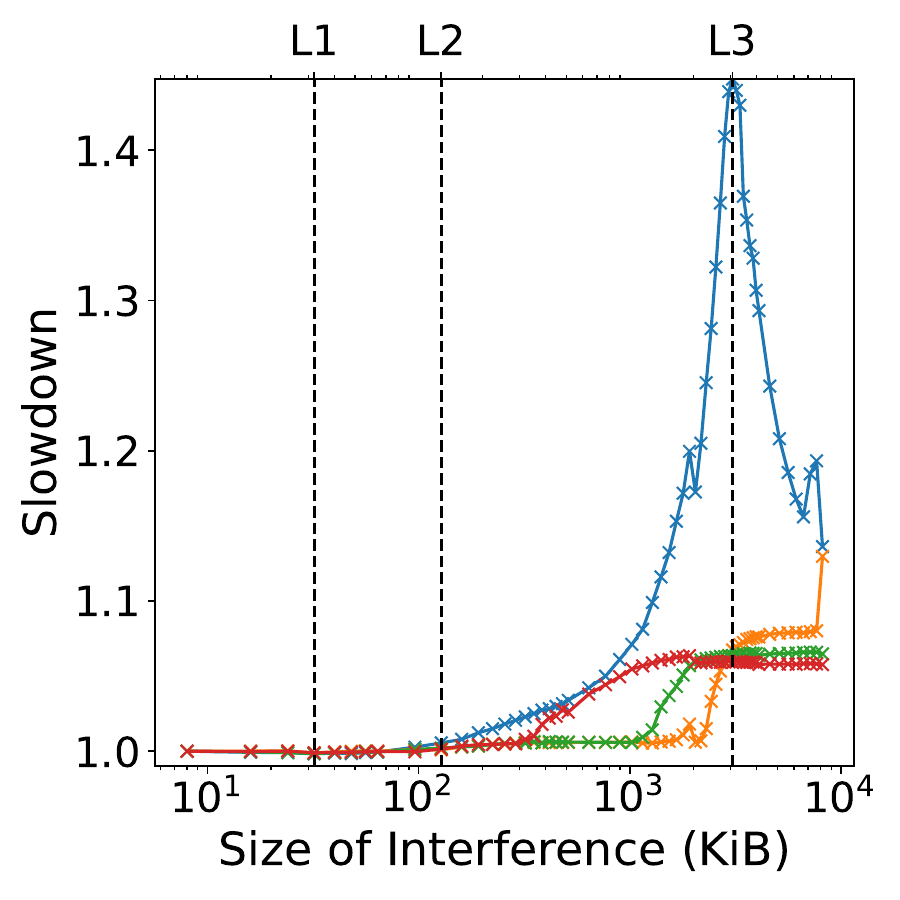}
        \captionsetup{justification=centering}
        \caption{RK3588: Way / \modifyInterf}
        \label{fig:rk3588-way-mser-modify}
    \end{subfigure}
    \begin{subfigure}{0.24\textwidth}
        \centering
        \includegraphics[width=\textwidth]{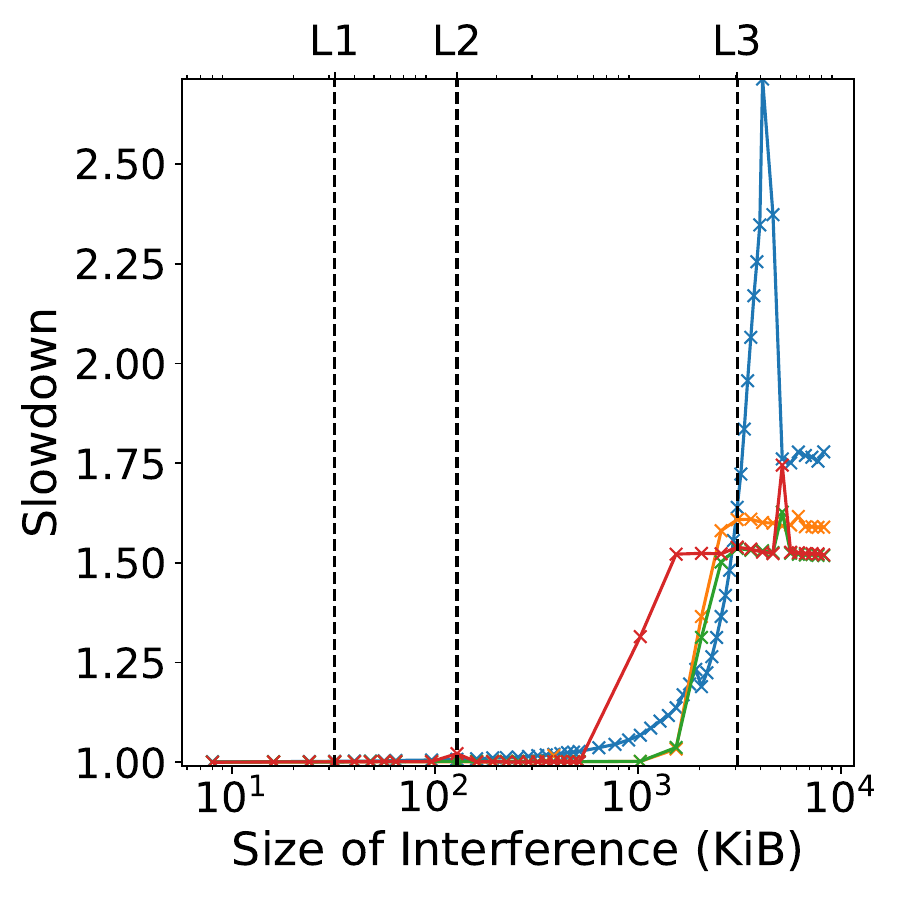}
        \captionsetup{justification=centering}
        \caption{RK3588: Set / \prefetchInterf}
        \label{fig:rk3588-set-mser-prefetch}
    \end{subfigure}
    \begin{subfigure}{0.24\textwidth}
        \centering
        \includegraphics[width=\textwidth]{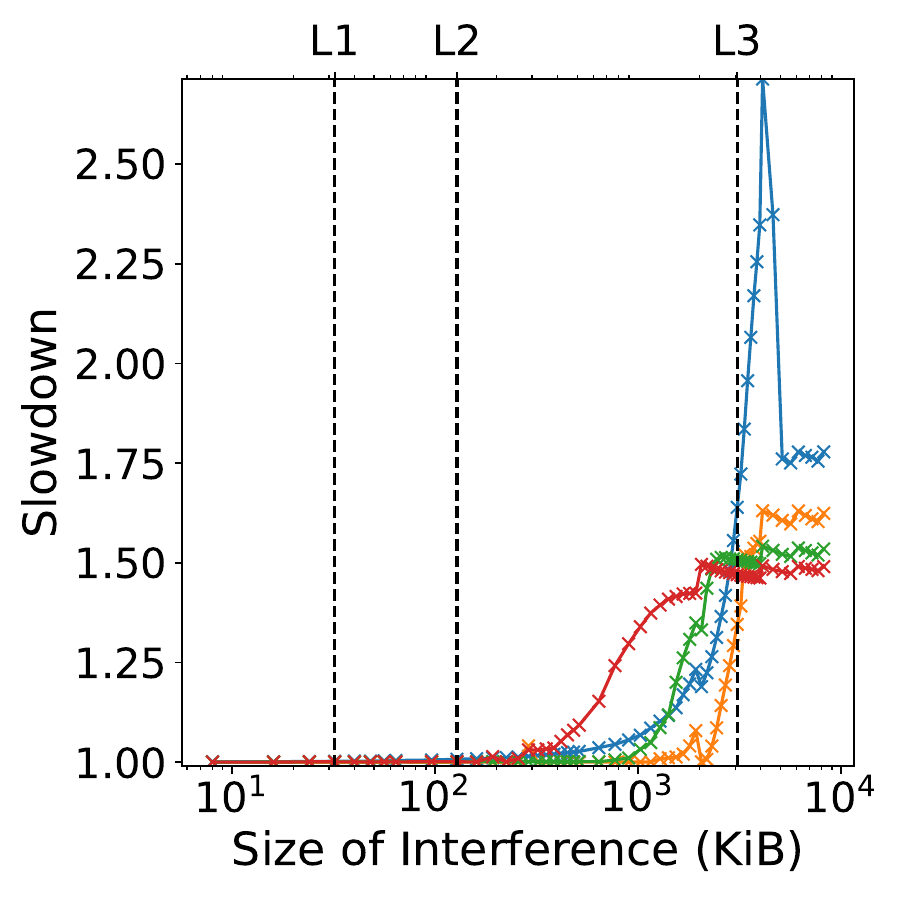}
        \captionsetup{justification=centering}
        \caption{RK3588: Way / \prefetchInterf}
        \label{fig:rk3588-way-mser-prefetch}
    \end{subfigure}

    \caption{Interference on \mser/\vga benchmark for RK3568 and RK3588 with \emph{Set} and \emph{Way} partitioning. Each plot reports the execution slowdown w.r.t the \emph{no-interference} case with increasing size (KiB) of \modifyInterf (a,b,e,f) and \prefetchInterf (c,d,g,h) interference. For each board, L1, L2 (if applicable), and L3 sizes are reported.}
    \label{fig:set-way-comparison-mser-vga}
\vspace{-1em}
\end{figure*}

%% file: figures/fig-no-partition.tex
\begin{figure}[htbp]
    \begin{subfigure}{0.49\columnwidth}
        \centering
        \includegraphics[width=\columnwidth]{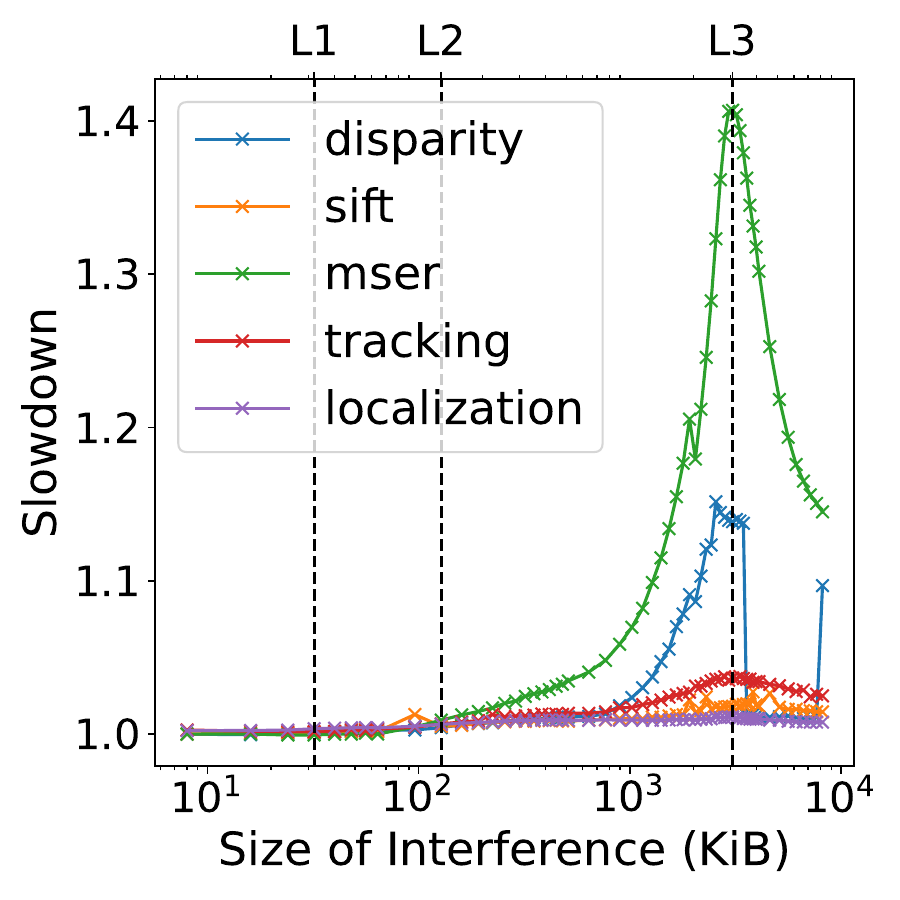}
        \captionsetup{justification=centering}
        \caption{Slowdown of different benchmarks}
        \label{fig:rk3588-benchmarks-no-partition} 
    \end{subfigure}
    \begin{subfigure}{0.49\columnwidth}
        \centering
        \includegraphics[width=\columnwidth]{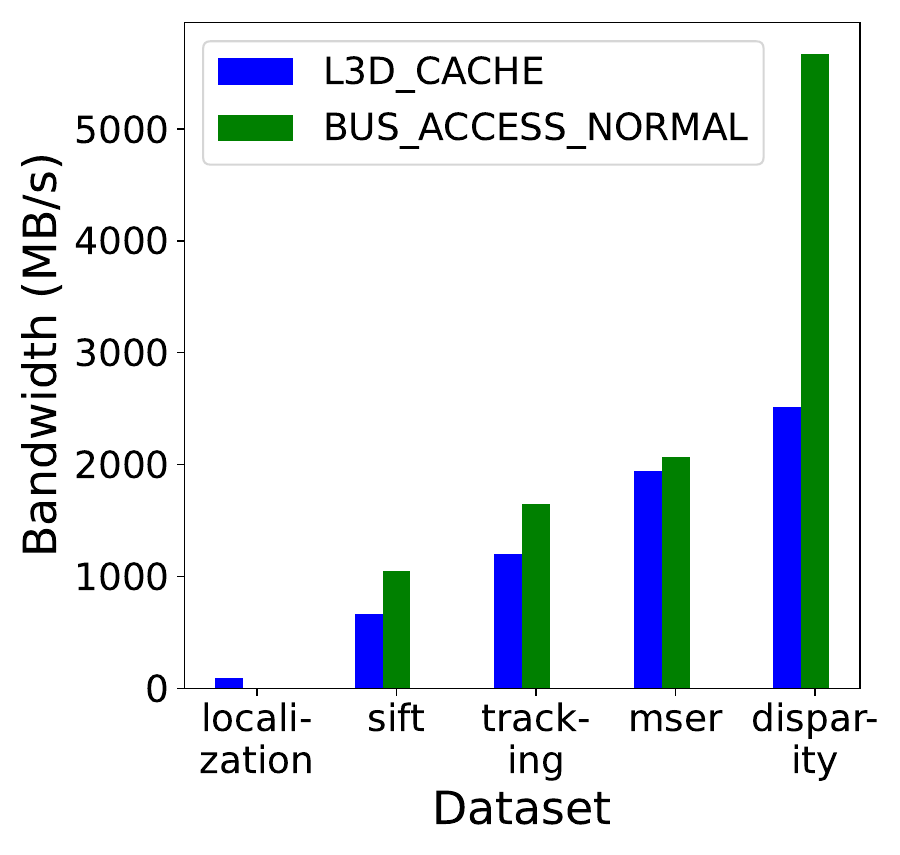}
        \captionsetup{justification=centering}
        \caption{Performance Counter Bandwidth}
        \label{fig:rk3588-perf-no-partition}
    \end{subfigure}

    \caption{Effect of \modifyInterf interference on different benchmarks/\vga dataset on the RK3588 platform without partitioning.}
    \label{fig:no-partition-benchmarks}
\vspace{-1em}
\end{figure}
%

%% file: figures/fig-bus-access.tex
\begin{figure}[t!]
    \centering
    \includegraphics[width=\columnwidth]{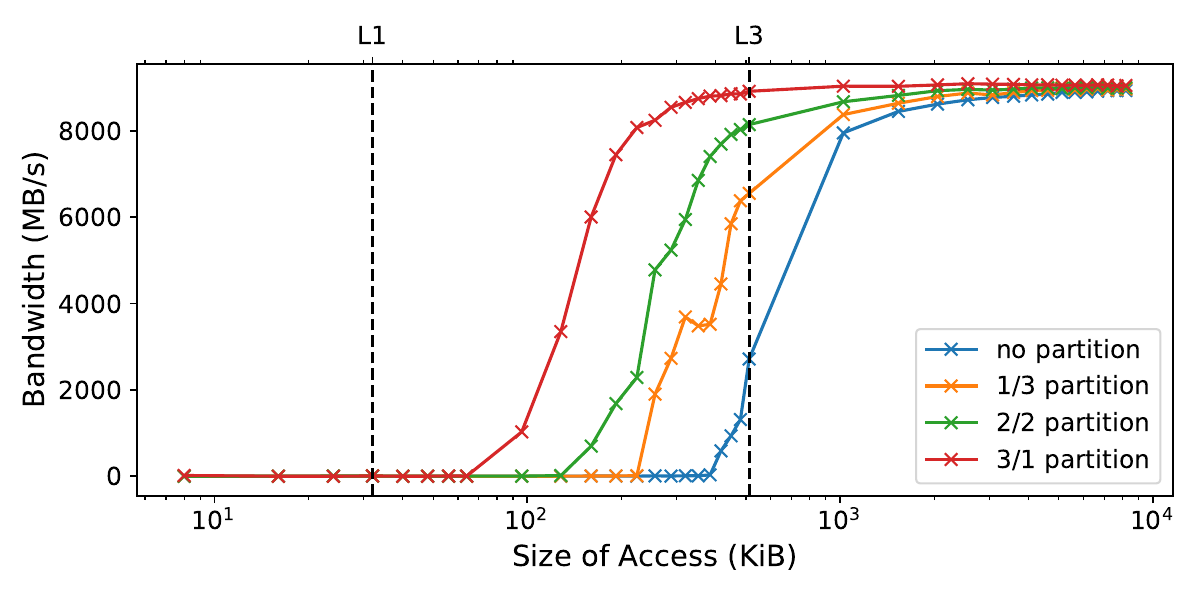}
    \caption{Bandwidth of \busaccessdsu as the size of \modifyInterf access in interference-bench increases on the RK3568 platform}
    \label{fig:64-modify-rk3568}
\end{figure}

%% file: figures/fig-l1-refill-counter.tex
\begin{figure}[!t]%
\centering%
\includegraphics[width=\columnwidth, clip]{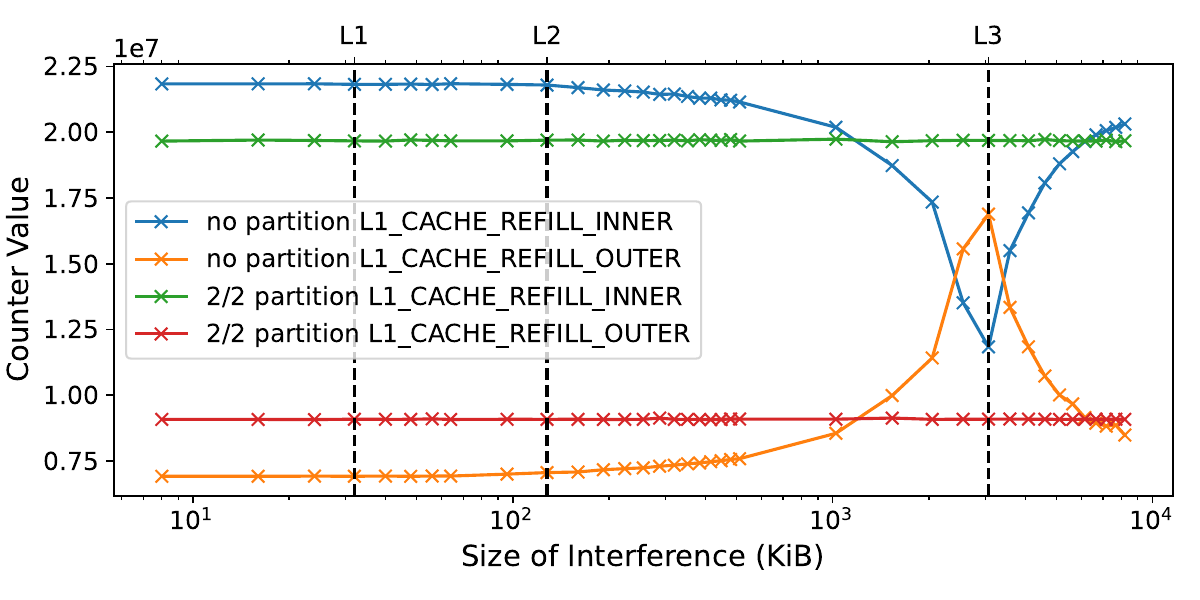}%
\caption{
	Value of Performance Counters in single execution of the \mser/\vga benchmark in the presence of \modifyInterf interference on RK3588 platform
}%
\label{fig:44-45-mser-modify}%
\vspace{-1em}
\end{figure}%

%% file: figures/fig-orin-way-disparity-read-cif.tex
\begin{figure*}[htbp]
    \begin{subfigure}{\textwidth}
        \centering
        \includegraphics[width=0.5\textwidth]{figures/set_subplot/legend.pdf}
    \end{subfigure}
    \centering
    \begin{subfigure}{0.32\textwidth}
        \centering
        \includegraphics[width=\textwidth]{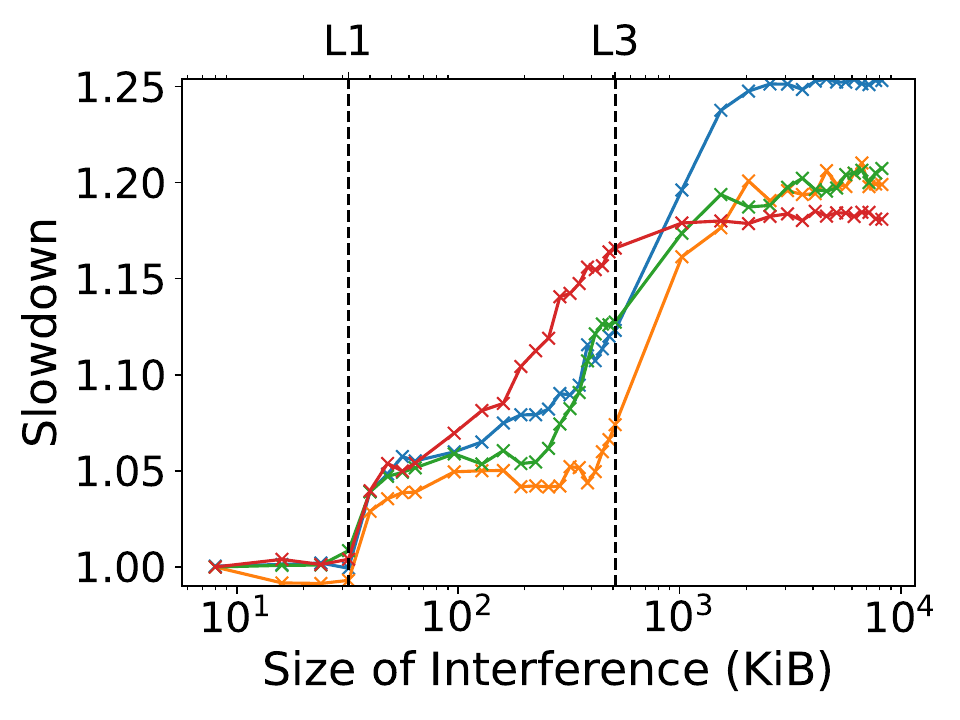}
        \captionsetup{justification=centering}
        \caption{RK3568}
        \label{fig:rk3568-way-disparity-read}
    \end{subfigure}
    \begin{subfigure}{0.32\textwidth}
        \centering
        \includegraphics[width=\textwidth]{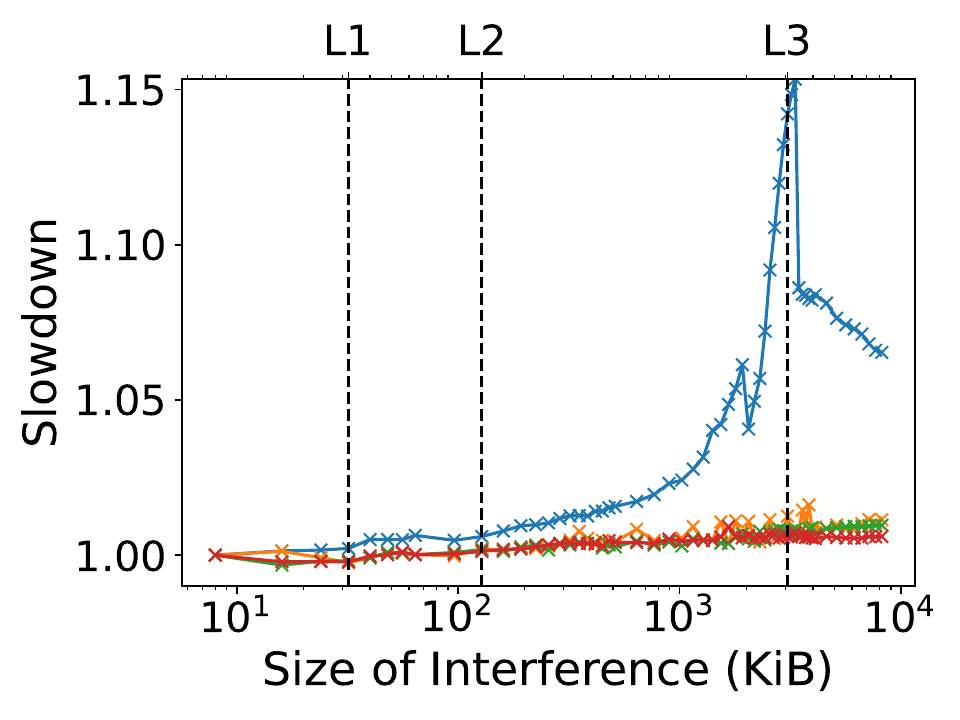}
        \captionsetup{justification=centering}
        \caption{RK3588}
        \label{fig:rk3588-way-disparity-read}
    \end{subfigure}    
    \begin{subfigure}{0.32\textwidth}
        \centering
        \includegraphics[width=\textwidth]{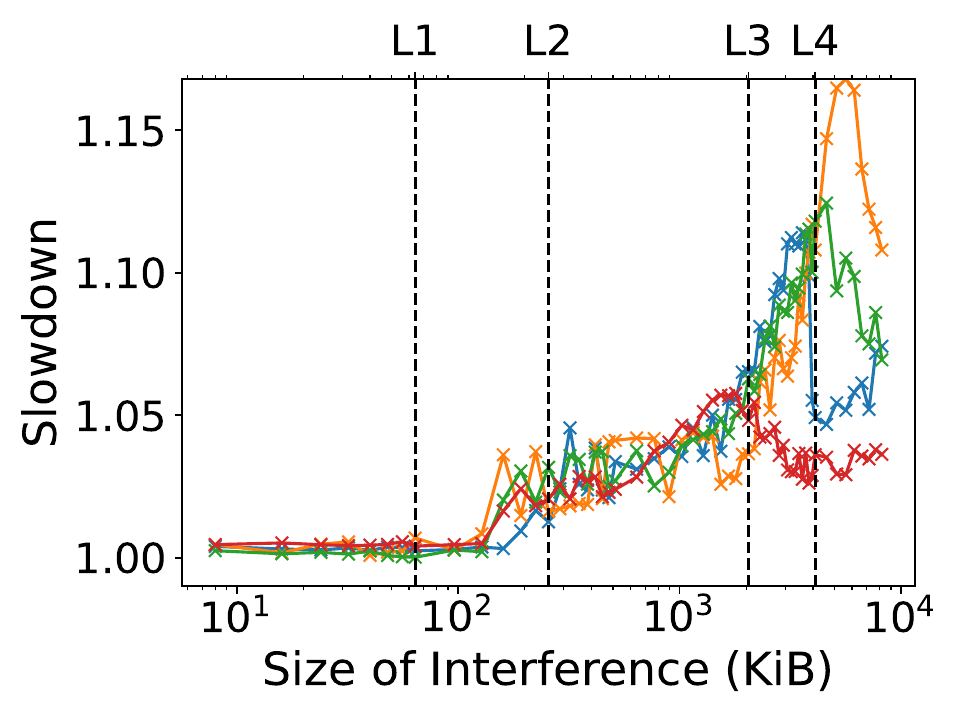}
        \captionsetup{justification=centering}
        \caption{Orin}
        \label{fig:orin-way-disparity-read}
    \end{subfigure}

    \caption{Slowdown of  \disparity/\cif under read interference on RK3568, RK3588, and Orin with way partitioning.}
    \label{fig:way-compare-orin}
 \vspace{-1em}
\end{figure*}

%% file: figures/fig-dsu-clusters.tex
\begin{figure}[!t]%
\centering%
\includegraphics[width=\columnwidth, clip]{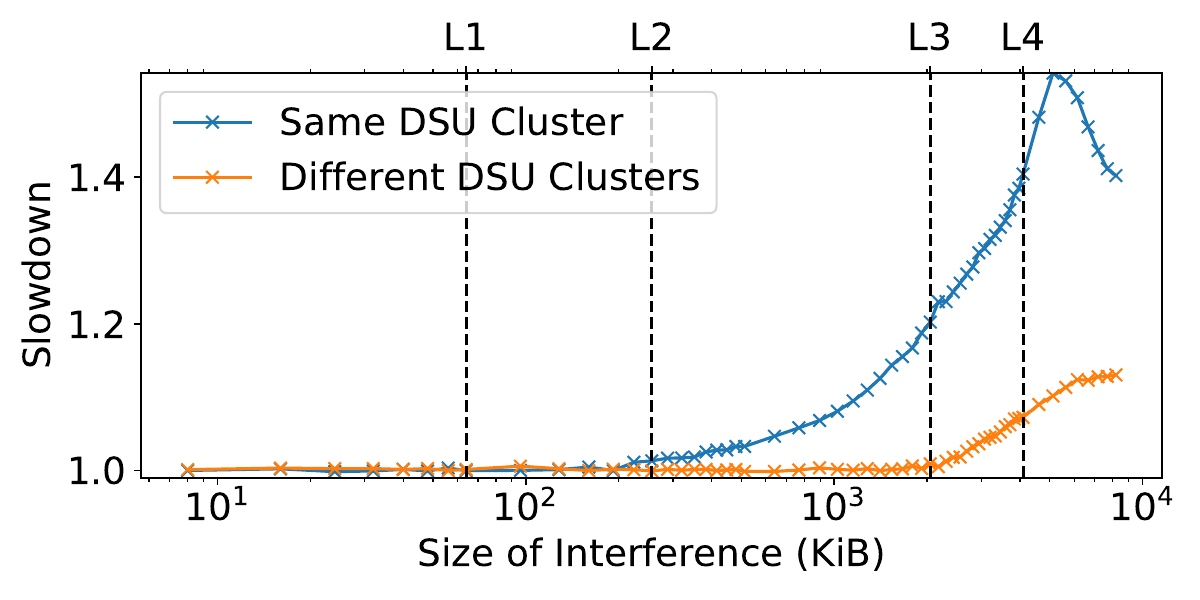}%
\caption{
    Slowdown of \mser/\vga on Orin under \modifyInterf interference running on same and different DSU clusters.
}
\label{fig:dsu-clusters}%
\vspace{-1em}
\end{figure}%

%% file: figures/fig-write-streaming.tex
\begin{figure*}[htbp]
    \begin{subfigure}{\textwidth}
        \centering
        \includegraphics[width=0.5\textwidth]{figures/set_subplot/legend.pdf}
    \end{subfigure}
    \centering
    \begin{subfigure}{0.32\textwidth}
        \centering
        \includegraphics[width=\textwidth]{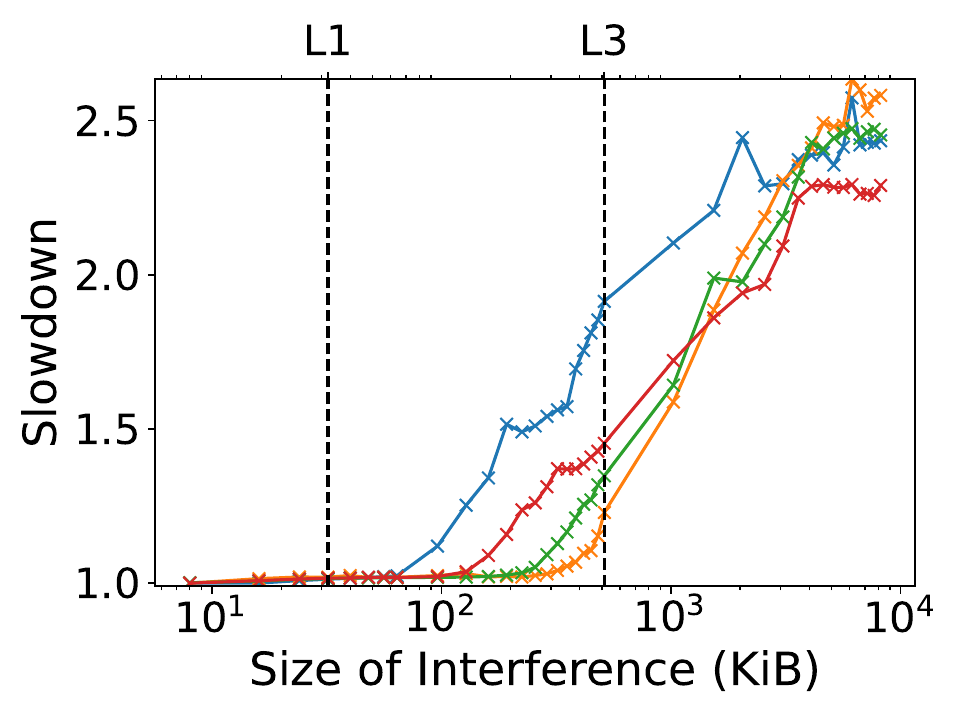}
        \captionsetup{justification=centering}
        \caption{RK3568}
        \label{fig:rk3568-set-disparity-write}
    \end{subfigure}
    \begin{subfigure}{0.32\textwidth}
        \centering
        \includegraphics[width=\textwidth]{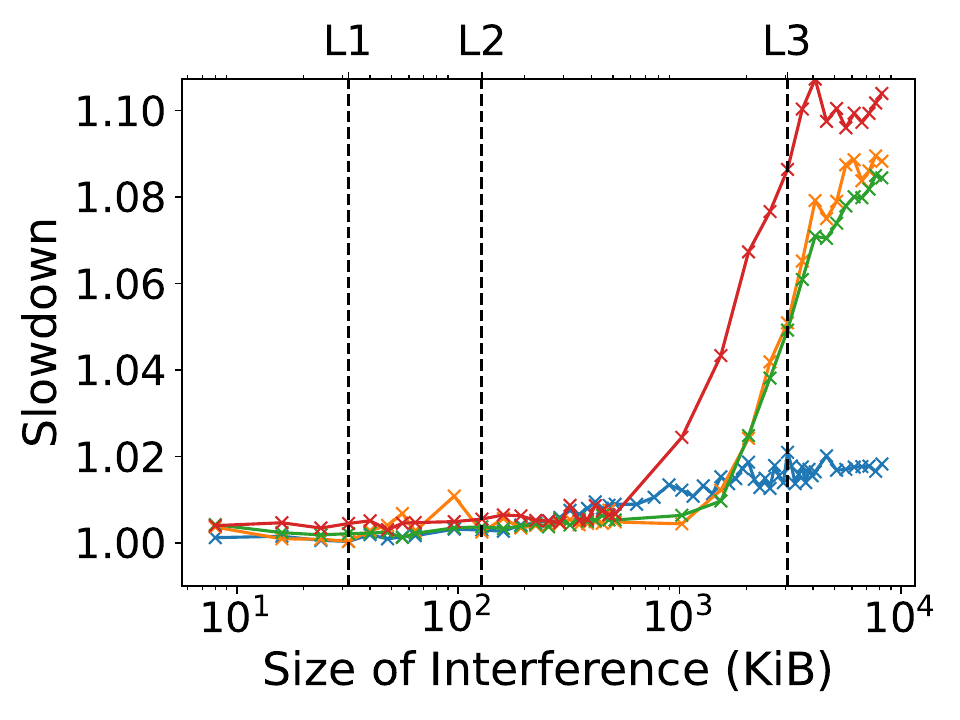}
        \captionsetup{justification=centering}
        \caption{RK3588}
        \label{fig:rk3588-set-disparity-write}
    \end{subfigure}    
    \begin{subfigure}{0.32\textwidth}
        \centering
        \includegraphics[width=\textwidth]{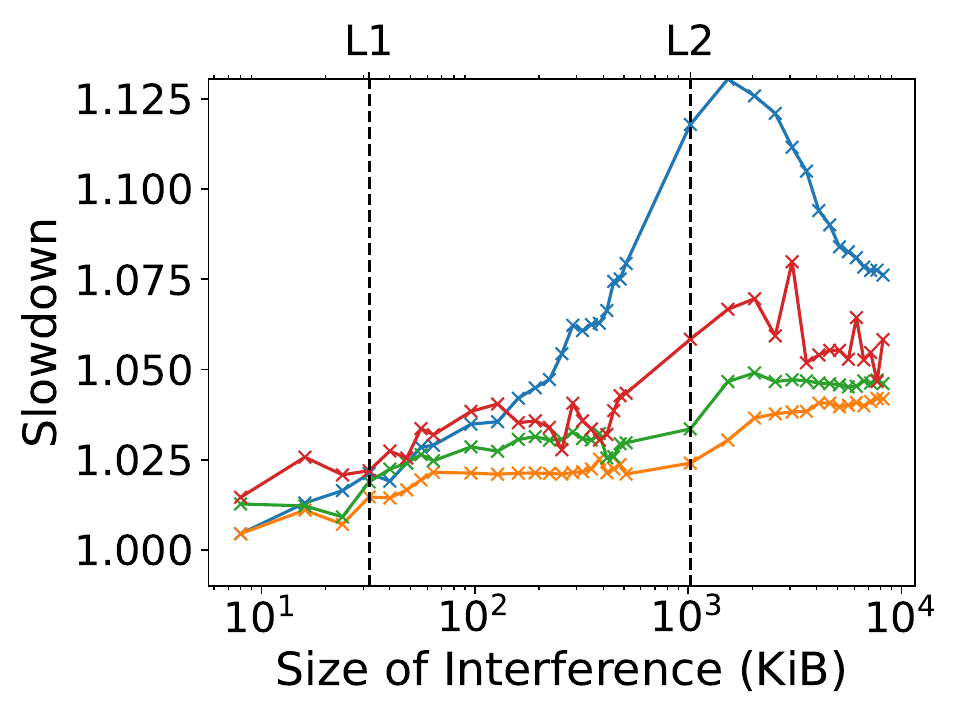}
        \captionsetup{justification=centering}
        \caption{ZCU102}
        \label{fig:zcu102-set-disparity-write}
    \end{subfigure}

    \caption{Slowdown of \disparity/\cif under \writeInterf interference on RK3568, RK3588, and ZCU102 with set partitioning.}
    \label{fig:set-compare-write-streaming}
 \vspace{-1em}
\end{figure*}

%% file: figures/fig-nowritestream.tex
\begin{figure}[htbp]
    \begin{subfigure}{\columnwidth}
        \centering
        \includegraphics[width=0.9\columnwidth]{figures/set_subplot/legend.pdf}
    \end{subfigure}
    \centering
    \begin{subfigure}{0.49\columnwidth}
        \centering
        \includegraphics[width=\columnwidth]{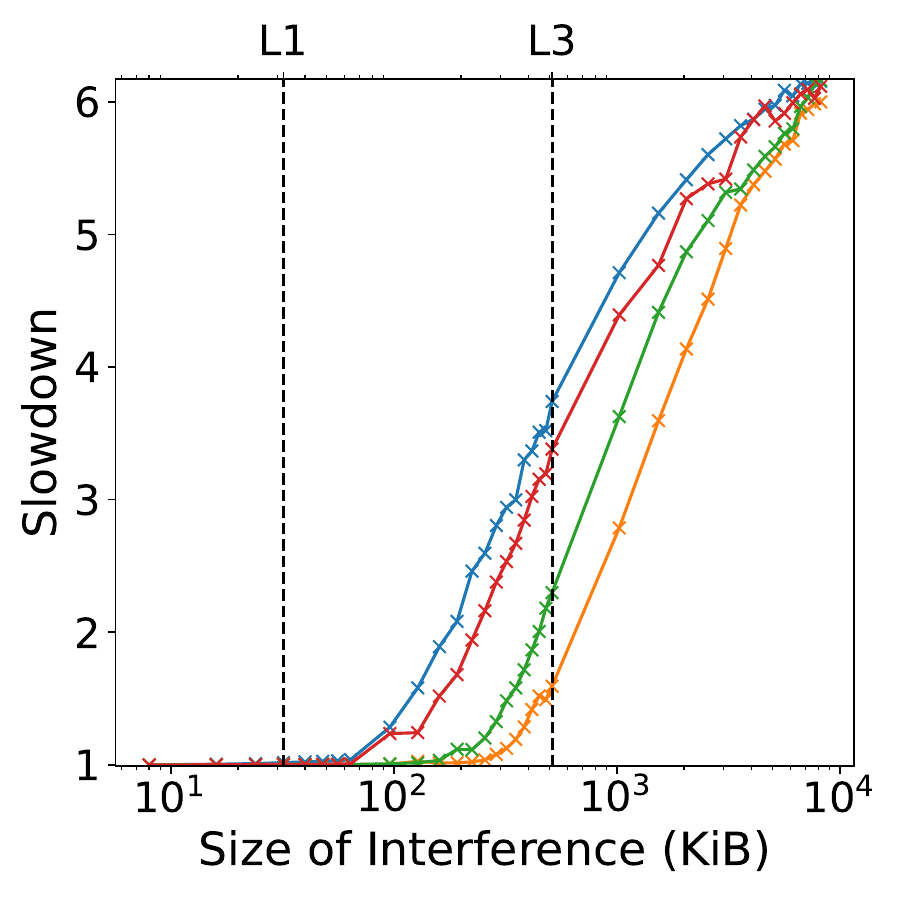}
        \captionsetup{justification=centering}
        \caption{Enabled}
        \label{fig:rk3568-way-mser-write-stream}
    \end{subfigure}
    \begin{subfigure}{0.49\columnwidth}
        \centering
        \includegraphics[width=\columnwidth]{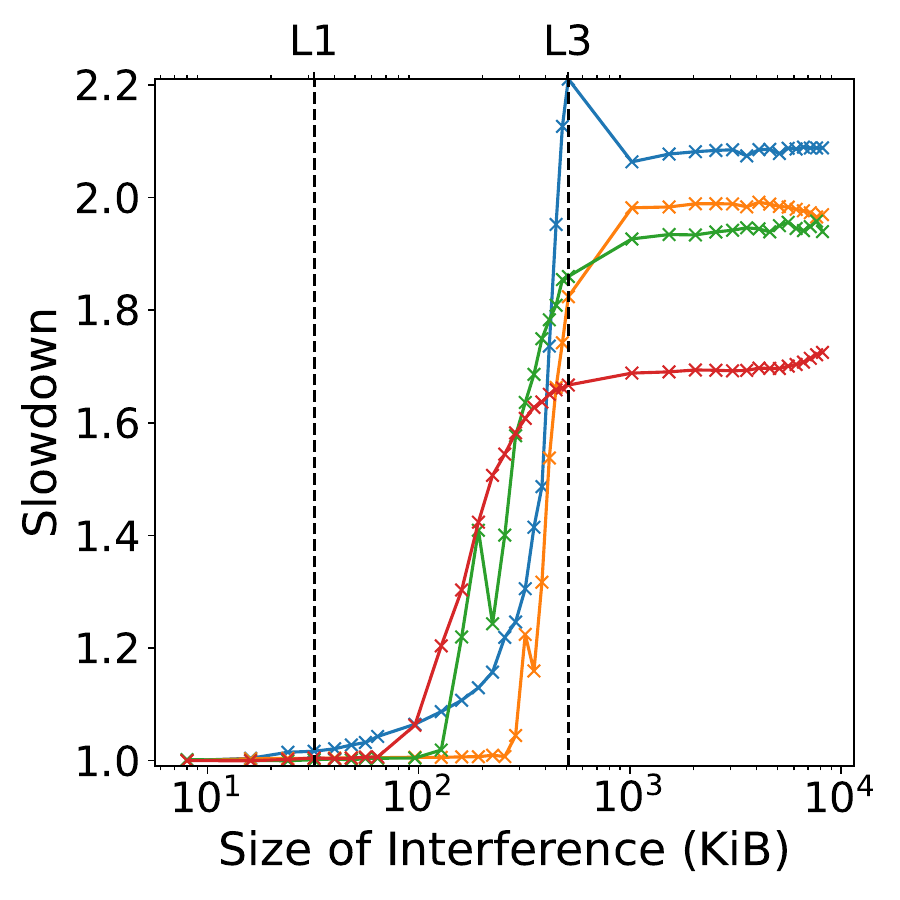}
        \captionsetup{justification=centering}
        \caption{Disabled}
        \label{fig:rk3568-way-mser-write-nostream}
    \end{subfigure}    

    \caption{Slowdown of \mser/\vga under write interference with \emph{write-streaming} mode a) Enabled and b) Disabled on RK3568}
    \label{fig:rk3568-write-streaming-mode}
 \vspace{-1em}
\end{figure}

%% file: figures/fig-l1-increase.tex
\begin{figure*}[htbp]
    \begin{subfigure}{\textwidth}
        \centering
        \includegraphics[width=0.5\textwidth]{figures/set_subplot/legend.pdf}
    \end{subfigure}
    \centering
    \begin{subfigure}{0.32\textwidth}
        \centering
        \includegraphics[width=\textwidth]{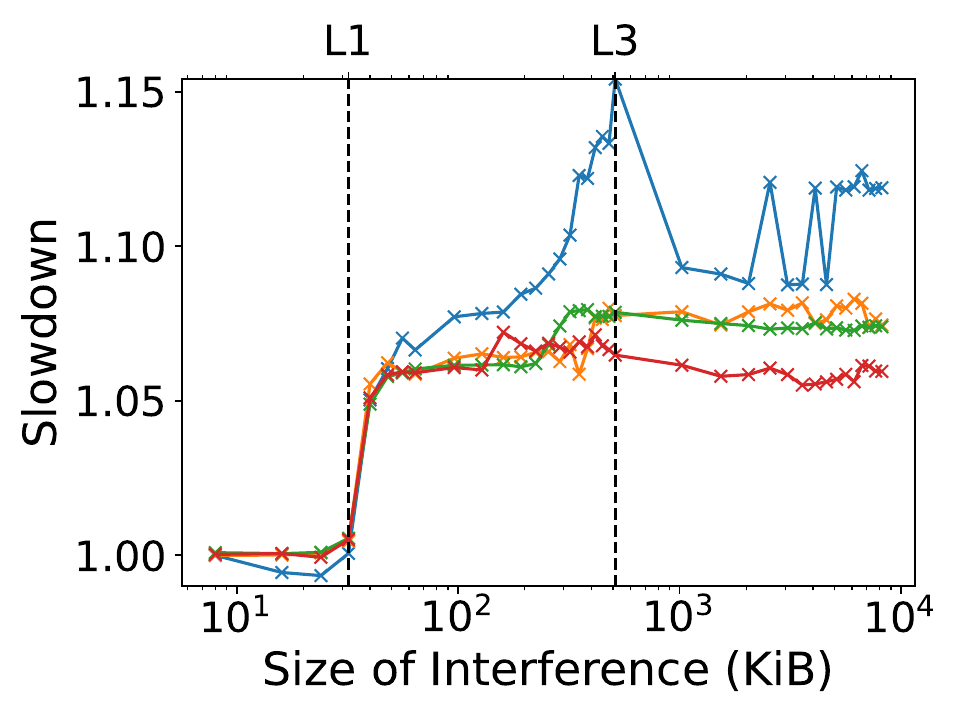}
        \captionsetup{justification=centering}
        \caption{RK3568}
        \label{fig:rk3568-set-disparity-modify}
    \end{subfigure}
    \begin{subfigure}{0.32\textwidth}
        \centering
        \includegraphics[width=\textwidth]{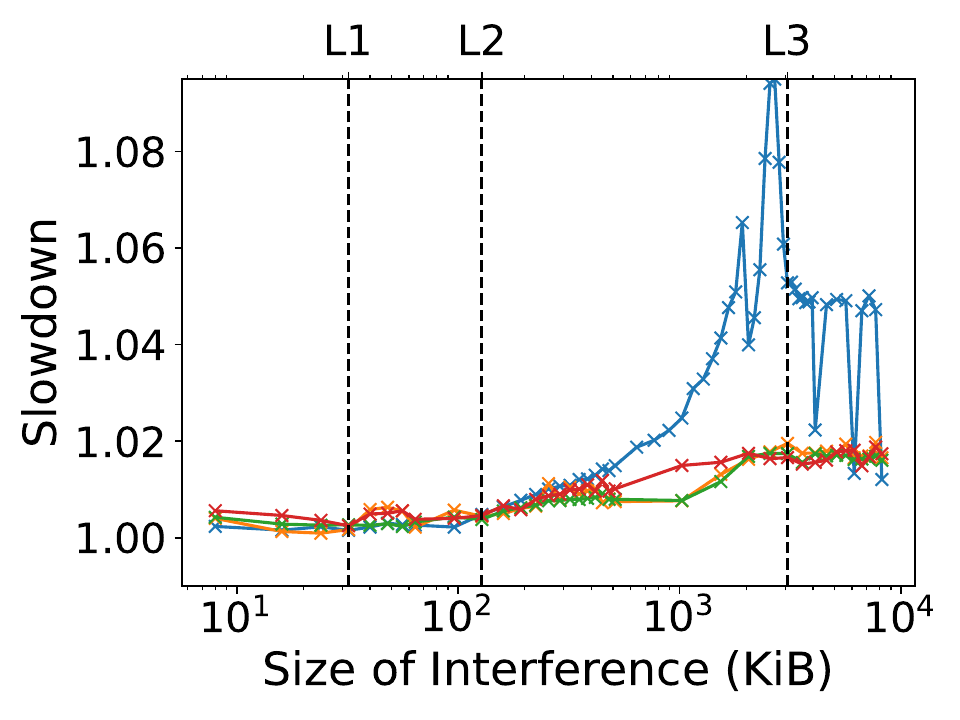}
        \captionsetup{justification=centering}
        \caption{RK3588}
        \label{fig:rk3588-set-disparity-modify}
    \end{subfigure}
    \begin{subfigure}{0.32\textwidth}
        \centering
        \includegraphics[width=\textwidth]{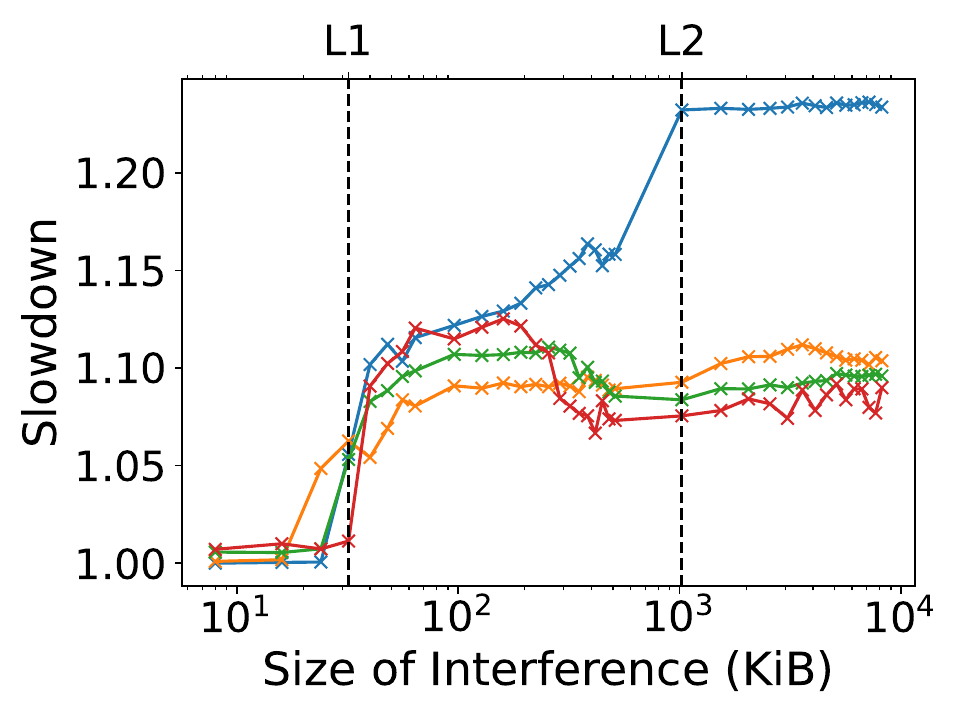}
        \captionsetup{justification=centering}
        \caption{ZCU102}
        \label{fig:zcu102-set-disparity-modify}
    \end{subfigure}    

    \caption{Slowdown of \disparity/\cif under \modifyInterf interference on RK3568, RK3588, and ZCU102 with set partitioning.}
    \label{fig:set-compare-l1-modify}
 \vspace{-1em}
\end{figure*}



%% file: figures/fig-prefetch.tex
\begin{figure}[htbp]
    \begin{subfigure}{\columnwidth}
        \centering
        \includegraphics[width=0.5\columnwidth, ]{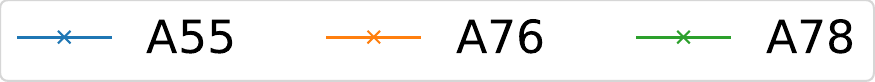}
        \label{fig:prefetch-legend-cpu} 
    \end{subfigure}
    \centering
    \begin{subfigure}{0.49\columnwidth}
        \centering
        \includegraphics[width=\columnwidth]{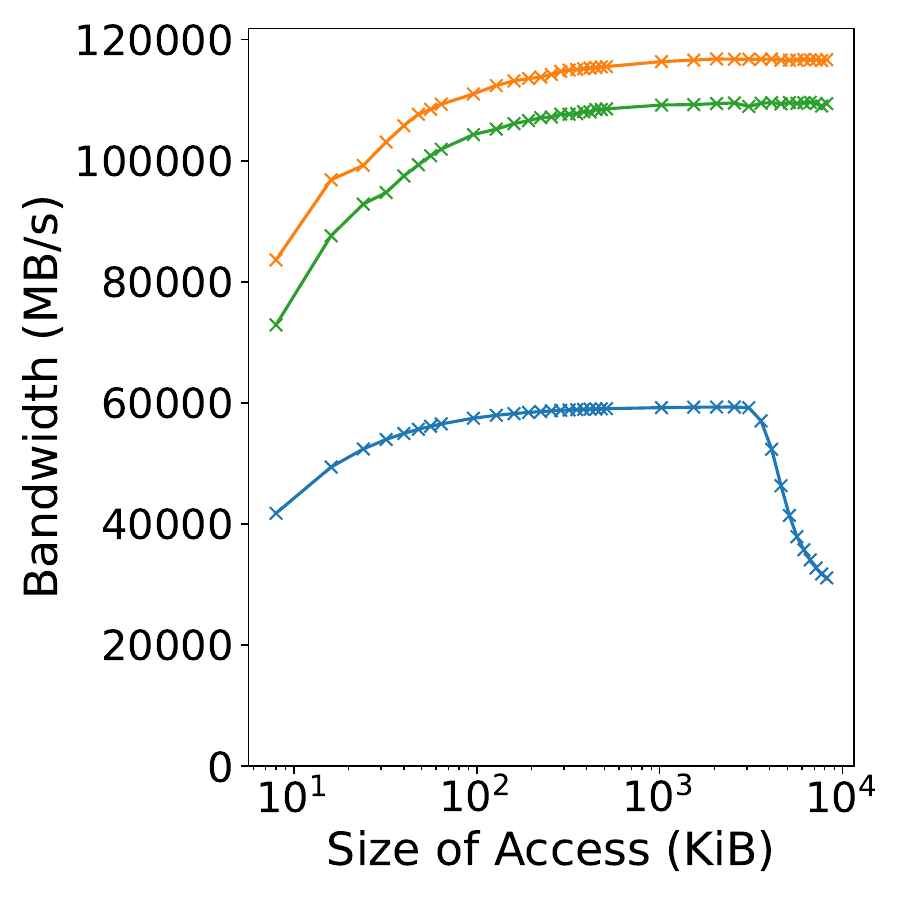}
        \captionsetup{justification=centering}
        \caption{Execution Time}
        \label{fig:prefetch-exec}
    \end{subfigure}
    \begin{subfigure}{0.49\columnwidth}
        \centering
        \includegraphics[width=\columnwidth]{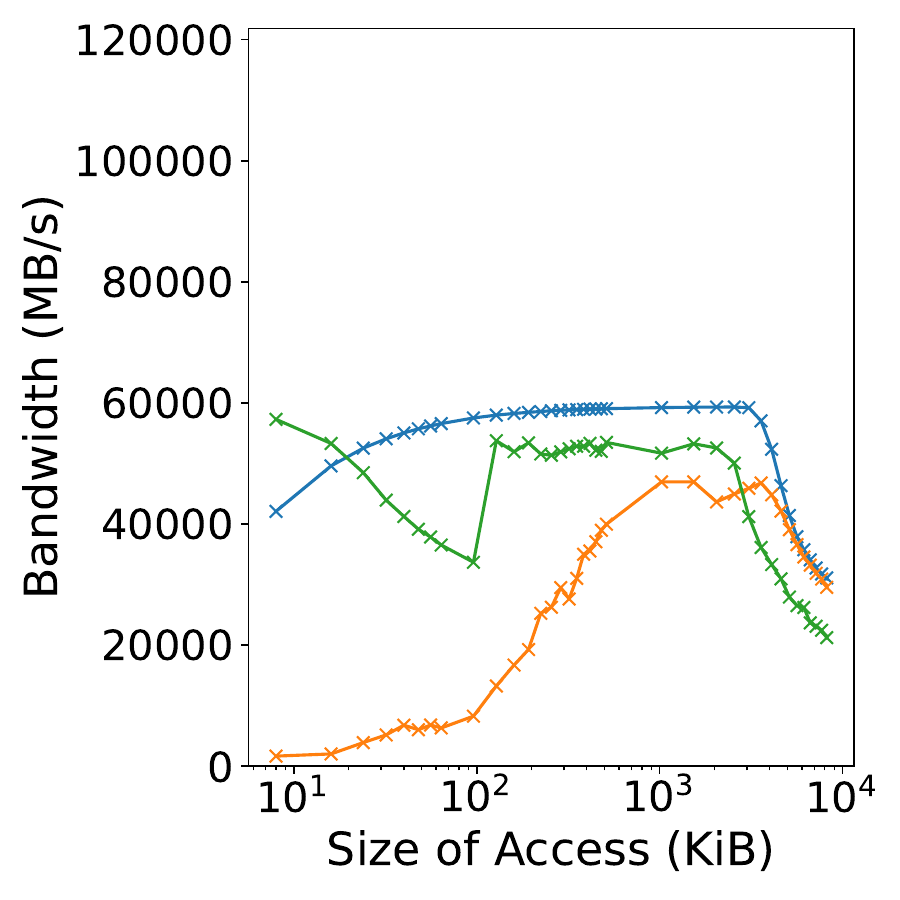}
        \captionsetup{justification=centering}
        \caption{Performance Counter}
        \label{fig:prefetch-pref}
    \end{subfigure}    

    \caption{Bandwidth of executing \prefetchInterf calculated from a) Execution time of program b) Performance Counters}
    \label{fig:prefetch-compare}
 \vspace{-1em}
\end{figure}

%% file: 7-conclusion.tex
\section{Conclusion and Future Directions}
\label{sec:conclusion}


This paper presented a comprehensive evaluation of Arm’s DSU technology for real-time cache partitioning across multiple COTS platforms. 
Through detailed testing, we examined the capabilities and limitations of Arm DSU’s way partitioning feature, comparing it with established set partitioning techniques such as \emph{page coloring}. 
Our findings show that way partitioning provides effective isolation with lower setup overhead, making it a practical option for reducing interference in real-time applications. 
Conversely, set partitioning offers finer control by allowing selective cache partitioning, not only at the last-level cache (LLC) but also at higher hierarchy levels (\eg L1, L2). 
This flexibility is particularly beneficial on platforms like NVIDIA Orin, where managing interference at both the L3 and L4 cache levels can improve performance and predictability.

Overall, this study identifies the trade-offs practitioners should consider when selecting a cache partitioning strategy for real-time systems. While way-partitioning offers ease of implementation and resilience against hardware constraints, its effectiveness depends on workload characteristics and system-level optimizations. Understanding platform-specific architectural behaviors—such as memory controller saturation, prefetching strategies, and write-streaming mechanisms, is needed for achieving predictable performance.

This study focuses exclusively on inter-core interference.
We defer to future work the effective integration
of cache partitioning at scheduling level to manage interference among different tasks.
Here, both way and set partitioning could be leveraged for task-level isolation, but their efficiency in this context requires further investigation. While way-partitioning can be dynamically adjusted by modifying the \code{CLUSTERPARTCR} register, the potential overhead of frequent changes has not been studied. Future work should assess whether this overhead is lower than the one incurred in set-partitioning.

%% file: 8-appendix.tex
\input{figures-appendix}

%% file: figures-appendix.tex
        \subsection{RK3568}
        \tableRKsixeight
        \clearpage

    \begin{figure}[H]
        \begin{subfigure}{\textwidth}
            \centering
            \includegraphics[width=0.5\textwidth]{figures/set_subplot/legend.pdf}
        \end{subfigure}
        \centering
        
        \begin{subfigure}{0.24\textwidth}
            \centering
            \includegraphics[width=\textwidth]{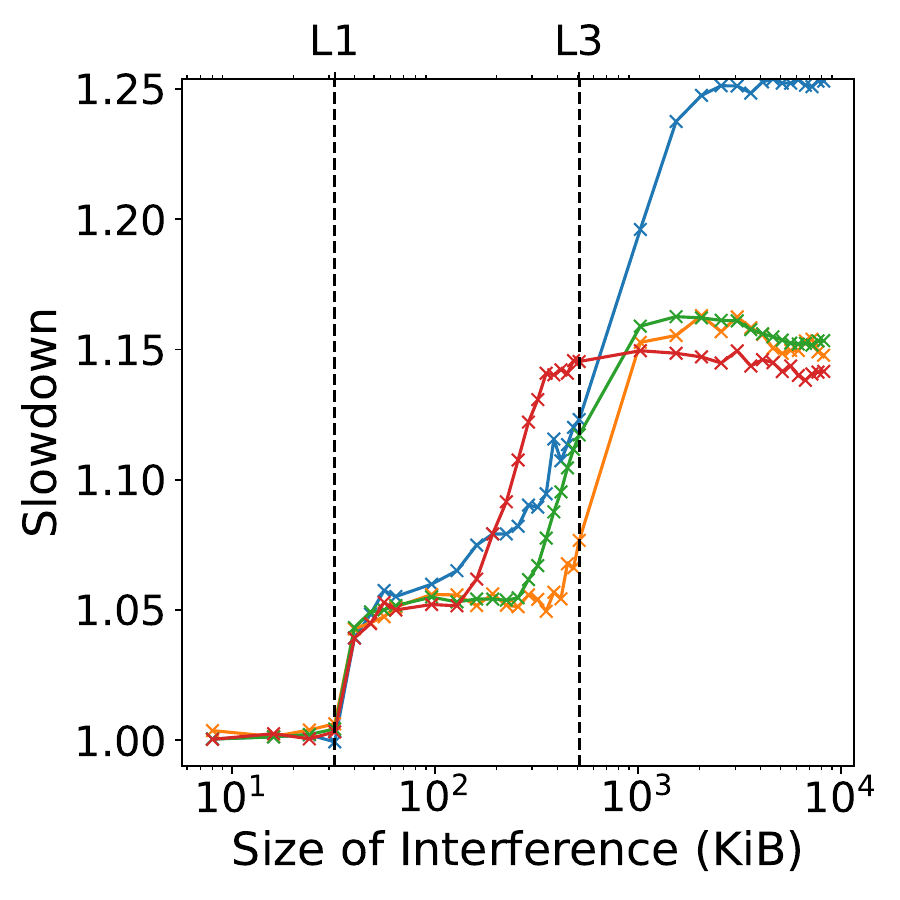}
            \captionsetup{justification=centering}
            \caption{rk3568: Set / Read}
            \label{fig:all-rk3568-set-disparity-read-cif}
        \end{subfigure}
        \hfill
        \begin{subfigure}{0.24\textwidth}
            \centering
            \includegraphics[width=\textwidth]{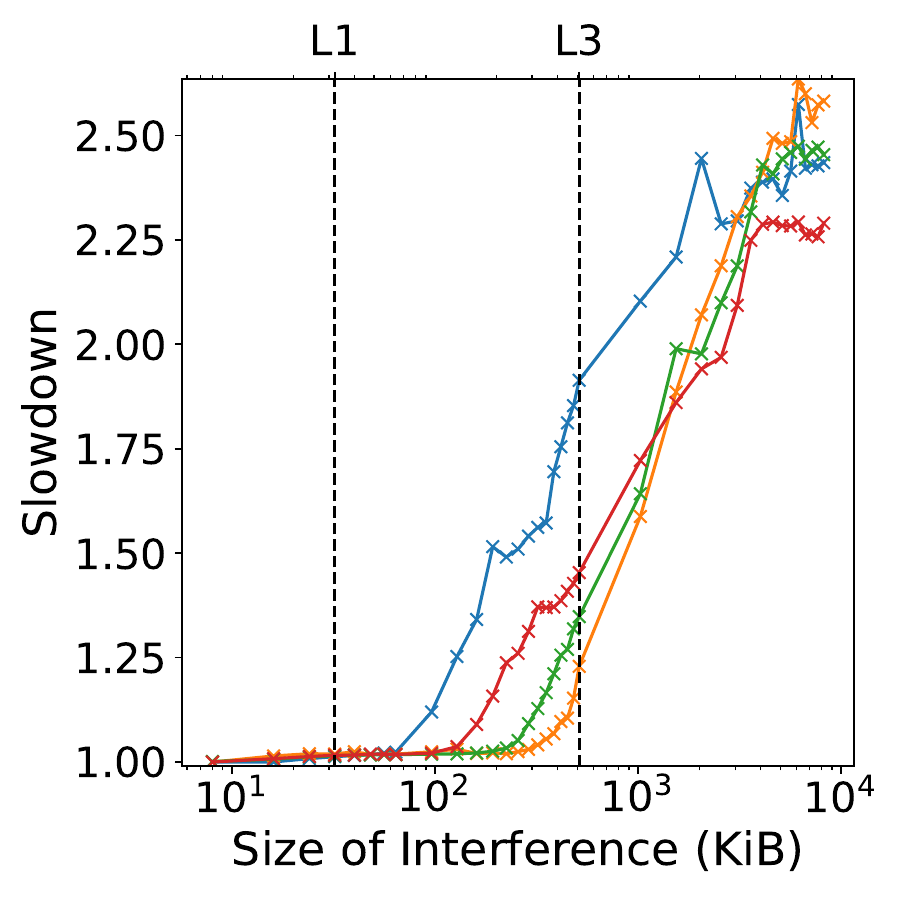}
            \captionsetup{justification=centering}
            \caption{rk3568: Set / Write}
            \label{fig:all-rk3568-set-disparity-write-cif}
        \end{subfigure}
        \hfill
        \begin{subfigure}{0.24\textwidth}
            \centering
            \includegraphics[width=\textwidth]{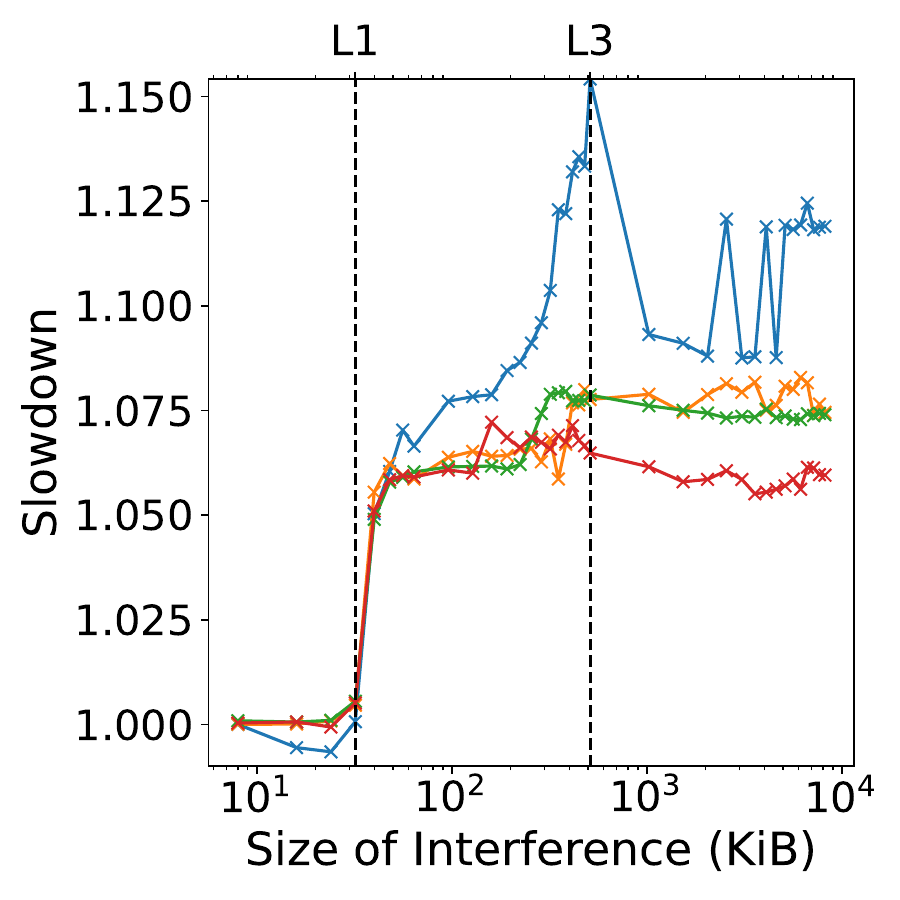}
            \captionsetup{justification=centering}
            \caption{rk3568: Set / Modify}
            \label{fig:all-rk3568-set-disparity-modify-cif}
        \end{subfigure}
        \hfill
        \begin{subfigure}{0.24\textwidth}
            \centering
            \includegraphics[width=\textwidth]{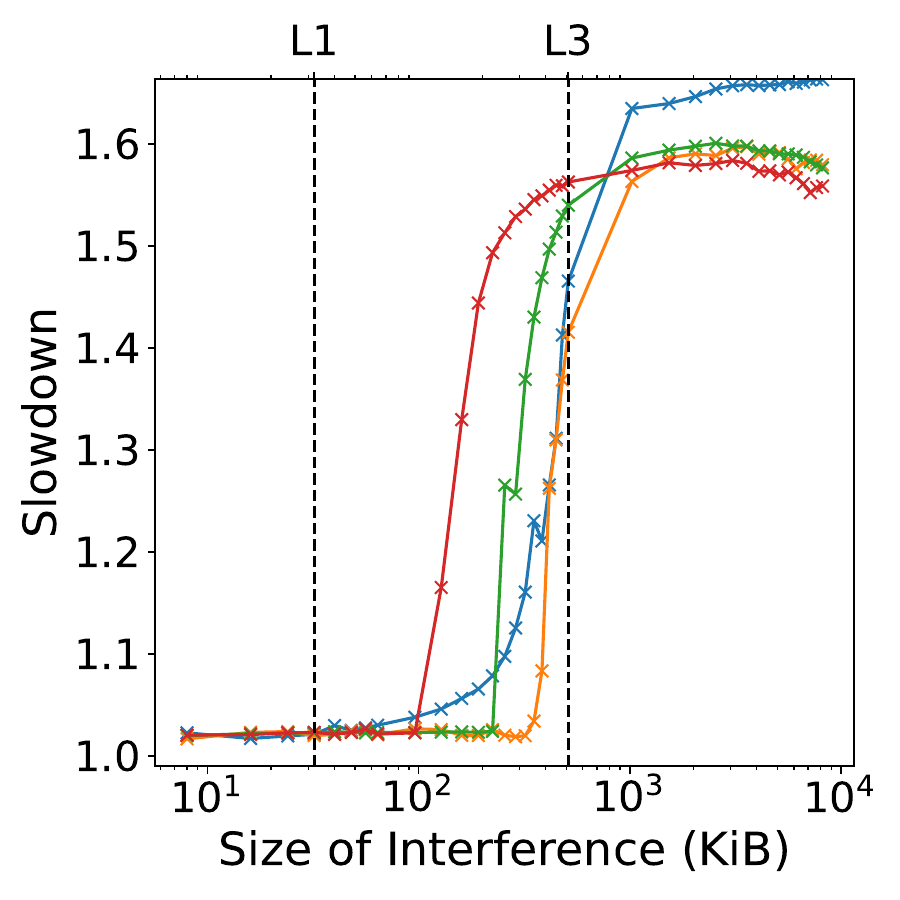}
            \captionsetup{justification=centering}
            \caption{rk3568: Set / Prefetch}
            \label{fig:all-rk3568-set-disparity-prefetch-cif}
        \end{subfigure}
        \hfill
        \begin{subfigure}{0.24\textwidth}
            \centering
            \includegraphics[width=\textwidth]{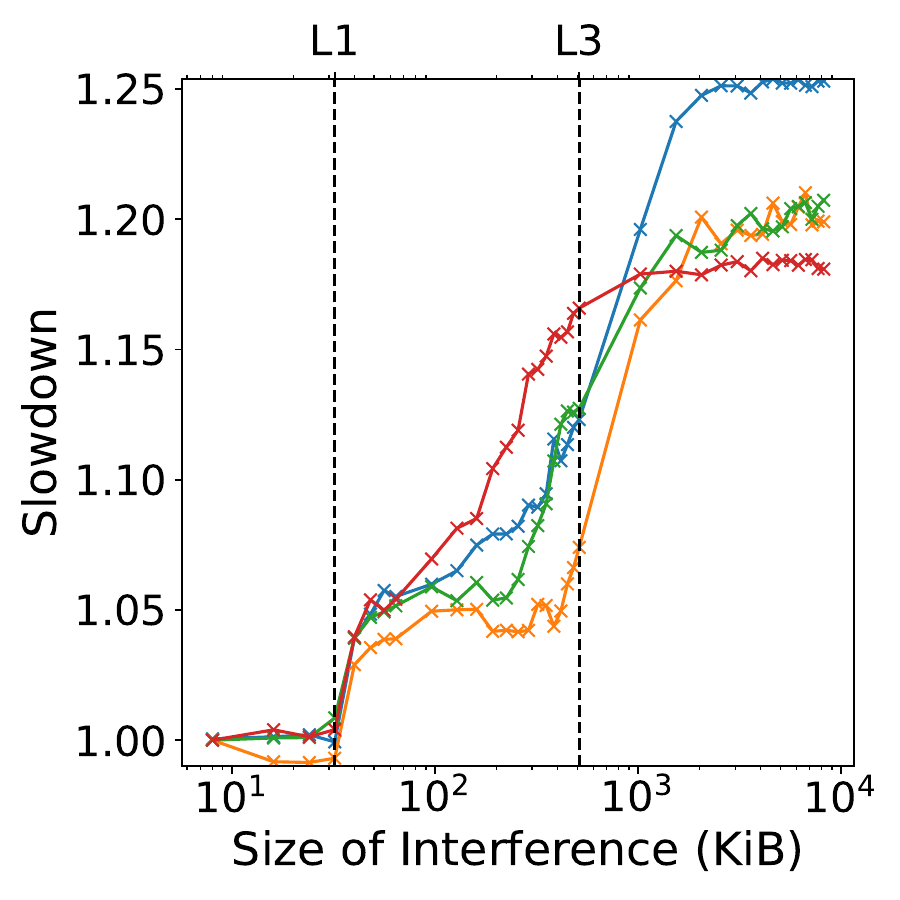}
            \captionsetup{justification=centering}
            \caption{rk3568: Way / Read}
            \label{fig:all-rk3568-way-disparity-read-cif}
        \end{subfigure}
        \hfill
        \begin{subfigure}{0.24\textwidth}
            \centering
            \includegraphics[width=\textwidth]{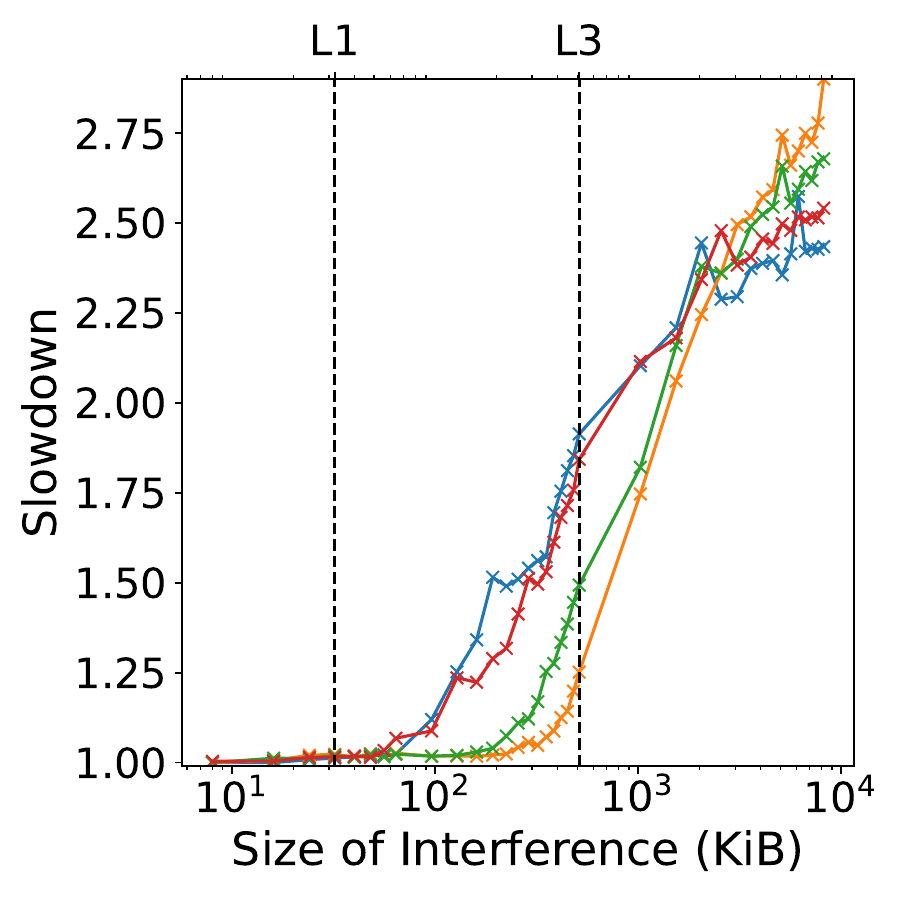}
            \captionsetup{justification=centering}
            \caption{rk3568: Way / Write}
            \label{fig:all-rk3568-way-disparity-write-cif}
        \end{subfigure}
        \hfill
        \begin{subfigure}{0.24\textwidth}
            \centering
            \includegraphics[width=\textwidth]{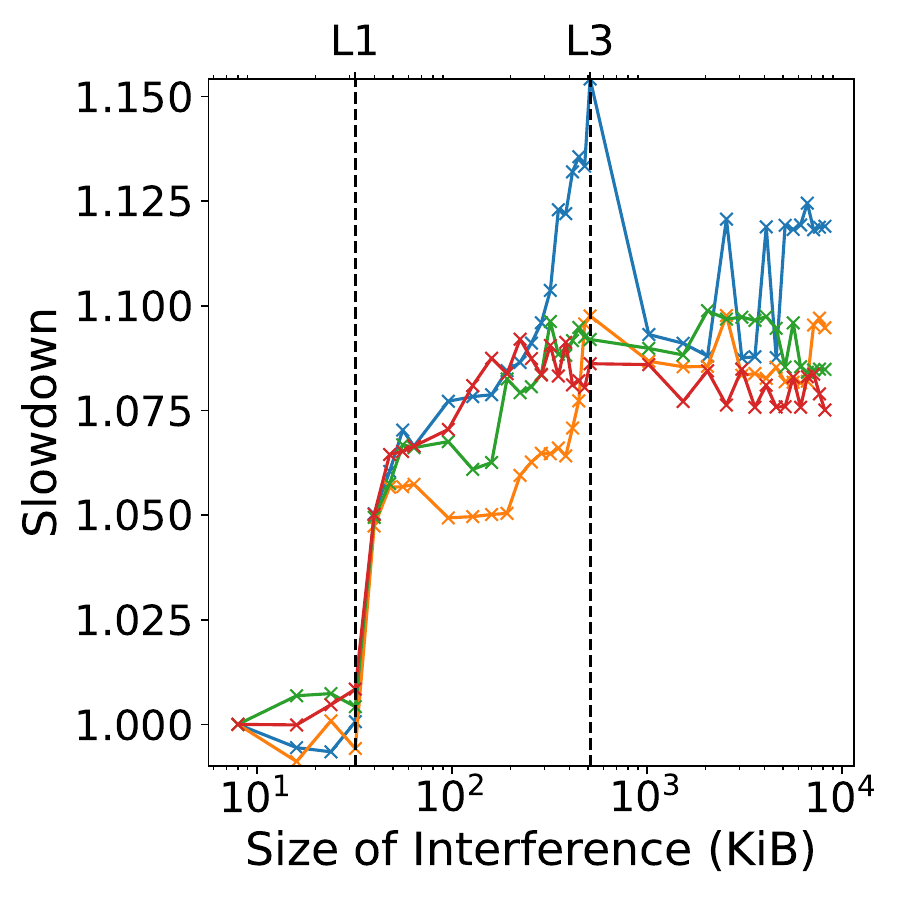}
            \captionsetup{justification=centering}
            \caption{rk3568: Way / Modify}
            \label{fig:all-rk3568-way-disparity-modify-cif}
        \end{subfigure}
        \hfill
        \begin{subfigure}{0.24\textwidth}
            \centering
            \includegraphics[width=\textwidth]{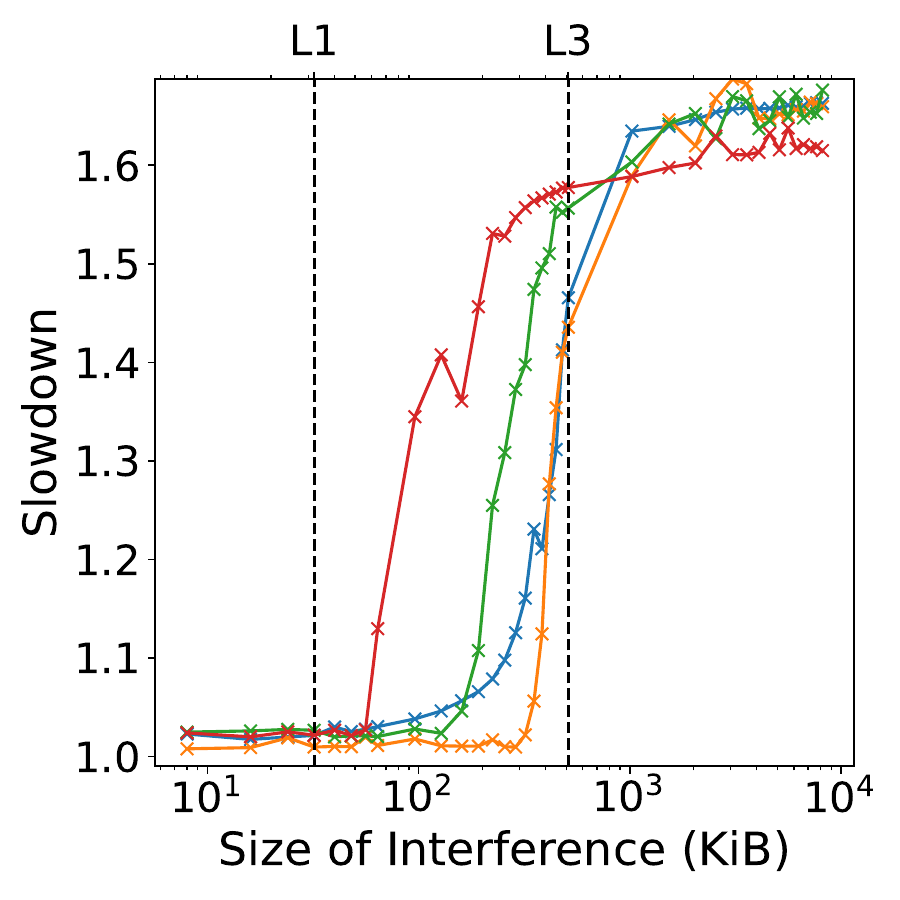}
            \captionsetup{justification=centering}
            \caption{rk3568: Way / Prefetch}
            \label{fig:all-rk3568-way-disparity-prefetch-cif}
        \end{subfigure}
        \hfill
        
        \caption{Execution Slowdown on \textit{'Disparity'} benchmark for \textit{'CIF'} dataset on \textit{'RK3568'} with Interferences and cache partitioning.}
        \label{fig:rk3568-disparity-cif}
    \end{figure}

    \begin{figure}[H]
        \begin{subfigure}{\textwidth}
            \centering
            \includegraphics[width=0.5\textwidth]{figures/set_subplot/legend.pdf}
        \end{subfigure}
        \centering
        
        \begin{subfigure}{0.24\textwidth}
            \centering
            \includegraphics[width=\textwidth]{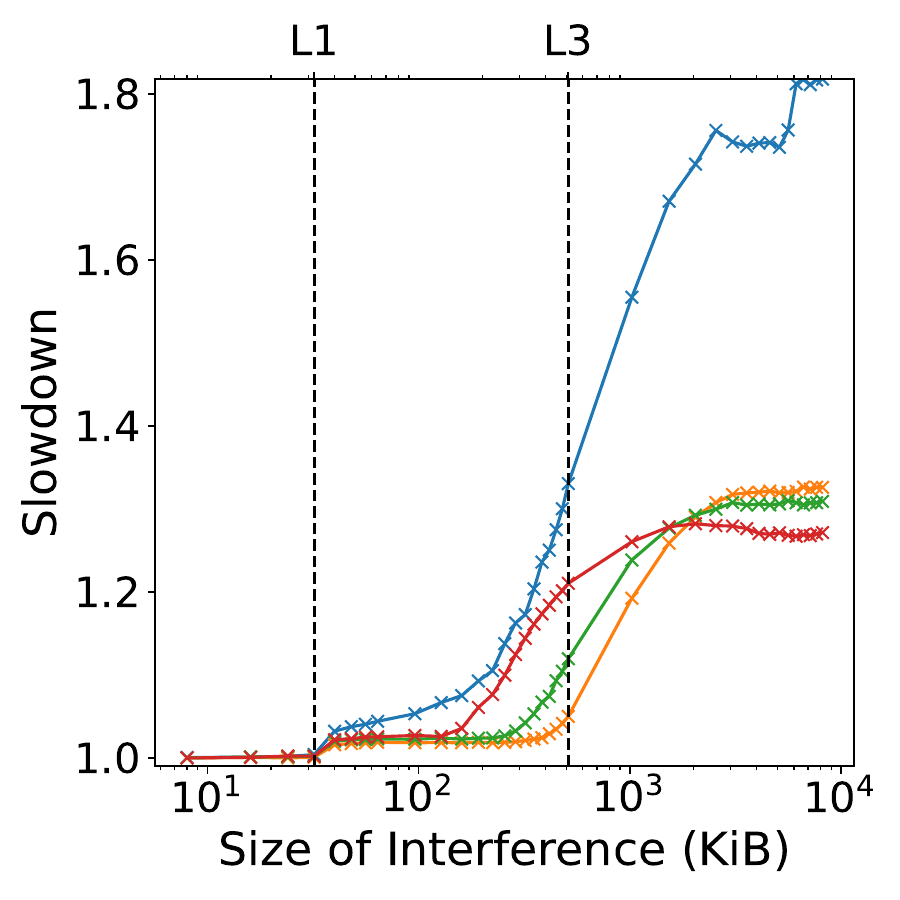}
            \captionsetup{justification=centering}
            \caption{rk3568: Set / Read}
            \label{fig:all-rk3568-set-mser-read-cif}
        \end{subfigure}
        \hfill
        \begin{subfigure}{0.24\textwidth}
            \centering
            \includegraphics[width=\textwidth]{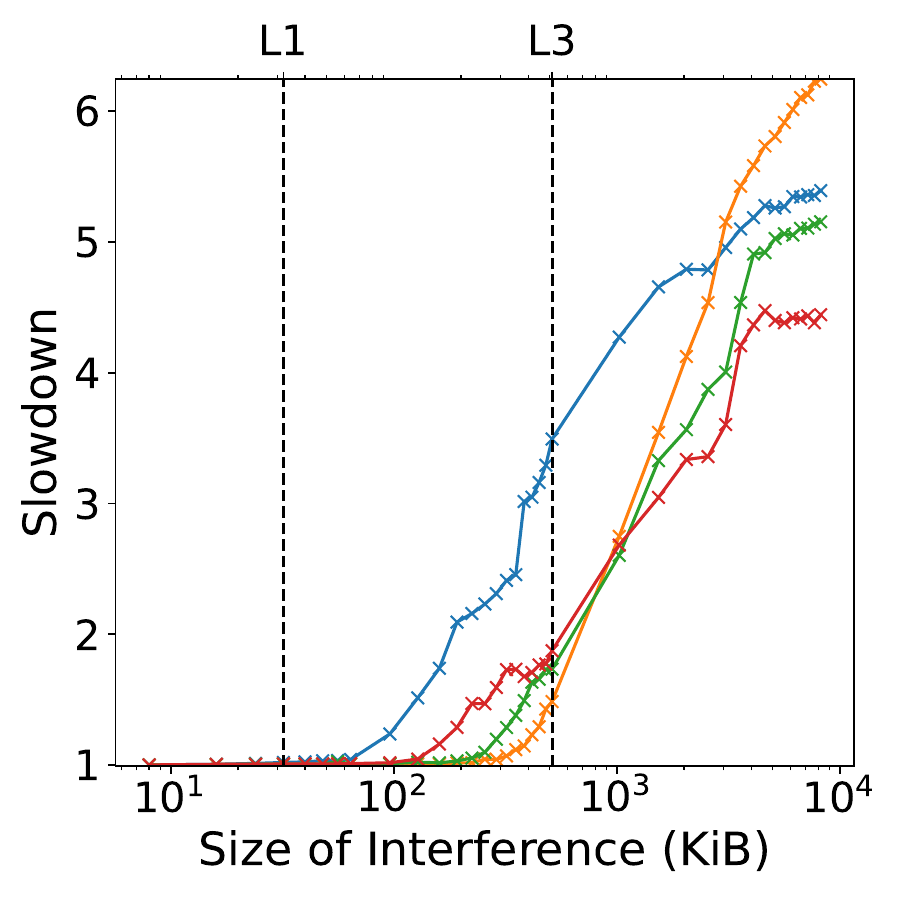}
            \captionsetup{justification=centering}
            \caption{rk3568: Set / Write}
            \label{fig:all-rk3568-set-mser-write-cif}
        \end{subfigure}
        \hfill
        \begin{subfigure}{0.24\textwidth}
            \centering
            \includegraphics[width=\textwidth]{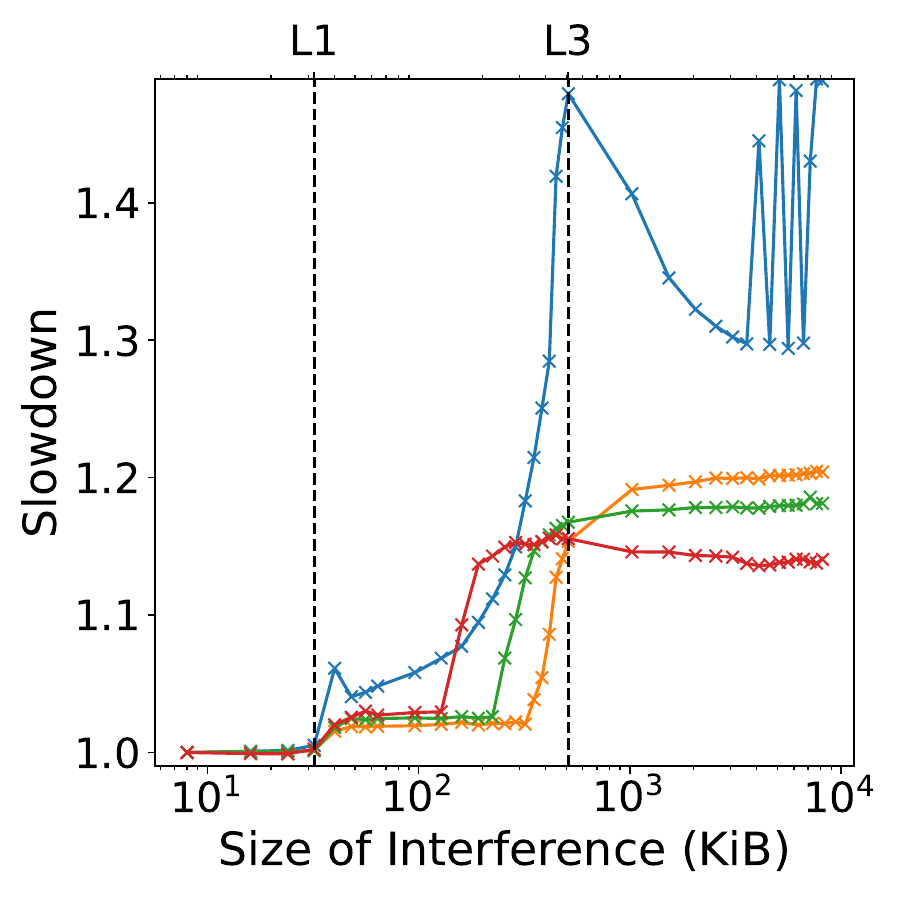}
            \captionsetup{justification=centering}
            \caption{rk3568: Set / Modify}
            \label{fig:all-rk3568-set-mser-modify-cif}
        \end{subfigure}
        \hfill
        \begin{subfigure}{0.24\textwidth}
            \centering
            \includegraphics[width=\textwidth]{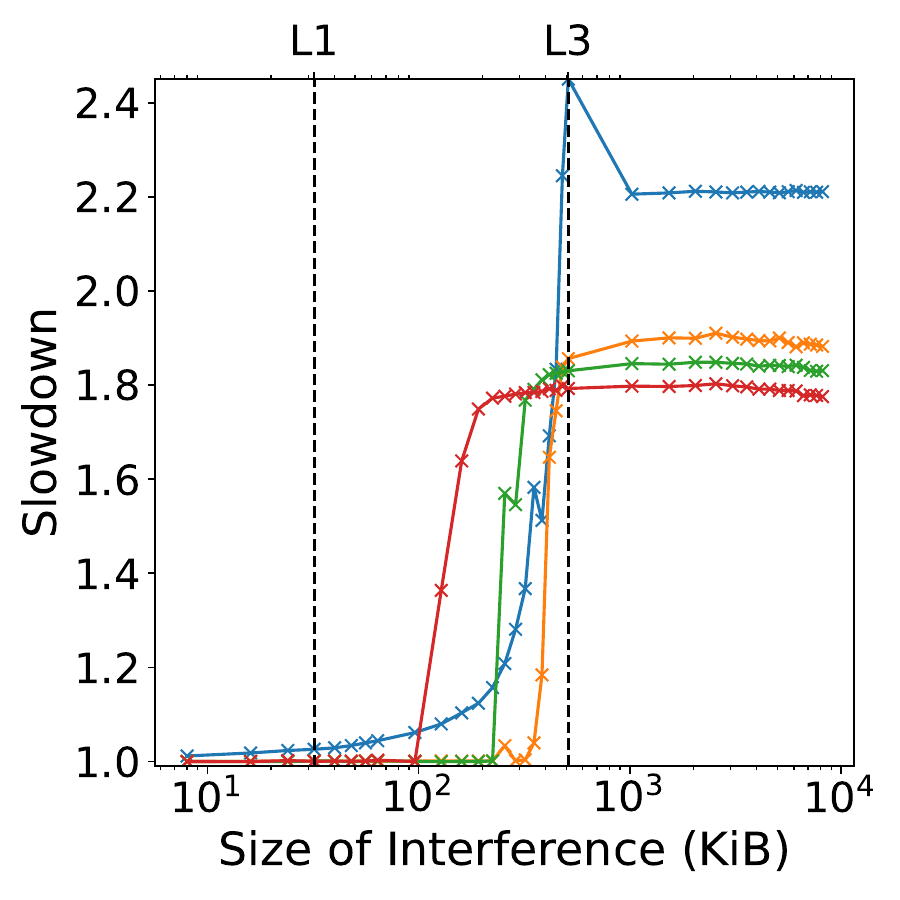}
            \captionsetup{justification=centering}
            \caption{rk3568: Set / Prefetch}
            \label{fig:all-rk3568-set-mser-prefetch-cif}
        \end{subfigure}
        \hfill
        \begin{subfigure}{0.24\textwidth}
            \centering
            \includegraphics[width=\textwidth]{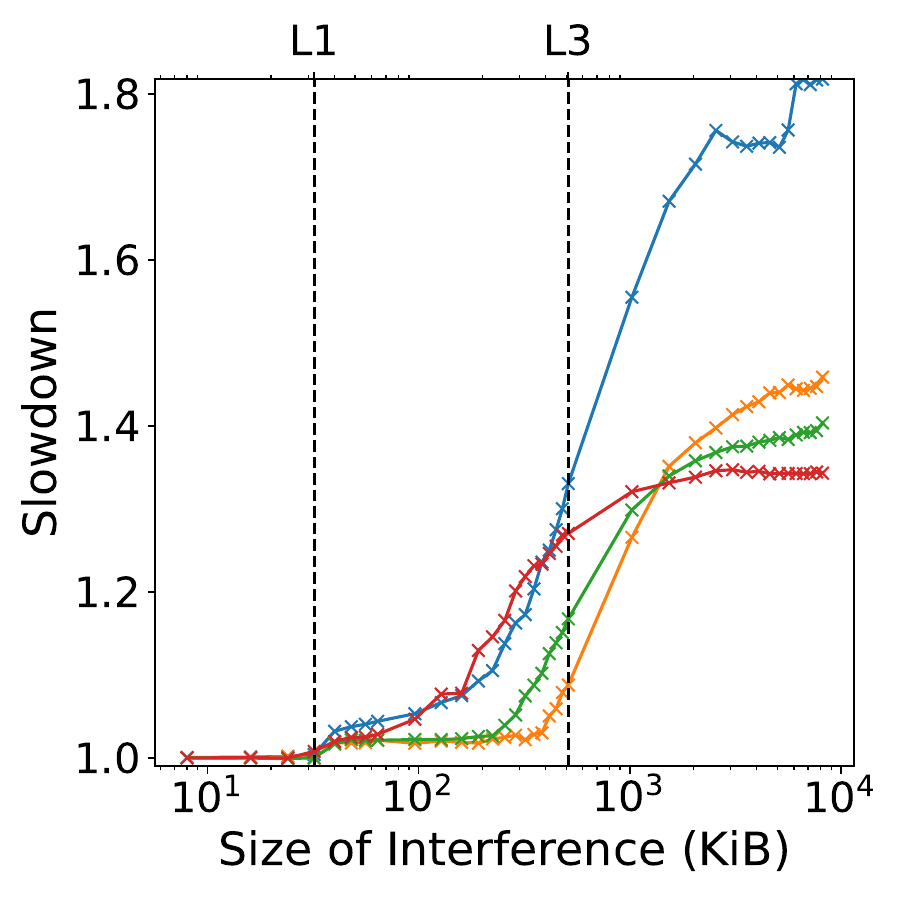}
            \captionsetup{justification=centering}
            \caption{rk3568: Way / Read}
            \label{fig:all-rk3568-way-mser-read-cif}
        \end{subfigure}
        \hfill
        \begin{subfigure}{0.24\textwidth}
            \centering
            \includegraphics[width=\textwidth]{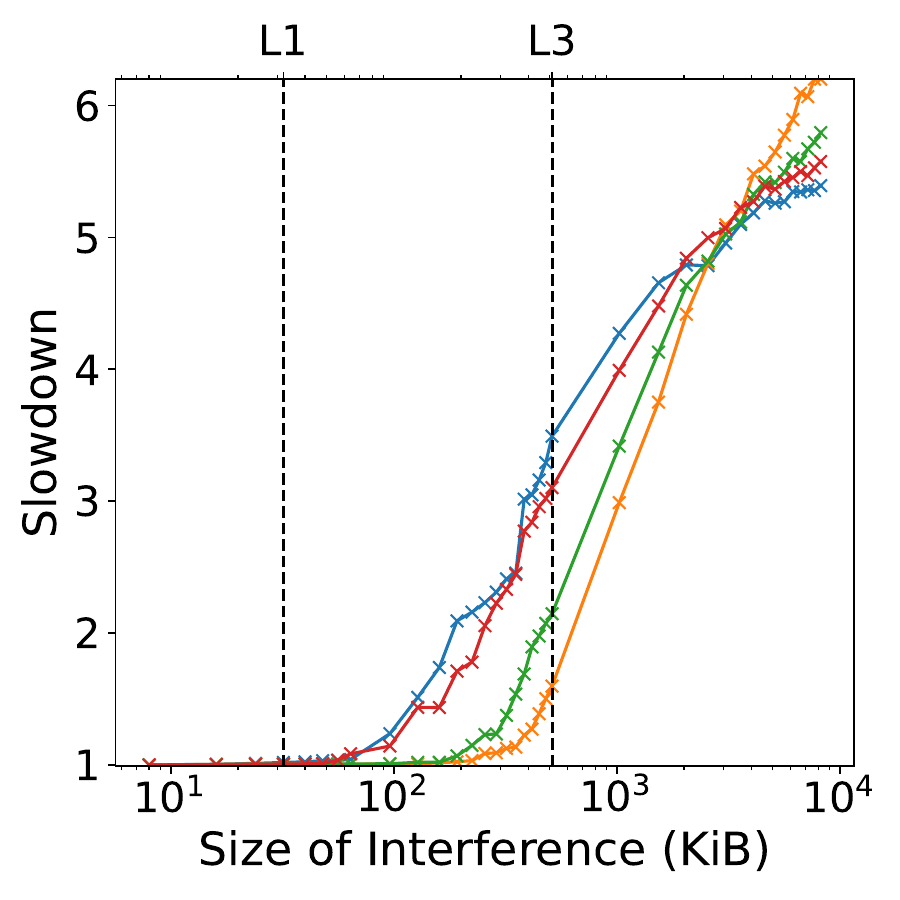}
            \captionsetup{justification=centering}
            \caption{rk3568: Way / Write}
            \label{fig:all-rk3568-way-mser-write-cif}
        \end{subfigure}
        \hfill
        \begin{subfigure}{0.24\textwidth}
            \centering
            \includegraphics[width=\textwidth]{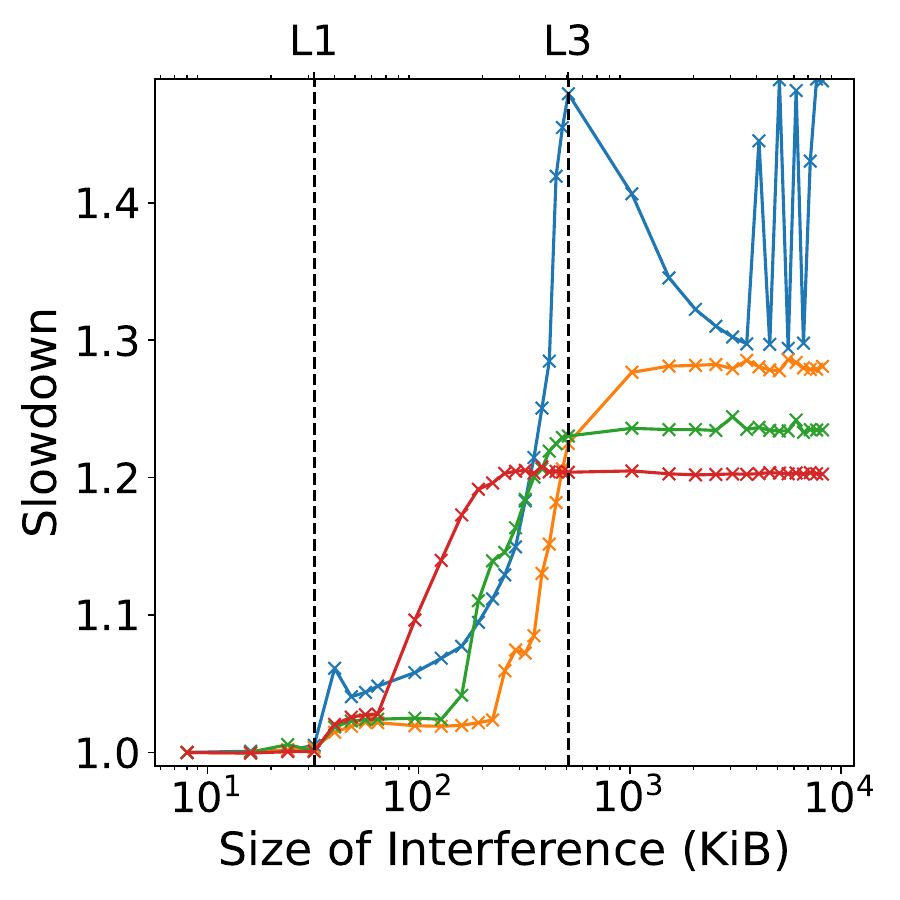}
            \captionsetup{justification=centering}
            \caption{rk3568: Way / Modify}
            \label{fig:all-rk3568-way-mser-modify-cif}
        \end{subfigure}
        \hfill
        \begin{subfigure}{0.24\textwidth}
            \centering
            \includegraphics[width=\textwidth]{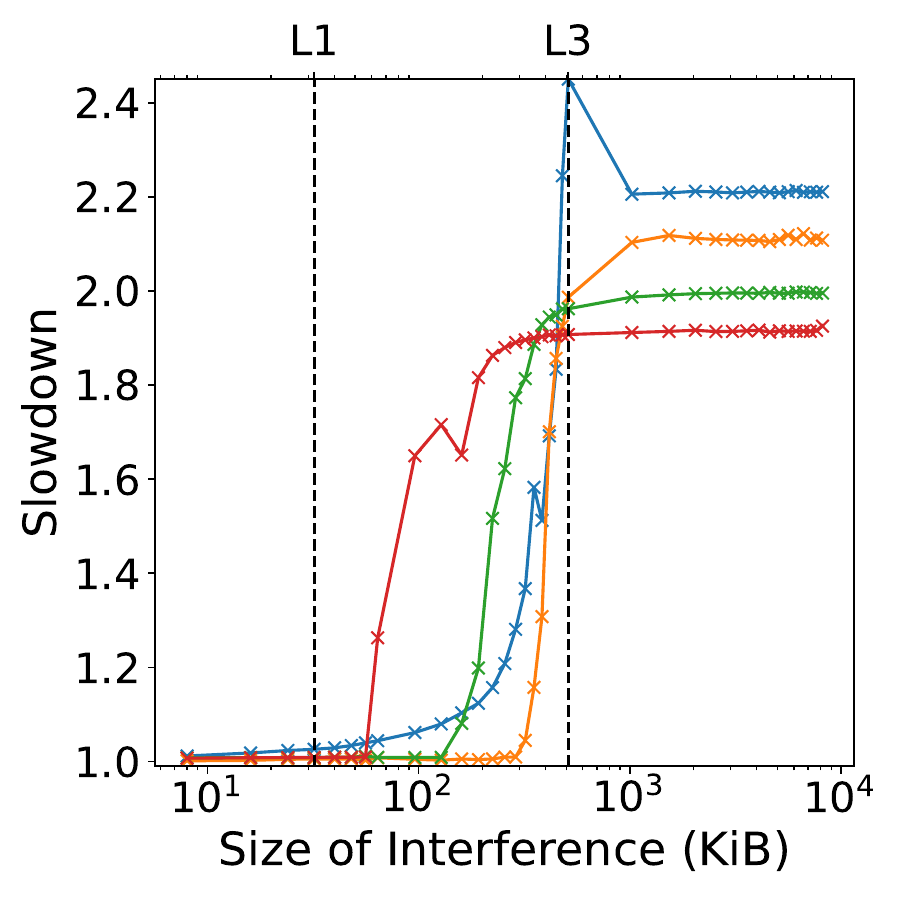}
            \captionsetup{justification=centering}
            \caption{rk3568: Way / Prefetch}
            \label{fig:all-rk3568-way-mser-prefetch-cif}
        \end{subfigure}
        \hfill
        
        \caption{Execution Slowdown on \textit{'Mser'} benchmark for \textit{'CIF'} dataset on \textit{'RK3568'} with Interferences and cache partitioning.}
        \label{fig:rk3568-mser-cif}
    \end{figure}

    \begin{figure}[H]
        \begin{subfigure}{\textwidth}
            \centering
            \includegraphics[width=0.5\textwidth]{figures/set_subplot/legend.pdf}
        \end{subfigure}
        \centering
        
        \begin{subfigure}{0.24\textwidth}
            \centering
            \includegraphics[width=\textwidth]{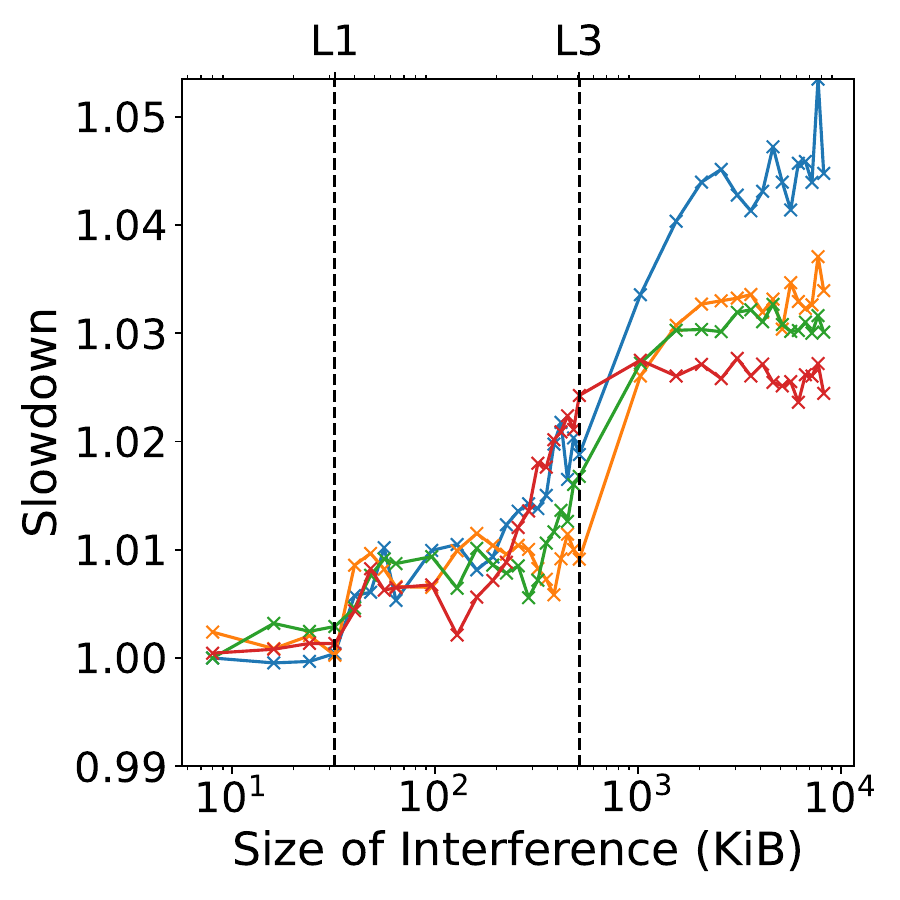}
            \captionsetup{justification=centering}
            \caption{rk3568: Set / Read}
            \label{fig:all-rk3568-set-tracking-read-cif}
        \end{subfigure}
        \hfill
        \begin{subfigure}{0.24\textwidth}
            \centering
            \includegraphics[width=\textwidth]{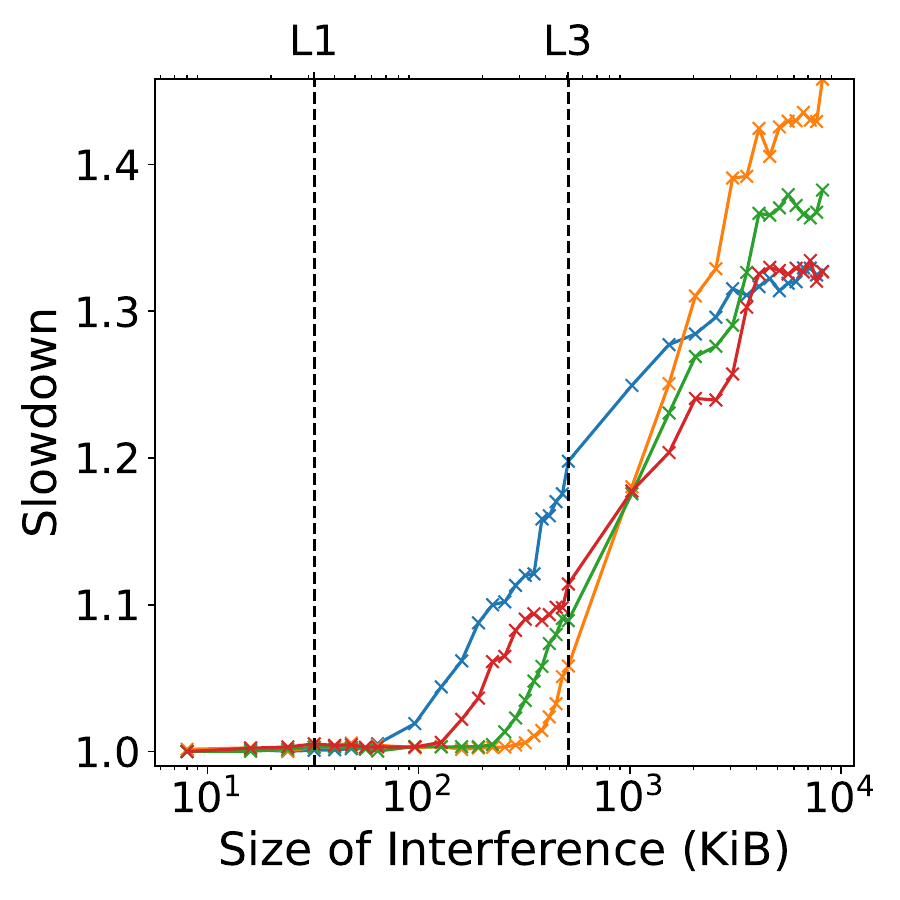}
            \captionsetup{justification=centering}
            \caption{rk3568: Set / Write}
            \label{fig:all-rk3568-set-tracking-write-cif}
        \end{subfigure}
        \hfill
        \begin{subfigure}{0.24\textwidth}
            \centering
            \includegraphics[width=\textwidth]{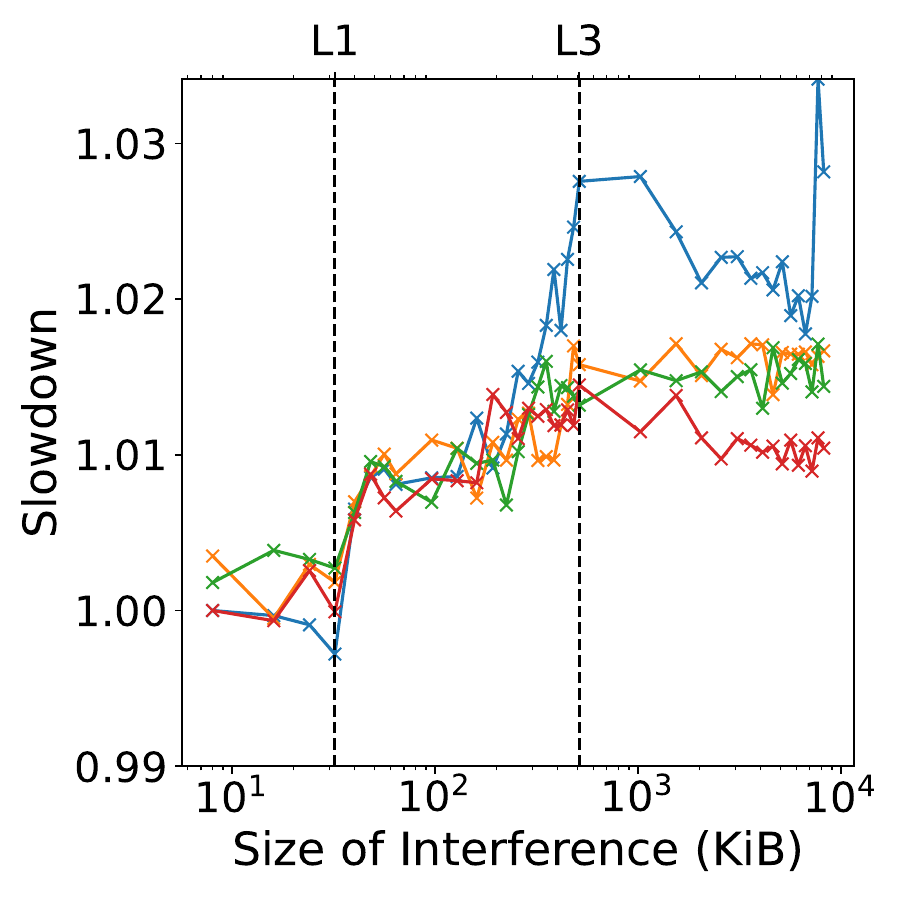}
            \captionsetup{justification=centering}
            \caption{rk3568: Set / Modify}
            \label{fig:all-rk3568-set-tracking-modify-cif}
        \end{subfigure}
        \hfill
        \begin{subfigure}{0.24\textwidth}
            \centering
            \includegraphics[width=\textwidth]{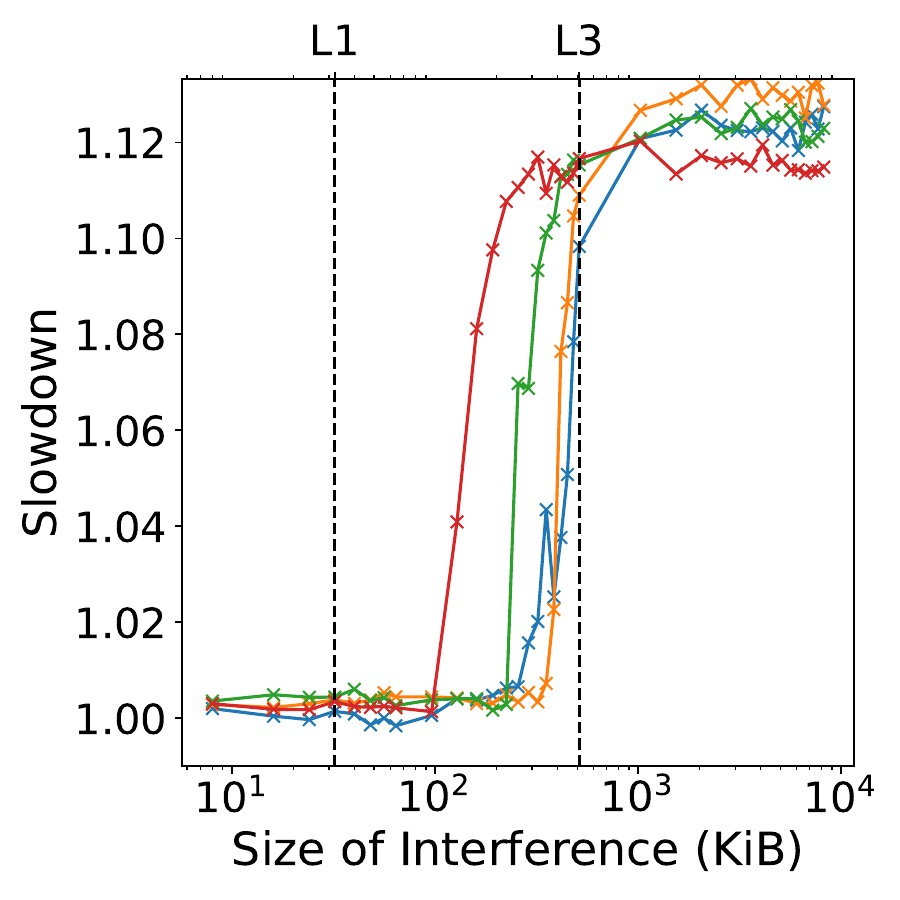}
            \captionsetup{justification=centering}
            \caption{rk3568: Set / Prefetch}
            \label{fig:all-rk3568-set-tracking-prefetch-cif}
        \end{subfigure}
        \hfill
        \begin{subfigure}{0.24\textwidth}
            \centering
            \includegraphics[width=\textwidth]{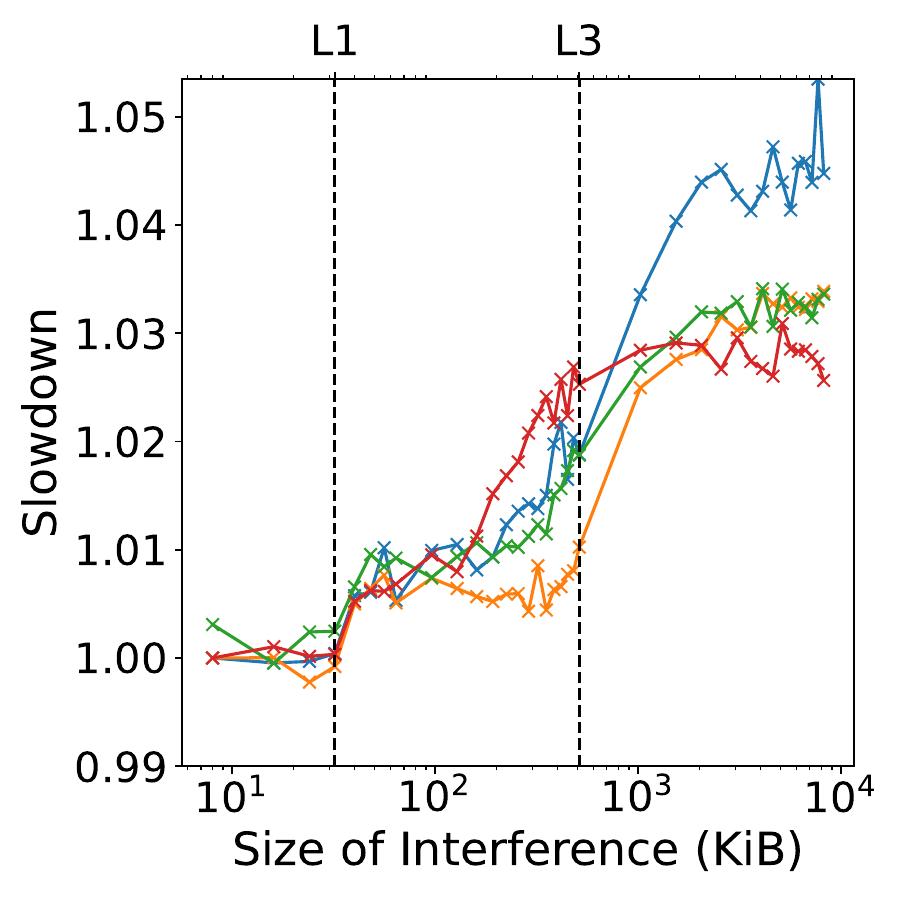}
            \captionsetup{justification=centering}
            \caption{rk3568: Way / Read}
            \label{fig:all-rk3568-way-tracking-read-cif}
        \end{subfigure}
        \hfill
        \begin{subfigure}{0.24\textwidth}
            \centering
            \includegraphics[width=\textwidth]{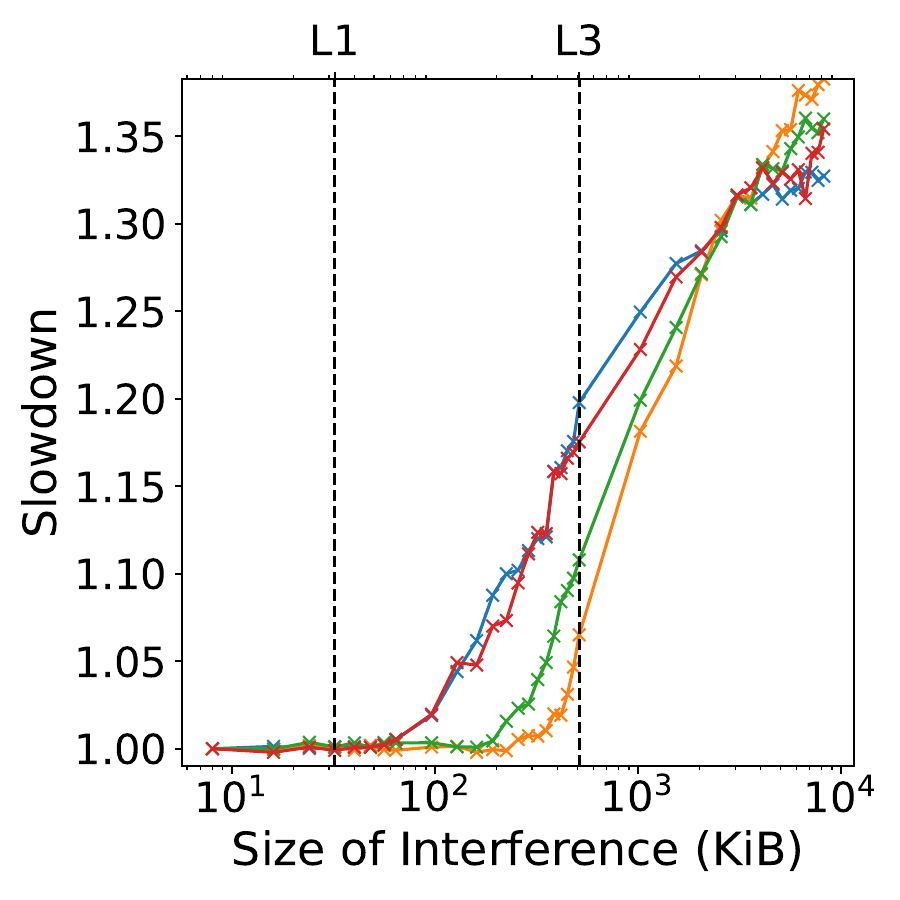}
            \captionsetup{justification=centering}
            \caption{rk3568: Way / Write}
            \label{fig:all-rk3568-way-tracking-write-cif}
        \end{subfigure}
        \hfill
        \begin{subfigure}{0.24\textwidth}
            \centering
            \includegraphics[width=\textwidth]{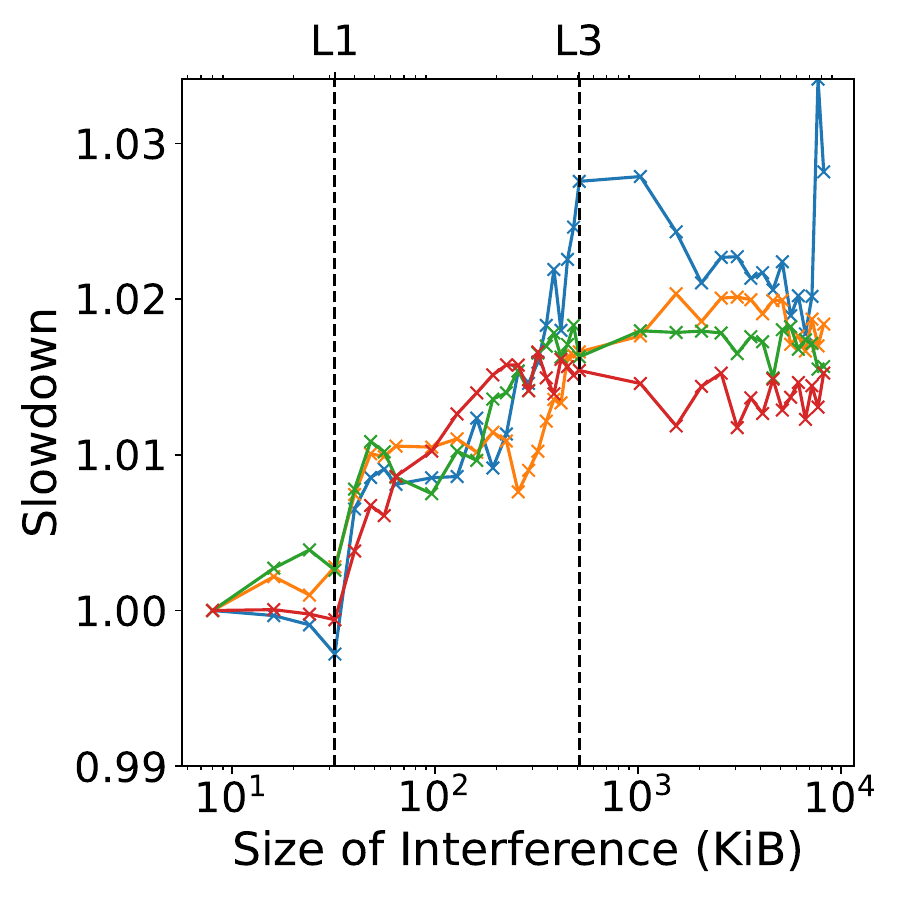}
            \captionsetup{justification=centering}
            \caption{rk3568: Way / Modify}
            \label{fig:all-rk3568-way-tracking-modify-cif}
        \end{subfigure}
        \hfill
        \begin{subfigure}{0.24\textwidth}
            \centering
            \includegraphics[width=\textwidth]{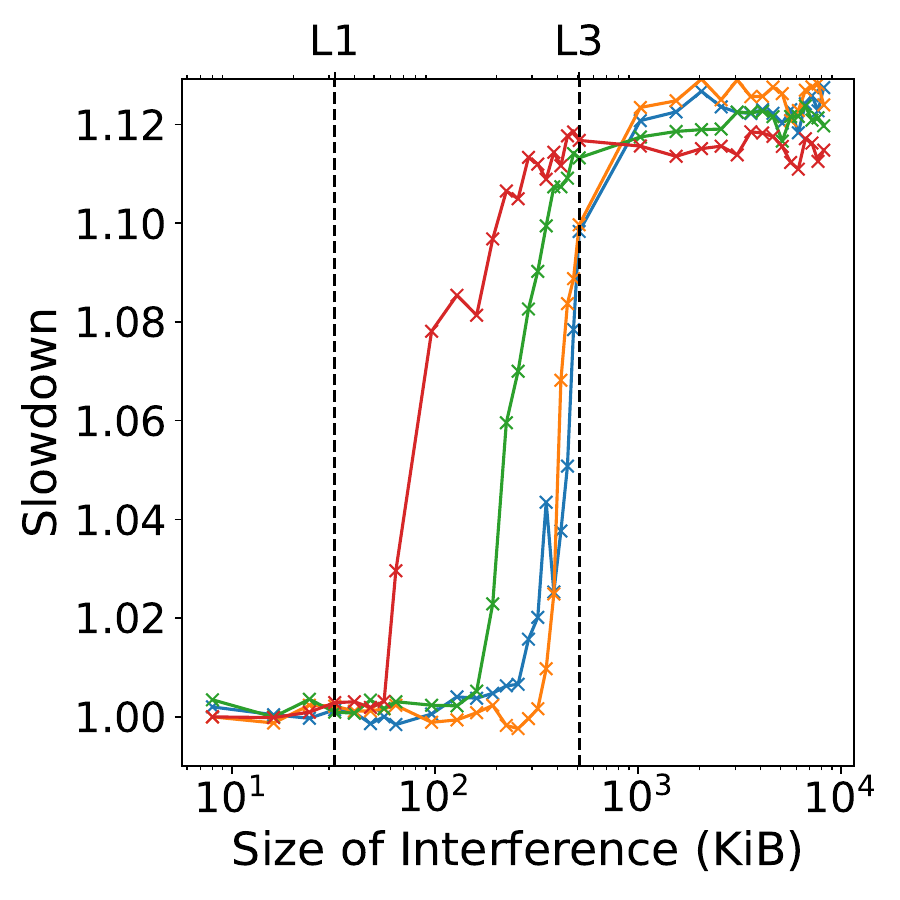}
            \captionsetup{justification=centering}
            \caption{rk3568: Way / Prefetch}
            \label{fig:all-rk3568-way-tracking-prefetch-cif}
        \end{subfigure}
        \hfill
        
        \caption{Execution Slowdown on \textit{'Tracking'} benchmark for \textit{'CIF'} dataset on \textit{'RK3568'} with Interferences and cache partitioning.}
        \label{fig:rk3568-tracking-cif}
    \end{figure}

    \begin{figure}[H]
        \begin{subfigure}{\textwidth}
            \centering
            \includegraphics[width=0.5\textwidth]{figures/set_subplot/legend.pdf}
        \end{subfigure}
        \centering
        
        \begin{subfigure}{0.24\textwidth}
            \centering
            \includegraphics[width=\textwidth]{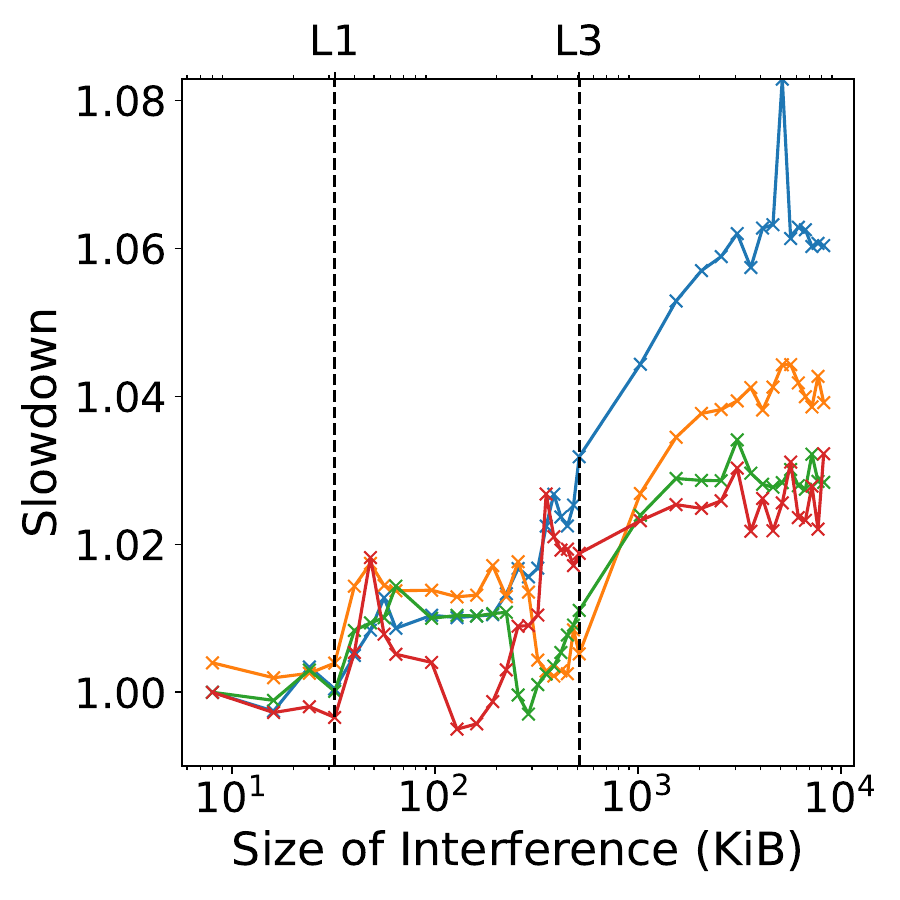}
            \captionsetup{justification=centering}
            \caption{rk3568: Set / Read}
            \label{fig:all-rk3568-set-sift-read-cif}
        \end{subfigure}
        \hfill
        \begin{subfigure}{0.24\textwidth}
            \centering
            \includegraphics[width=\textwidth]{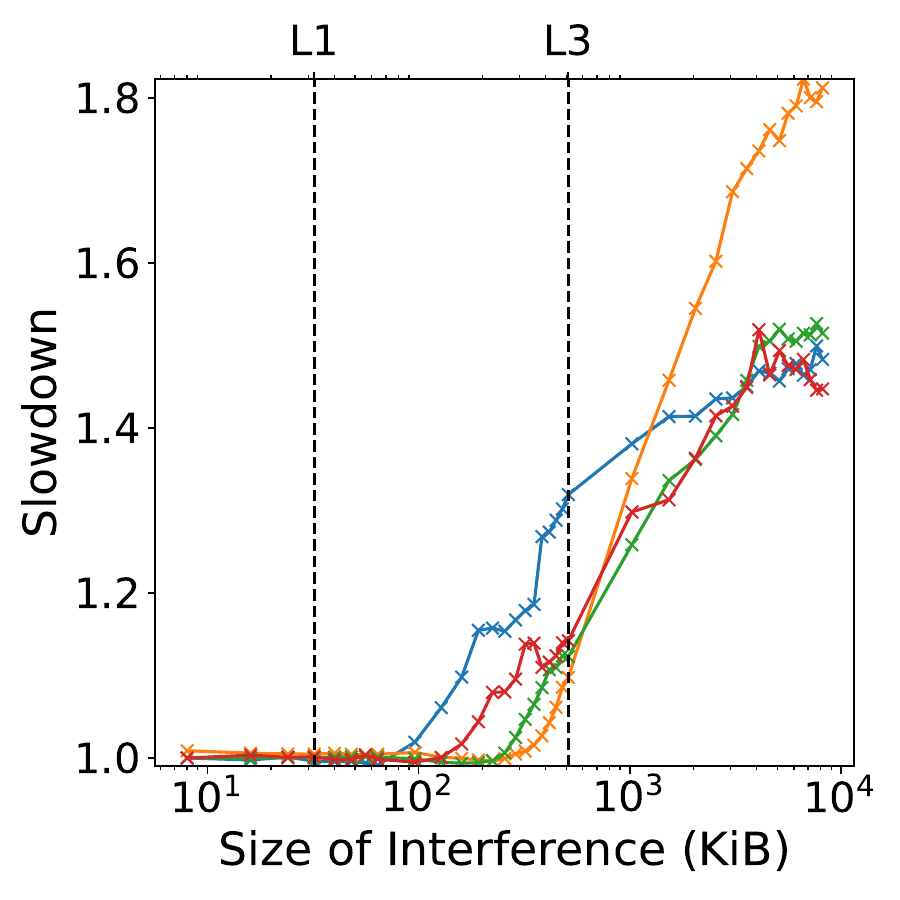}
            \captionsetup{justification=centering}
            \caption{rk3568: Set / Write}
            \label{fig:all-rk3568-set-sift-write-cif}
        \end{subfigure}
        \hfill
        \begin{subfigure}{0.24\textwidth}
            \centering
            \includegraphics[width=\textwidth]{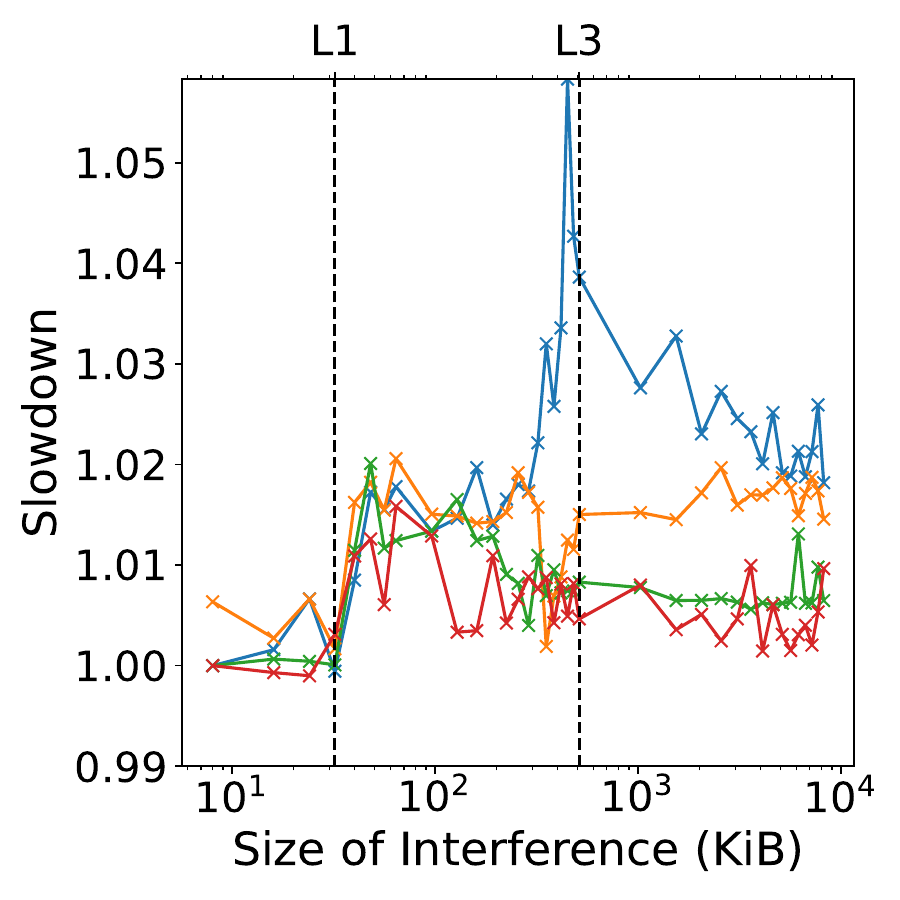}
            \captionsetup{justification=centering}
            \caption{rk3568: Set / Modify}
            \label{fig:all-rk3568-set-sift-modify-cif}
        \end{subfigure}
        \hfill
        \begin{subfigure}{0.24\textwidth}
            \centering
            \includegraphics[width=\textwidth]{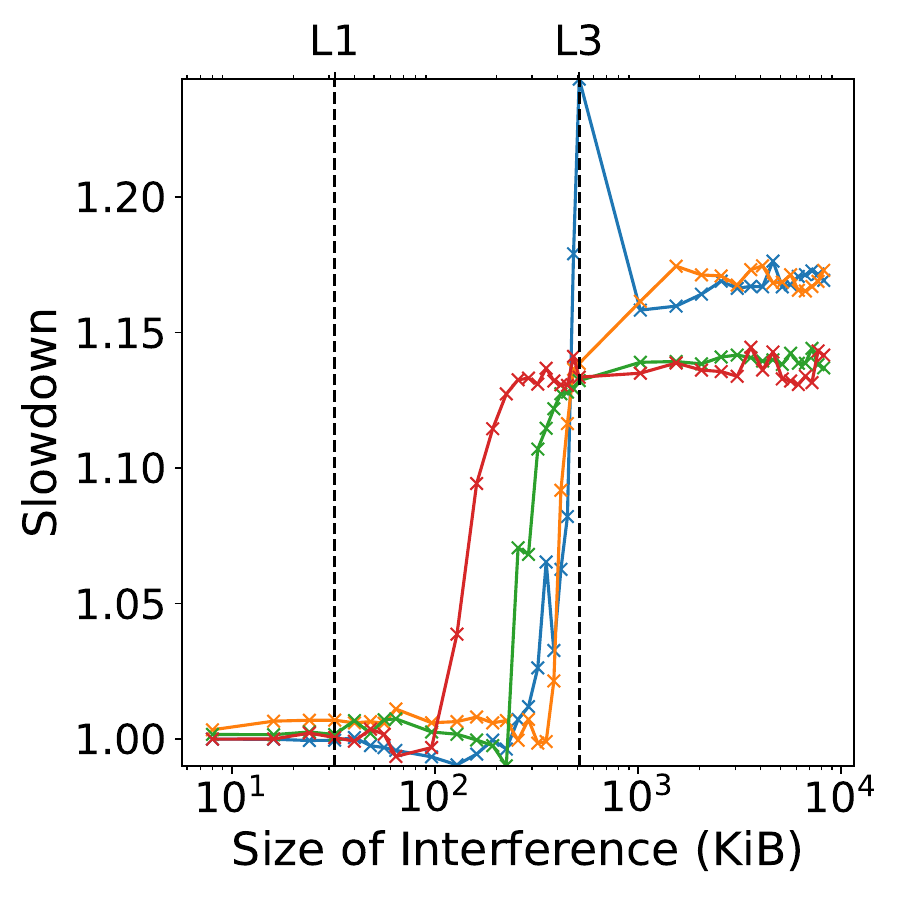}
            \captionsetup{justification=centering}
            \caption{rk3568: Set / Prefetch}
            \label{fig:all-rk3568-set-sift-prefetch-cif}
        \end{subfigure}
        \hfill
        \begin{subfigure}{0.24\textwidth}
            \centering
            \includegraphics[width=\textwidth]{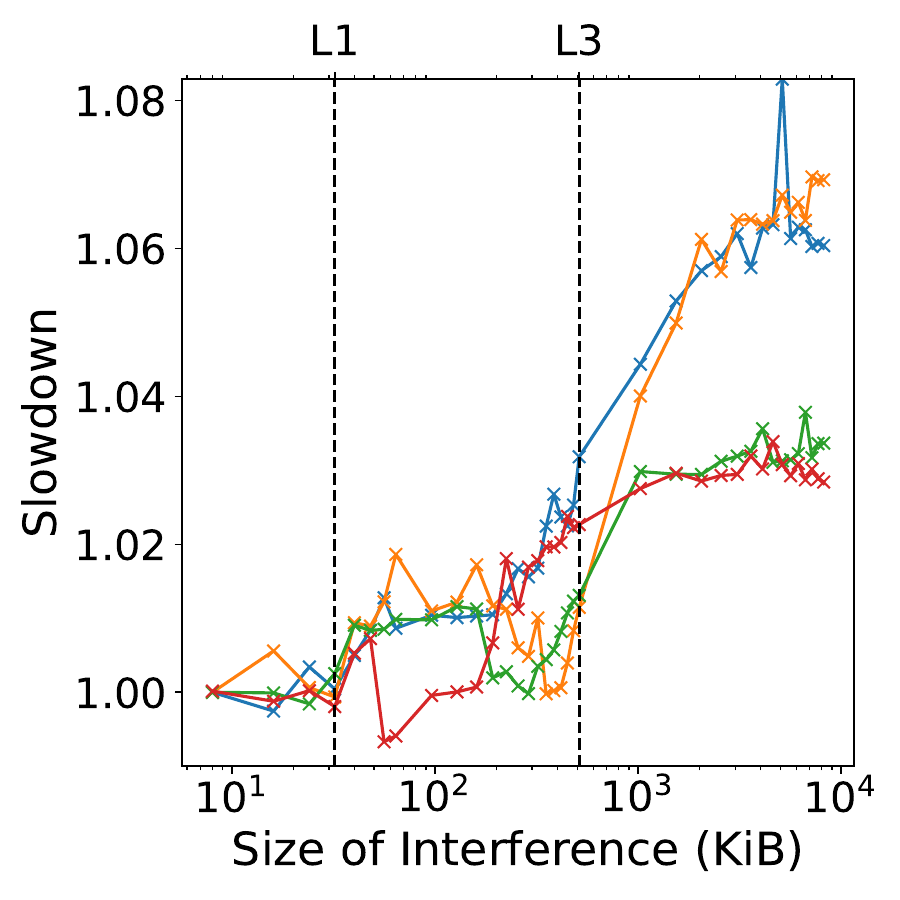}
            \captionsetup{justification=centering}
            \caption{rk3568: Way / Read}
            \label{fig:all-rk3568-way-sift-read-cif}
        \end{subfigure}
        \hfill
        \begin{subfigure}{0.24\textwidth}
            \centering
            \includegraphics[width=\textwidth]{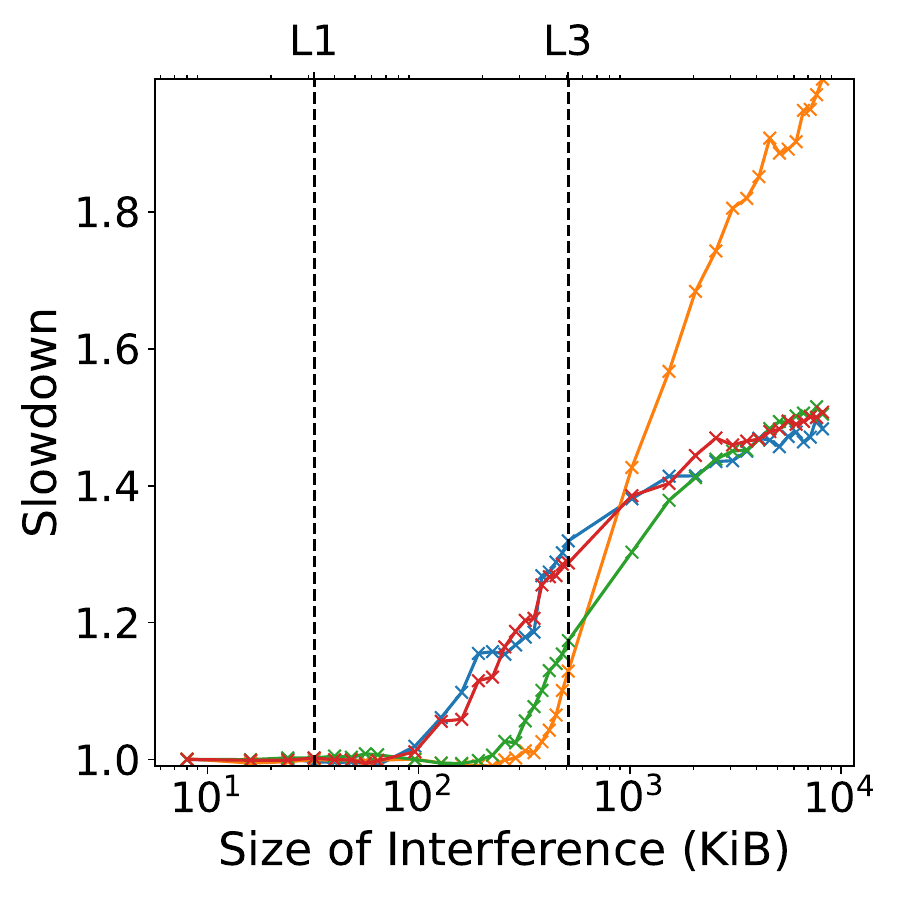}
            \captionsetup{justification=centering}
            \caption{rk3568: Way / Write}
            \label{fig:all-rk3568-way-sift-write-cif}
        \end{subfigure}
        \hfill
        \begin{subfigure}{0.24\textwidth}
            \centering
            \includegraphics[width=\textwidth]{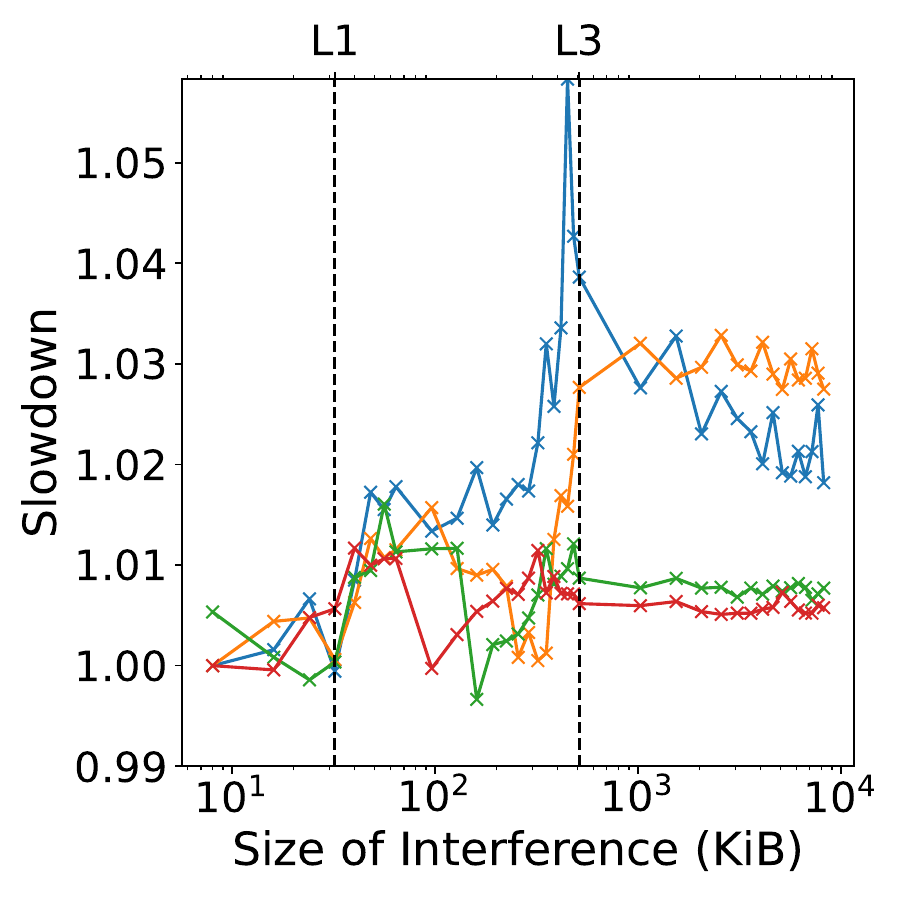}
            \captionsetup{justification=centering}
            \caption{rk3568: Way / Modify}
            \label{fig:all-rk3568-way-sift-modify-cif}
        \end{subfigure}
        \hfill
        \begin{subfigure}{0.24\textwidth}
            \centering
            \includegraphics[width=\textwidth]{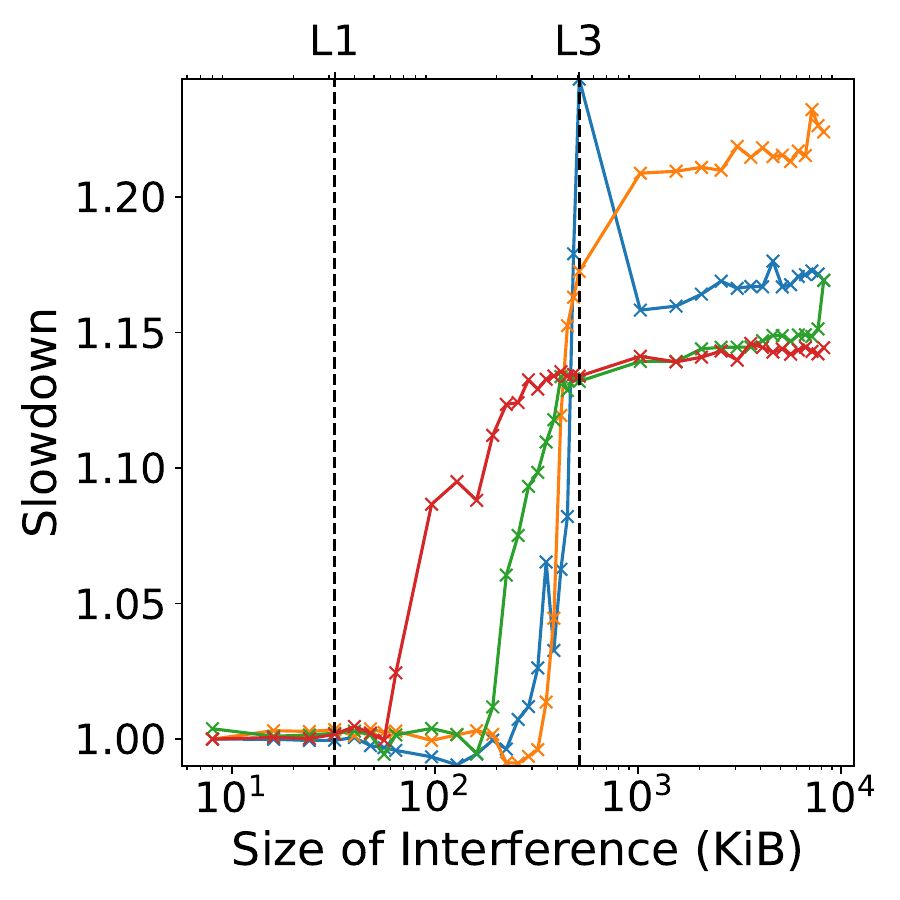}
            \captionsetup{justification=centering}
            \caption{rk3568: Way / Prefetch}
            \label{fig:all-rk3568-way-sift-prefetch-cif}
        \end{subfigure}
        \hfill
        
        \caption{Execution Slowdown on \textit{'Sift'} benchmark for \textit{'CIF'} dataset on \textit{'RK3568'} with Interferences and cache partitioning.}
        \label{fig:rk3568-sift-cif}
    \end{figure}

    \begin{figure}[H]
        \begin{subfigure}{\textwidth}
            \centering
            \includegraphics[width=0.5\textwidth]{figures/set_subplot/legend.pdf}
        \end{subfigure}
        \centering
        
        \begin{subfigure}{0.24\textwidth}
            \centering
            \includegraphics[width=\textwidth]{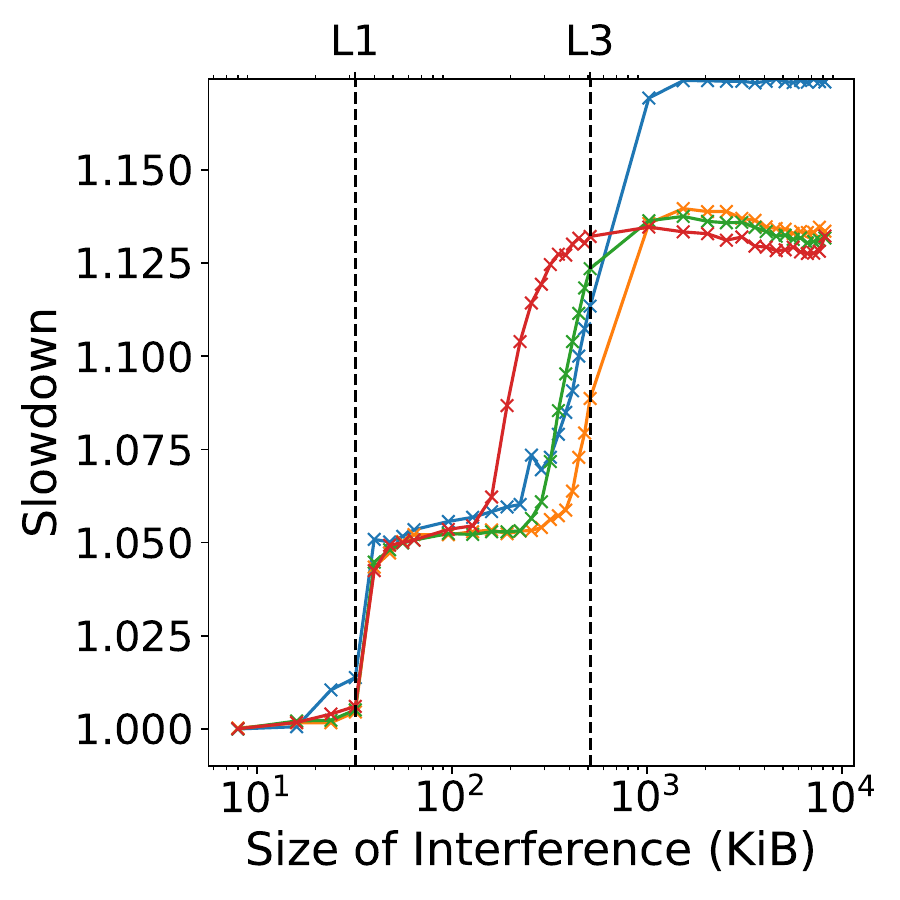}
            \captionsetup{justification=centering}
            \caption{rk3568: Set / Read}
            \label{fig:all-rk3568-set-disparity-read-vga}
        \end{subfigure}
        \hfill
        \begin{subfigure}{0.24\textwidth}
            \centering
            \includegraphics[width=\textwidth]{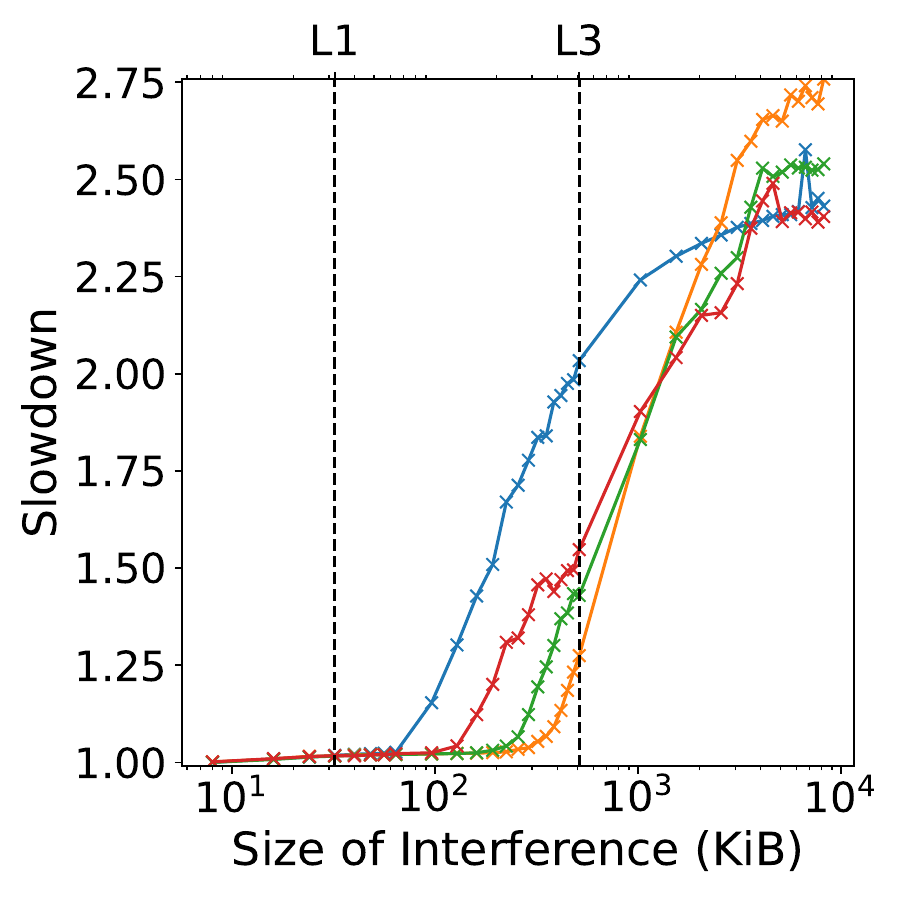}
            \captionsetup{justification=centering}
            \caption{rk3568: Set / Write}
            \label{fig:all-rk3568-set-disparity-write-vga}
        \end{subfigure}
        \hfill
        \begin{subfigure}{0.24\textwidth}
            \centering
            \includegraphics[width=\textwidth]{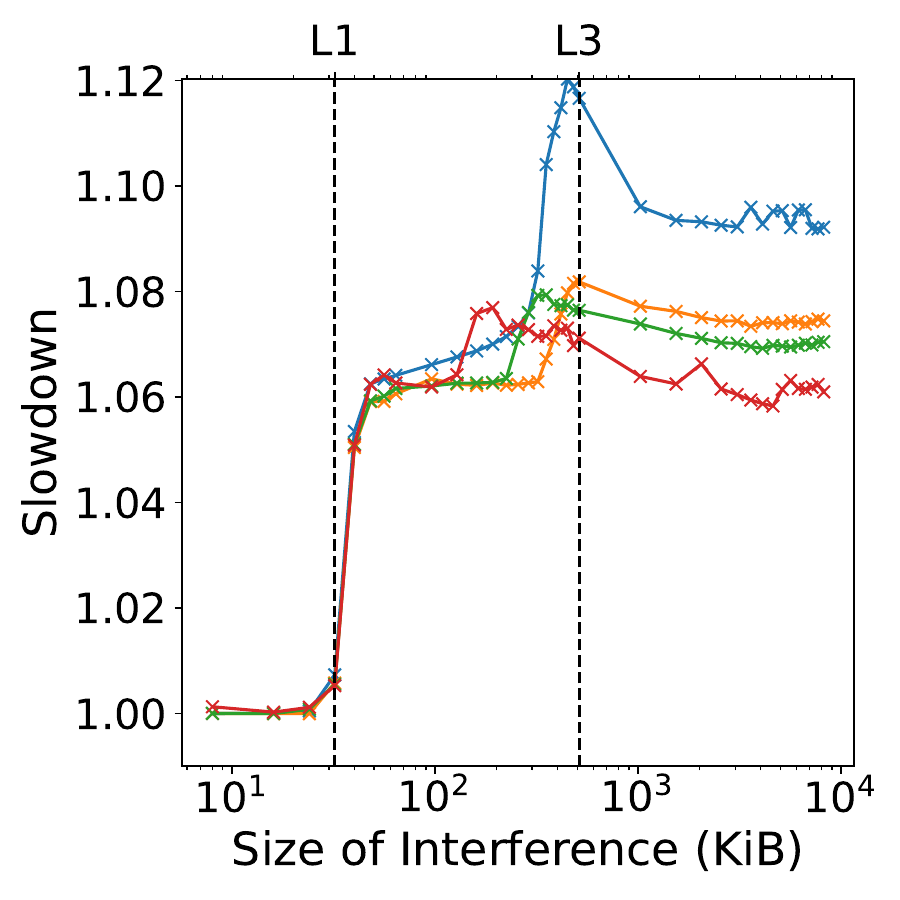}
            \captionsetup{justification=centering}
            \caption{rk3568: Set / Modify}
            \label{fig:all-rk3568-set-disparity-modify-vga}
        \end{subfigure}
        \hfill
        \begin{subfigure}{0.24\textwidth}
            \centering
            \includegraphics[width=\textwidth]{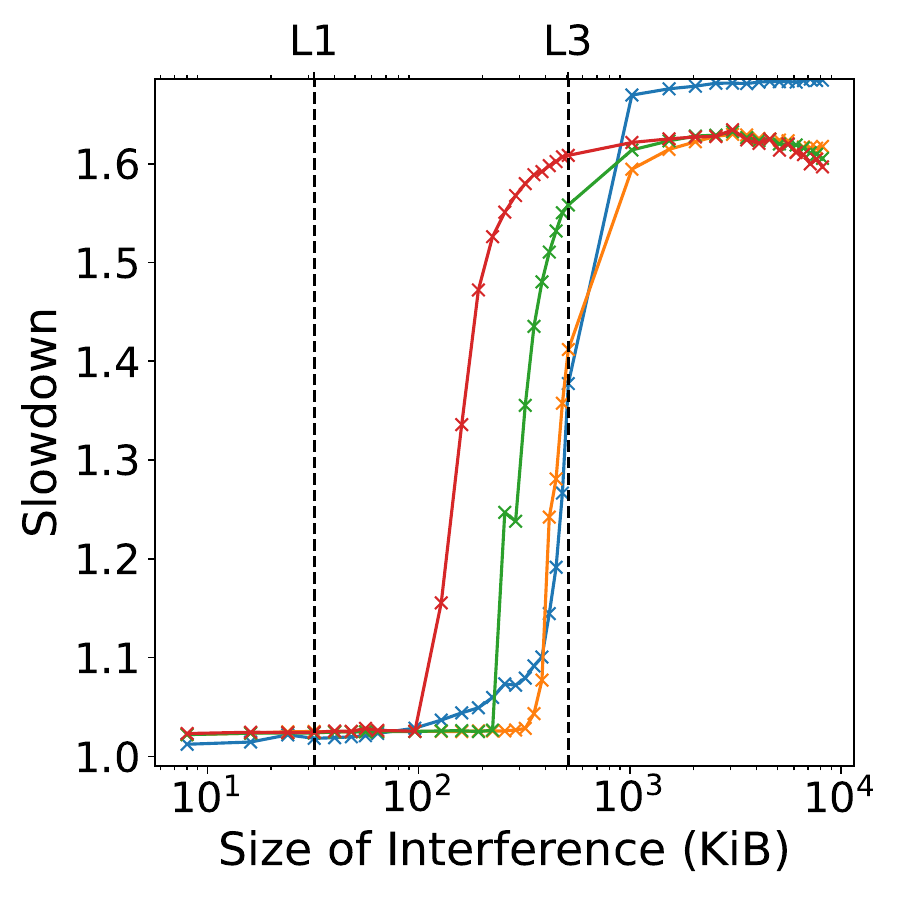}
            \captionsetup{justification=centering}
            \caption{rk3568: Set / Prefetch}
            \label{fig:all-rk3568-set-disparity-prefetch-vga}
        \end{subfigure}
        \hfill
        \begin{subfigure}{0.24\textwidth}
            \centering
            \includegraphics[width=\textwidth]{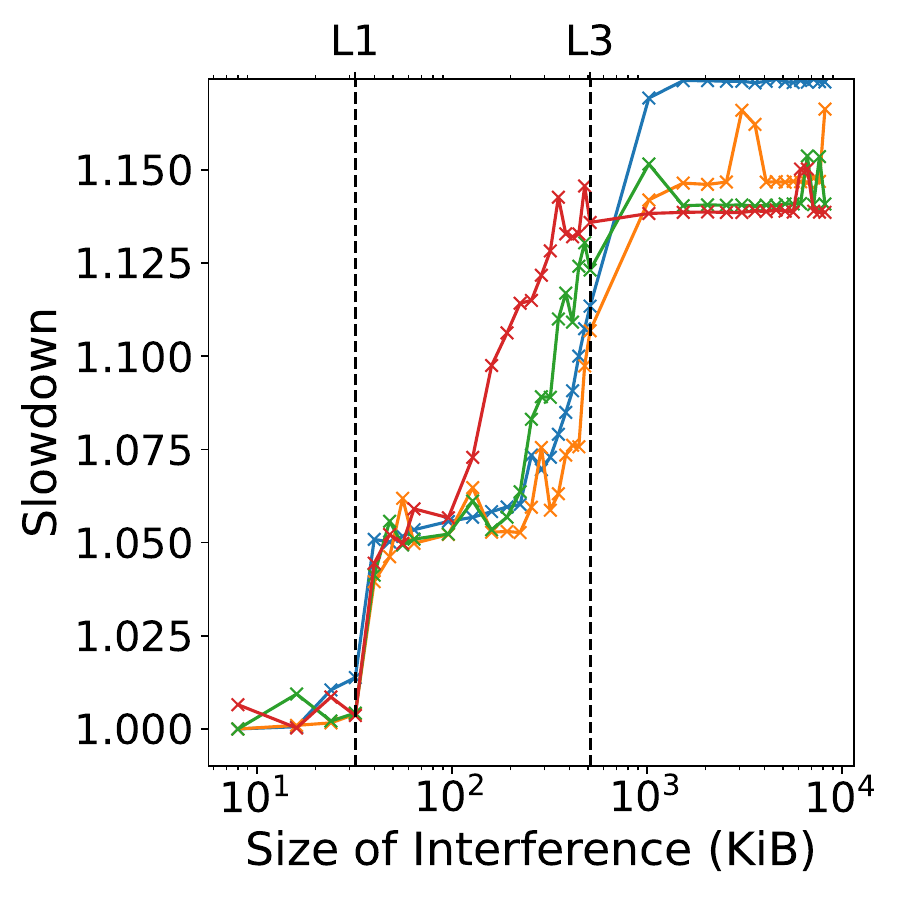}
            \captionsetup{justification=centering}
            \caption{rk3568: Way / Read}
            \label{fig:all-rk3568-way-disparity-read-vga}
        \end{subfigure}
        \hfill
        \begin{subfigure}{0.24\textwidth}
            \centering
            \includegraphics[width=\textwidth]{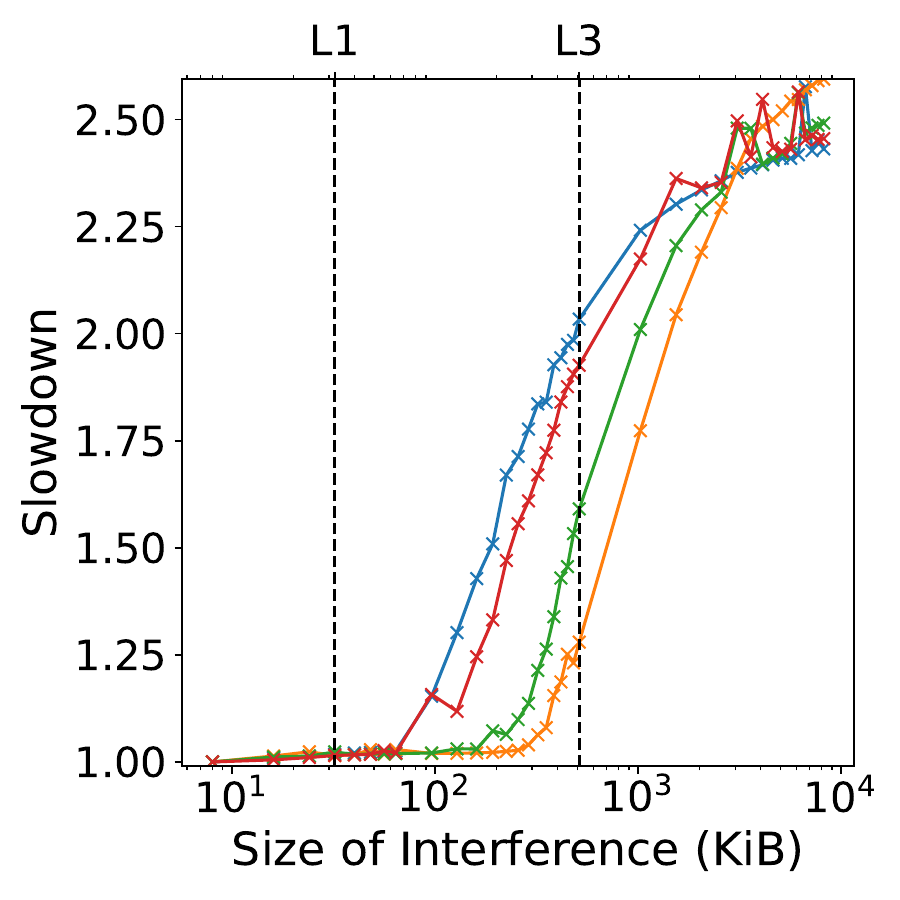}
            \captionsetup{justification=centering}
            \caption{rk3568: Way / Write}
            \label{fig:all-rk3568-way-disparity-write-vga}
        \end{subfigure}
        \hfill
        \begin{subfigure}{0.24\textwidth}
            \centering
            \includegraphics[width=\textwidth]{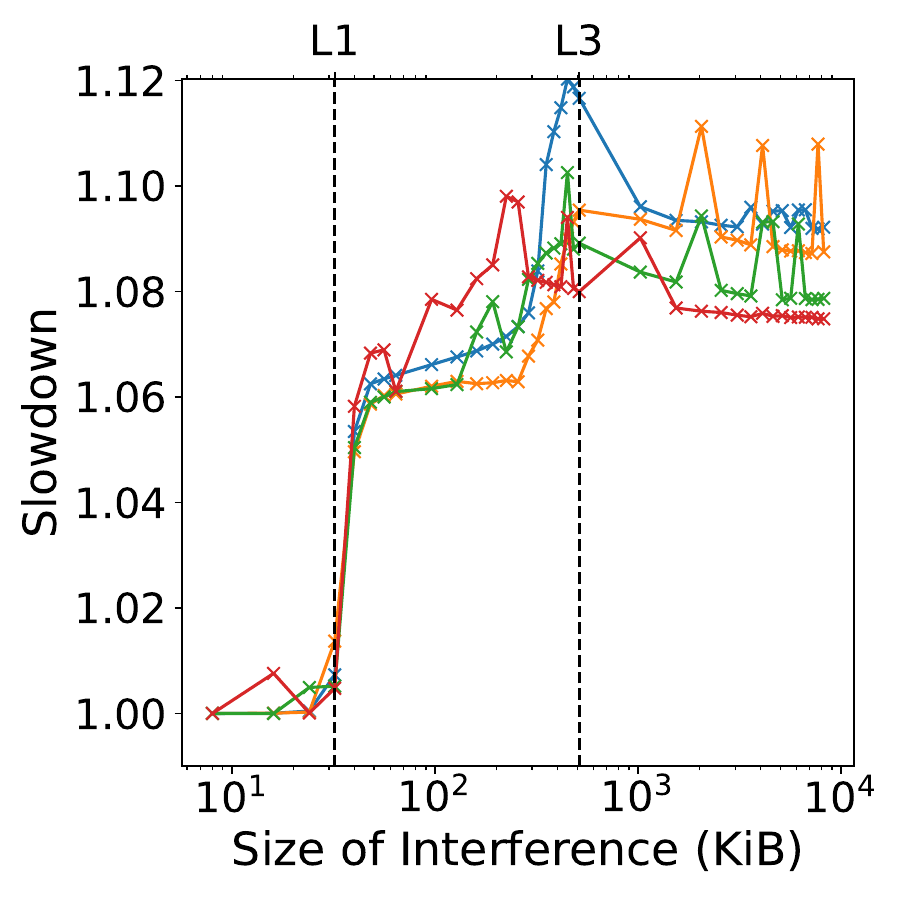}
            \captionsetup{justification=centering}
            \caption{rk3568: Way / Modify}
            \label{fig:all-rk3568-way-disparity-modify-vga}
        \end{subfigure}
        \hfill
        \begin{subfigure}{0.24\textwidth}
            \centering
            \includegraphics[width=\textwidth]{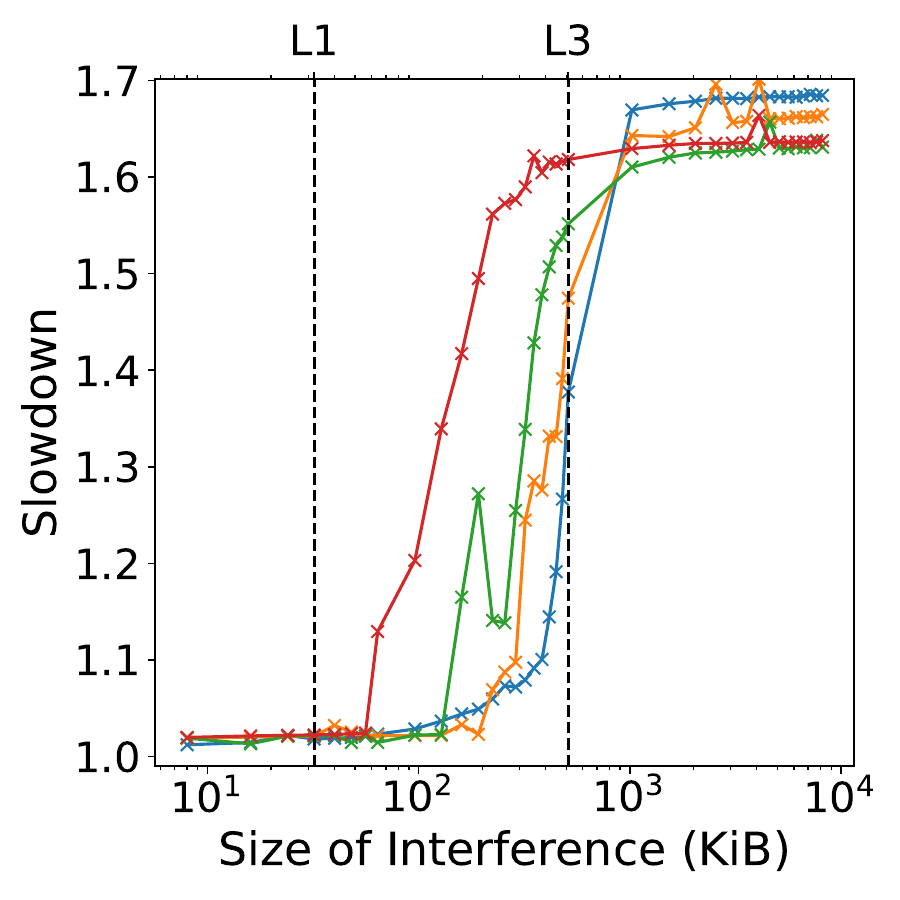}
            \captionsetup{justification=centering}
            \caption{rk3568: Way / Prefetch}
            \label{fig:all-rk3568-way-disparity-prefetch-vga}
        \end{subfigure}
        \hfill
        
        \caption{Execution Slowdown on \textit{'Disparity'} benchmark for \textit{'VGA'} dataset on \textit{'RK3568'} with Interferences and cache partitioning.}
        \label{fig:rk3568-disparity-vga}
    \end{figure}

    \begin{figure}[H]
        \begin{subfigure}{\textwidth}
            \centering
            \includegraphics[width=0.5\textwidth]{figures/set_subplot/legend.pdf}
        \end{subfigure}
        \centering
        
        \begin{subfigure}{0.24\textwidth}
            \centering
            \includegraphics[width=\textwidth]{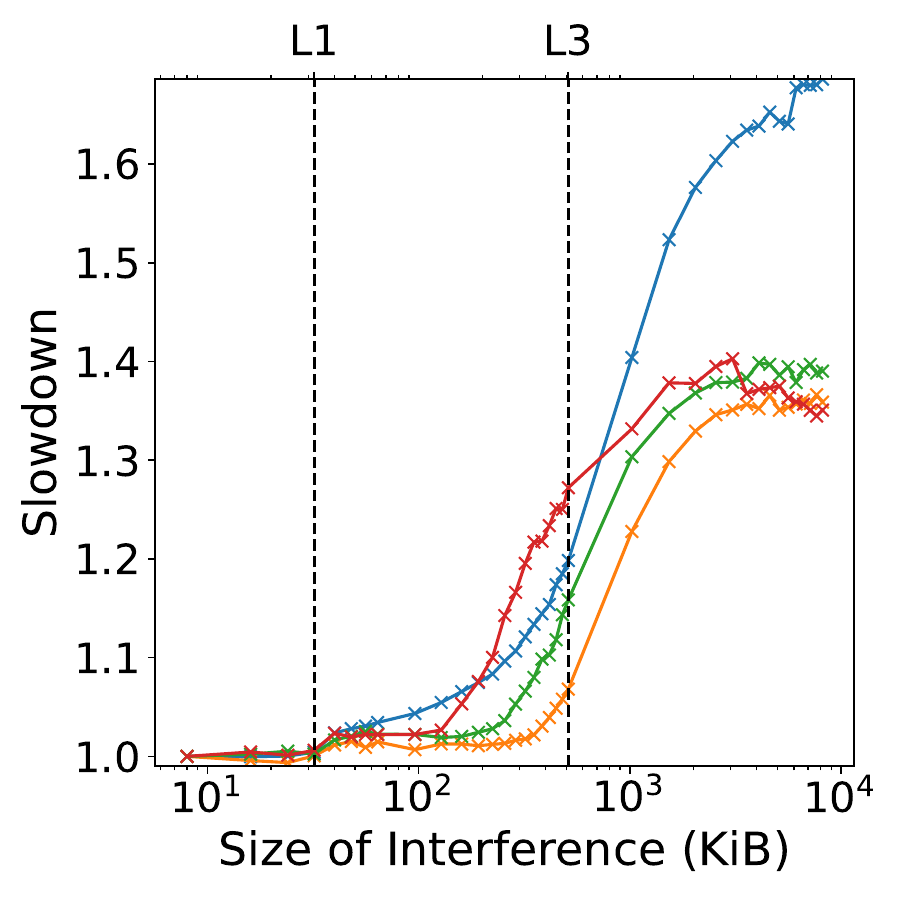}
            \captionsetup{justification=centering}
            \caption{rk3568: Set / Read}
            \label{fig:all-rk3568-set-mser-read-vga}
        \end{subfigure}
        \hfill
        \begin{subfigure}{0.24\textwidth}
            \centering
            \includegraphics[width=\textwidth]{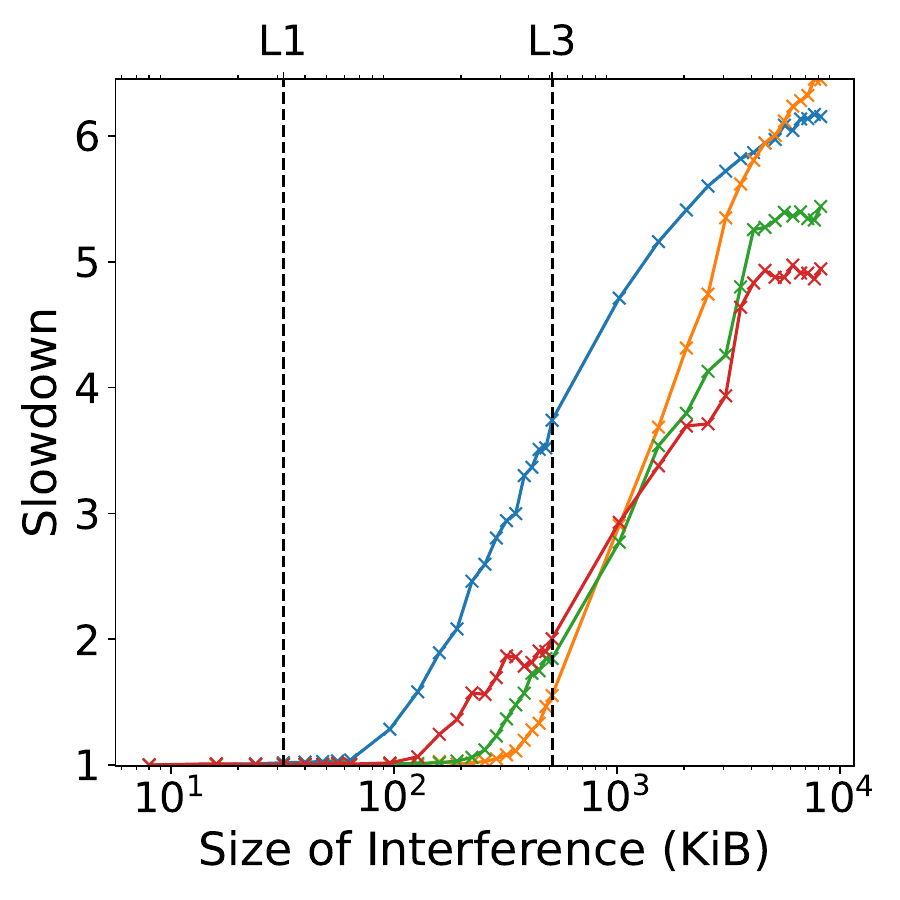}
            \captionsetup{justification=centering}
            \caption{rk3568: Set / Write}
            \label{fig:all-rk3568-set-mser-write-vga}
        \end{subfigure}
        \hfill
        \begin{subfigure}{0.24\textwidth}
            \centering
            \includegraphics[width=\textwidth]{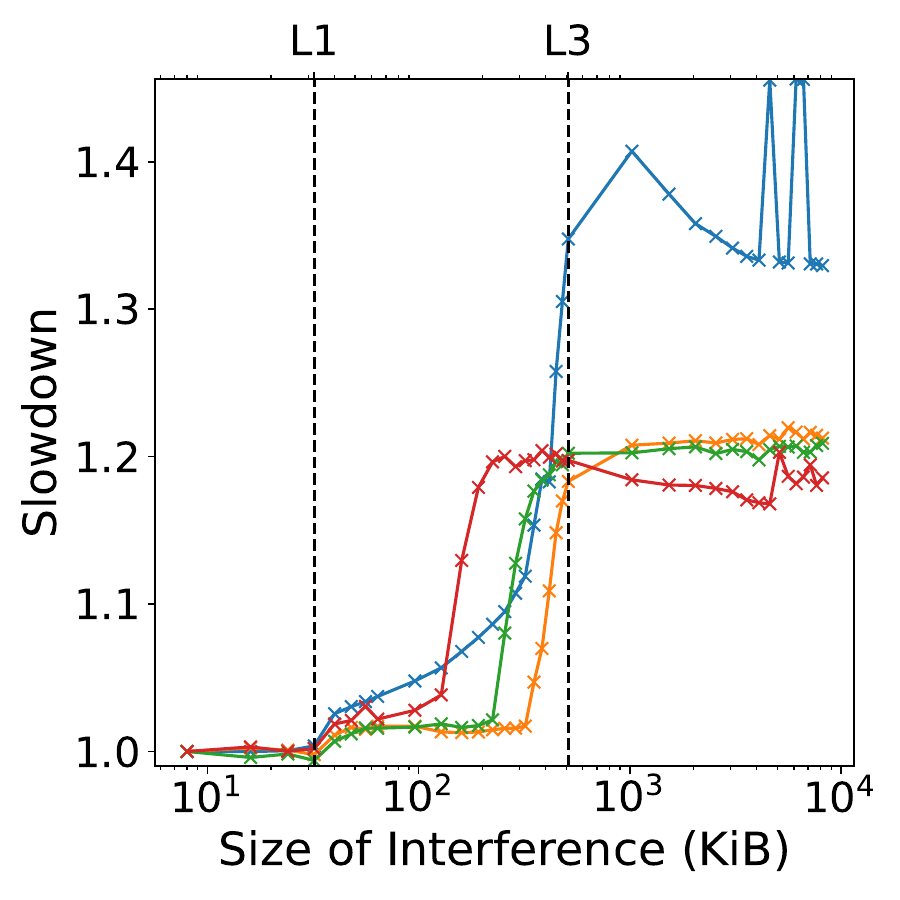}
            \captionsetup{justification=centering}
            \caption{rk3568: Set / Modify}
            \label{fig:all-rk3568-set-mser-modify-vga}
        \end{subfigure}
        \hfill
        \begin{subfigure}{0.24\textwidth}
            \centering
            \includegraphics[width=\textwidth]{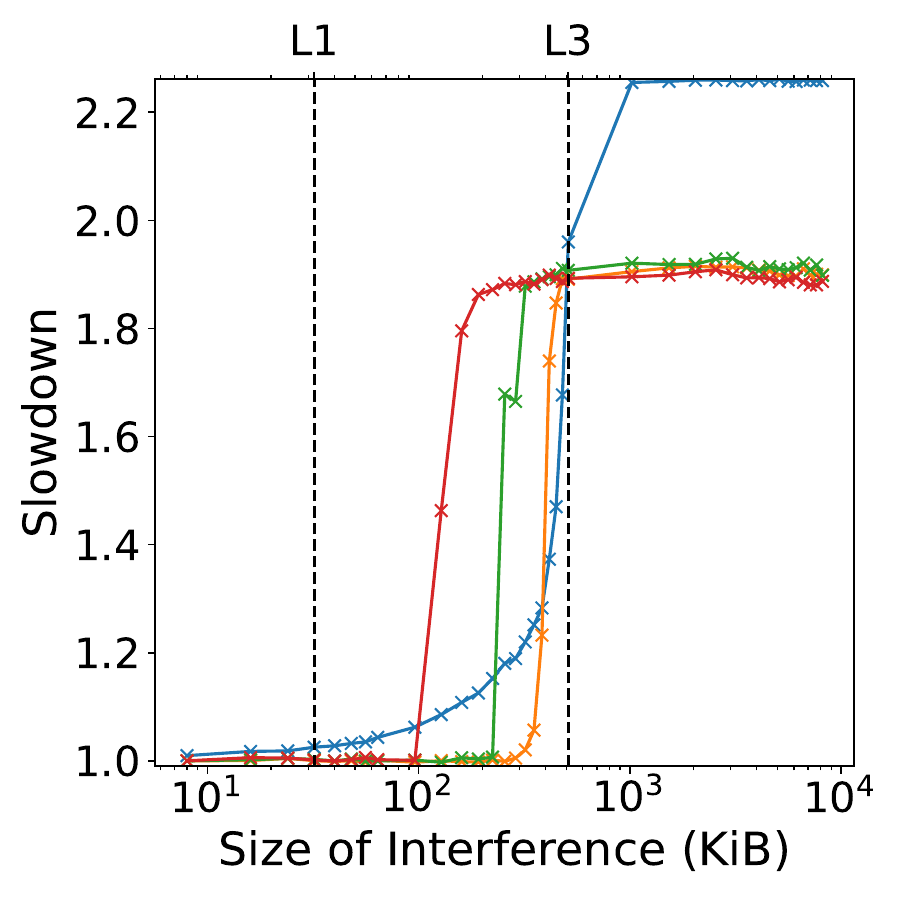}
            \captionsetup{justification=centering}
            \caption{rk3568: Set / Prefetch}
            \label{fig:all-rk3568-set-mser-prefetch-vga}
        \end{subfigure}
        \hfill
        \begin{subfigure}{0.24\textwidth}
            \centering
            \includegraphics[width=\textwidth]{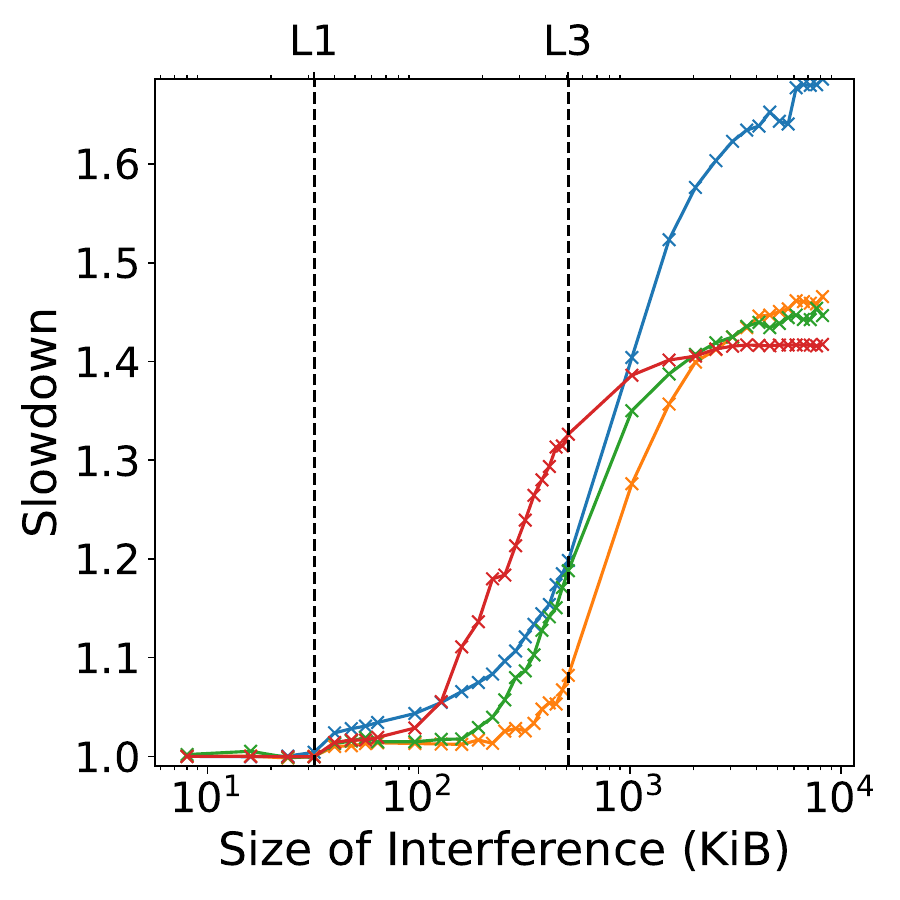}
            \captionsetup{justification=centering}
            \caption{rk3568: Way / Read}
            \label{fig:all-rk3568-way-mser-read-vga}
        \end{subfigure}
        \hfill
        \begin{subfigure}{0.24\textwidth}
            \centering
            \includegraphics[width=\textwidth]{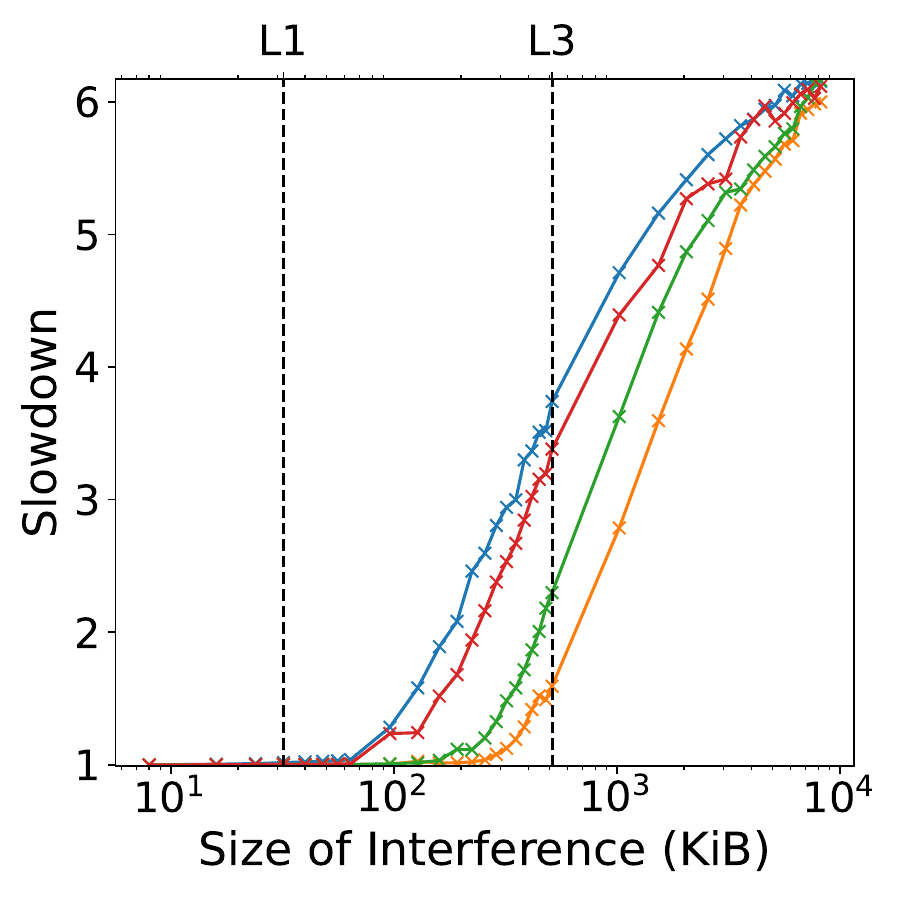}
            \captionsetup{justification=centering}
            \caption{rk3568: Way / Write}
            \label{fig:all-rk3568-way-mser-write-vga}
        \end{subfigure}
        \hfill
        \begin{subfigure}{0.24\textwidth}
            \centering
            \includegraphics[width=\textwidth]{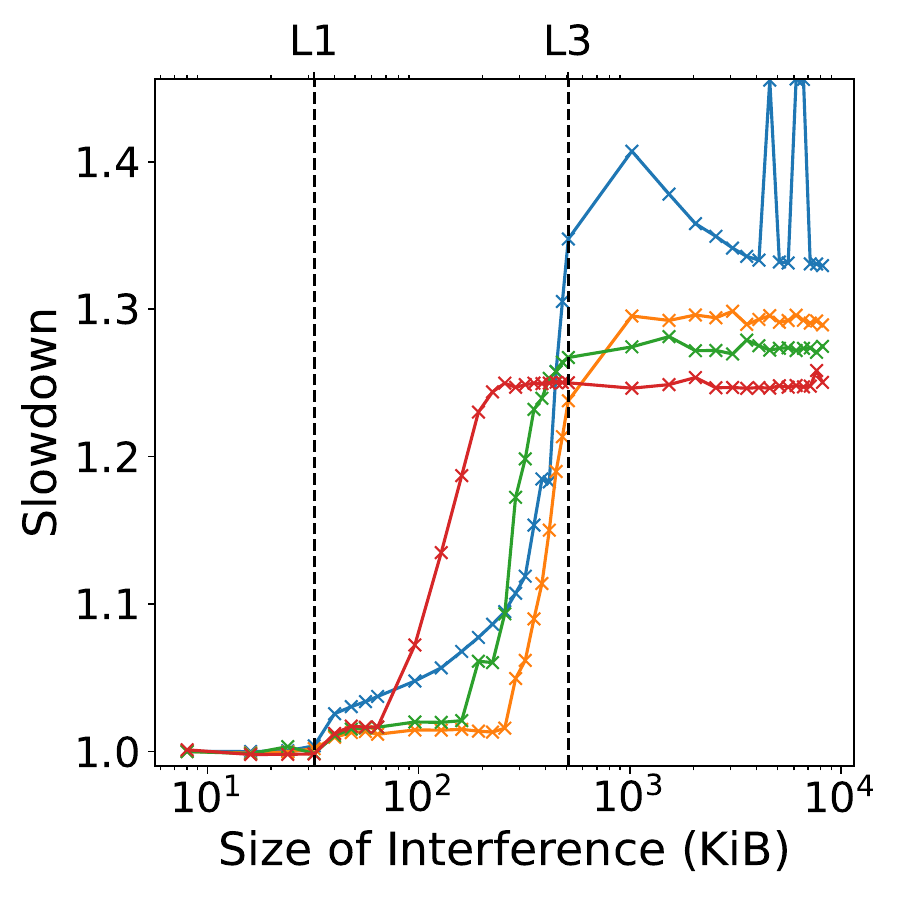}
            \captionsetup{justification=centering}
            \caption{rk3568: Way / Modify}
            \label{fig:all-rk3568-way-mser-modify-vga}
        \end{subfigure}
        \hfill
        \begin{subfigure}{0.24\textwidth}
            \centering
            \includegraphics[width=\textwidth]{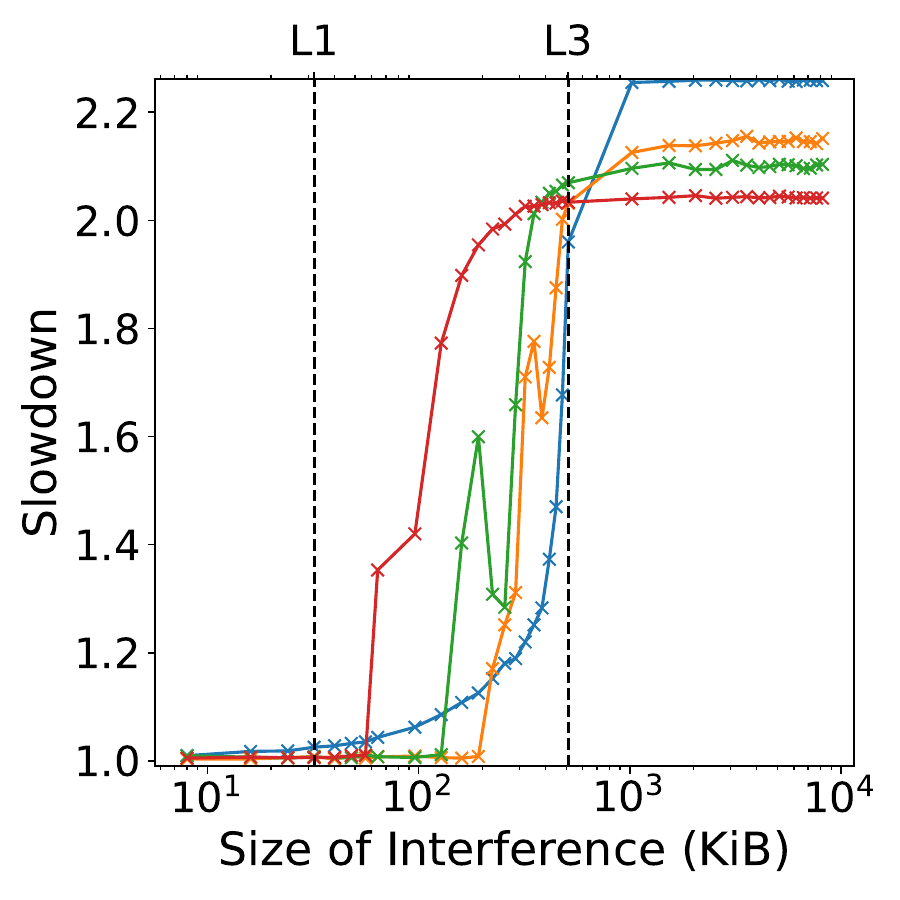}
            \captionsetup{justification=centering}
            \caption{rk3568: Way / Prefetch}
            \label{fig:all-rk3568-way-mser-prefetch-vga}
        \end{subfigure}
        \hfill
        
        \caption{Execution Slowdown on \textit{'Mser'} benchmark for \textit{'VGA'} dataset on \textit{'RK3568'} with Interferences and cache partitioning.}
        \label{fig:rk3568-mser-vga}
    \end{figure}

    \begin{figure}[H]
        \begin{subfigure}{\textwidth}
            \centering
            \includegraphics[width=0.5\textwidth]{figures/set_subplot/legend.pdf}
        \end{subfigure}
        \centering
        
        \begin{subfigure}{0.24\textwidth}
            \centering
            \includegraphics[width=\textwidth]{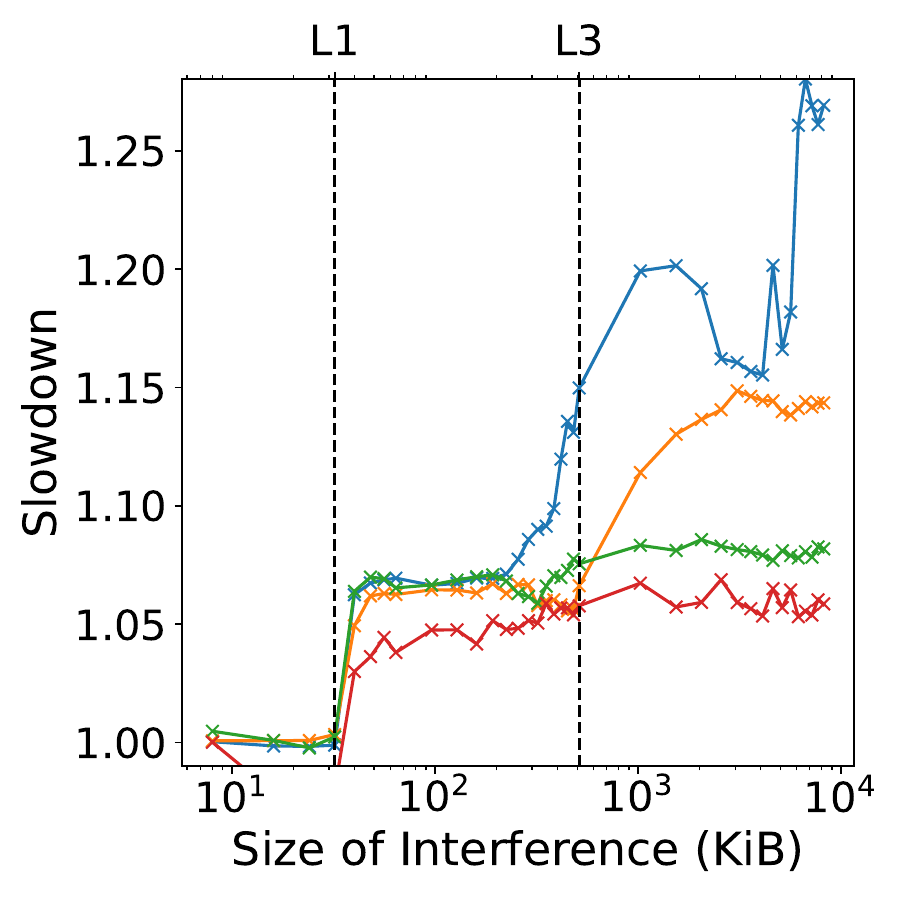}
            \captionsetup{justification=centering}
            \caption{rk3568: Set / Read}
            \label{fig:all-rk3568-set-tracking-read-vga}
        \end{subfigure}
        \hfill
        \begin{subfigure}{0.24\textwidth}
            \centering
            \includegraphics[width=\textwidth]{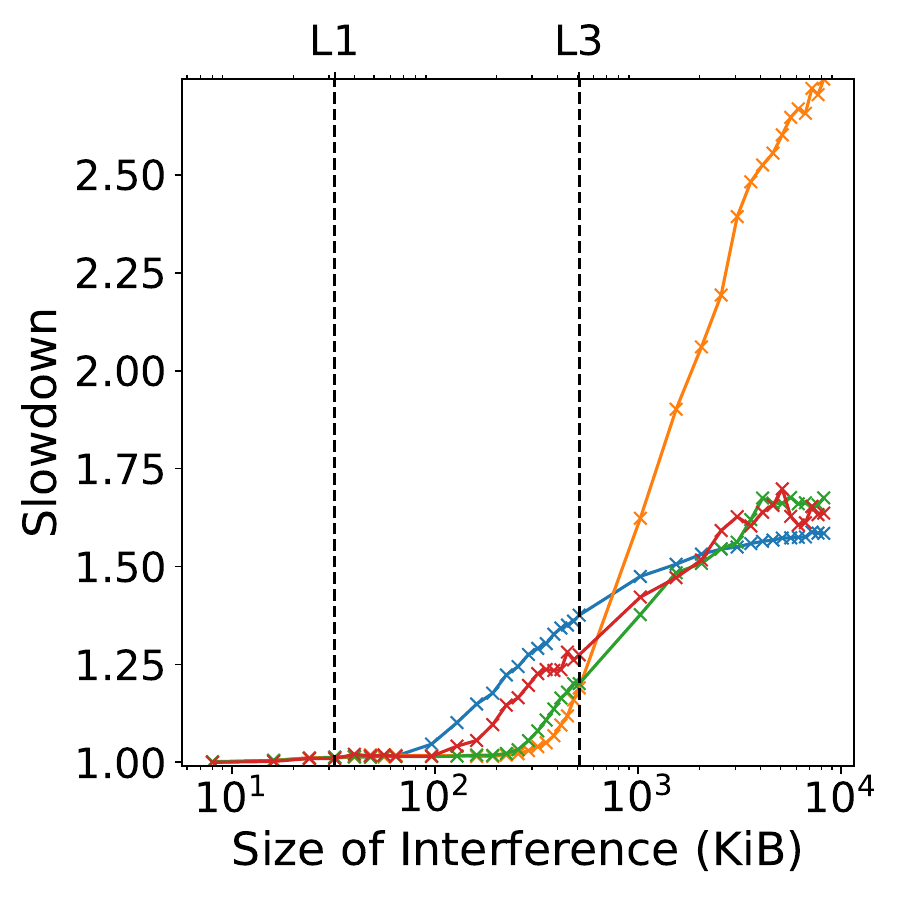}
            \captionsetup{justification=centering}
            \caption{rk3568: Set / Write}
            \label{fig:all-rk3568-set-tracking-write-vga}
        \end{subfigure}
        \hfill
        \begin{subfigure}{0.24\textwidth}
            \centering
            \includegraphics[width=\textwidth]{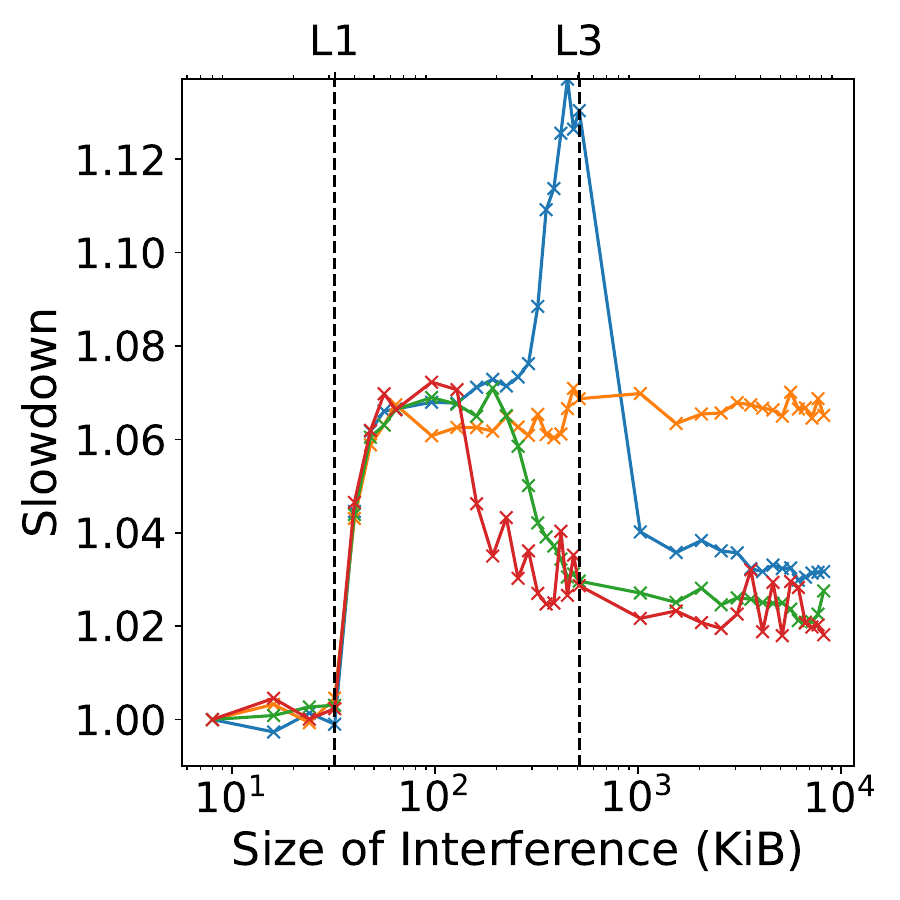}
            \captionsetup{justification=centering}
            \caption{rk3568: Set / Modify}
            \label{fig:all-rk3568-set-tracking-modify-vga}
        \end{subfigure}
        \hfill
        \begin{subfigure}{0.24\textwidth}
            \centering
            \includegraphics[width=\textwidth]{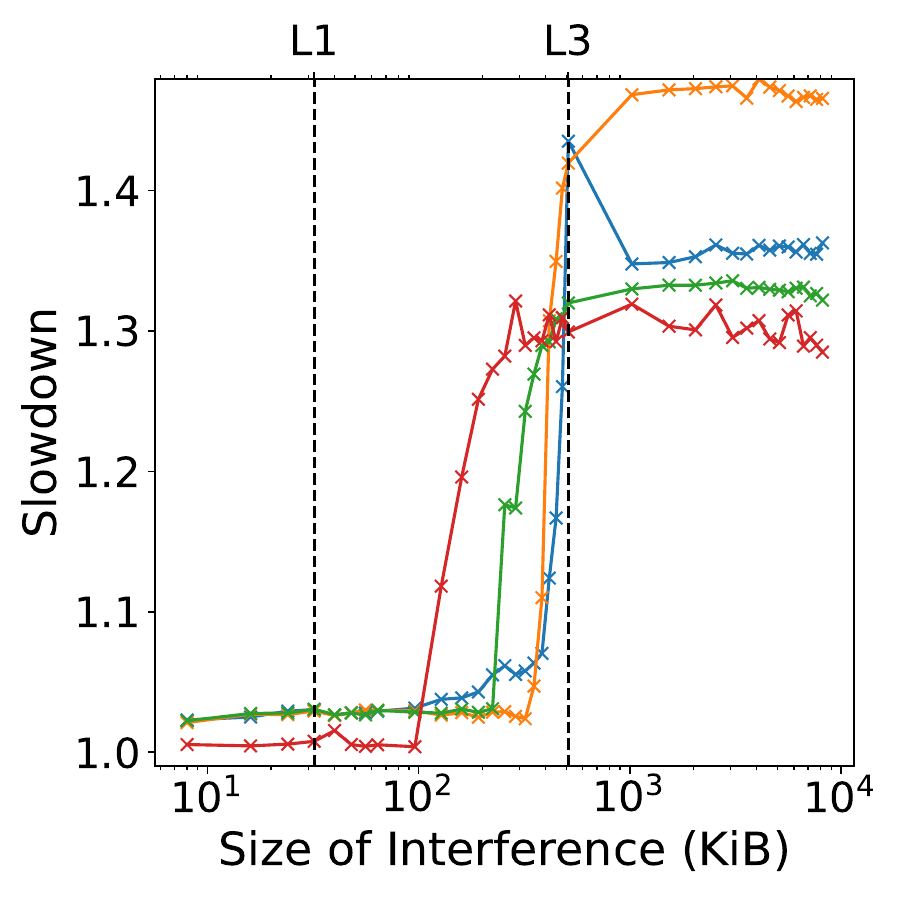}
            \captionsetup{justification=centering}
            \caption{rk3568: Set / Prefetch}
            \label{fig:all-rk3568-set-tracking-prefetch-vga}
        \end{subfigure}
        \hfill
        \begin{subfigure}{0.24\textwidth}
            \centering
            \includegraphics[width=\textwidth]{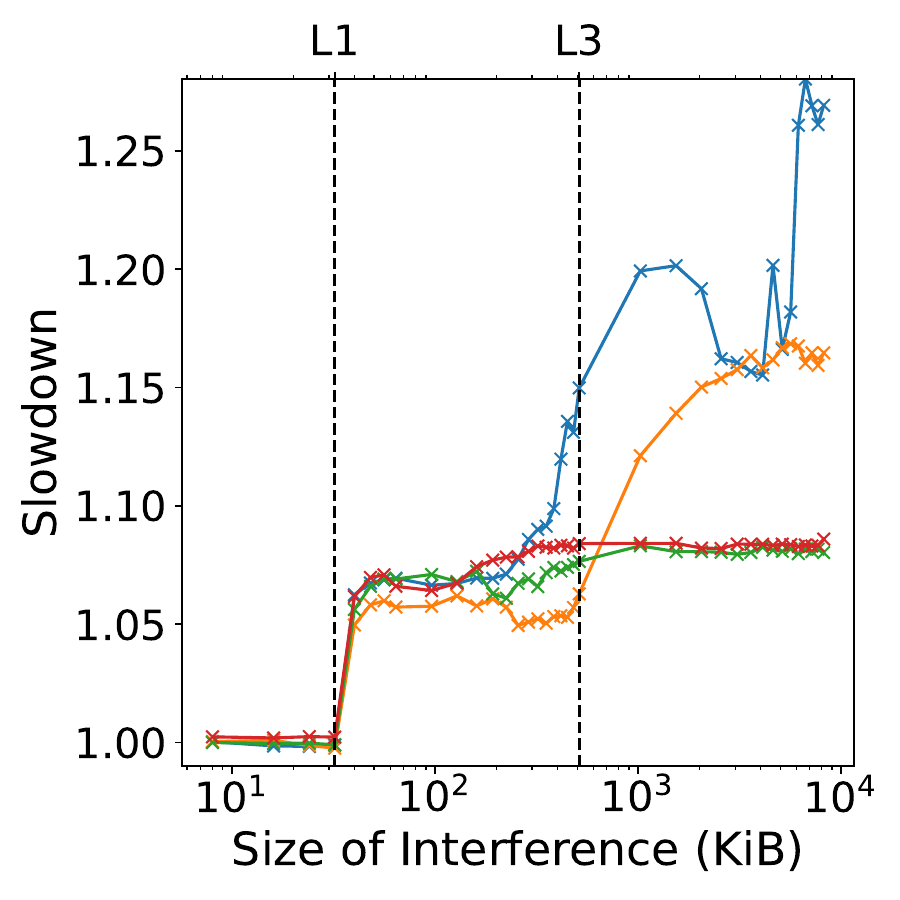}
            \captionsetup{justification=centering}
            \caption{rk3568: Way / Read}
            \label{fig:all-rk3568-way-tracking-read-vga}
        \end{subfigure}
        \hfill
        \begin{subfigure}{0.24\textwidth}
            \centering
            \includegraphics[width=\textwidth]{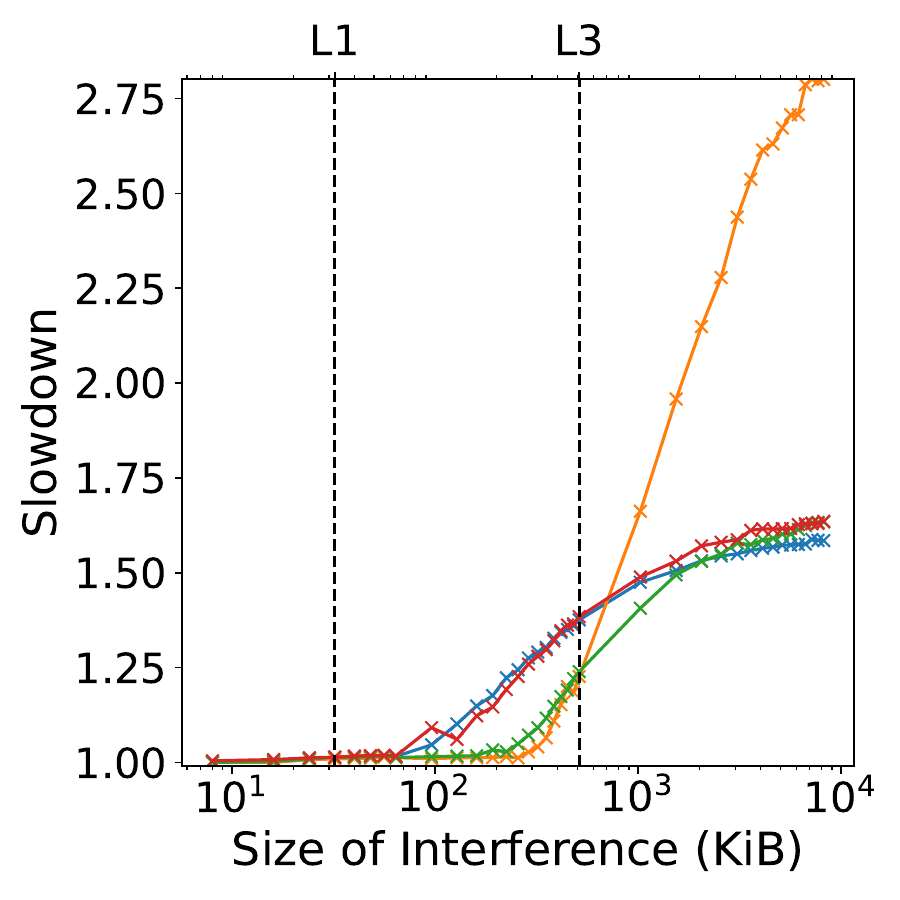}
            \captionsetup{justification=centering}
            \caption{rk3568: Way / Write}
            \label{fig:all-rk3568-way-tracking-write-vga}
        \end{subfigure}
        \hfill
        \begin{subfigure}{0.24\textwidth}
            \centering
            \includegraphics[width=\textwidth]{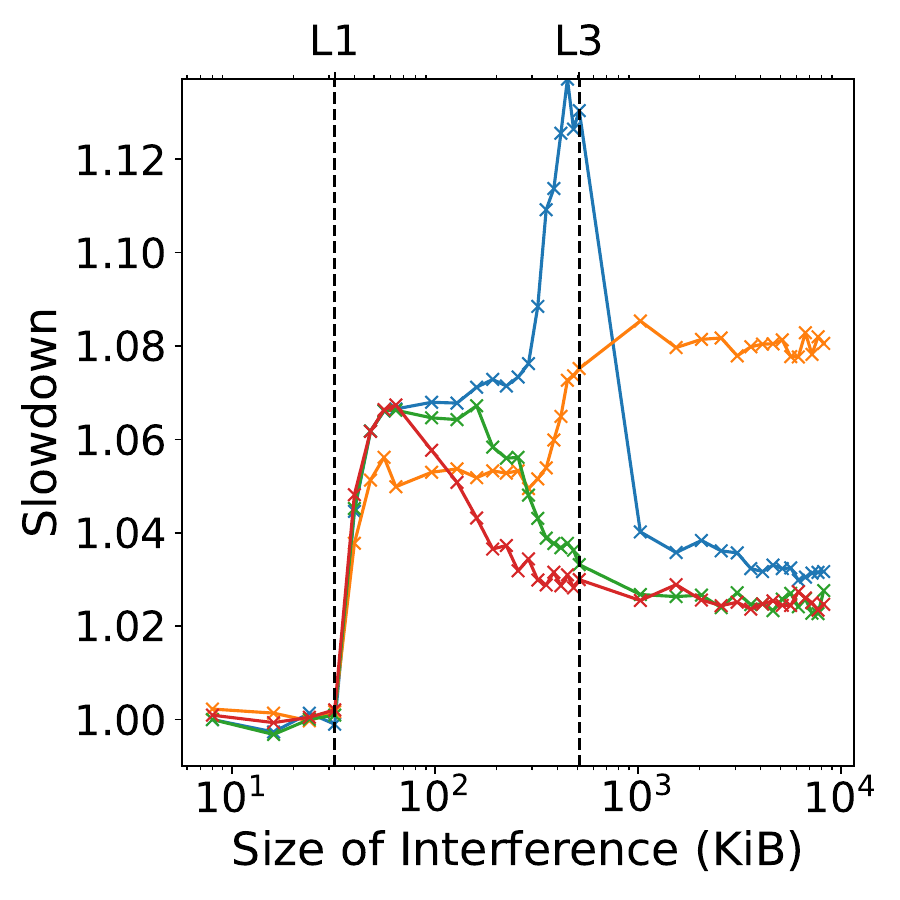}
            \captionsetup{justification=centering}
            \caption{rk3568: Way / Modify}
            \label{fig:all-rk3568-way-tracking-modify-vga}
        \end{subfigure}
        \hfill
        \begin{subfigure}{0.24\textwidth}
            \centering
            \includegraphics[width=\textwidth]{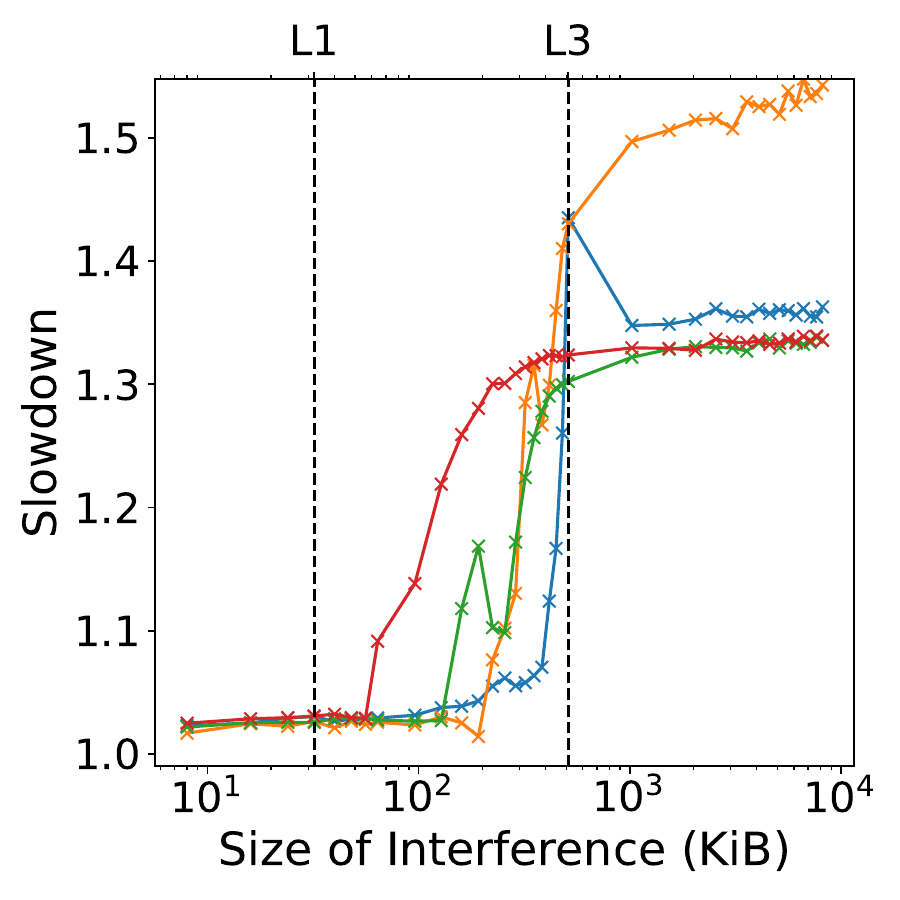}
            \captionsetup{justification=centering}
            \caption{rk3568: Way / Prefetch}
            \label{fig:all-rk3568-way-tracking-prefetch-vga}
        \end{subfigure}
        \hfill
        
        \caption{Execution Slowdown on \textit{'Tracking'} benchmark for \textit{'VGA'} dataset on \textit{'RK3568'} with Interferences and cache partitioning.}
        \label{fig:rk3568-tracking-vga}
    \end{figure}

    \begin{figure}[H]
        \begin{subfigure}{\textwidth}
            \centering
            \includegraphics[width=0.5\textwidth]{figures/set_subplot/legend.pdf}
        \end{subfigure}
        \centering
        
        \begin{subfigure}{0.24\textwidth}
            \centering
            \includegraphics[width=\textwidth]{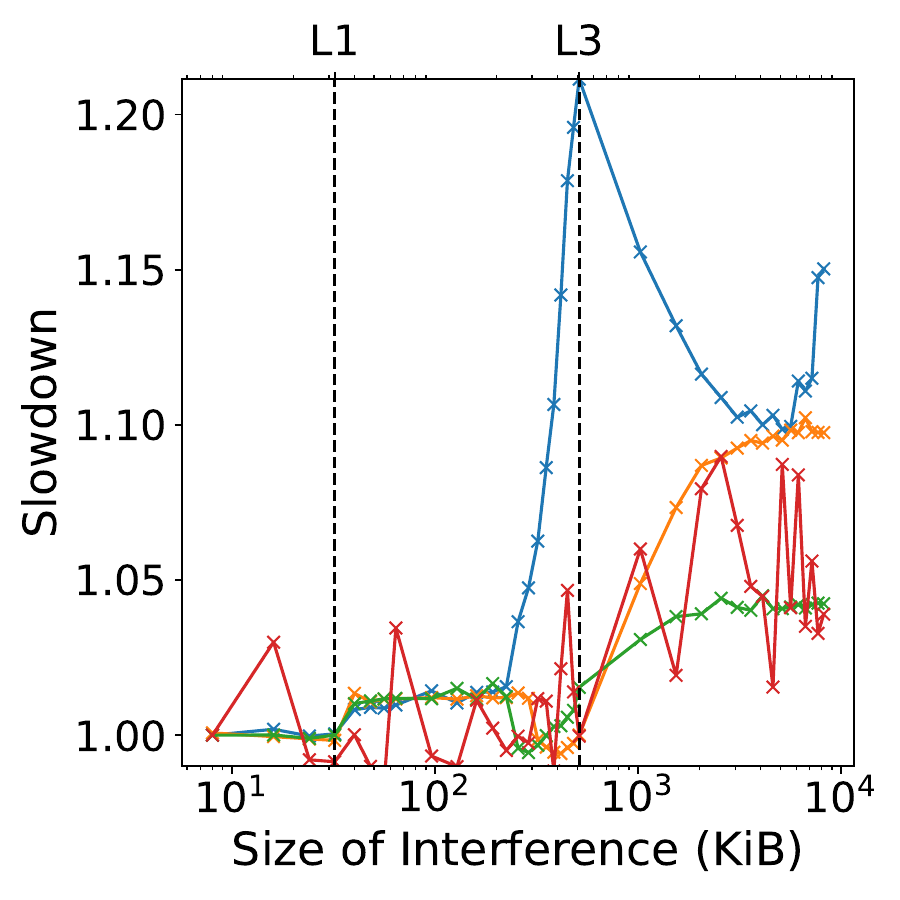}
            \captionsetup{justification=centering}
            \caption{rk3568: Set / Read}
            \label{fig:all-rk3568-set-sift-read-vga}
        \end{subfigure}
        \hfill
        \begin{subfigure}{0.24\textwidth}
            \centering
            \includegraphics[width=\textwidth]{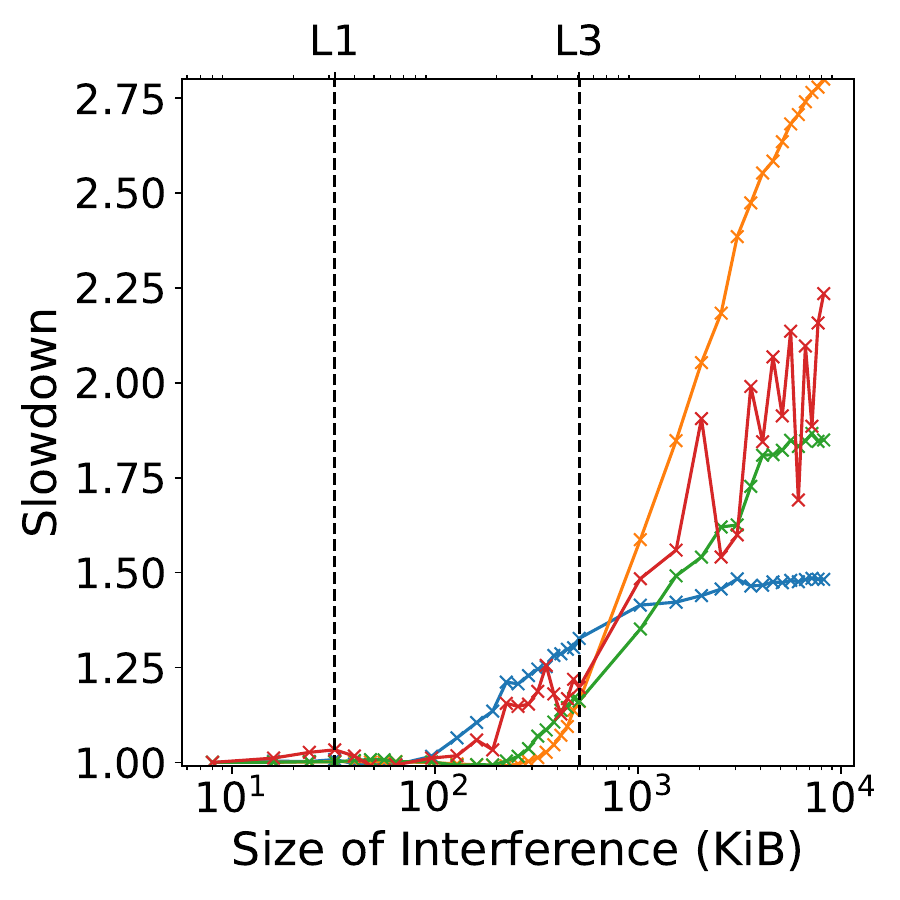}
            \captionsetup{justification=centering}
            \caption{rk3568: Set / Write}
            \label{fig:all-rk3568-set-sift-write-vga}
        \end{subfigure}
        \hfill
        \begin{subfigure}{0.24\textwidth}
            \centering
            \includegraphics[width=\textwidth]{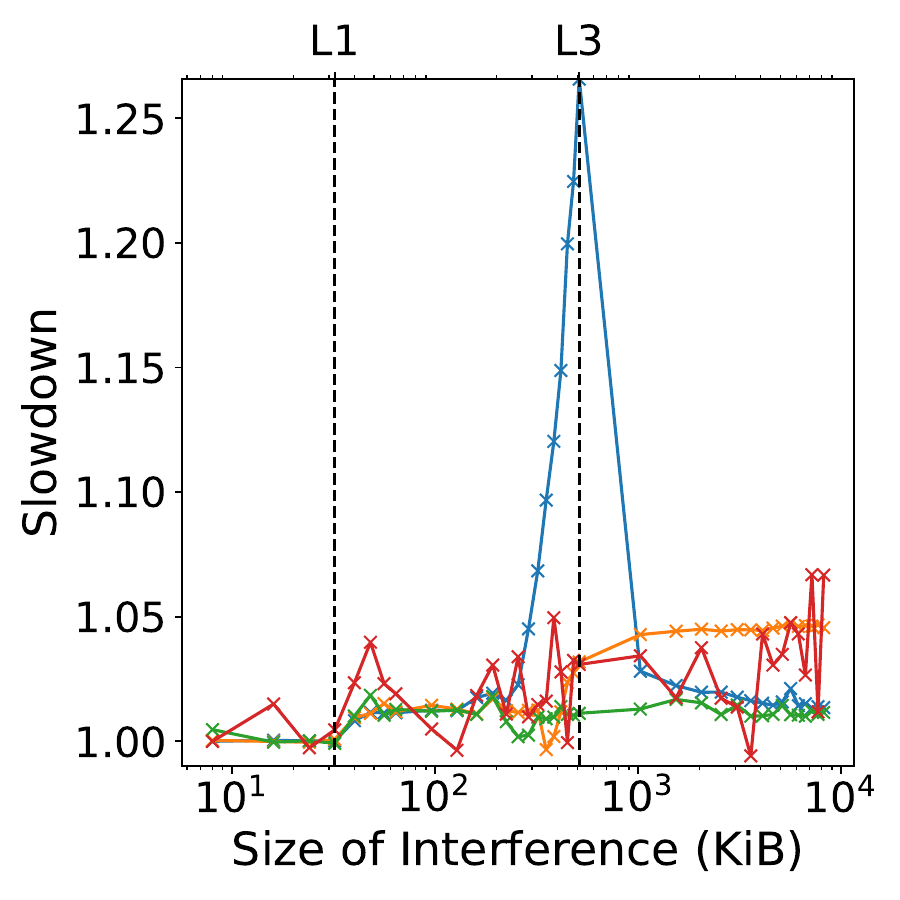}
            \captionsetup{justification=centering}
            \caption{rk3568: Set / Modify}
            \label{fig:all-rk3568-set-sift-modify-vga}
        \end{subfigure}
        \hfill
        \begin{subfigure}{0.24\textwidth}
            \centering
            \includegraphics[width=\textwidth]{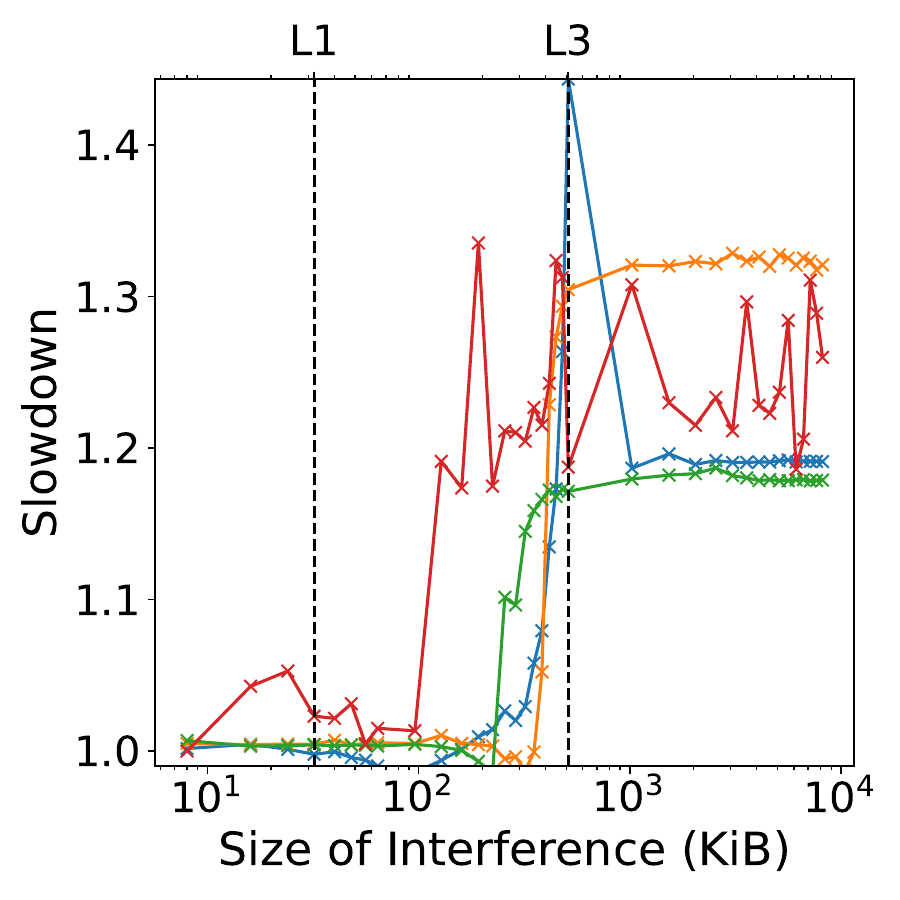}
            \captionsetup{justification=centering}
            \caption{rk3568: Set / Prefetch}
            \label{fig:all-rk3568-set-sift-prefetch-vga}
        \end{subfigure}
        \hfill
        \begin{subfigure}{0.24\textwidth}
            \centering
            \includegraphics[width=\textwidth]{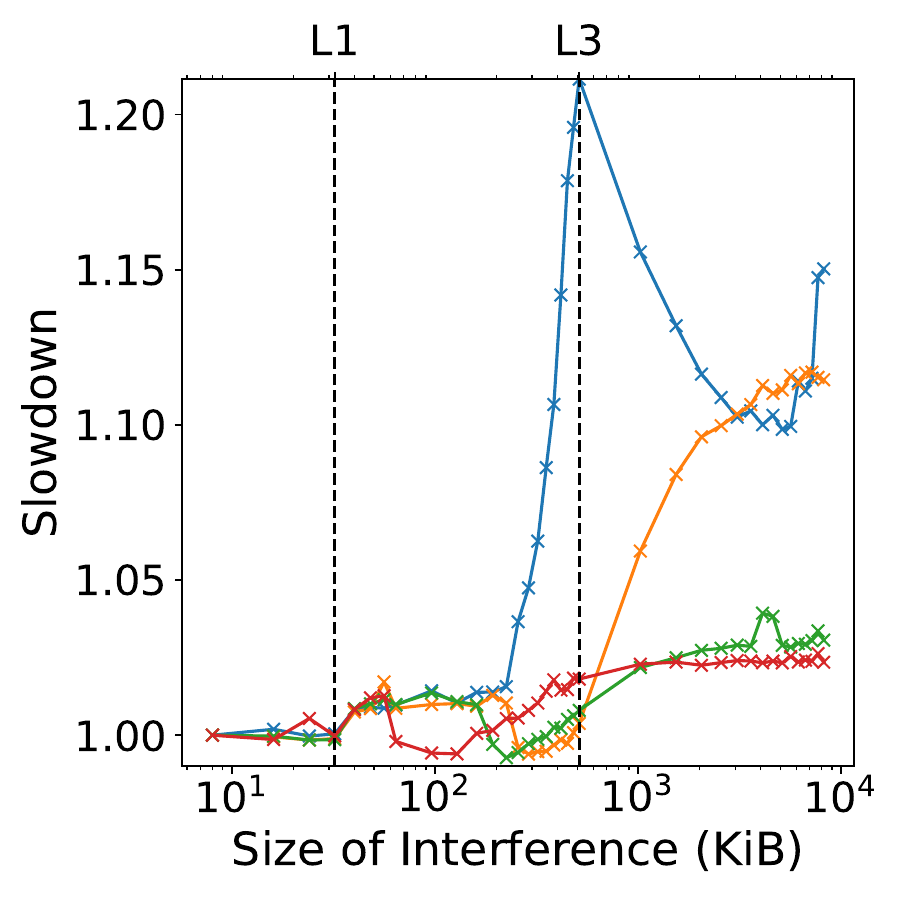}
            \captionsetup{justification=centering}
            \caption{rk3568: Way / Read}
            \label{fig:all-rk3568-way-sift-read-vga}
        \end{subfigure}
        \hfill
        \begin{subfigure}{0.24\textwidth}
            \centering
            \includegraphics[width=\textwidth]{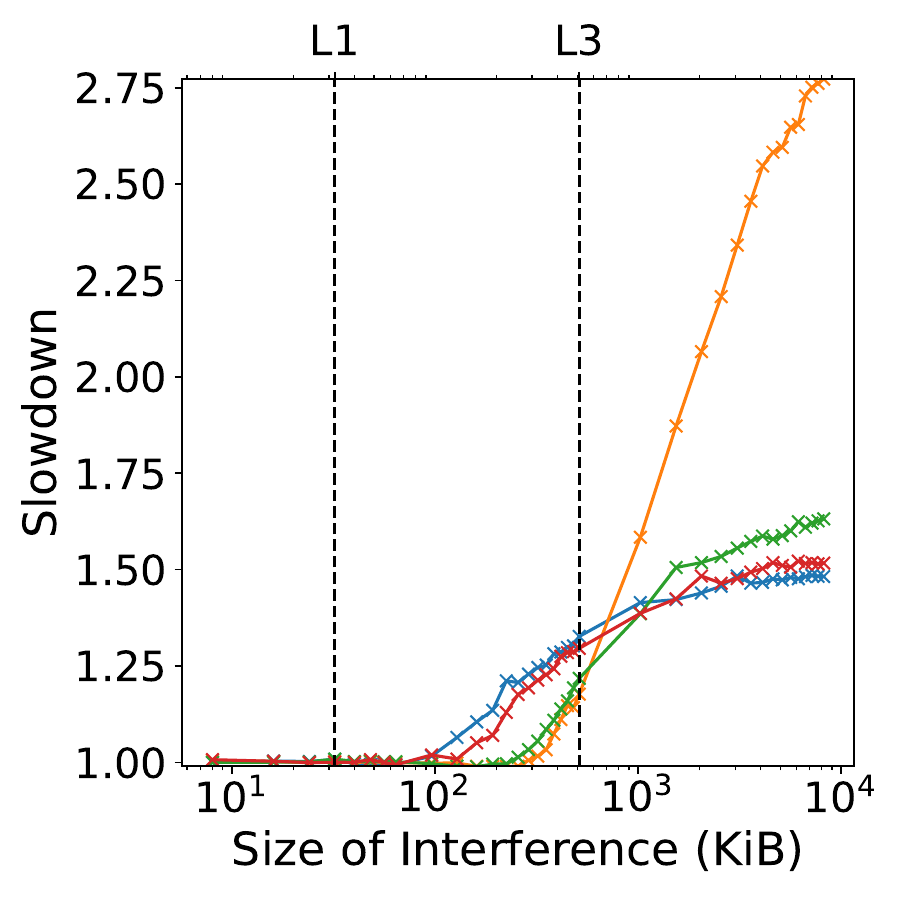}
            \captionsetup{justification=centering}
            \caption{rk3568: Way / Write}
            \label{fig:all-rk3568-way-sift-write-vga}
        \end{subfigure}
        \hfill
        \begin{subfigure}{0.24\textwidth}
            \centering
            \includegraphics[width=\textwidth]{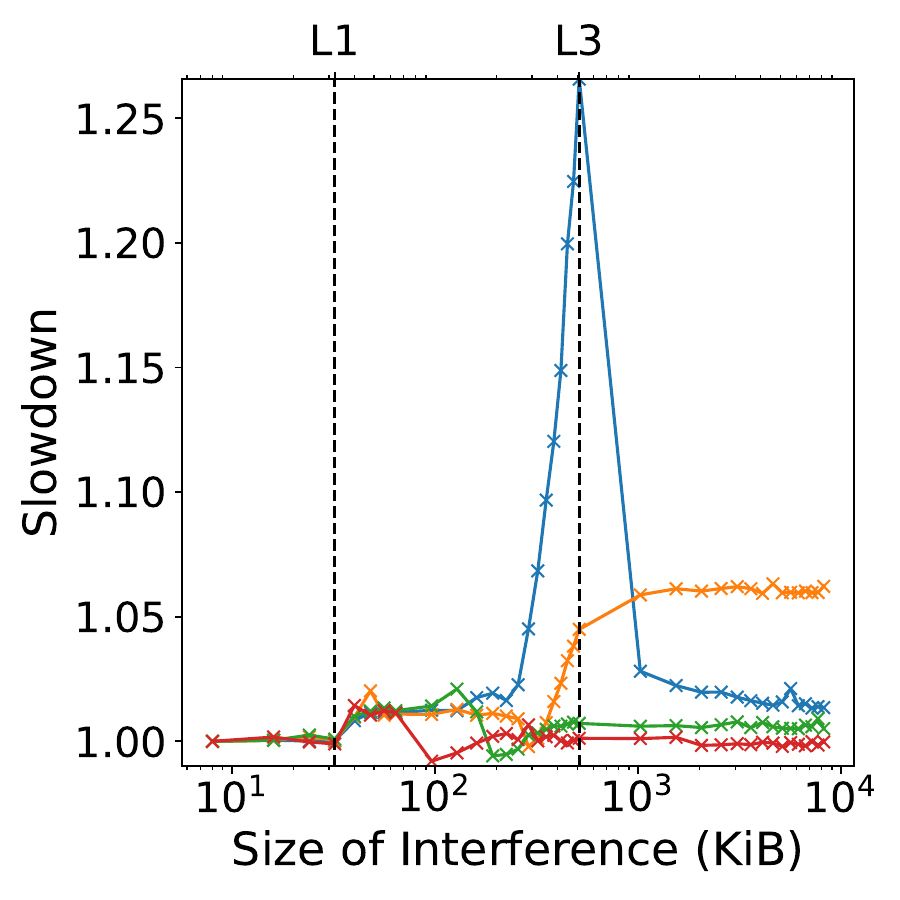}
            \captionsetup{justification=centering}
            \caption{rk3568: Way / Modify}
            \label{fig:all-rk3568-way-sift-modify-vga}
        \end{subfigure}
        \hfill
        \begin{subfigure}{0.24\textwidth}
            \centering
            \includegraphics[width=\textwidth]{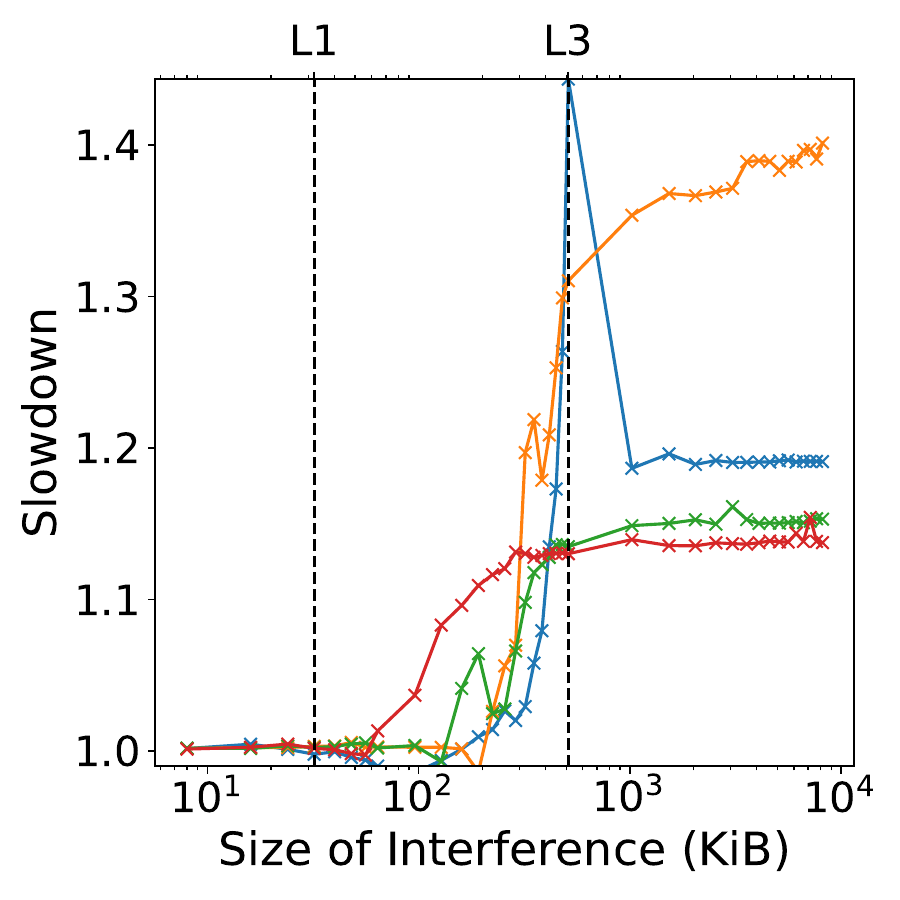}
            \captionsetup{justification=centering}
            \caption{rk3568: Way / Prefetch}
            \label{fig:all-rk3568-way-sift-prefetch-vga}
        \end{subfigure}
        \hfill
        
        \caption{Execution Slowdown on \textit{'Sift'} benchmark for \textit{'VGA'} dataset on \textit{'RK3568'} with Interferences and cache partitioning.}
        \label{fig:rk3568-sift-vga}
    \end{figure}

        \clearpage

        \subsection{RK3588}
        \tableRKeighteight
        \clearpage

    \begin{figure}[H]
        \begin{subfigure}{\textwidth}
            \centering
            \includegraphics[width=0.5\textwidth]{figures/set_subplot/legend.pdf}
        \end{subfigure}
        \centering
        
        \begin{subfigure}{0.24\textwidth}
            \centering
            \includegraphics[width=\textwidth]{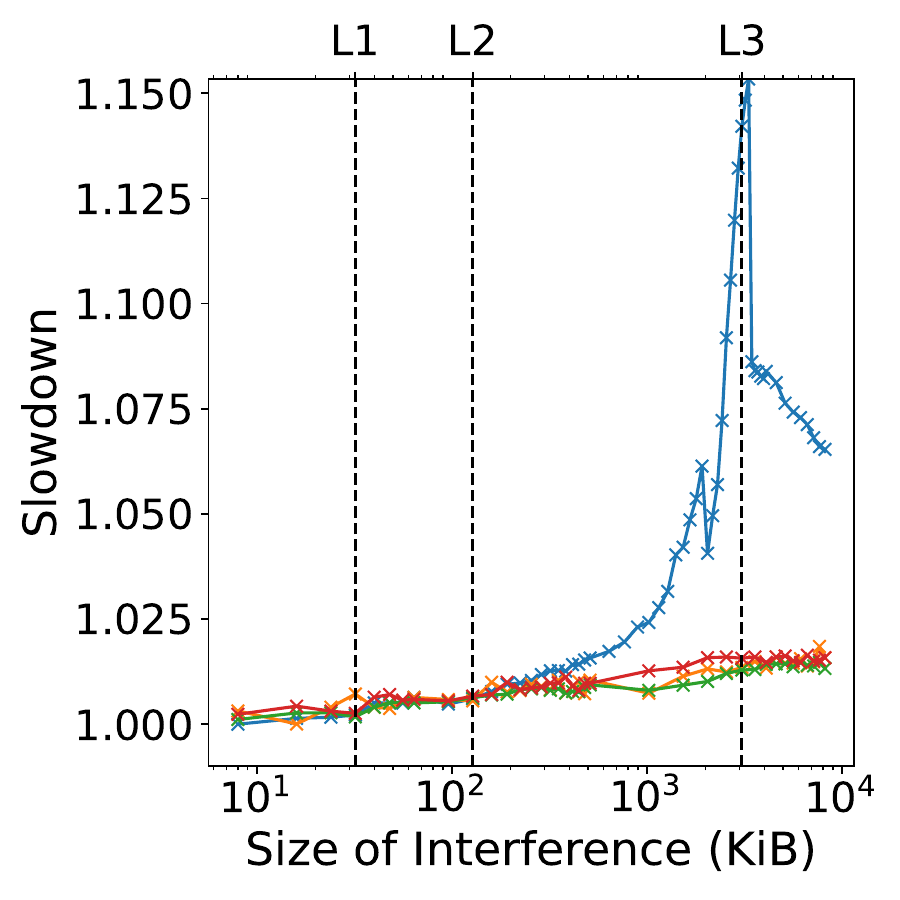}
            \captionsetup{justification=centering}
            \caption{rk3588: Set / Read}
            \label{fig:all-rk3588-set-disparity-read-cif}
        \end{subfigure}
        \hfill
        \begin{subfigure}{0.24\textwidth}
            \centering
            \includegraphics[width=\textwidth]{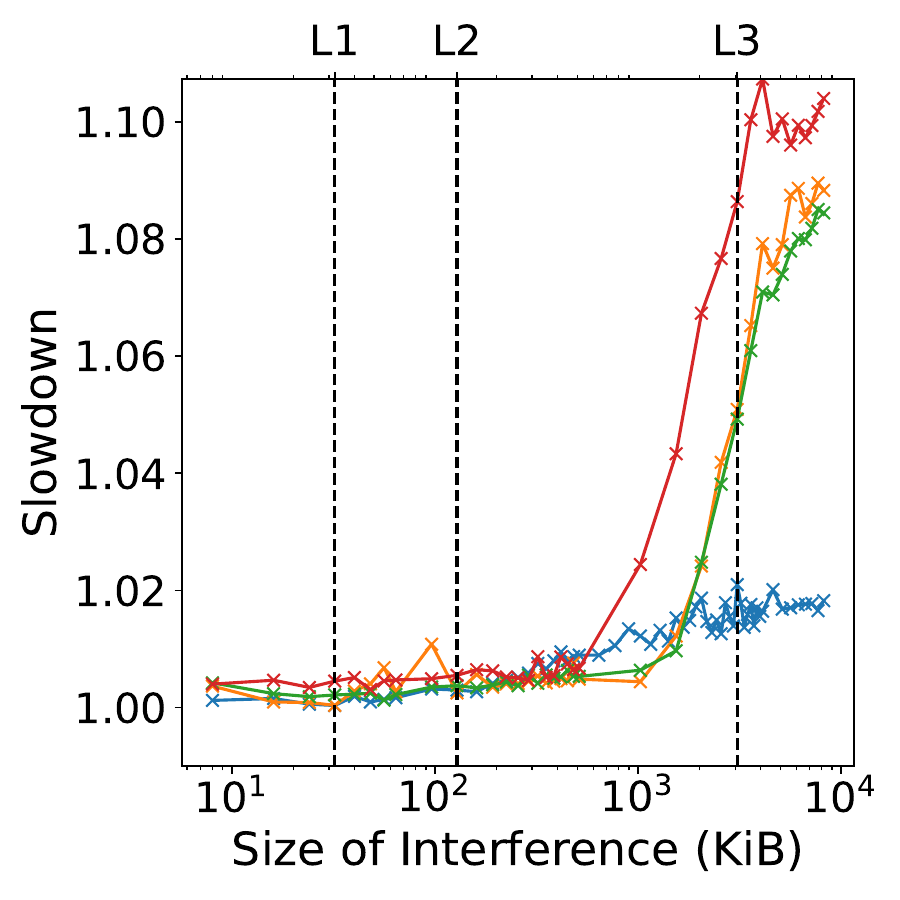}
            \captionsetup{justification=centering}
            \caption{rk3588: Set / Write}
            \label{fig:all-rk3588-set-disparity-write-cif}
        \end{subfigure}
        \hfill
        \begin{subfigure}{0.24\textwidth}
            \centering
            \includegraphics[width=\textwidth]{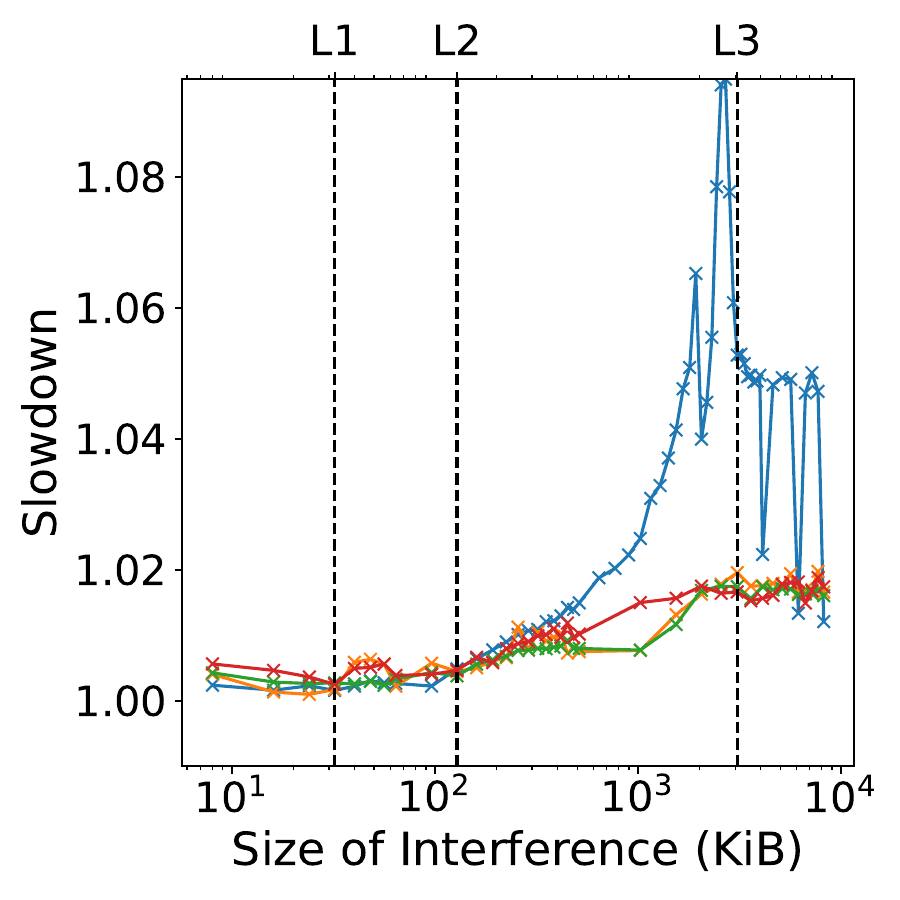}
            \captionsetup{justification=centering}
            \caption{rk3588: Set / Modify}
            \label{fig:all-rk3588-set-disparity-modify-cif}
        \end{subfigure}
        \hfill
        \begin{subfigure}{0.24\textwidth}
            \centering
            \includegraphics[width=\textwidth]{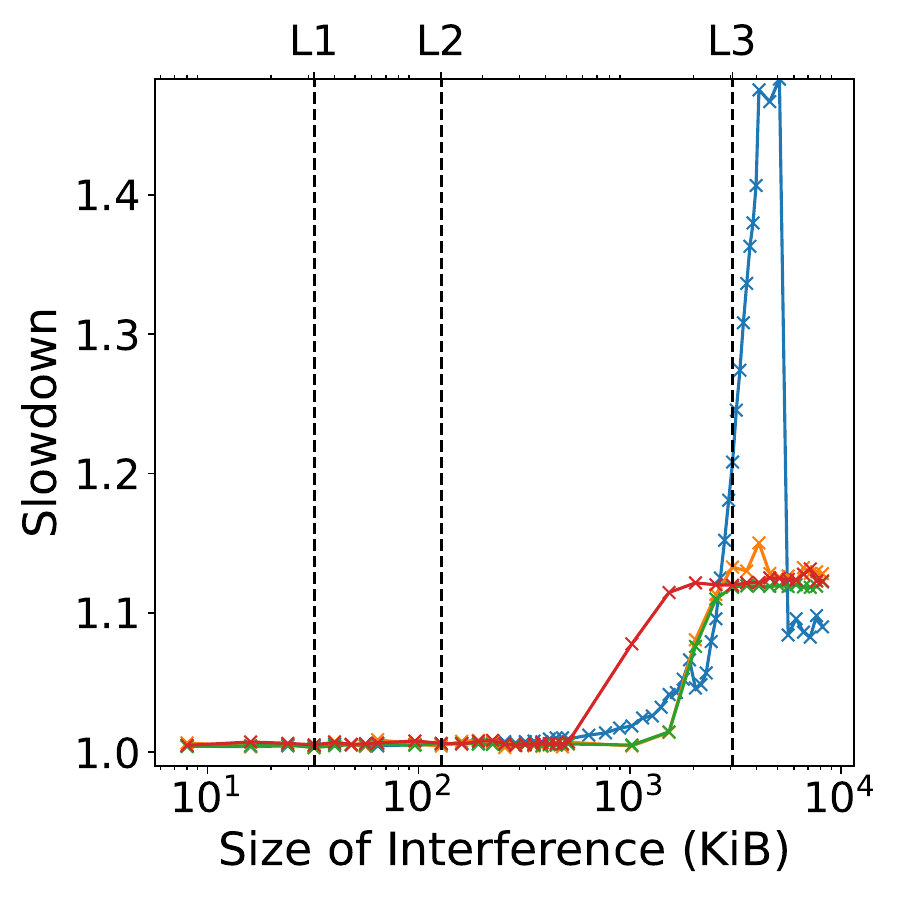}
            \captionsetup{justification=centering}
            \caption{rk3588: Set / Prefetch}
            \label{fig:all-rk3588-set-disparity-prefetch-cif}
        \end{subfigure}
        \hfill
        \begin{subfigure}{0.24\textwidth}
            \centering
            \includegraphics[width=\textwidth]{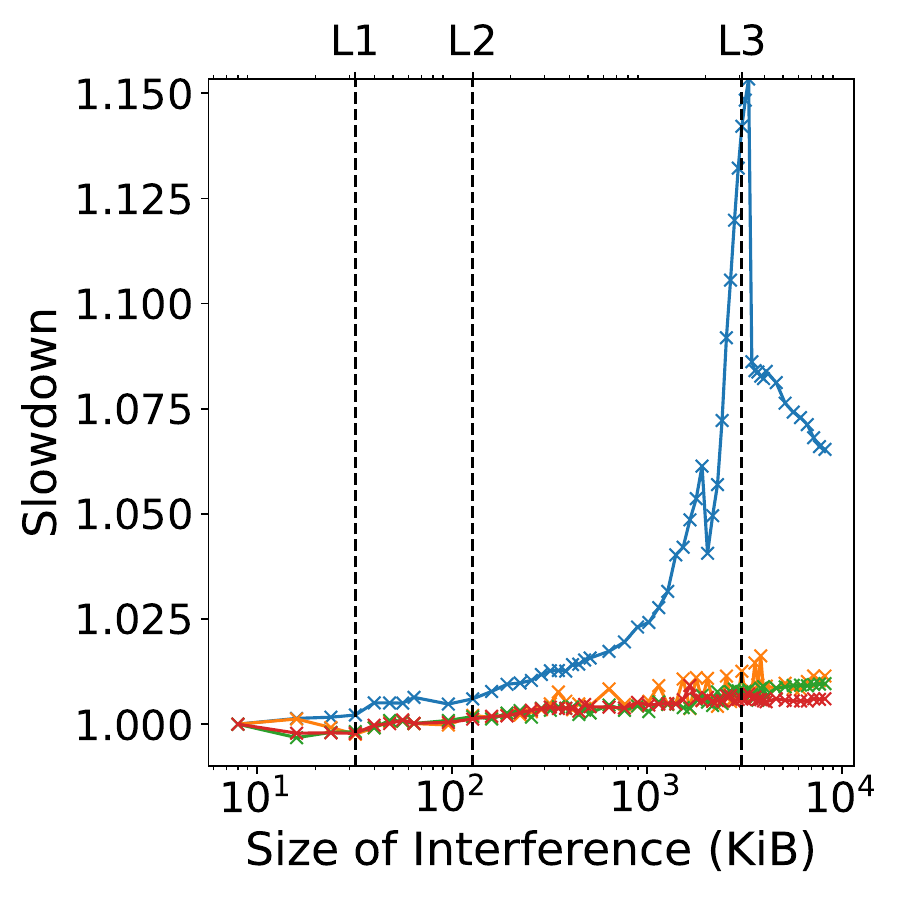}
            \captionsetup{justification=centering}
            \caption{rk3588: Way / Read}
            \label{fig:all-rk3588-way-disparity-read-cif}
        \end{subfigure}
        \hfill
        \begin{subfigure}{0.24\textwidth}
            \centering
            \includegraphics[width=\textwidth]{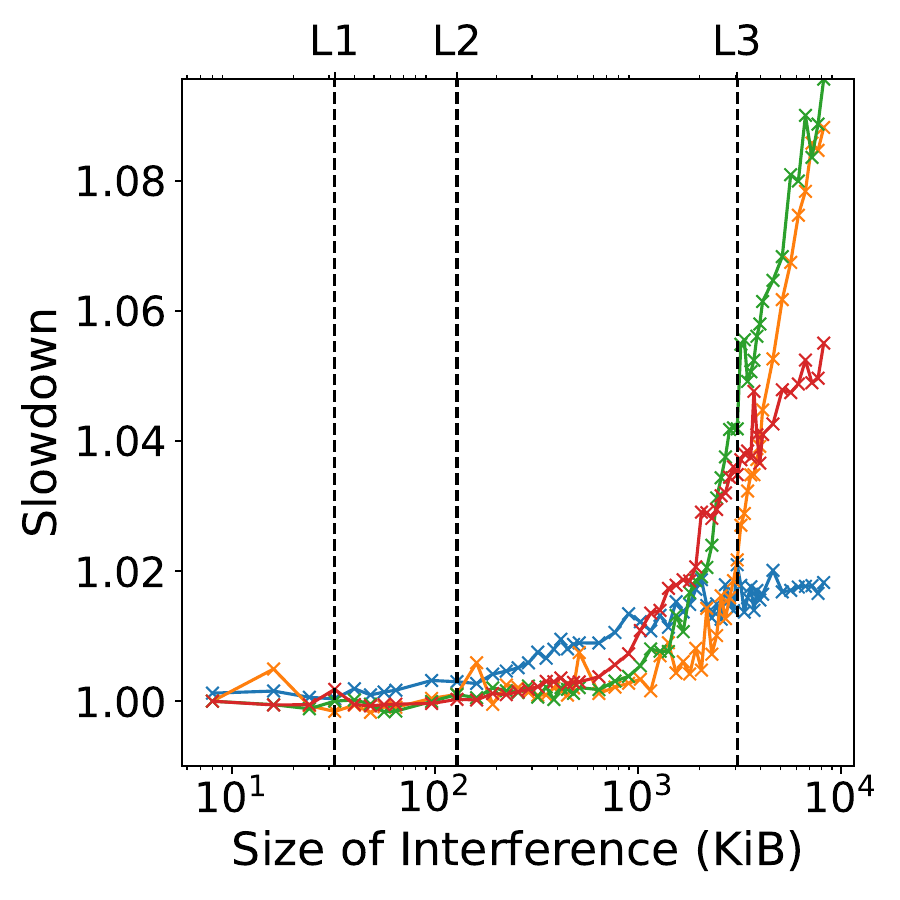}
            \captionsetup{justification=centering}
            \caption{rk3588: Way / Write}
            \label{fig:all-rk3588-way-disparity-write-cif}
        \end{subfigure}
        \hfill
        \begin{subfigure}{0.24\textwidth}
            \centering
            \includegraphics[width=\textwidth]{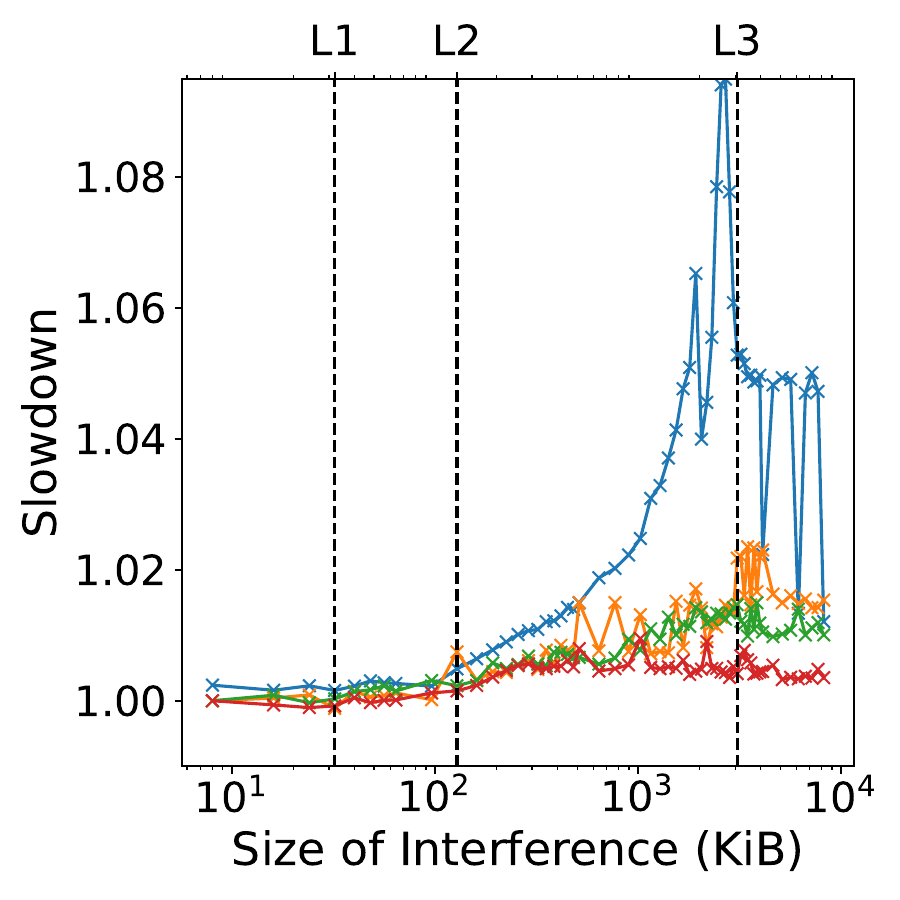}
            \captionsetup{justification=centering}
            \caption{rk3588: Way / Modify}
            \label{fig:all-rk3588-way-disparity-modify-cif}
        \end{subfigure}
        \hfill
        \begin{subfigure}{0.24\textwidth}
            \centering
            \includegraphics[width=\textwidth]{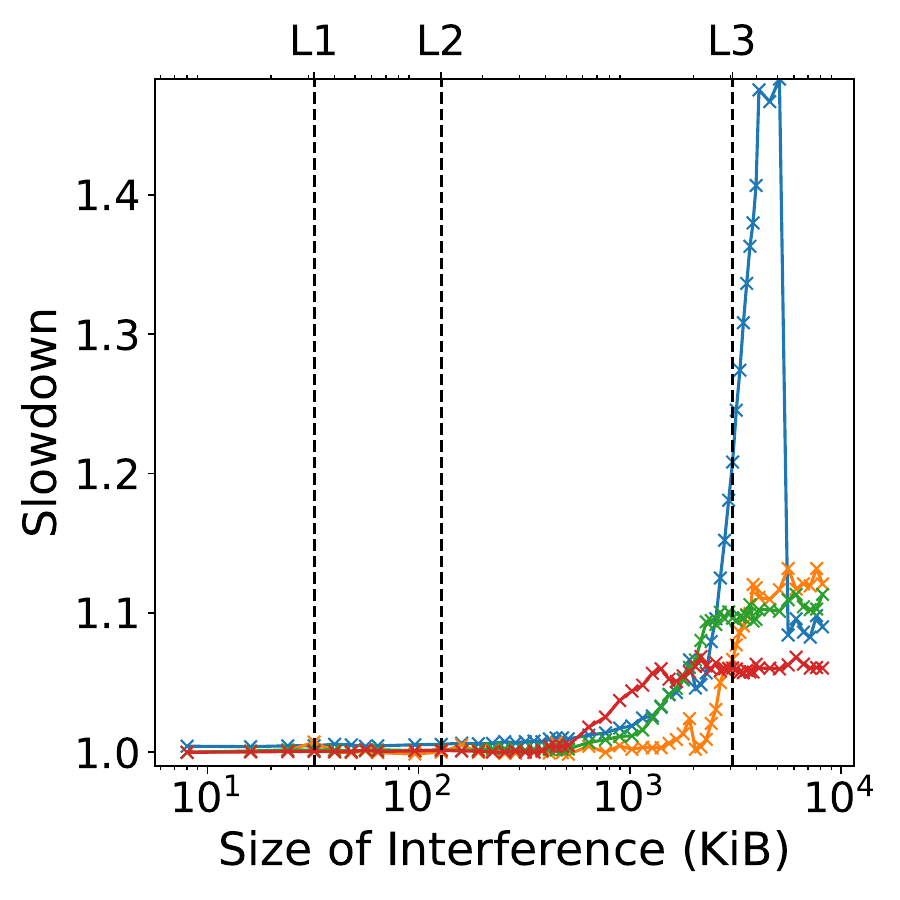}
            \captionsetup{justification=centering}
            \caption{rk3588: Way / Prefetch}
            \label{fig:all-rk3588-way-disparity-prefetch-cif}
        \end{subfigure}
        \hfill
        
        \caption{Execution Slowdown on \textit{'Disparity'} benchmark for \textit{'CIF'} dataset on \textit{'RK3588'} with Interferences and cache partitioning.}
        \label{fig:rk3588-disparity-cif}
    \end{figure}

    \begin{figure}[H]
        \begin{subfigure}{\textwidth}
            \centering
            \includegraphics[width=0.5\textwidth]{figures/set_subplot/legend.pdf}
        \end{subfigure}
        \centering
        
        \begin{subfigure}{0.24\textwidth}
            \centering
            \includegraphics[width=\textwidth]{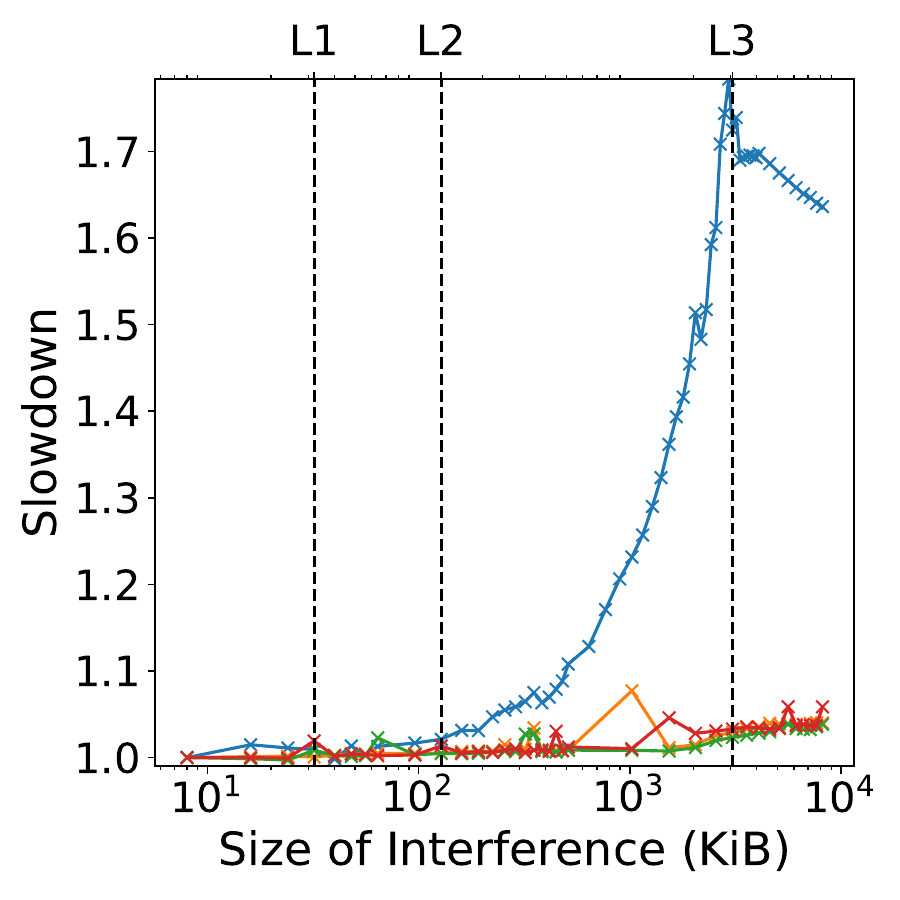}
            \captionsetup{justification=centering}
            \caption{rk3588: Set / Read}
            \label{fig:all-rk3588-set-mser-read-cif}
        \end{subfigure}
        \hfill
        \begin{subfigure}{0.24\textwidth}
            \centering
            \includegraphics[width=\textwidth]{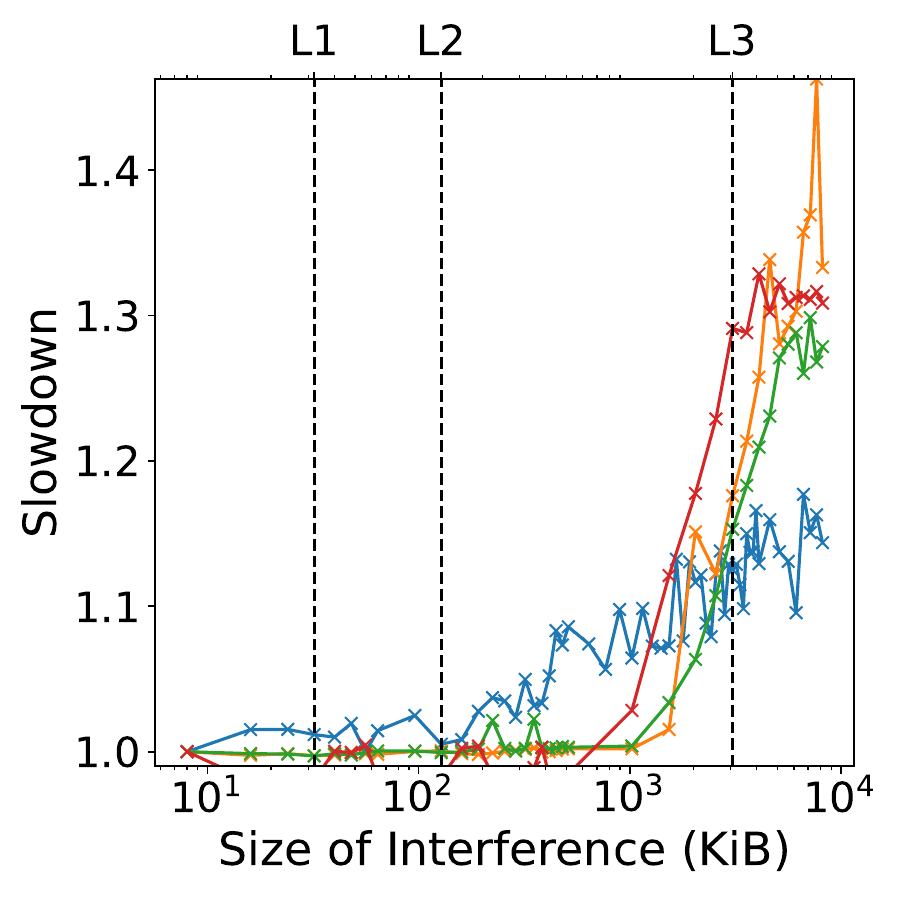}
            \captionsetup{justification=centering}
            \caption{rk3588: Set / Write}
            \label{fig:all-rk3588-set-mser-write-cif}
        \end{subfigure}
        \hfill
        \begin{subfigure}{0.24\textwidth}
            \centering
            \includegraphics[width=\textwidth]{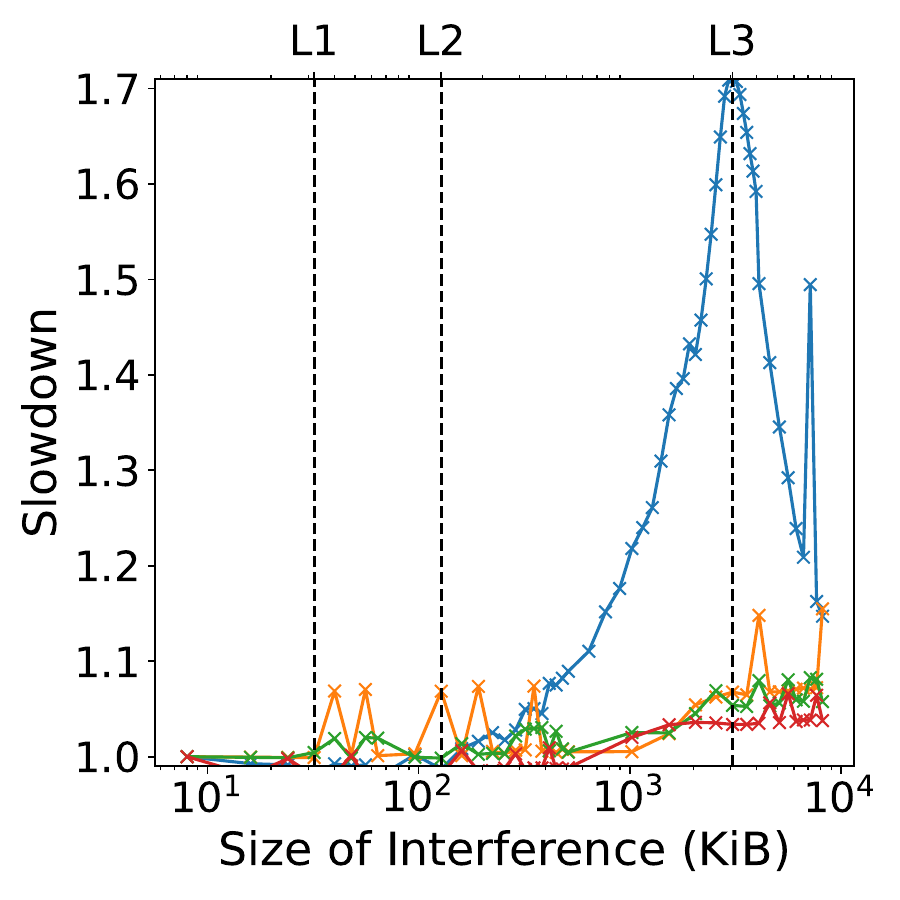}
            \captionsetup{justification=centering}
            \caption{rk3588: Set / Modify}
            \label{fig:all-rk3588-set-mser-modify-cif}
        \end{subfigure}
        \hfill
        \begin{subfigure}{0.24\textwidth}
            \centering
            \includegraphics[width=\textwidth]{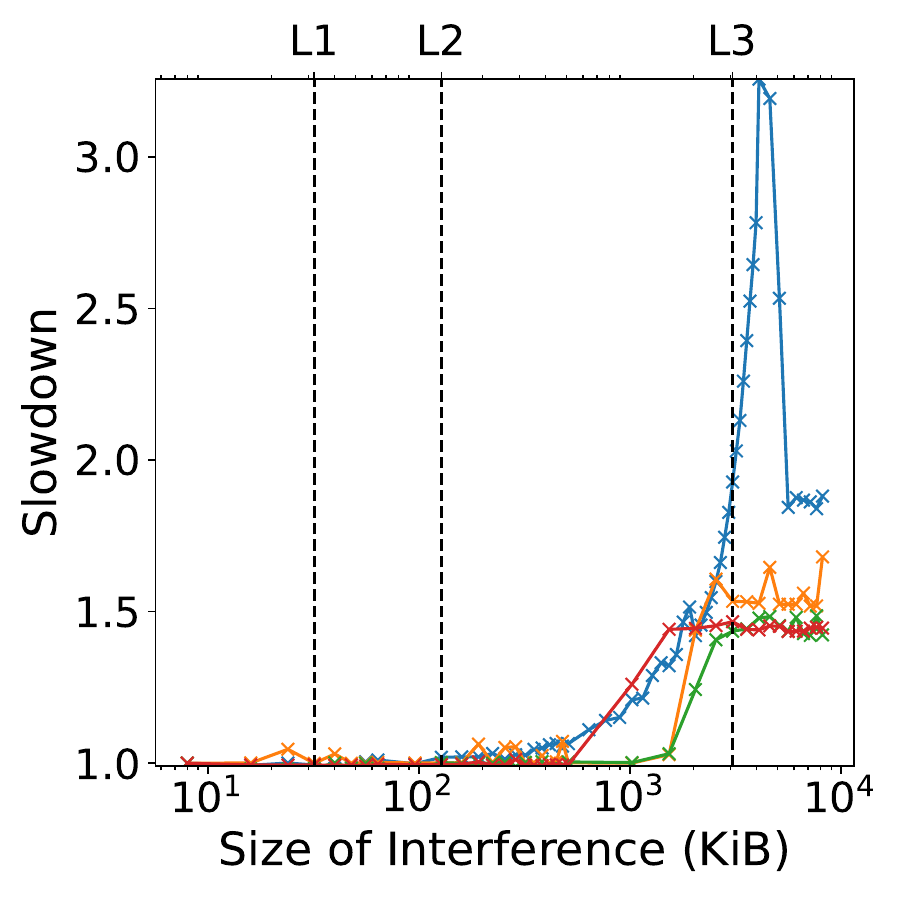}
            \captionsetup{justification=centering}
            \caption{rk3588: Set / Prefetch}
            \label{fig:all-rk3588-set-mser-prefetch-cif}
        \end{subfigure}
        \hfill
        \begin{subfigure}{0.24\textwidth}
            \centering
            \includegraphics[width=\textwidth]{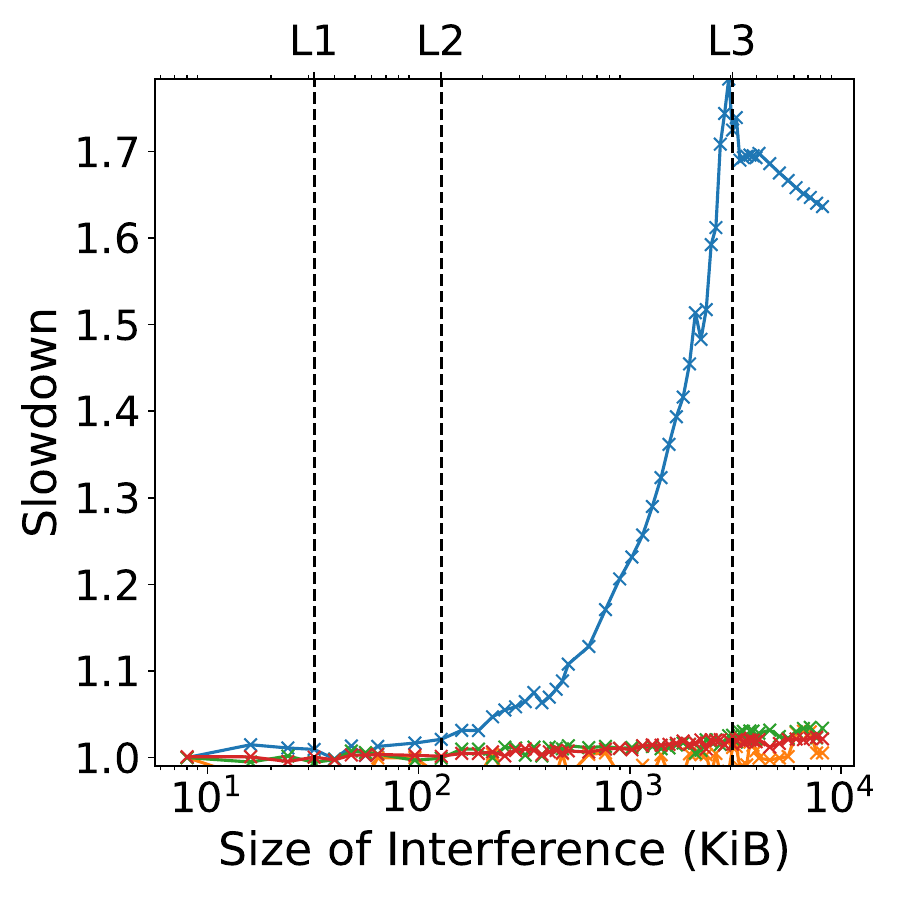}
            \captionsetup{justification=centering}
            \caption{rk3588: Way / Read}
            \label{fig:all-rk3588-way-mser-read-cif}
        \end{subfigure}
        \hfill
        \begin{subfigure}{0.24\textwidth}
            \centering
            \includegraphics[width=\textwidth]{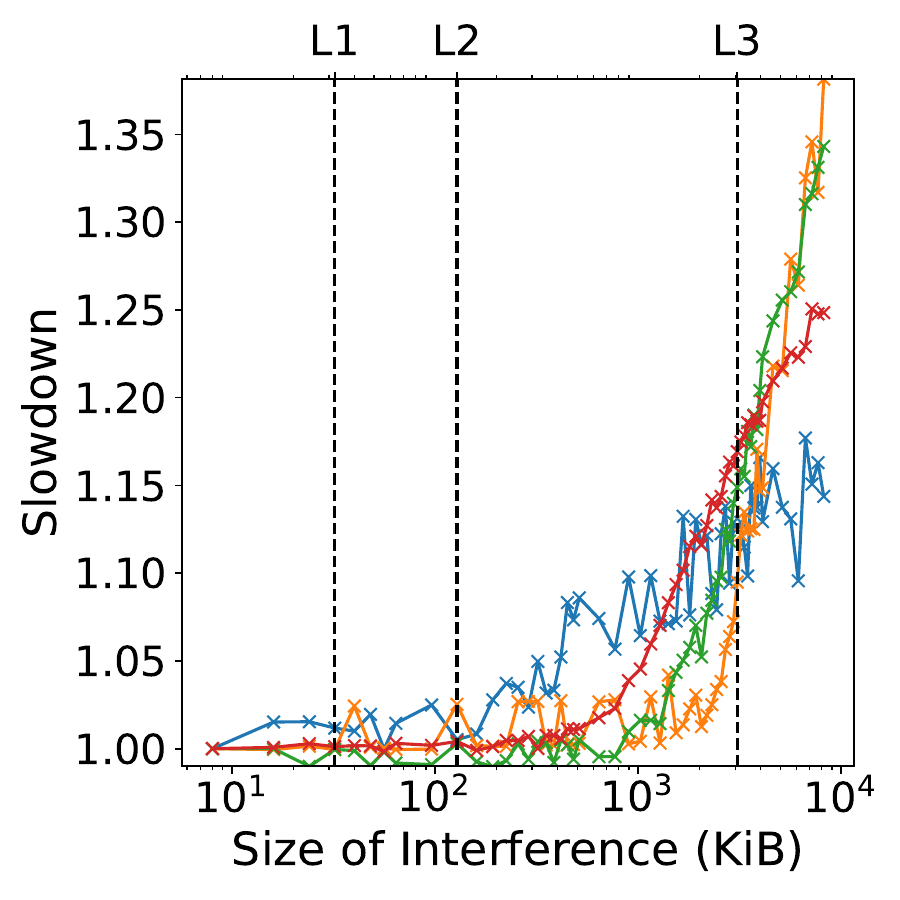}
            \captionsetup{justification=centering}
            \caption{rk3588: Way / Write}
            \label{fig:all-rk3588-way-mser-write-cif}
        \end{subfigure}
        \hfill
        \begin{subfigure}{0.24\textwidth}
            \centering
            \includegraphics[width=\textwidth]{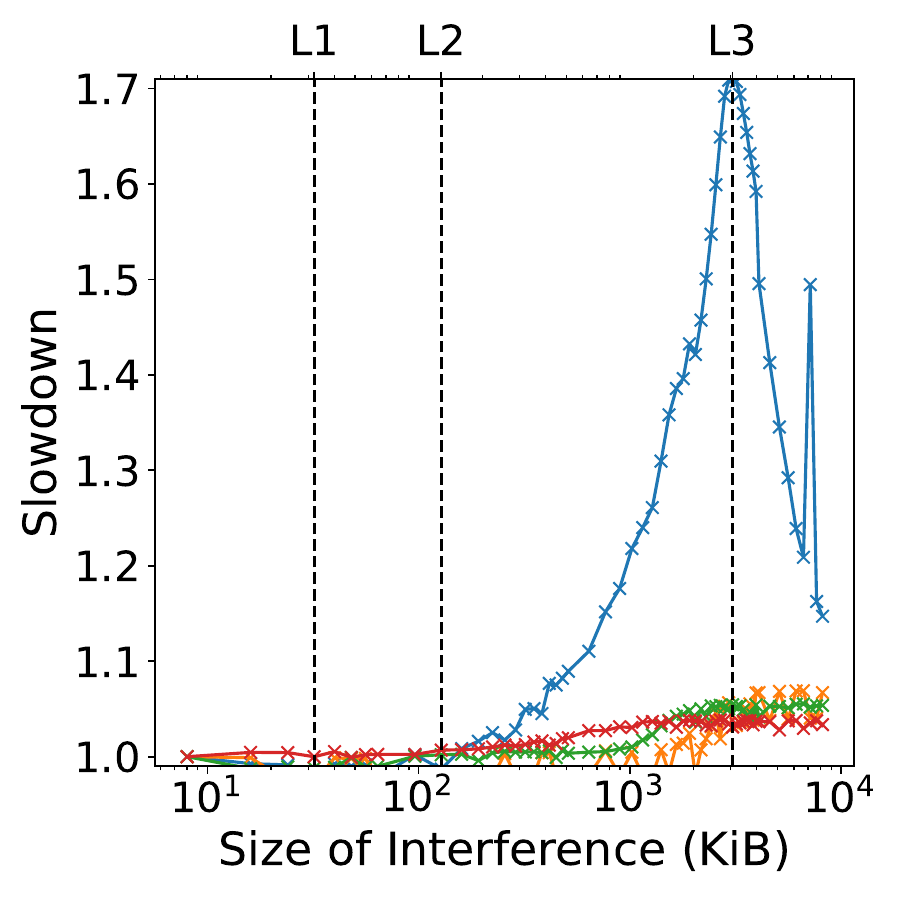}
            \captionsetup{justification=centering}
            \caption{rk3588: Way / Modify}
            \label{fig:all-rk3588-way-mser-modify-cif}
        \end{subfigure}
        \hfill
        \begin{subfigure}{0.24\textwidth}
            \centering
            \includegraphics[width=\textwidth]{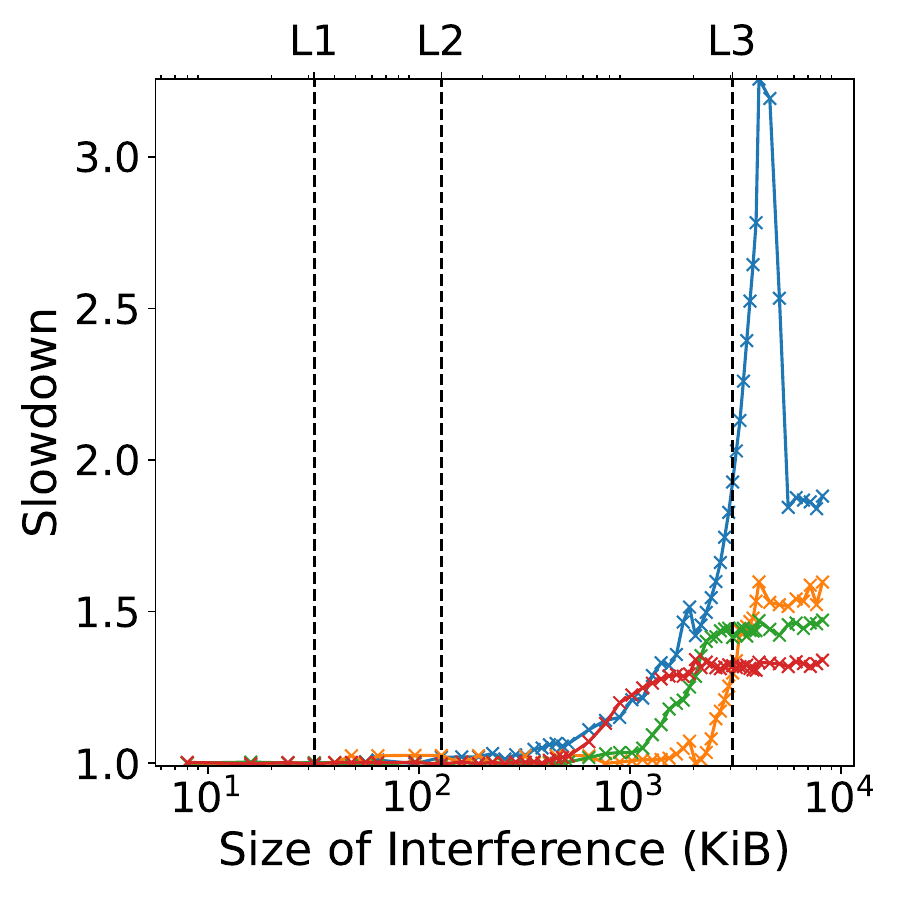}
            \captionsetup{justification=centering}
            \caption{rk3588: Way / Prefetch}
            \label{fig:all-rk3588-way-mser-prefetch-cif}
        \end{subfigure}
        \hfill
        
        \caption{Execution Slowdown on \textit{'Mser'} benchmark for \textit{'CIF'} dataset on \textit{'RK3588'} with Interferences and cache partitioning.}
        \label{fig:rk3588-mser-cif}
    \end{figure}

    \begin{figure}[H]
        \begin{subfigure}{\textwidth}
            \centering
            \includegraphics[width=0.5\textwidth]{figures/set_subplot/legend.pdf}
        \end{subfigure}
        \centering
        
        \begin{subfigure}{0.24\textwidth}
            \centering
            \includegraphics[width=\textwidth]{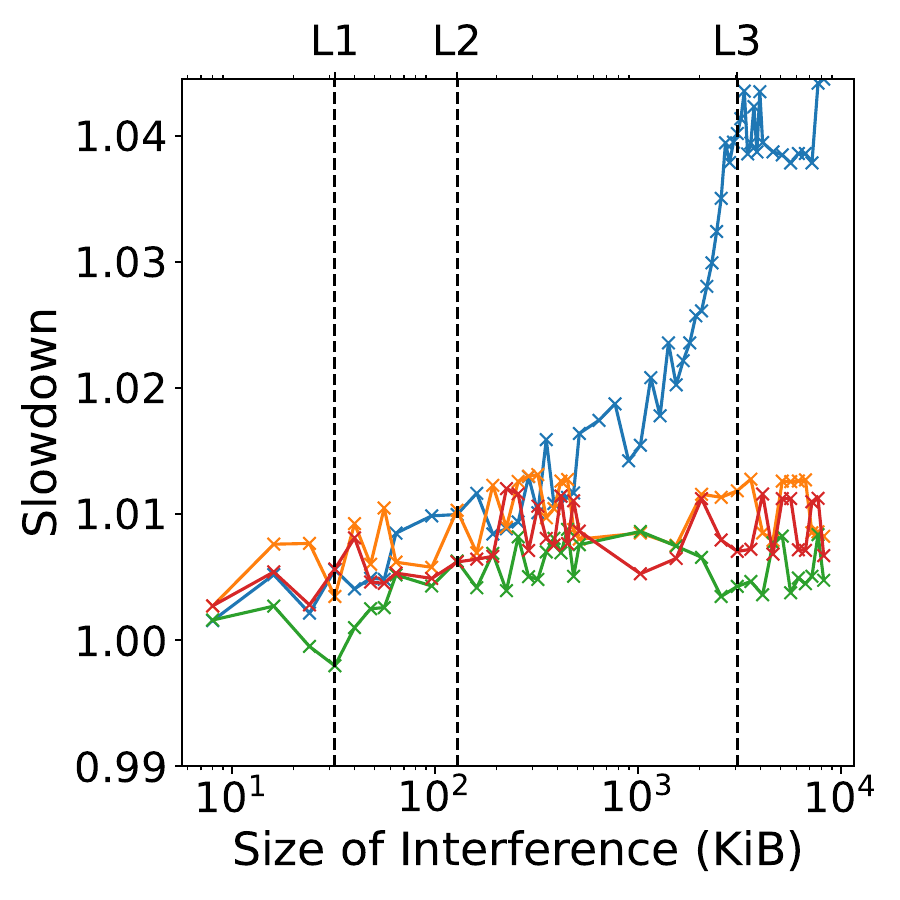}
            \captionsetup{justification=centering}
            \caption{rk3588: Set / Read}
            \label{fig:all-rk3588-set-tracking-read-cif}
        \end{subfigure}
        \hfill
        \begin{subfigure}{0.24\textwidth}
            \centering
            \includegraphics[width=\textwidth]{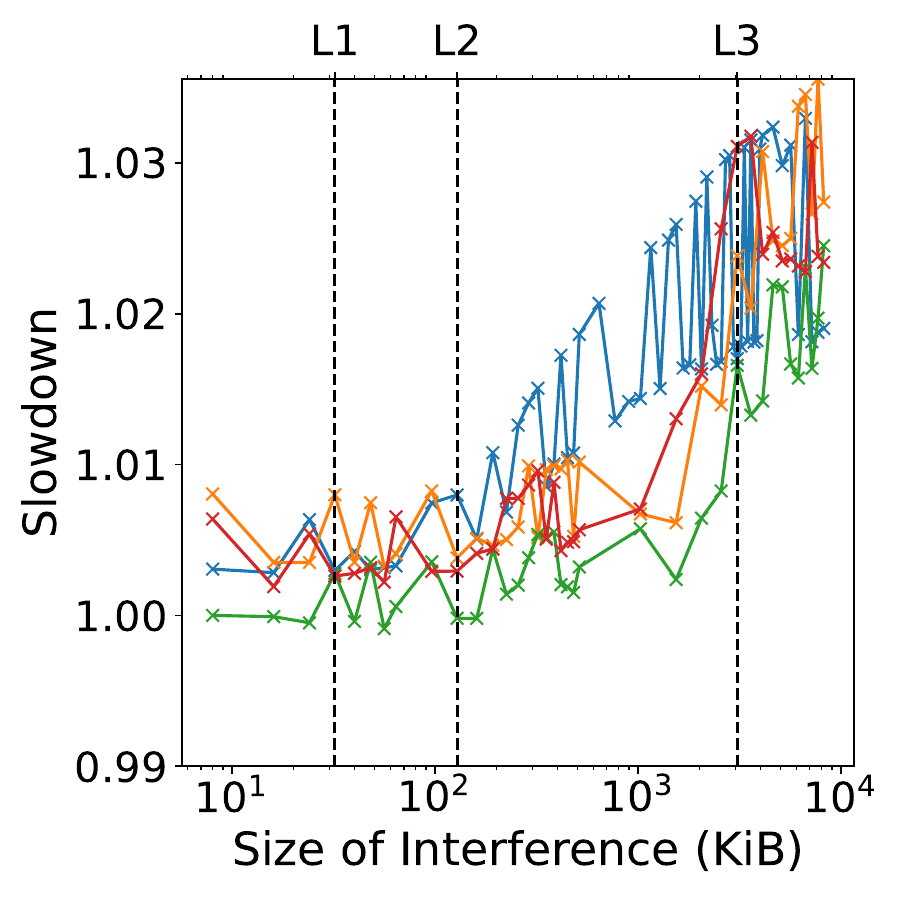}
            \captionsetup{justification=centering}
            \caption{rk3588: Set / Write}
            \label{fig:all-rk3588-set-tracking-write-cif}
        \end{subfigure}
        \hfill
        \begin{subfigure}{0.24\textwidth}
            \centering
            \includegraphics[width=\textwidth]{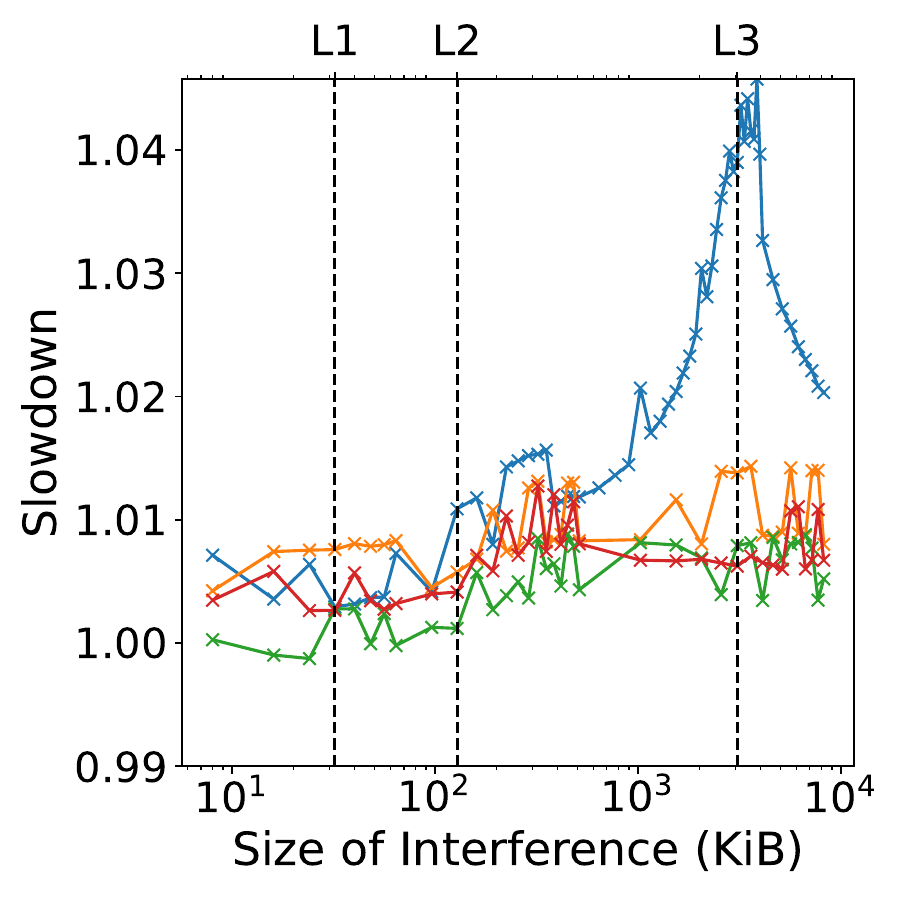}
            \captionsetup{justification=centering}
            \caption{rk3588: Set / Modify}
            \label{fig:all-rk3588-set-tracking-modify-cif}
        \end{subfigure}
        \hfill
        \begin{subfigure}{0.24\textwidth}
            \centering
            \includegraphics[width=\textwidth]{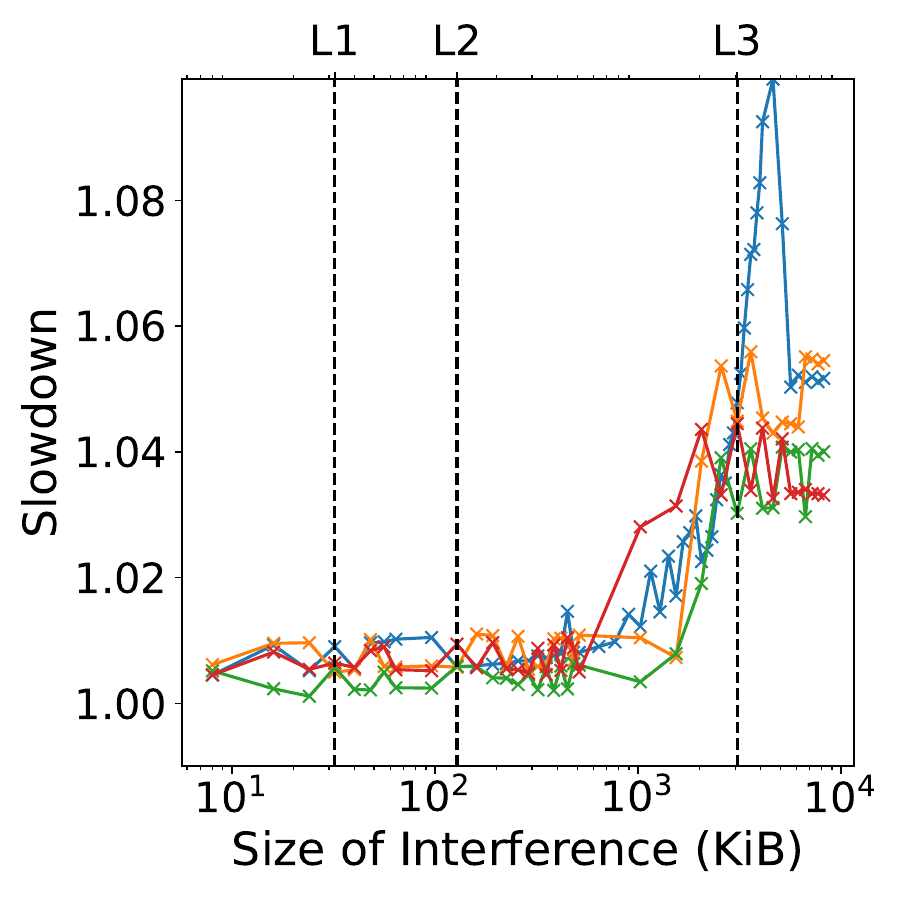}
            \captionsetup{justification=centering}
            \caption{rk3588: Set / Prefetch}
            \label{fig:all-rk3588-set-tracking-prefetch-cif}
        \end{subfigure}
        \hfill
        \begin{subfigure}{0.24\textwidth}
            \centering
            \includegraphics[width=\textwidth]{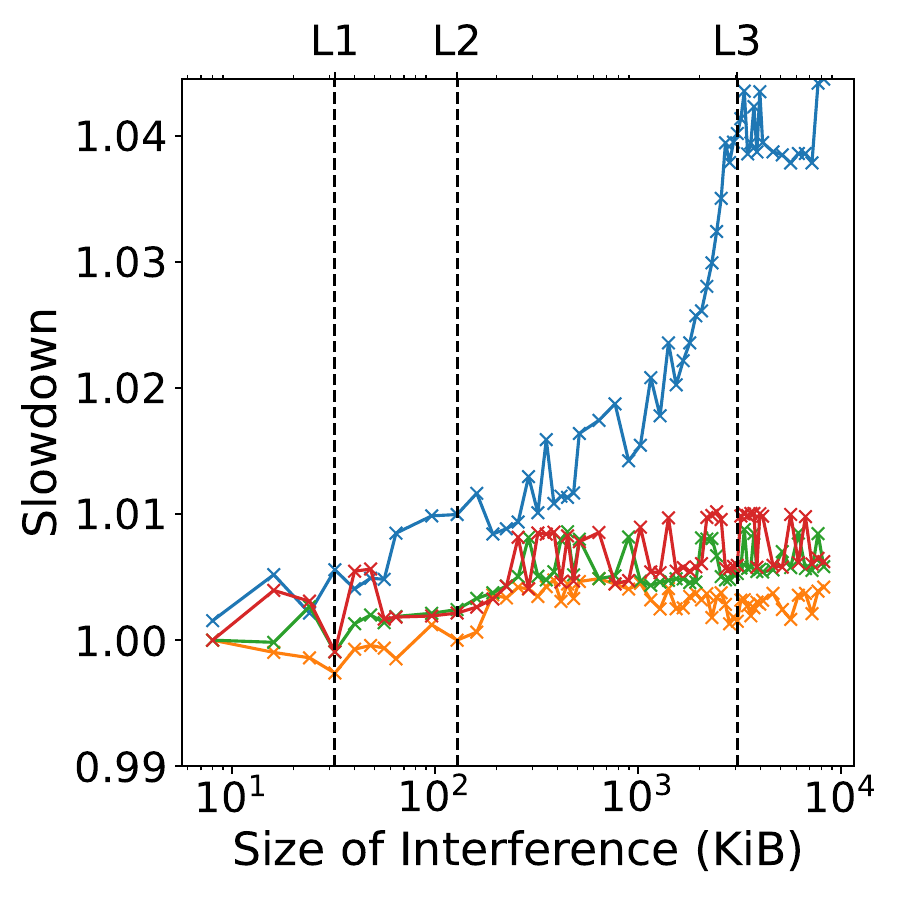}
            \captionsetup{justification=centering}
            \caption{rk3588: Way / Read}
            \label{fig:all-rk3588-way-tracking-read-cif}
        \end{subfigure}
        \hfill
        \begin{subfigure}{0.24\textwidth}
            \centering
            \includegraphics[width=\textwidth]{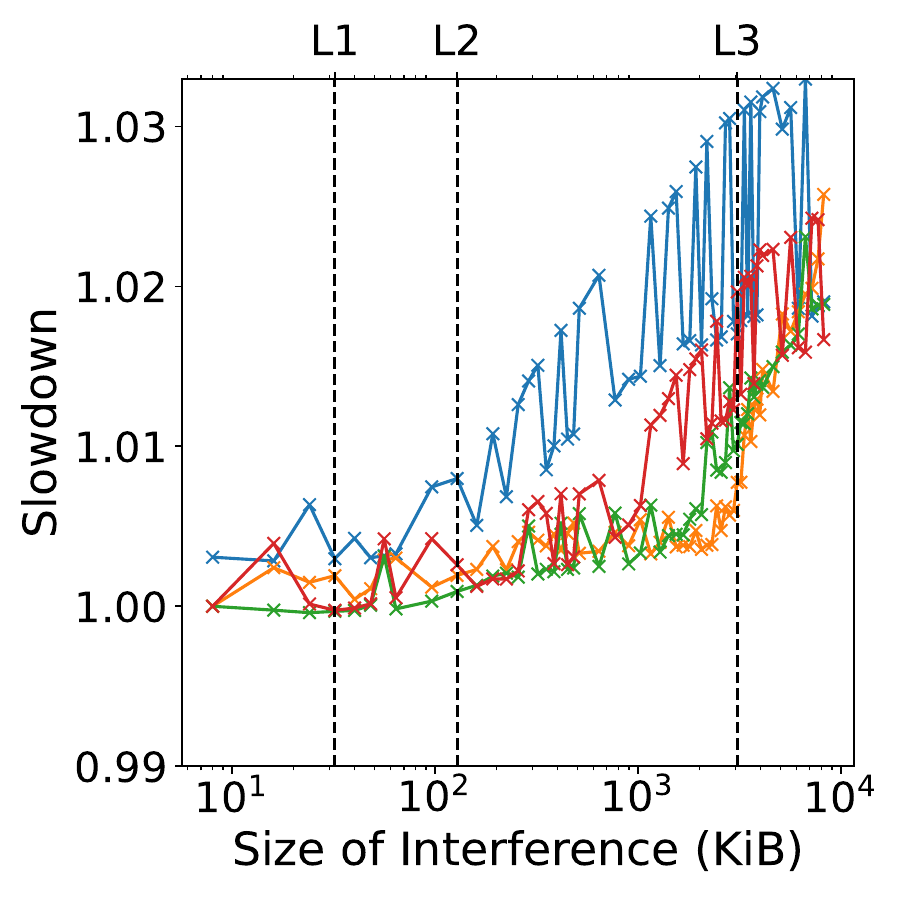}
            \captionsetup{justification=centering}
            \caption{rk3588: Way / Write}
            \label{fig:all-rk3588-way-tracking-write-cif}
        \end{subfigure}
        \hfill
        \begin{subfigure}{0.24\textwidth}
            \centering
            \includegraphics[width=\textwidth]{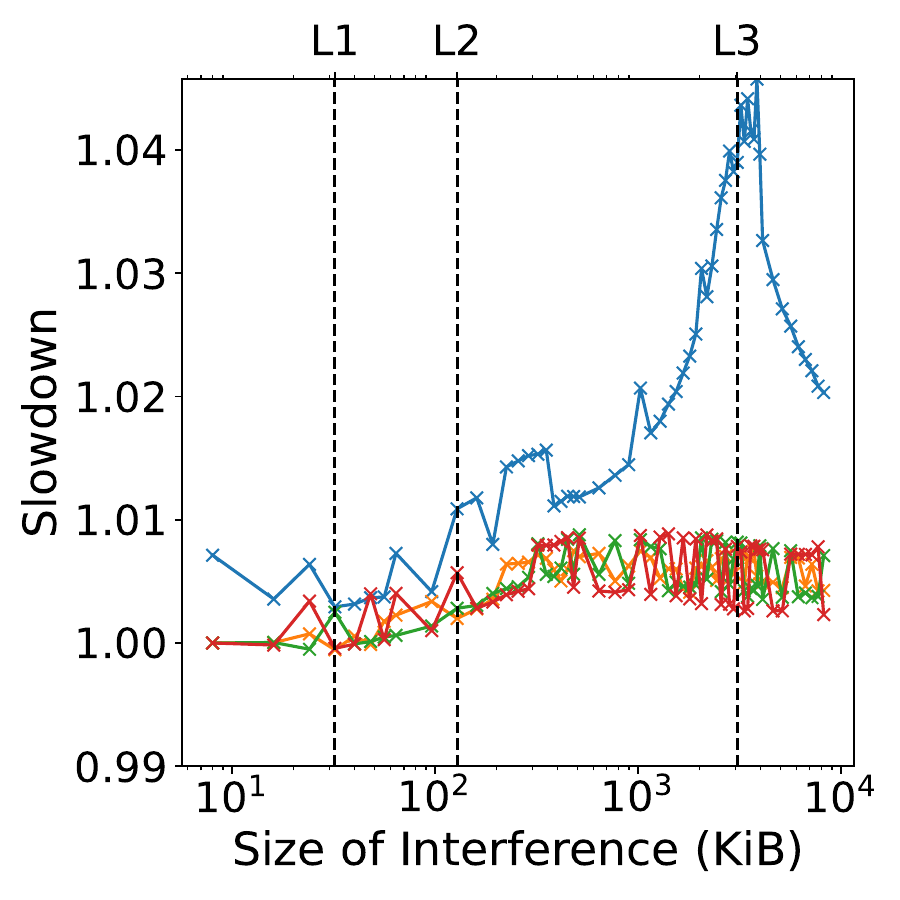}
            \captionsetup{justification=centering}
            \caption{rk3588: Way / Modify}
            \label{fig:all-rk3588-way-tracking-modify-cif}
        \end{subfigure}
        \hfill
        \begin{subfigure}{0.24\textwidth}
            \centering
            \includegraphics[width=\textwidth]{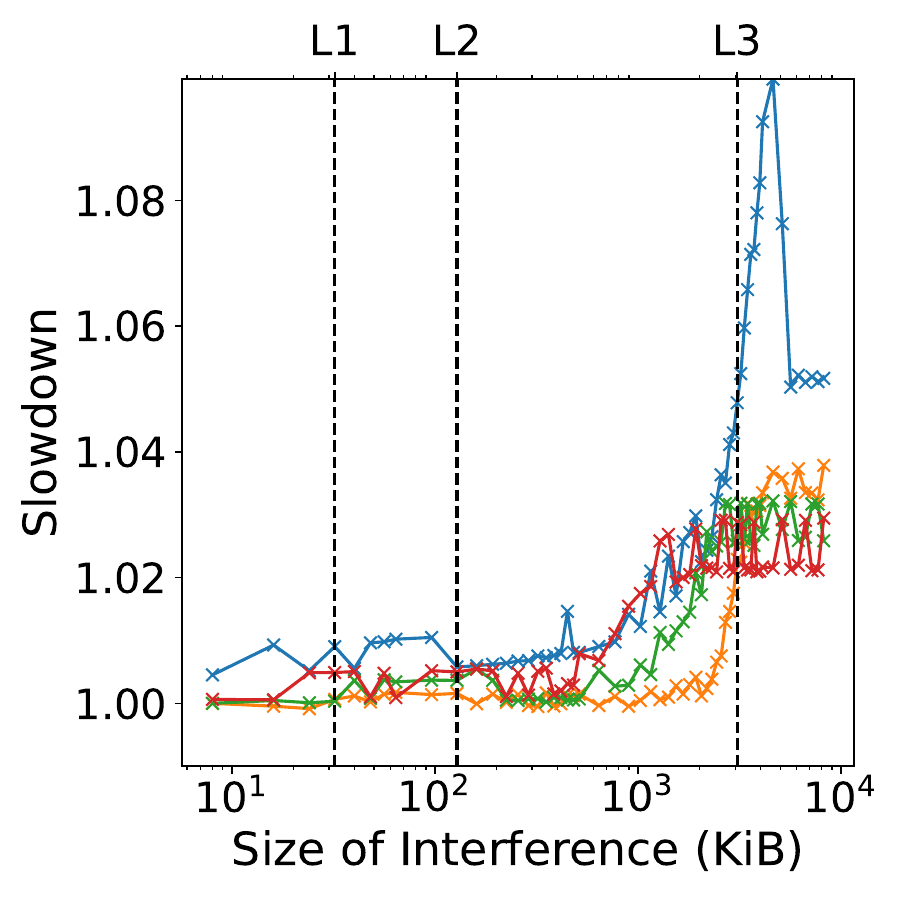}
            \captionsetup{justification=centering}
            \caption{rk3588: Way / Prefetch}
            \label{fig:all-rk3588-way-tracking-prefetch-cif}
        \end{subfigure}
        \hfill
        
        \caption{Execution Slowdown on \textit{'Tracking'} benchmark for \textit{'CIF'} dataset on \textit{'RK3588'} with Interferences and cache partitioning.}
        \label{fig:rk3588-tracking-cif}
    \end{figure}

    \begin{figure}[H]
        \begin{subfigure}{\textwidth}
            \centering
            \includegraphics[width=0.5\textwidth]{figures/set_subplot/legend.pdf}
        \end{subfigure}
        \centering
        
        \begin{subfigure}{0.24\textwidth}
            \centering
            \includegraphics[width=\textwidth]{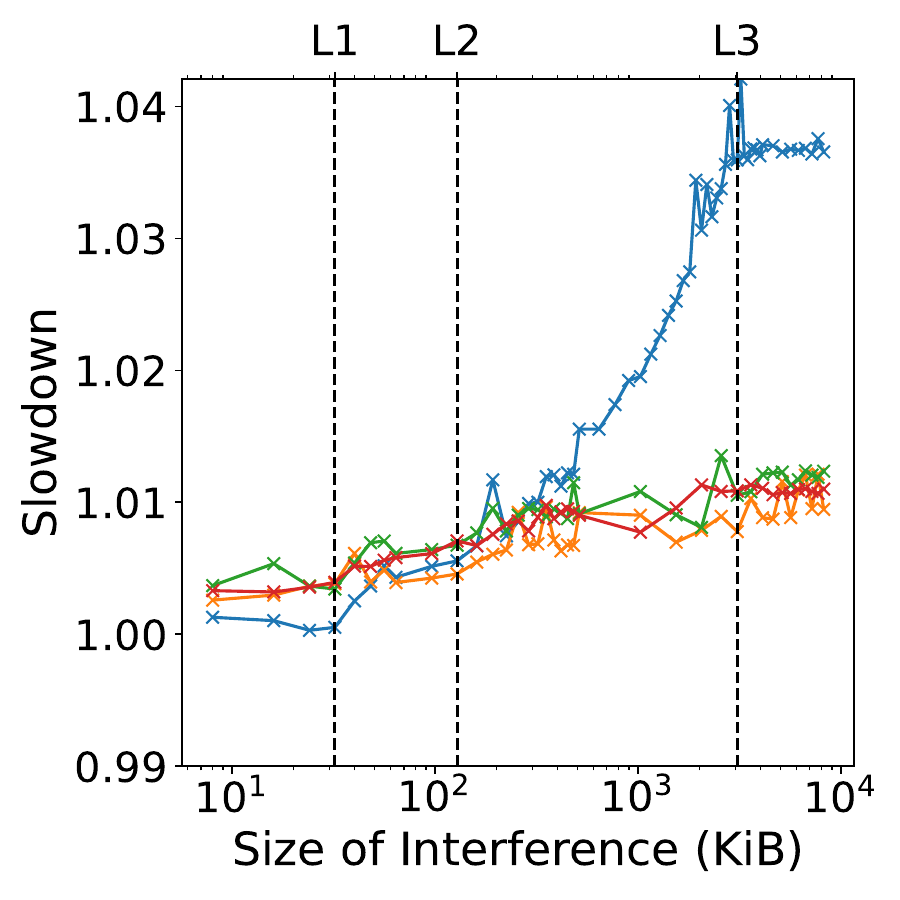}
            \captionsetup{justification=centering}
            \caption{rk3588: Set / Read}
            \label{fig:all-rk3588-set-sift-read-cif}
        \end{subfigure}
        \hfill
        \begin{subfigure}{0.24\textwidth}
            \centering
            \includegraphics[width=\textwidth]{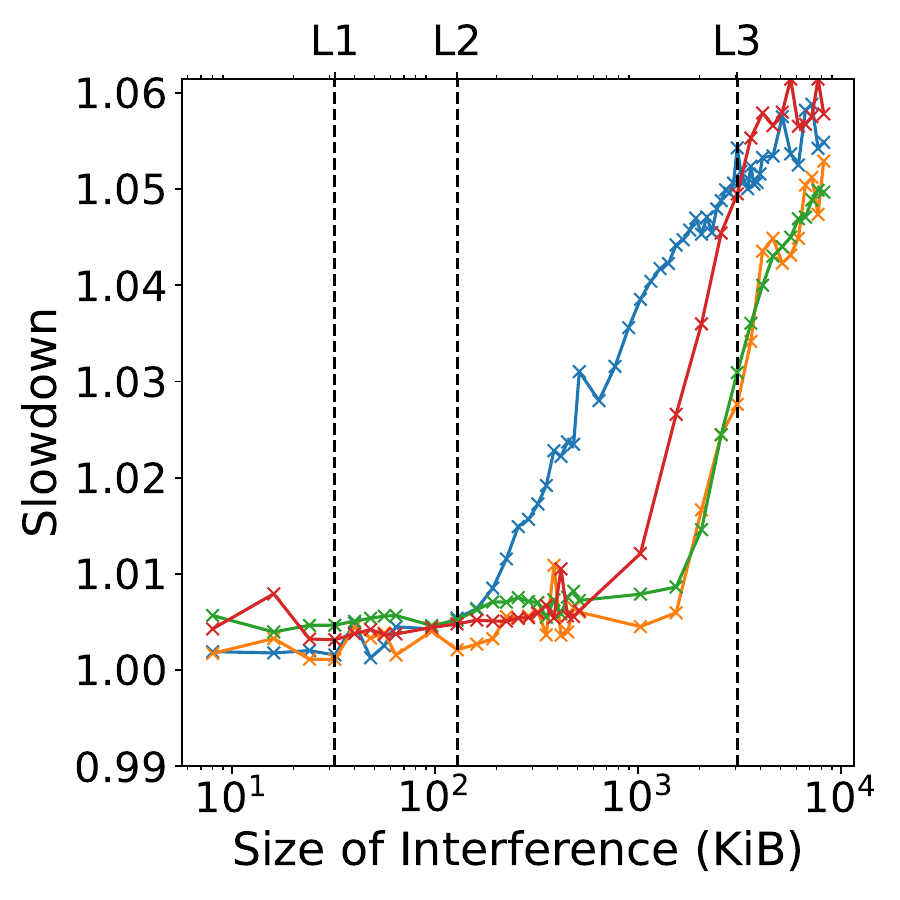}
            \captionsetup{justification=centering}
            \caption{rk3588: Set / Write}
            \label{fig:all-rk3588-set-sift-write-cif}
        \end{subfigure}
        \hfill
        \begin{subfigure}{0.24\textwidth}
            \centering
            \includegraphics[width=\textwidth]{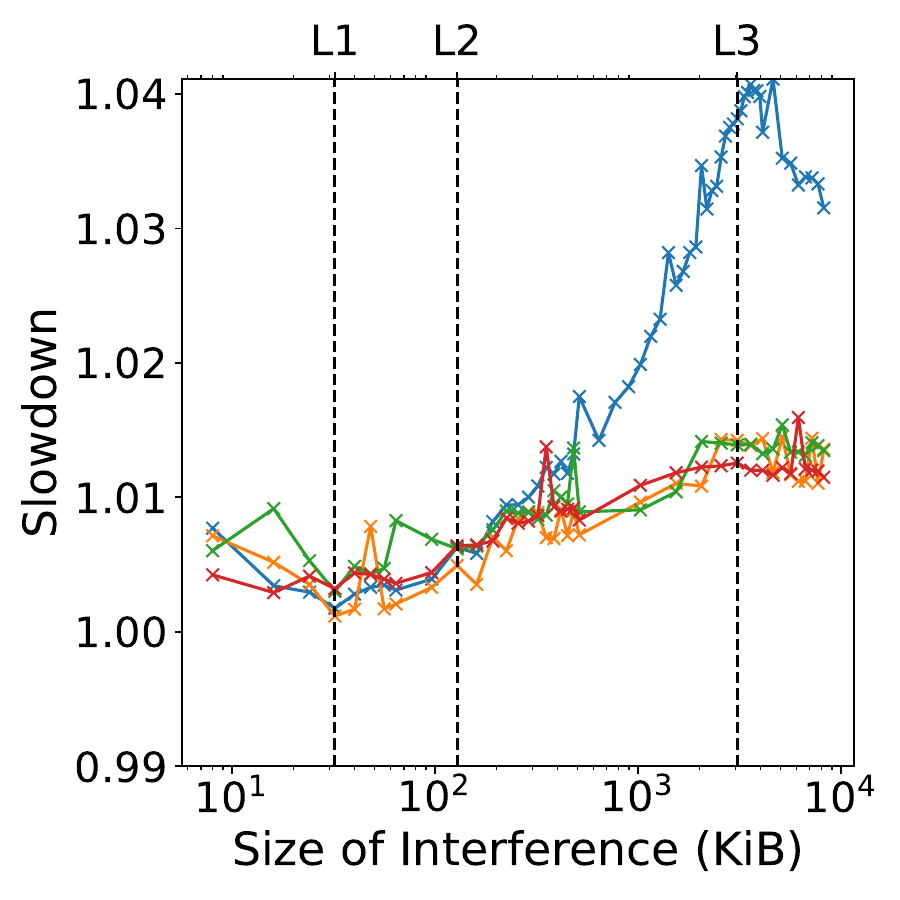}
            \captionsetup{justification=centering}
            \caption{rk3588: Set / Modify}
            \label{fig:all-rk3588-set-sift-modify-cif}
        \end{subfigure}
        \hfill
        \begin{subfigure}{0.24\textwidth}
            \centering
            \includegraphics[width=\textwidth]{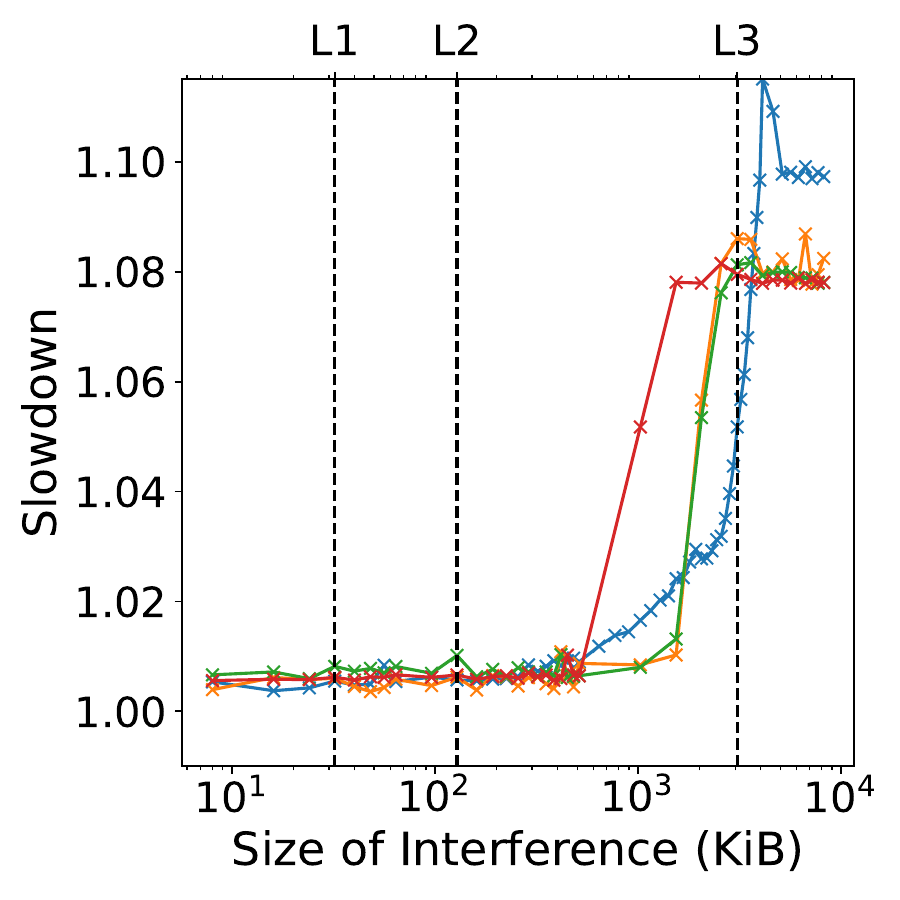}
            \captionsetup{justification=centering}
            \caption{rk3588: Set / Prefetch}
            \label{fig:all-rk3588-set-sift-prefetch-cif}
        \end{subfigure}
        \hfill
        \begin{subfigure}{0.24\textwidth}
            \centering
            \includegraphics[width=\textwidth]{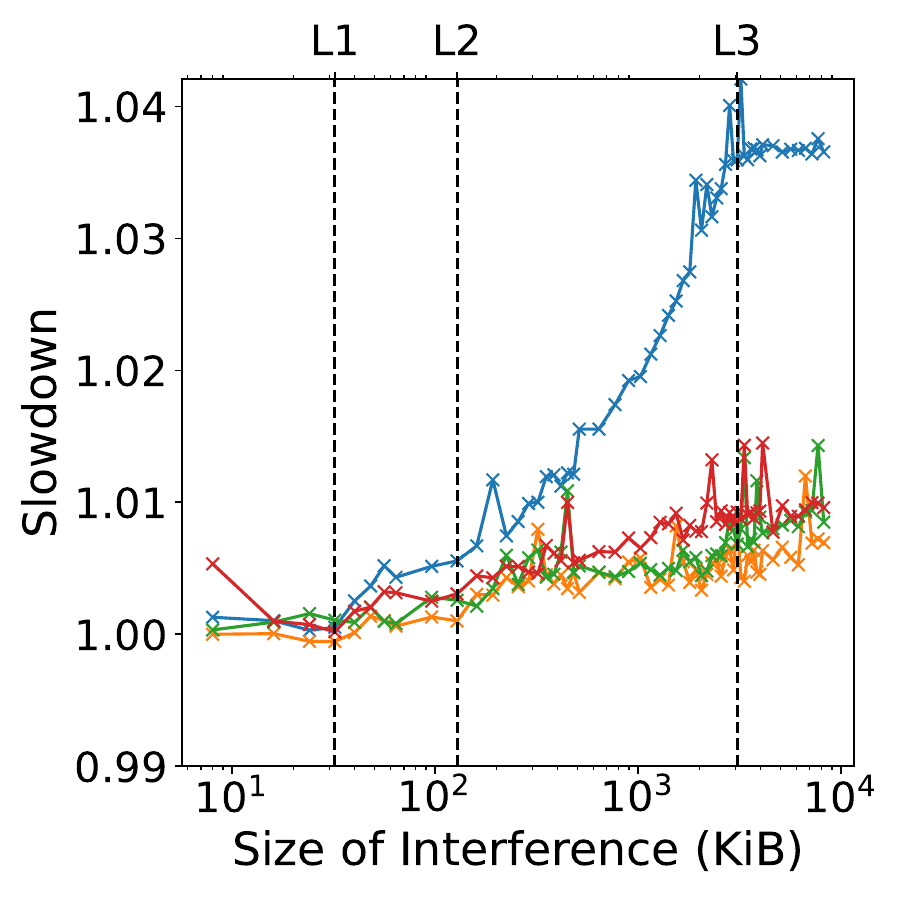}
            \captionsetup{justification=centering}
            \caption{rk3588: Way / Read}
            \label{fig:all-rk3588-way-sift-read-cif}
        \end{subfigure}
        \hfill
        \begin{subfigure}{0.24\textwidth}
            \centering
            \includegraphics[width=\textwidth]{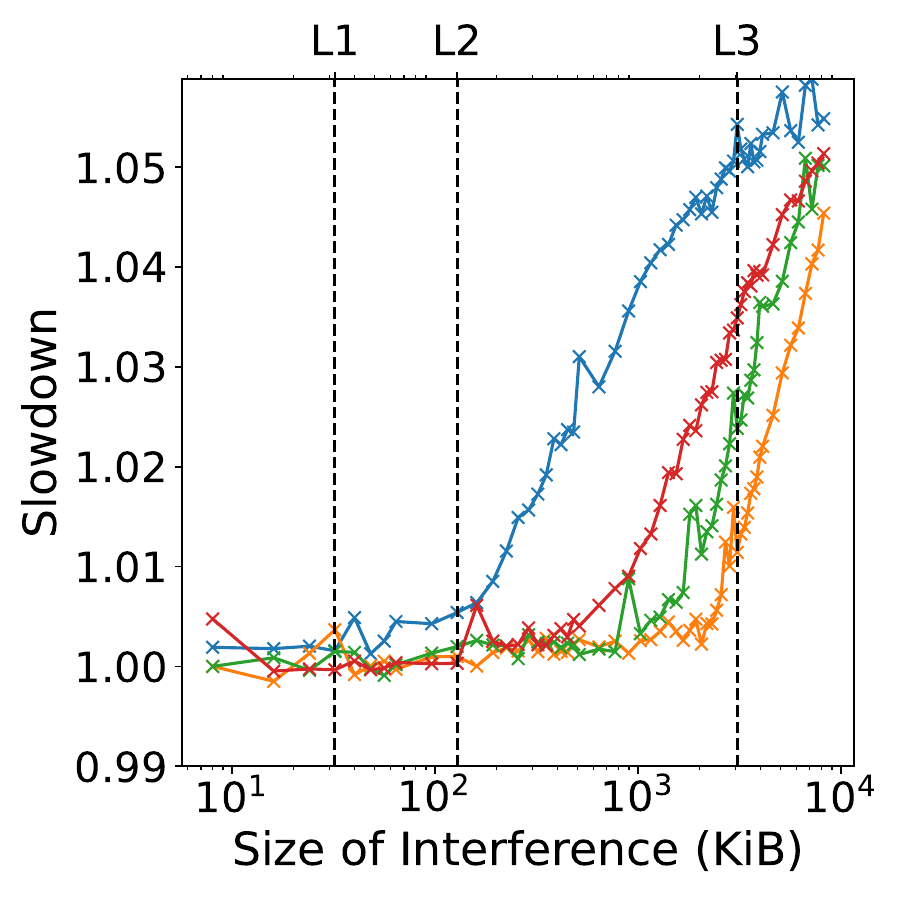}
            \captionsetup{justification=centering}
            \caption{rk3588: Way / Write}
            \label{fig:all-rk3588-way-sift-write-cif}
        \end{subfigure}
        \hfill
        \begin{subfigure}{0.24\textwidth}
            \centering
            \includegraphics[width=\textwidth]{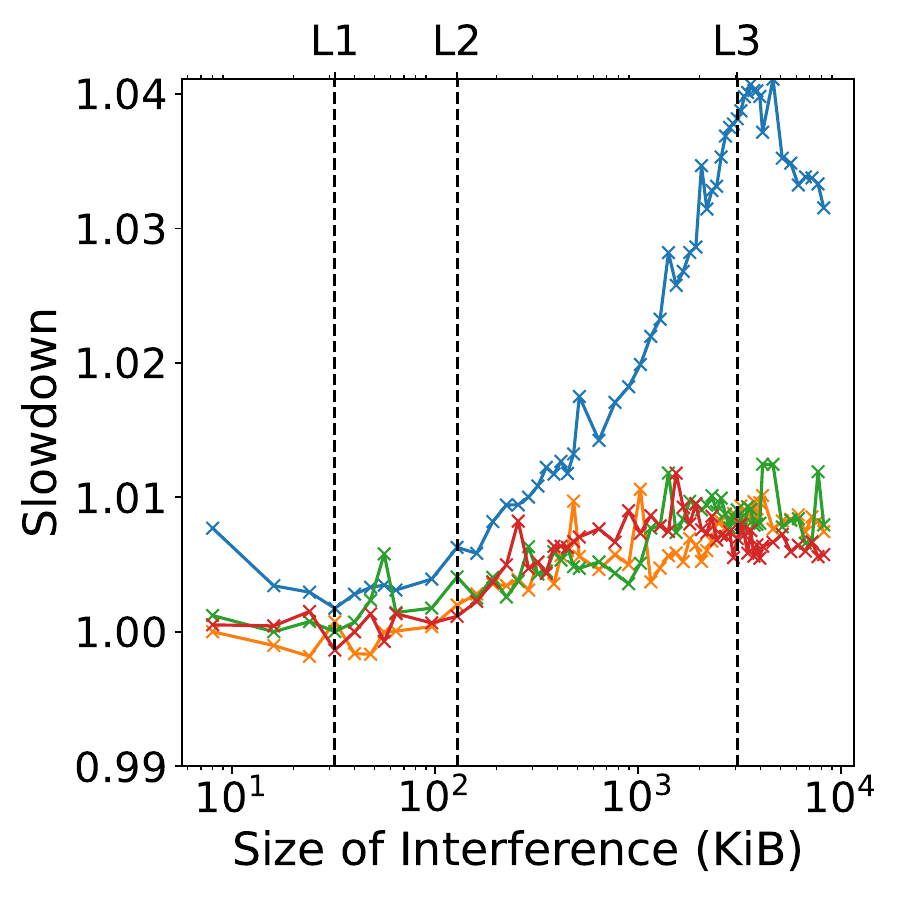}
            \captionsetup{justification=centering}
            \caption{rk3588: Way / Modify}
            \label{fig:all-rk3588-way-sift-modify-cif}
        \end{subfigure}
        \hfill
        \begin{subfigure}{0.24\textwidth}
            \centering
            \includegraphics[width=\textwidth]{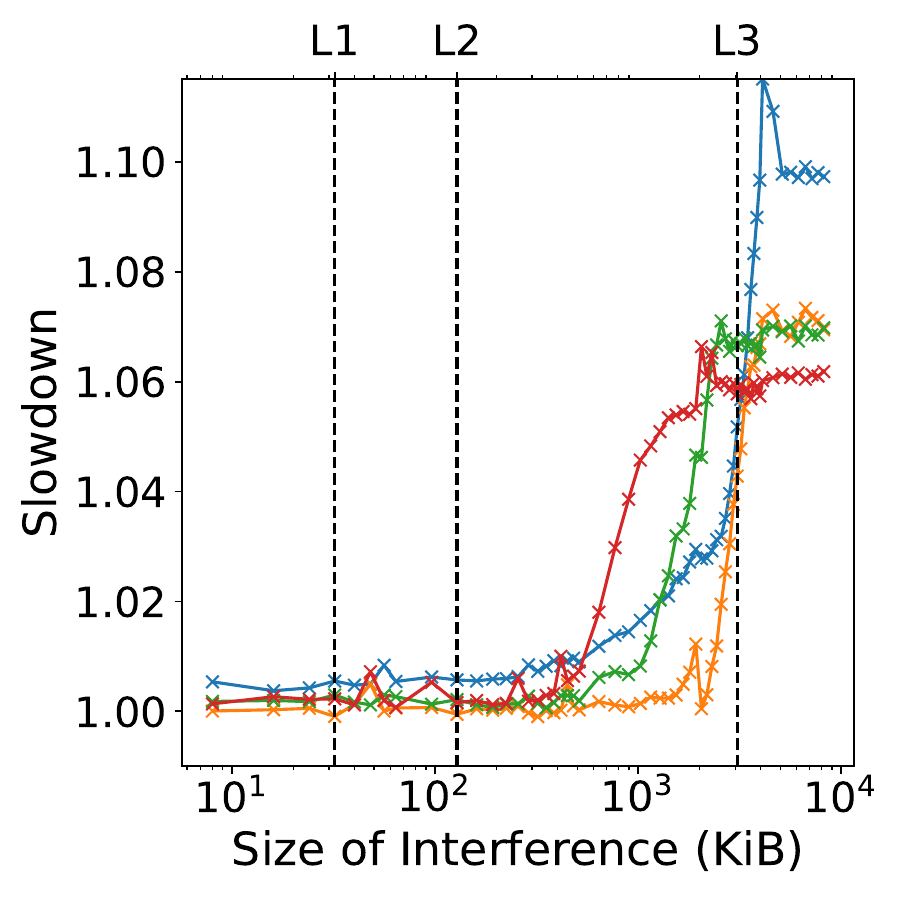}
            \captionsetup{justification=centering}
            \caption{rk3588: Way / Prefetch}
            \label{fig:all-rk3588-way-sift-prefetch-cif}
        \end{subfigure}
        \hfill
        
        \caption{Execution Slowdown on \textit{'Sift'} benchmark for \textit{'CIF'} dataset on \textit{'RK3588'} with Interferences and cache partitioning.}
        \label{fig:rk3588-sift-cif}
    \end{figure}

    \begin{figure}[H]
        \begin{subfigure}{\textwidth}
            \centering
            \includegraphics[width=0.5\textwidth]{figures/set_subplot/legend.pdf}
        \end{subfigure}
        \centering
        
        \begin{subfigure}{0.24\textwidth}
            \centering
            \includegraphics[width=\textwidth]{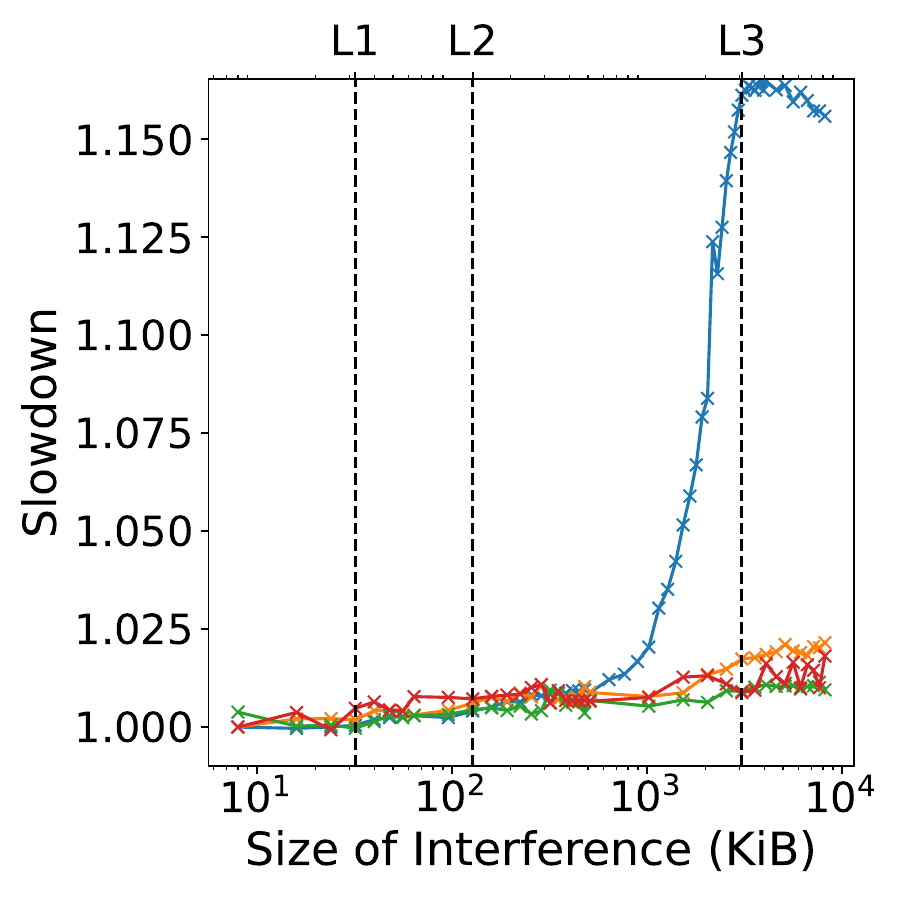}
            \captionsetup{justification=centering}
            \caption{rk3588: Set / Read}
            \label{fig:all-rk3588-set-disparity-read-vga}
        \end{subfigure}
        \hfill
        \begin{subfigure}{0.24\textwidth}
            \centering
            \includegraphics[width=\textwidth]{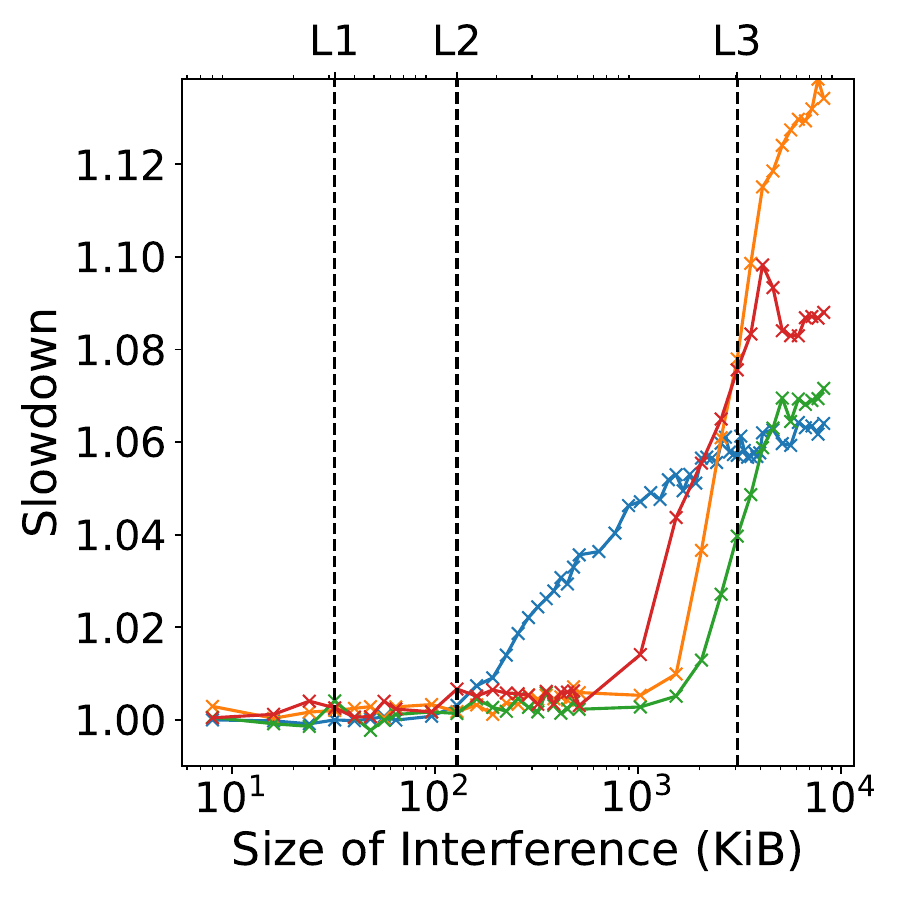}
            \captionsetup{justification=centering}
            \caption{rk3588: Set / Write}
            \label{fig:all-rk3588-set-disparity-write-vga}
        \end{subfigure}
        \hfill
        \begin{subfigure}{0.24\textwidth}
            \centering
            \includegraphics[width=\textwidth]{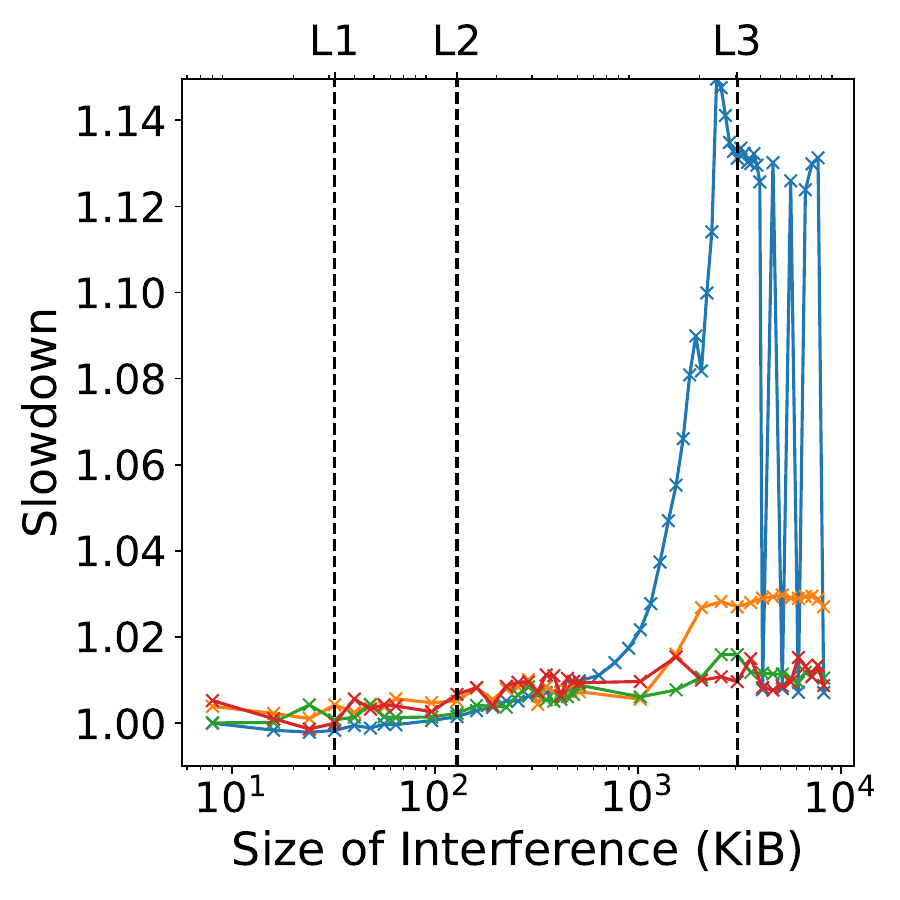}
            \captionsetup{justification=centering}
            \caption{rk3588: Set / Modify}
            \label{fig:all-rk3588-set-disparity-modify-vga}
        \end{subfigure}
        \hfill
        \begin{subfigure}{0.24\textwidth}
            \centering
            \includegraphics[width=\textwidth]{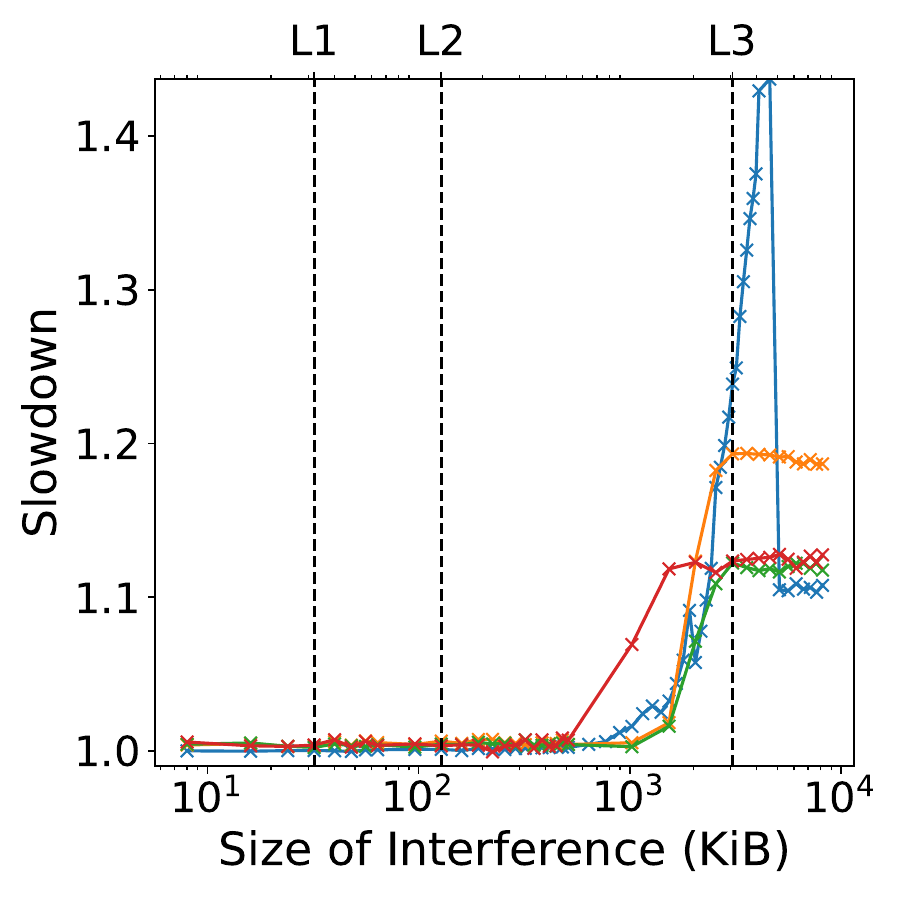}
            \captionsetup{justification=centering}
            \caption{rk3588: Set / Prefetch}
            \label{fig:all-rk3588-set-disparity-prefetch-vga}
        \end{subfigure}
        \hfill
        \begin{subfigure}{0.24\textwidth}
            \centering
            \includegraphics[width=\textwidth]{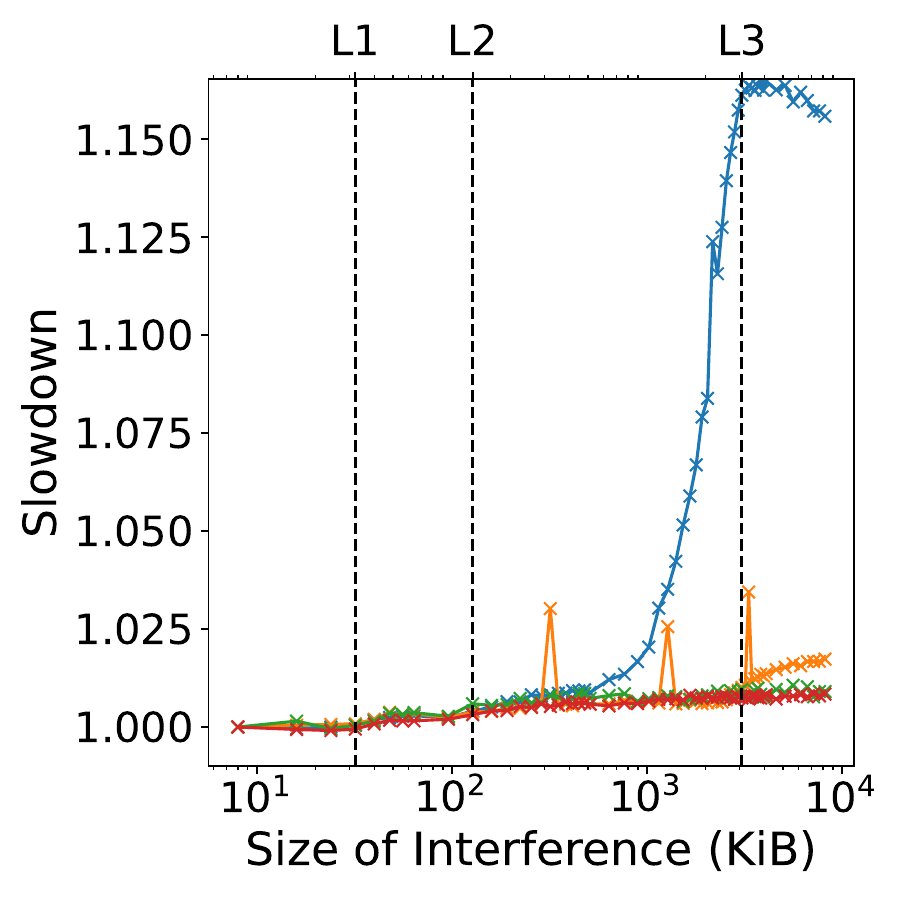}
            \captionsetup{justification=centering}
            \caption{rk3588: Way / Read}
            \label{fig:all-rk3588-way-disparity-read-vga}
        \end{subfigure}
        \hfill
        \begin{subfigure}{0.24\textwidth}
            \centering
            \includegraphics[width=\textwidth]{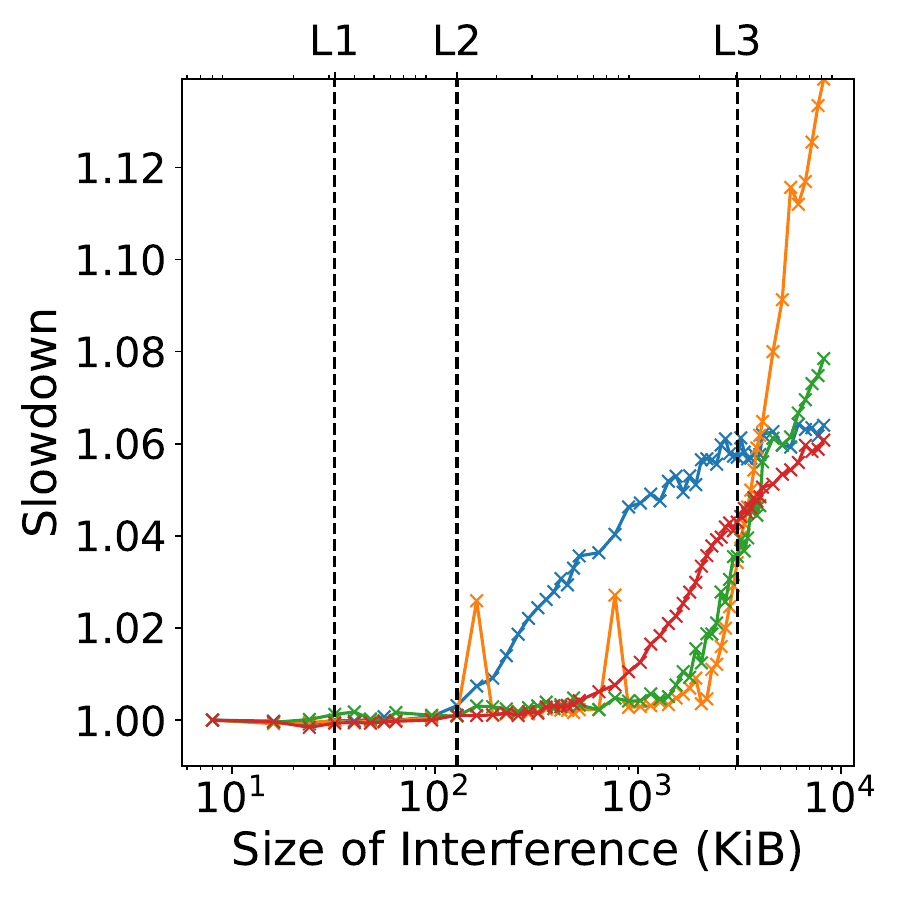}
            \captionsetup{justification=centering}
            \caption{rk3588: Way / Write}
            \label{fig:all-rk3588-way-disparity-write-vga}
        \end{subfigure}
        \hfill
        \begin{subfigure}{0.24\textwidth}
            \centering
            \includegraphics[width=\textwidth]{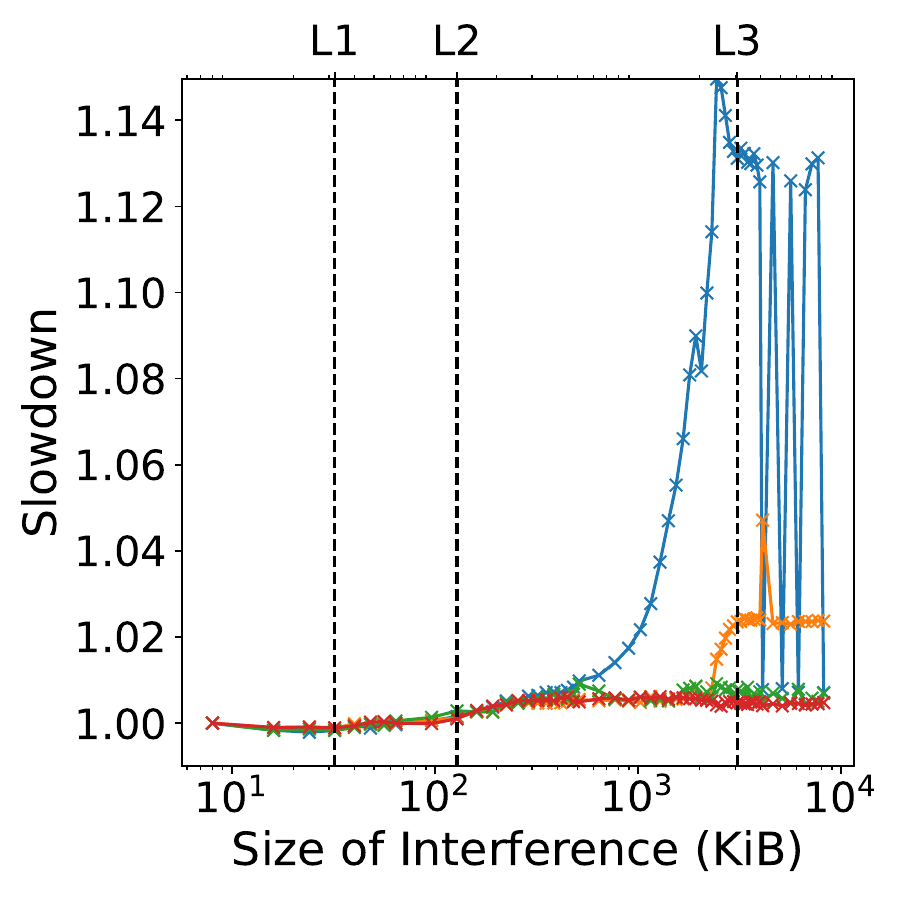}
            \captionsetup{justification=centering}
            \caption{rk3588: Way / Modify}
            \label{fig:all-rk3588-way-disparity-modify-vga}
        \end{subfigure}
        \hfill
        \begin{subfigure}{0.24\textwidth}
            \centering
            \includegraphics[width=\textwidth]{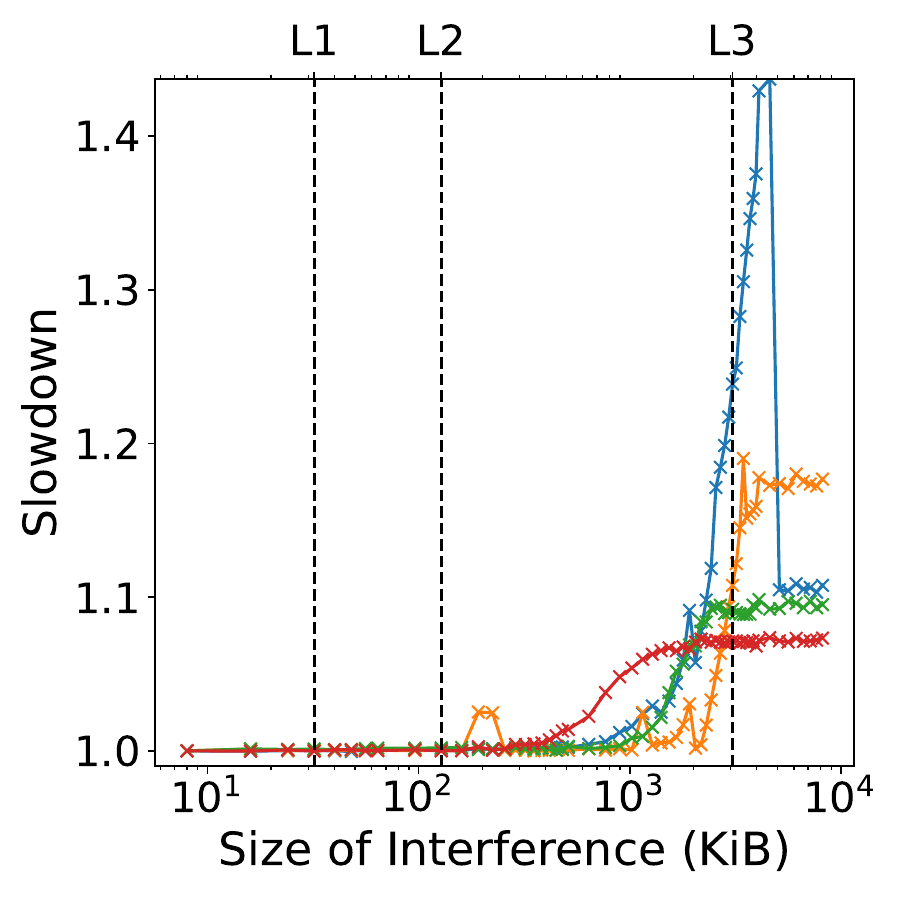}
            \captionsetup{justification=centering}
            \caption{rk3588: Way / Prefetch}
            \label{fig:all-rk3588-way-disparity-prefetch-vga}
        \end{subfigure}
        \hfill
        
        \caption{Execution Slowdown on \textit{'Disparity'} benchmark for \textit{'VGA'} dataset on \textit{'RK3588'} with Interferences and cache partitioning.}
        \label{fig:rk3588-disparity-vga}
    \end{figure}

    \begin{figure}[H]
        \begin{subfigure}{\textwidth}
            \centering
            \includegraphics[width=0.5\textwidth]{figures/set_subplot/legend.pdf}
        \end{subfigure}
        \centering
        
        \begin{subfigure}{0.24\textwidth}
            \centering
            \includegraphics[width=\textwidth]{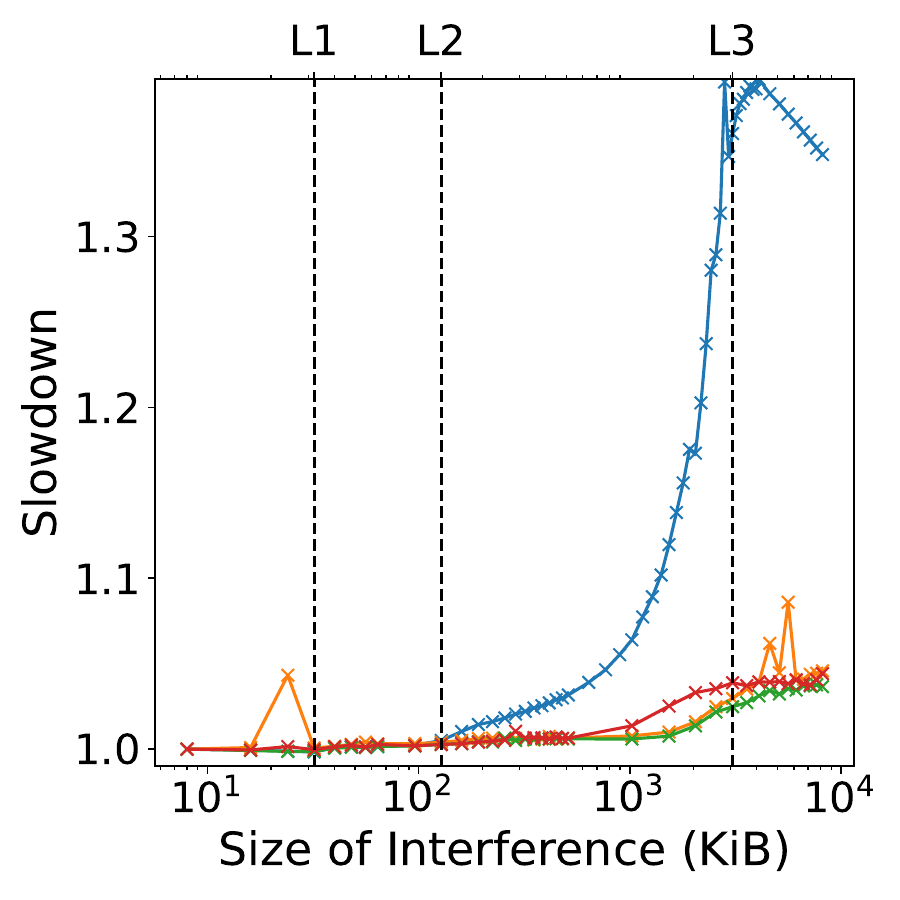}
            \captionsetup{justification=centering}
            \caption{rk3588: Set / Read}
            \label{fig:all-rk3588-set-mser-read-vga}
        \end{subfigure}
        \hfill
        \begin{subfigure}{0.24\textwidth}
            \centering
            \includegraphics[width=\textwidth]{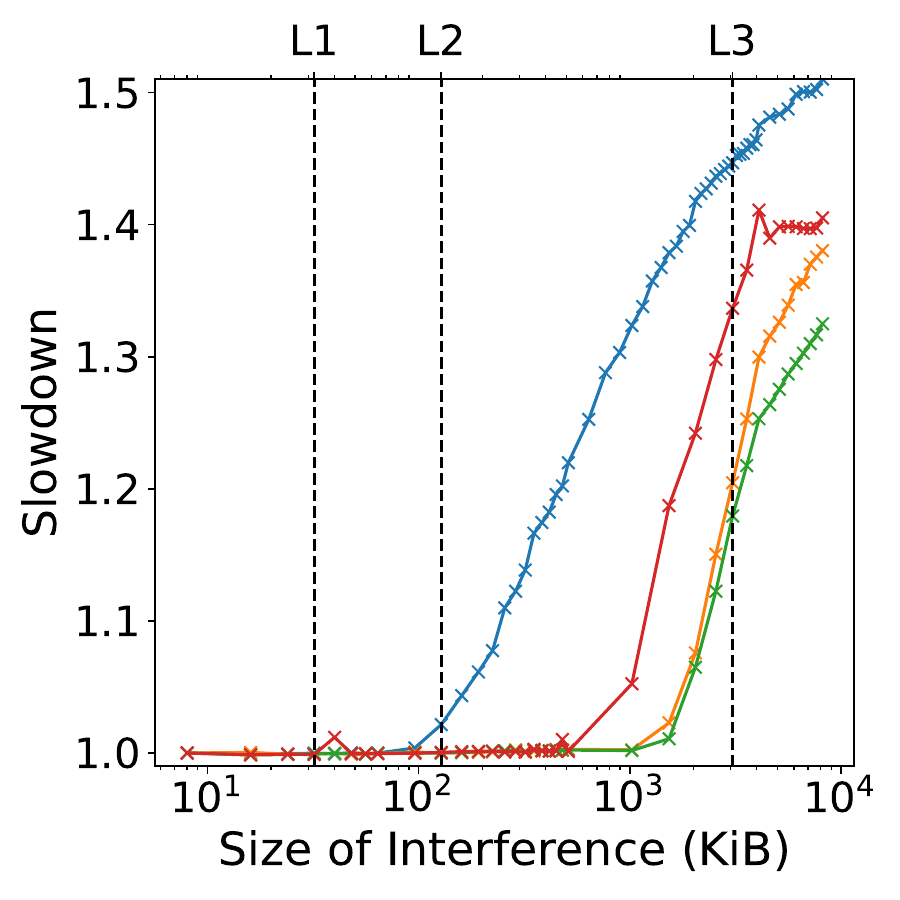}
            \captionsetup{justification=centering}
            \caption{rk3588: Set / Write}
            \label{fig:all-rk3588-set-mser-write-vga}
        \end{subfigure}
        \hfill
        \begin{subfigure}{0.24\textwidth}
            \centering
            \includegraphics[width=\textwidth]{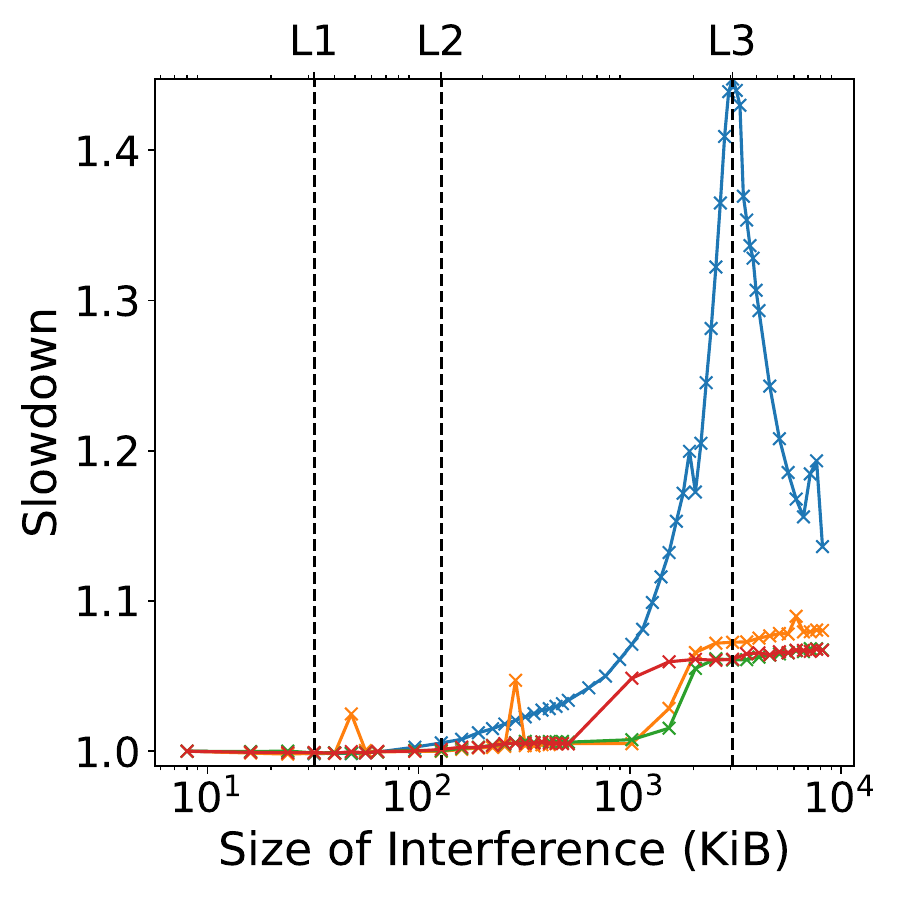}
            \captionsetup{justification=centering}
            \caption{rk3588: Set / Modify}
            \label{fig:all-rk3588-set-mser-modify-vga}
        \end{subfigure}
        \hfill
        \begin{subfigure}{0.24\textwidth}
            \centering
            \includegraphics[width=\textwidth]{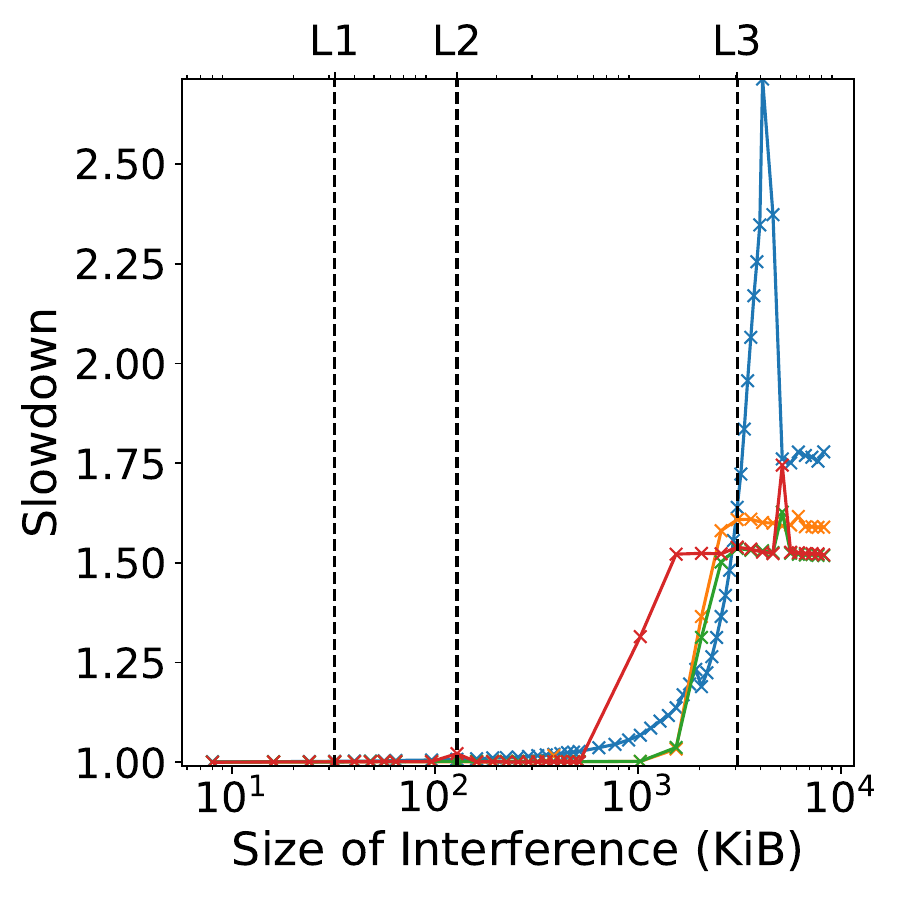}
            \captionsetup{justification=centering}
            \caption{rk3588: Set / Prefetch}
            \label{fig:all-rk3588-set-mser-prefetch-vga}
        \end{subfigure}
        \hfill
        \begin{subfigure}{0.24\textwidth}
            \centering
            \includegraphics[width=\textwidth]{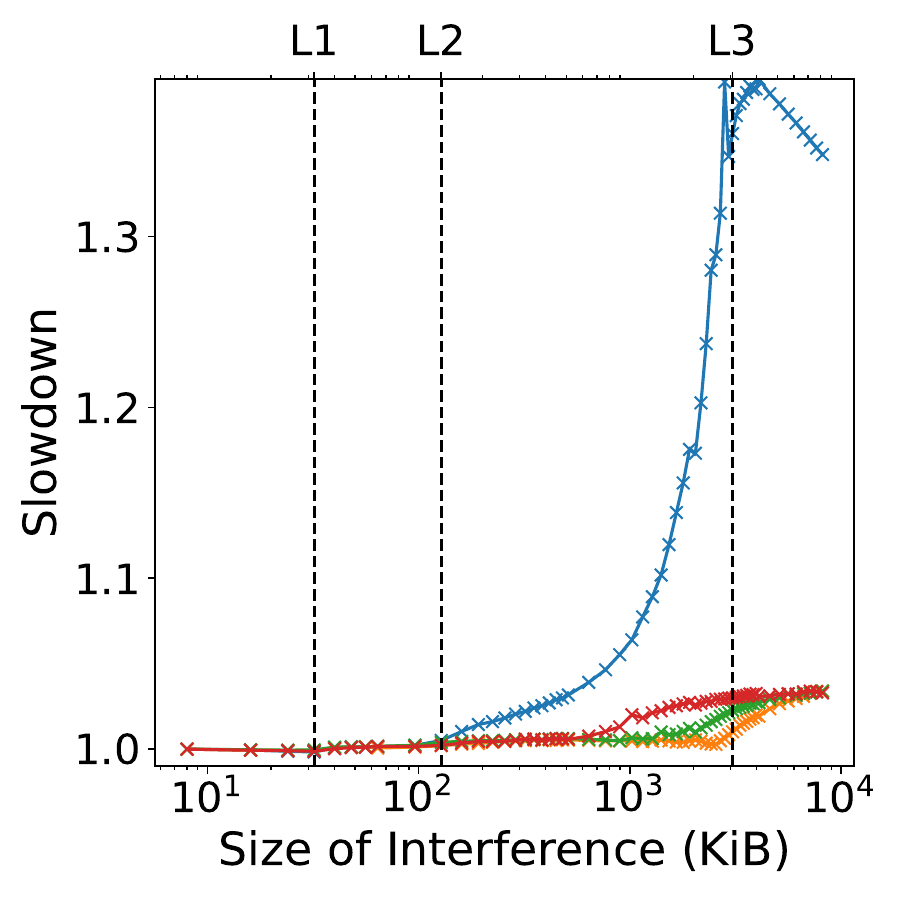}
            \captionsetup{justification=centering}
            \caption{rk3588: Way / Read}
            \label{fig:all-rk3588-way-mser-read-vga}
        \end{subfigure}
        \hfill
        \begin{subfigure}{0.24\textwidth}
            \centering
            \includegraphics[width=\textwidth]{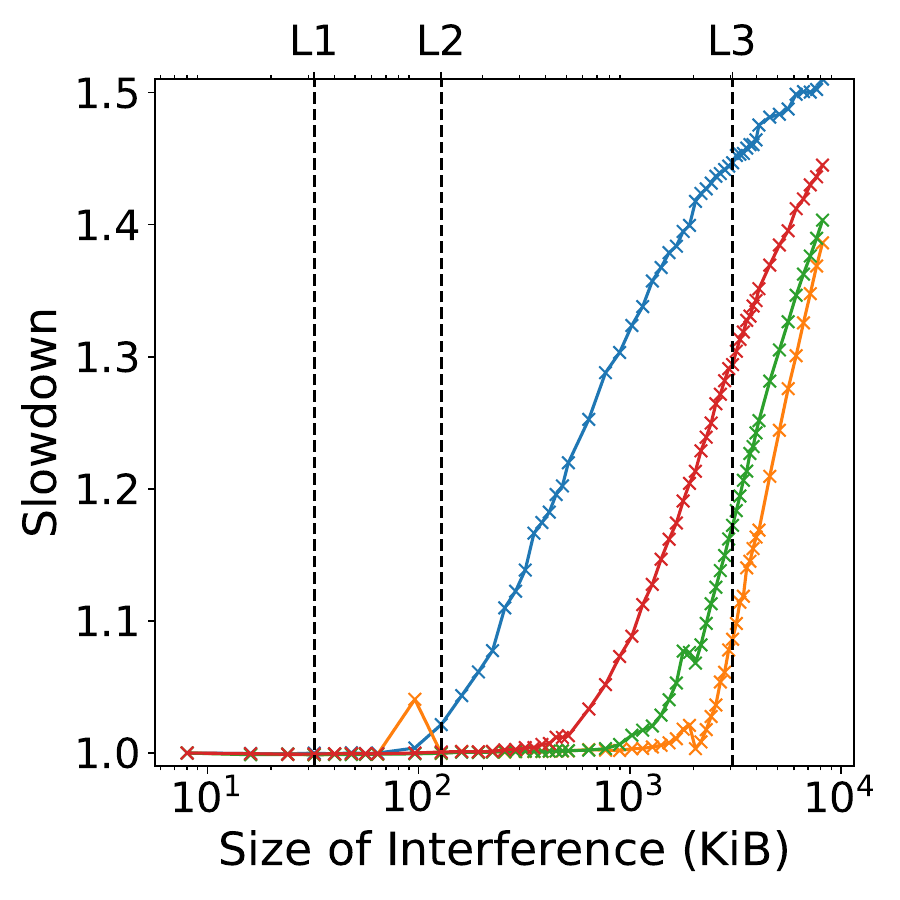}
            \captionsetup{justification=centering}
            \caption{rk3588: Way / Write}
            \label{fig:all-rk3588-way-mser-write-vga}
        \end{subfigure}
        \hfill
        \begin{subfigure}{0.24\textwidth}
            \centering
            \includegraphics[width=\textwidth]{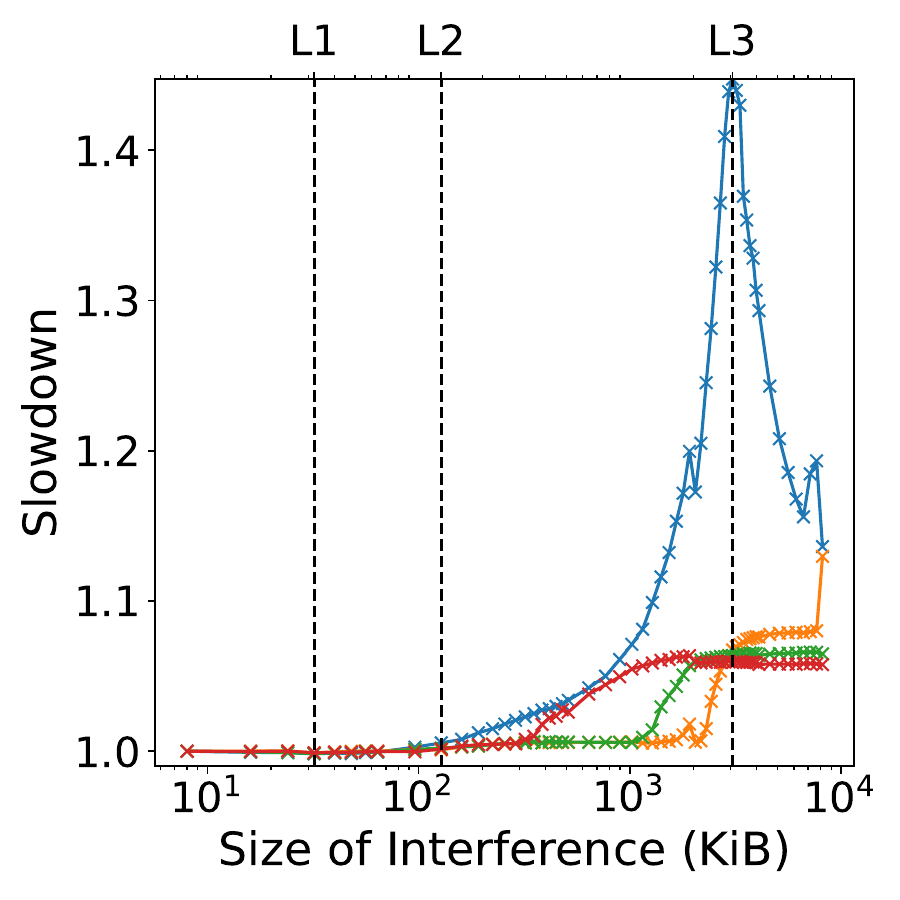}
            \captionsetup{justification=centering}
            \caption{rk3588: Way / Modify}
            \label{fig:all-rk3588-way-mser-modify-vga}
        \end{subfigure}
        \hfill
        \begin{subfigure}{0.24\textwidth}
            \centering
            \includegraphics[width=\textwidth]{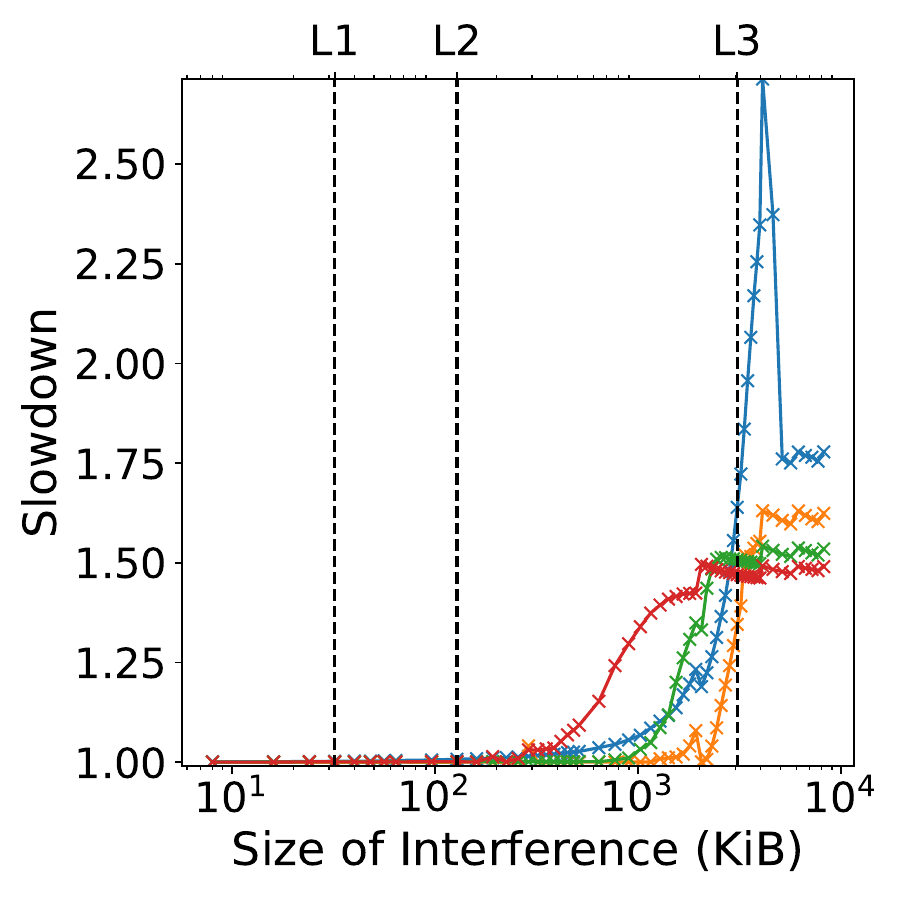}
            \captionsetup{justification=centering}
            \caption{rk3588: Way / Prefetch}
            \label{fig:all-rk3588-way-mser-prefetch-vga}
        \end{subfigure}
        \hfill
        
        \caption{Execution Slowdown on \textit{'Mser'} benchmark for \textit{'VGA'} dataset on \textit{'RK3588'} with Interferences and cache partitioning.}
        \label{fig:rk3588-mser-vga}
    \end{figure}

    \begin{figure}[H]
        \begin{subfigure}{\textwidth}
            \centering
            \includegraphics[width=0.5\textwidth]{figures/set_subplot/legend.pdf}
        \end{subfigure}
        \centering
        
        \begin{subfigure}{0.24\textwidth}
            \centering
            \includegraphics[width=\textwidth]{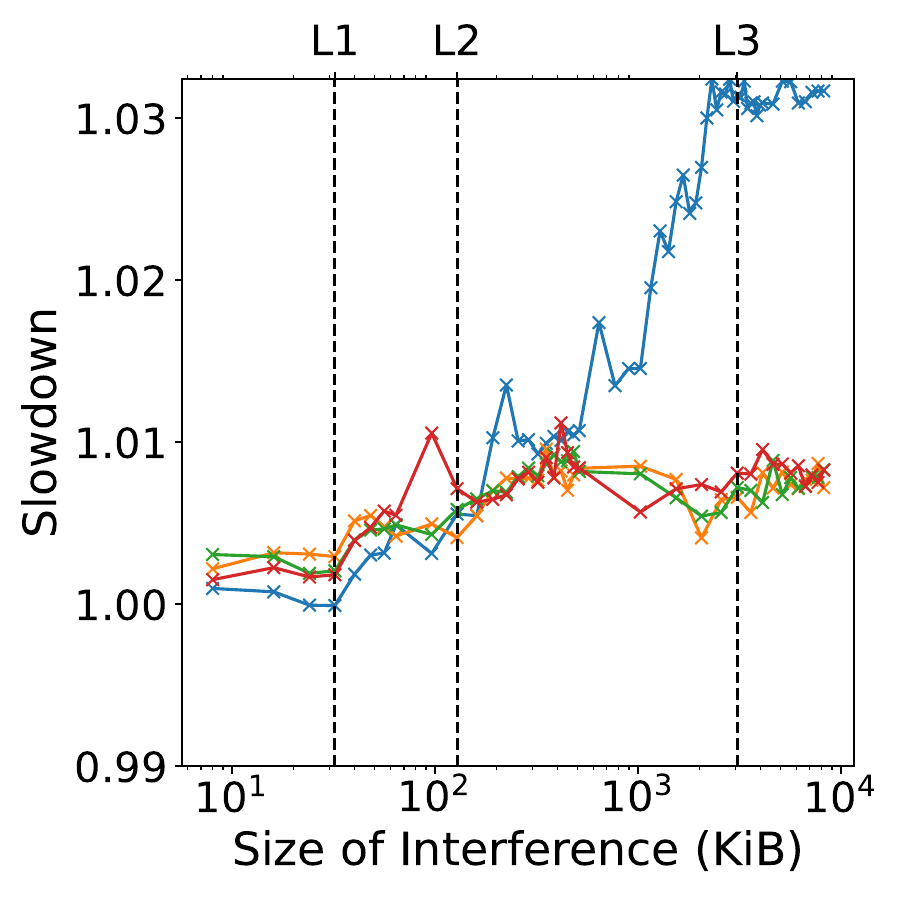}
            \captionsetup{justification=centering}
            \caption{rk3588: Set / Read}
            \label{fig:all-rk3588-set-tracking-read-vga}
        \end{subfigure}
        \hfill
        \begin{subfigure}{0.24\textwidth}
            \centering
            \includegraphics[width=\textwidth]{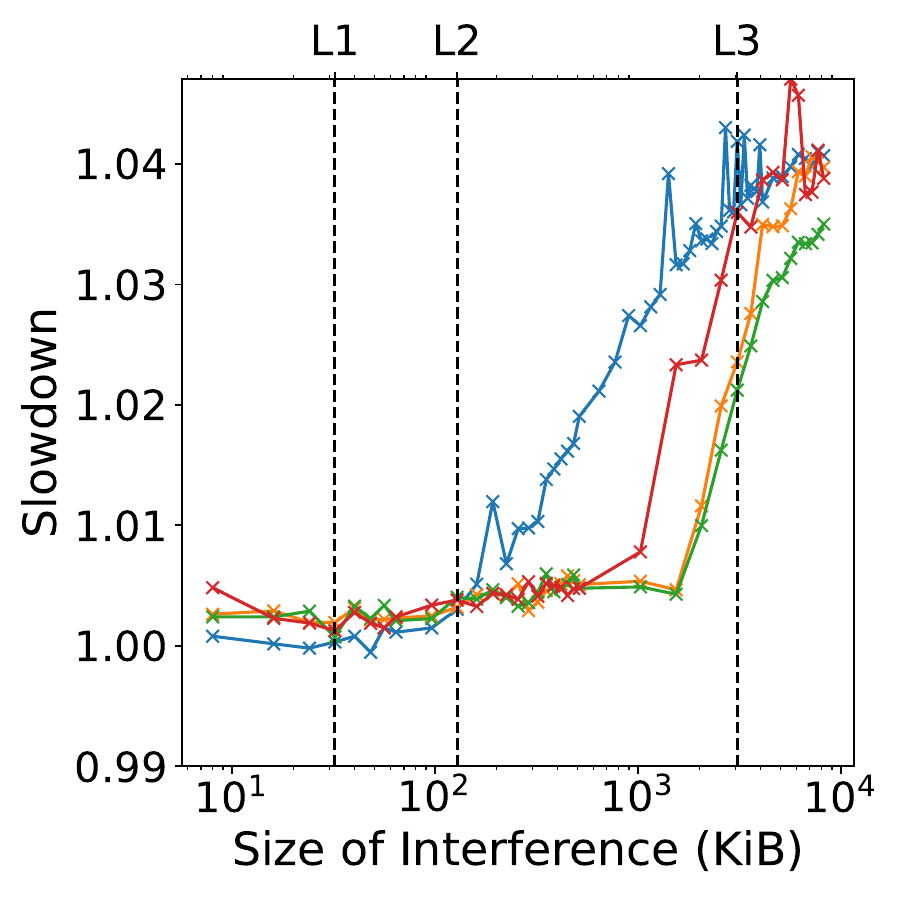}
            \captionsetup{justification=centering}
            \caption{rk3588: Set / Write}
            \label{fig:all-rk3588-set-tracking-write-vga}
        \end{subfigure}
        \hfill
        \begin{subfigure}{0.24\textwidth}
            \centering
            \includegraphics[width=\textwidth]{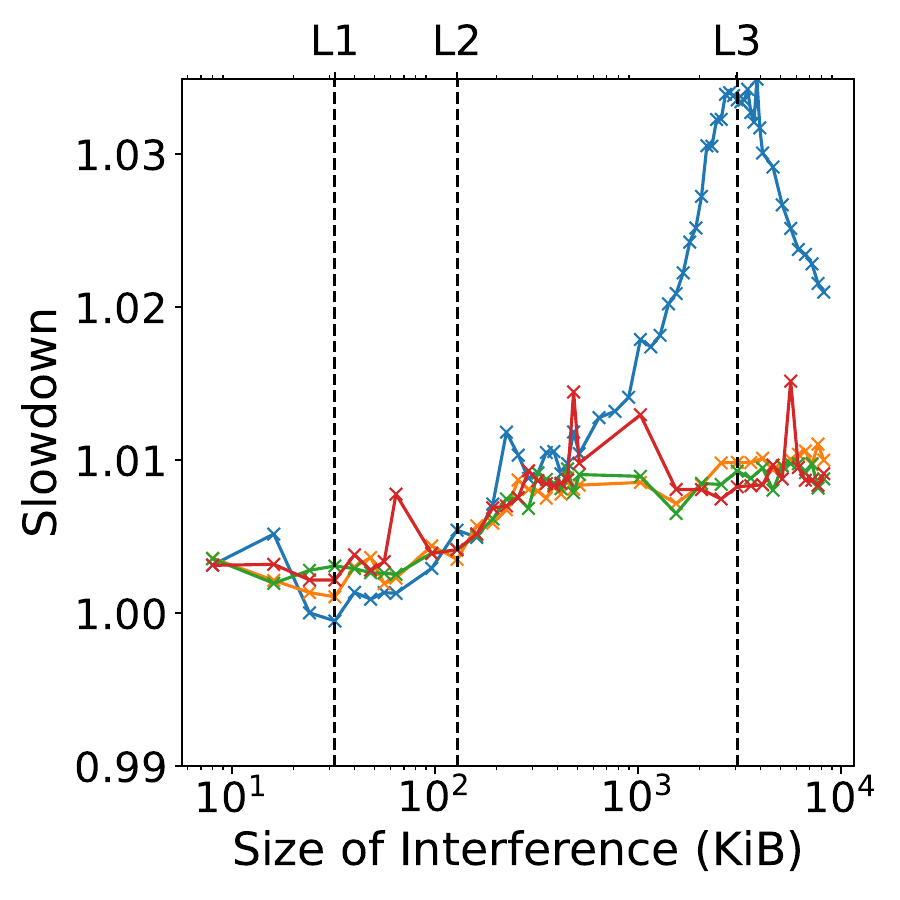}
            \captionsetup{justification=centering}
            \caption{rk3588: Set / Modify}
            \label{fig:all-rk3588-set-tracking-modify-vga}
        \end{subfigure}
        \hfill
        \begin{subfigure}{0.24\textwidth}
            \centering
            \includegraphics[width=\textwidth]{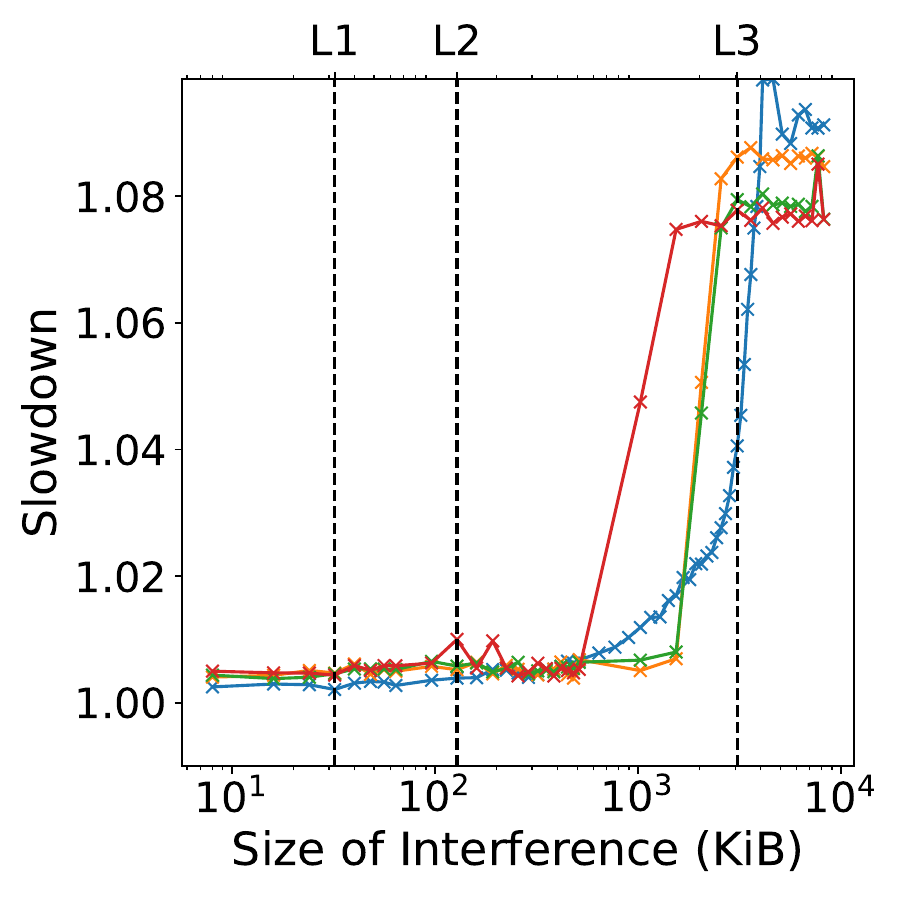}
            \captionsetup{justification=centering}
            \caption{rk3588: Set / Prefetch}
            \label{fig:all-rk3588-set-tracking-prefetch-vga}
        \end{subfigure}
        \hfill
        \begin{subfigure}{0.24\textwidth}
            \centering
            \includegraphics[width=\textwidth]{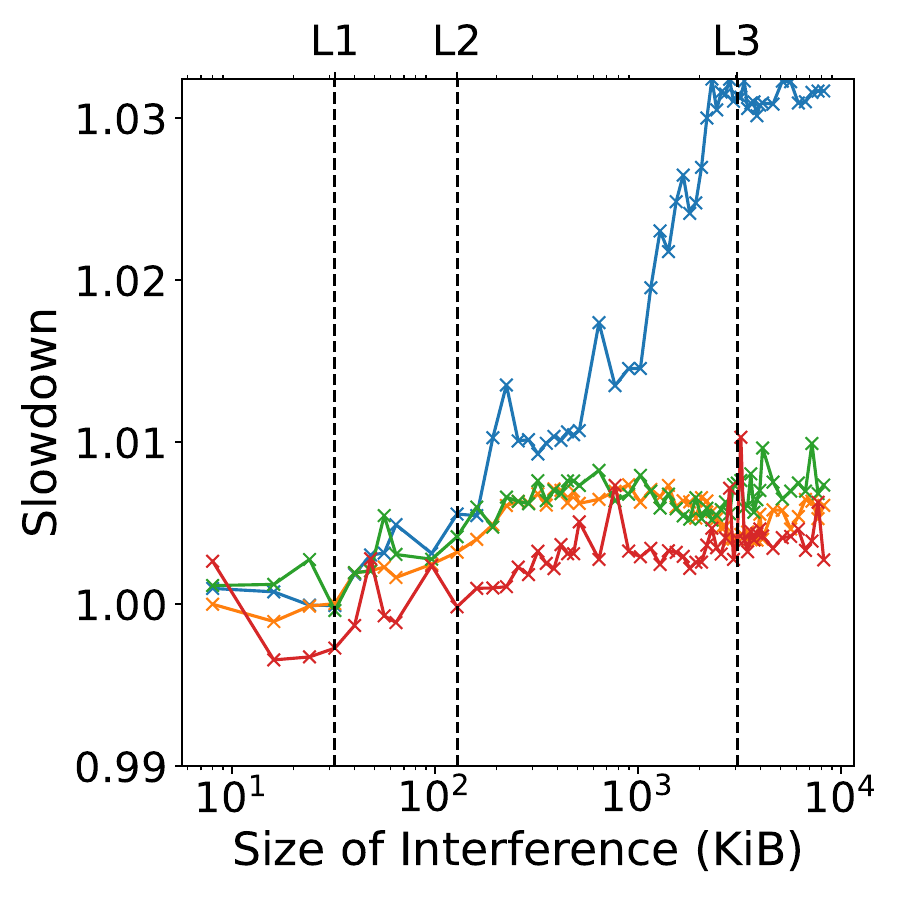}
            \captionsetup{justification=centering}
            \caption{rk3588: Way / Read}
            \label{fig:all-rk3588-way-tracking-read-vga}
        \end{subfigure}
        \hfill
        \begin{subfigure}{0.24\textwidth}
            \centering
            \includegraphics[width=\textwidth]{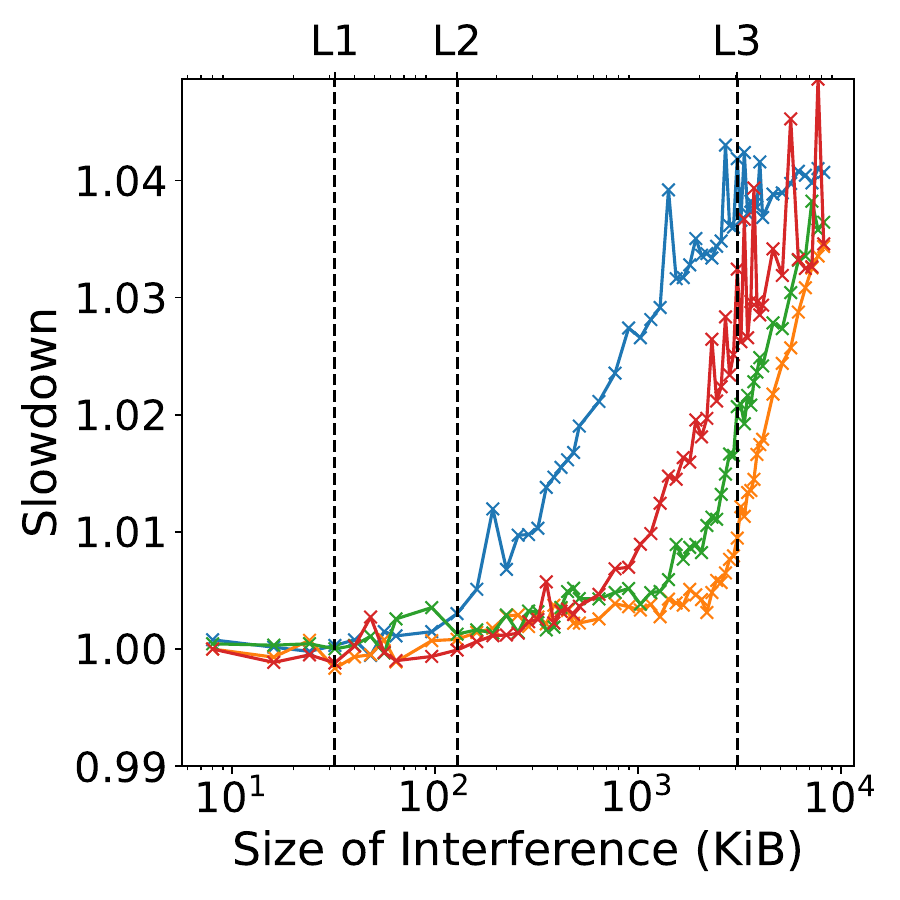}
            \captionsetup{justification=centering}
            \caption{rk3588: Way / Write}
            \label{fig:all-rk3588-way-tracking-write-vga}
        \end{subfigure}
        \hfill
        \begin{subfigure}{0.24\textwidth}
            \centering
            \includegraphics[width=\textwidth]{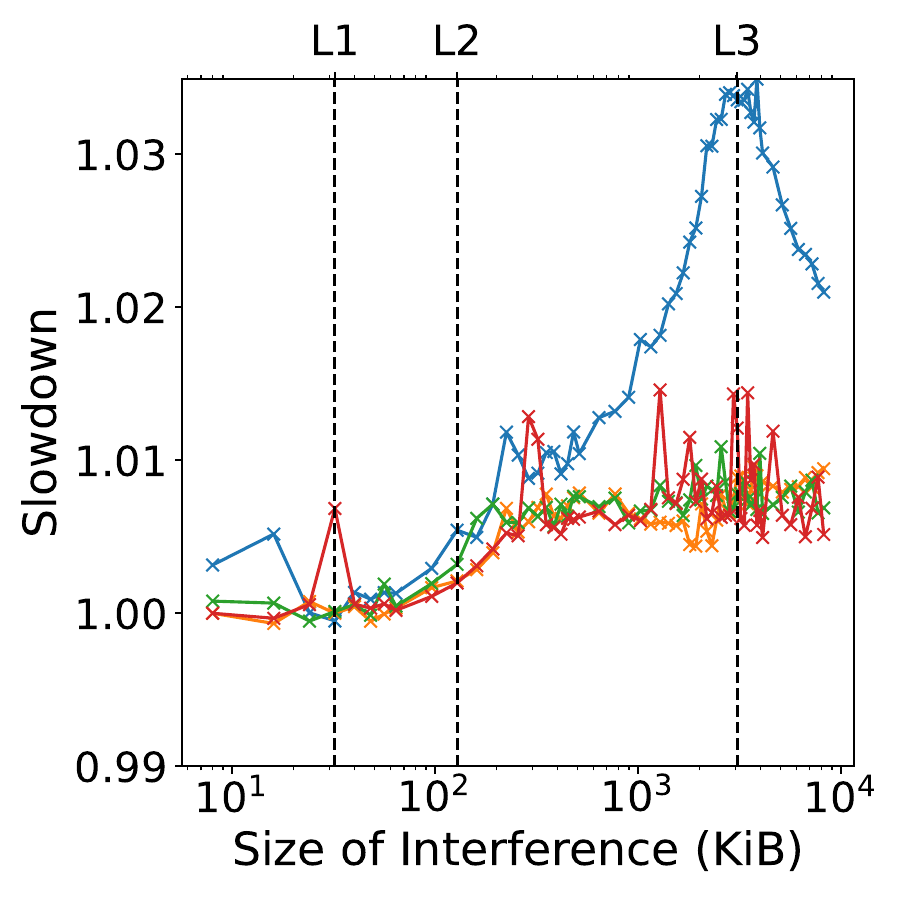}
            \captionsetup{justification=centering}
            \caption{rk3588: Way / Modify}
            \label{fig:all-rk3588-way-tracking-modify-vga}
        \end{subfigure}
        \hfill
        \begin{subfigure}{0.24\textwidth}
            \centering
            \includegraphics[width=\textwidth]{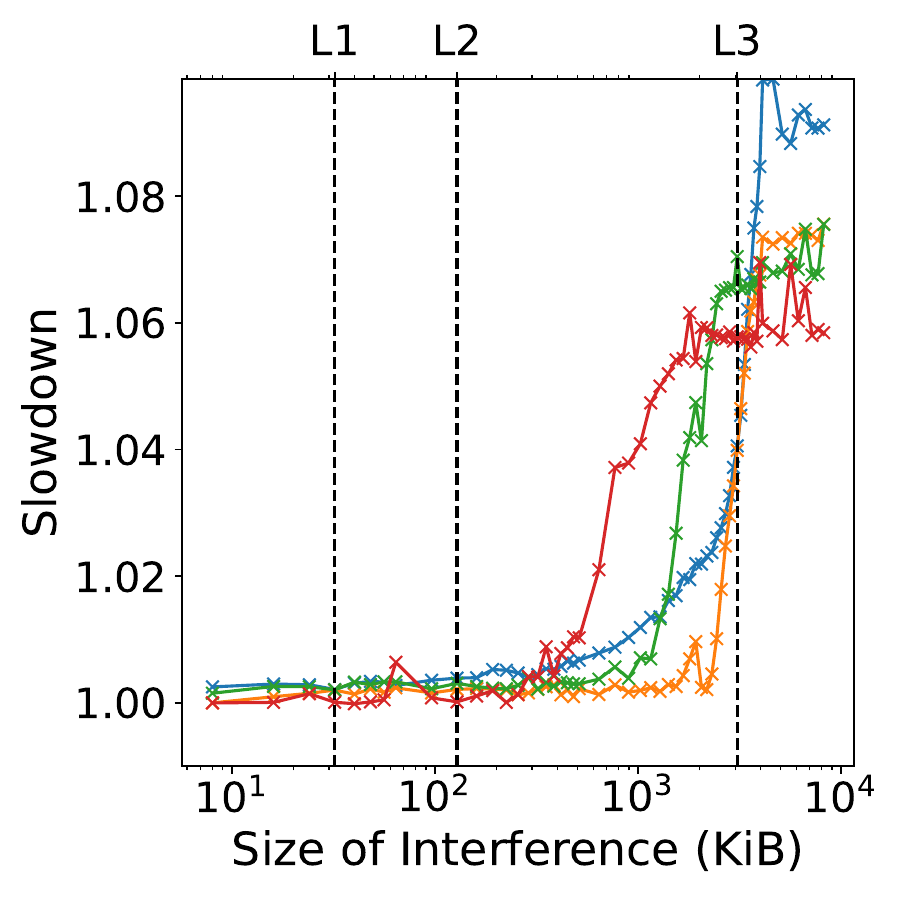}
            \captionsetup{justification=centering}
            \caption{rk3588: Way / Prefetch}
            \label{fig:all-rk3588-way-tracking-prefetch-vga}
        \end{subfigure}
        \hfill
        
        \caption{Execution Slowdown on \textit{'Tracking'} benchmark for \textit{'VGA'} dataset on \textit{'RK3588'} with Interferences and cache partitioning.}
        \label{fig:rk3588-tracking-vga}
    \end{figure}

    \begin{figure}[H]
        \begin{subfigure}{\textwidth}
            \centering
            \includegraphics[width=0.5\textwidth]{figures/set_subplot/legend.pdf}
        \end{subfigure}
        \centering
        
        \begin{subfigure}{0.24\textwidth}
            \centering
            \includegraphics[width=\textwidth]{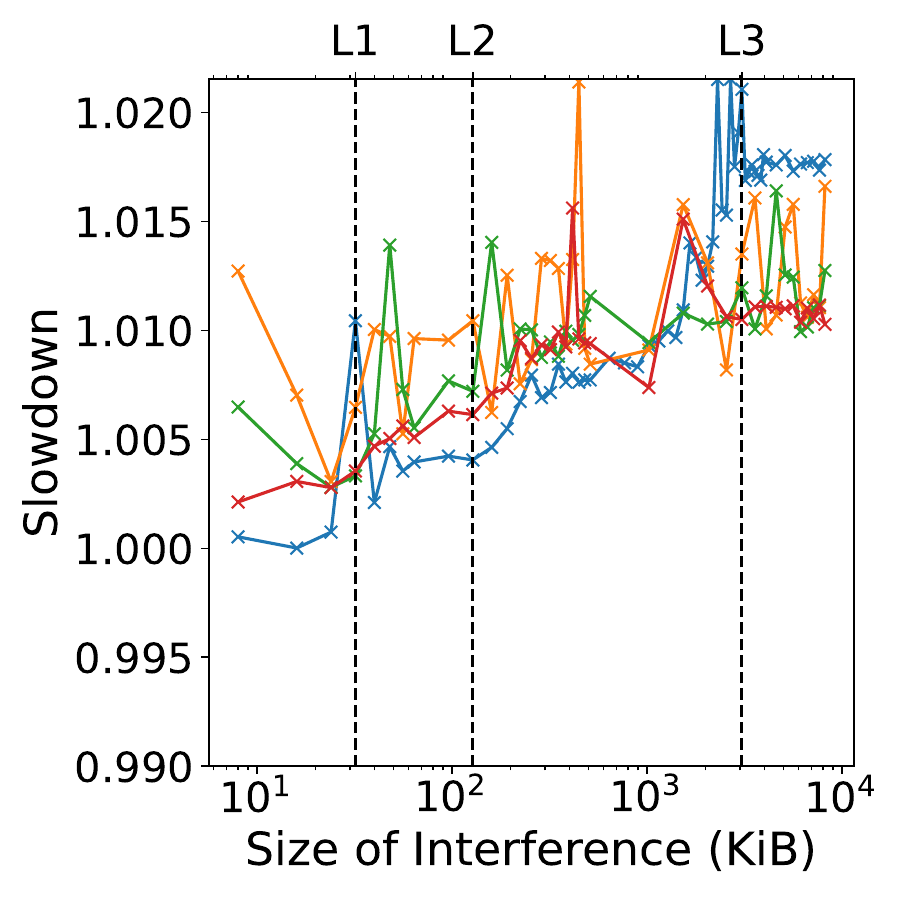}
            \captionsetup{justification=centering}
            \caption{rk3588: Set / Read}
            \label{fig:all-rk3588-set-sift-read-vga}
        \end{subfigure}
        \hfill
        \begin{subfigure}{0.24\textwidth}
            \centering
            \includegraphics[width=\textwidth]{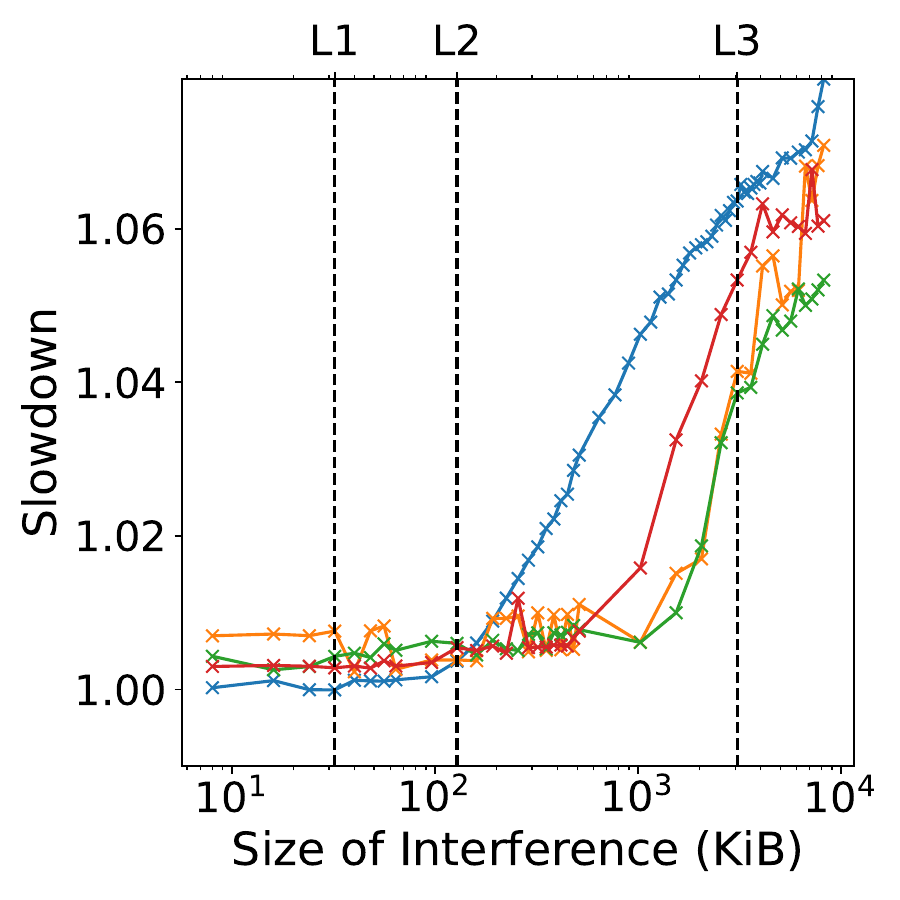}
            \captionsetup{justification=centering}
            \caption{rk3588: Set / Write}
            \label{fig:all-rk3588-set-sift-write-vga}
        \end{subfigure}
        \hfill
        \begin{subfigure}{0.24\textwidth}
            \centering
            \includegraphics[width=\textwidth]{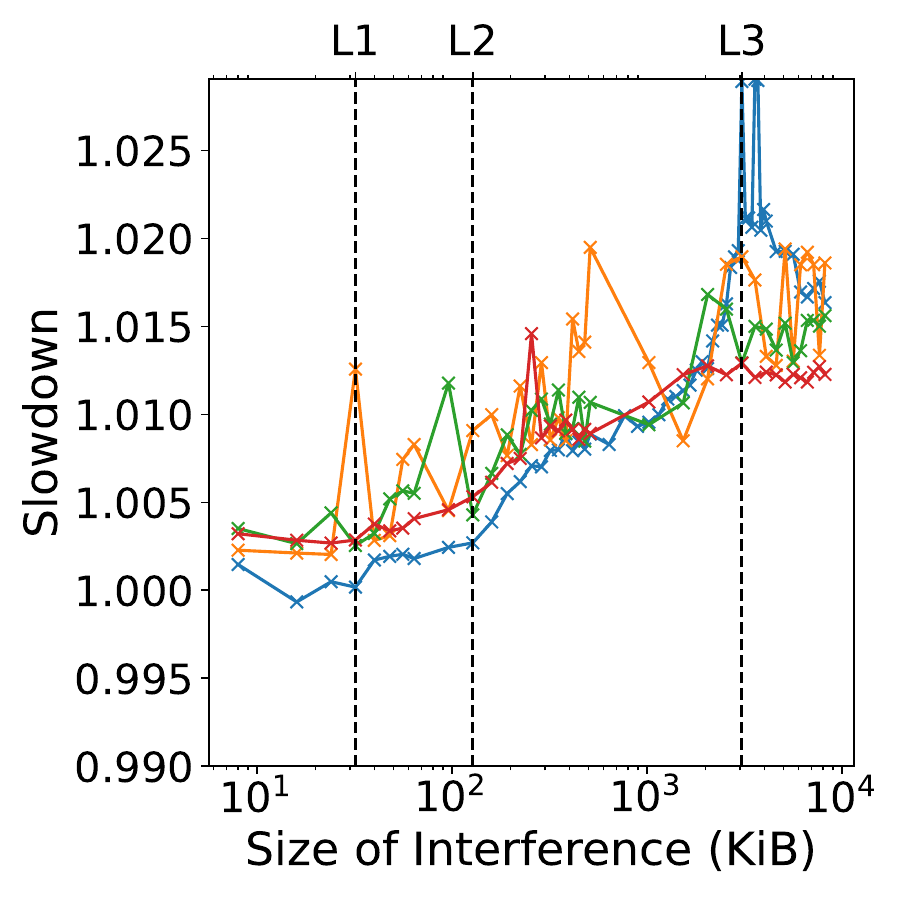}
            \captionsetup{justification=centering}
            \caption{rk3588: Set / Modify}
            \label{fig:all-rk3588-set-sift-modify-vga}
        \end{subfigure}
        \hfill
        \begin{subfigure}{0.24\textwidth}
            \centering
            \includegraphics[width=\textwidth]{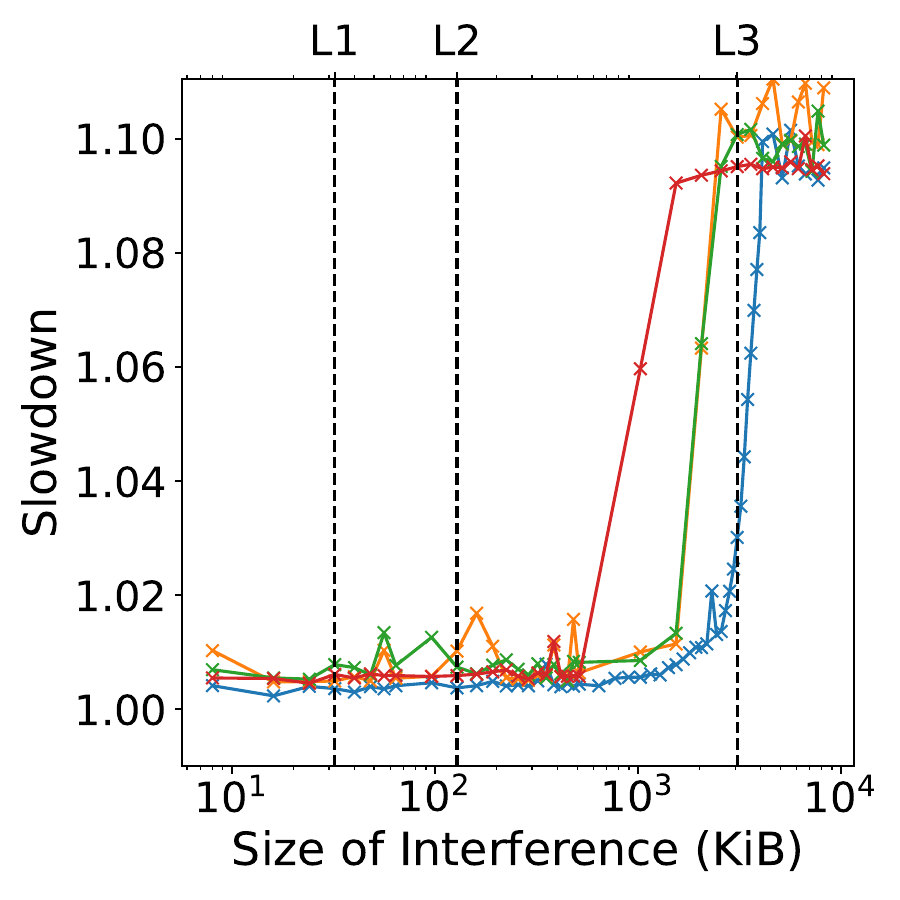}
            \captionsetup{justification=centering}
            \caption{rk3588: Set / Prefetch}
            \label{fig:all-rk3588-set-sift-prefetch-vga}
        \end{subfigure}
        \hfill
        \begin{subfigure}{0.24\textwidth}
            \centering
            \includegraphics[width=\textwidth]{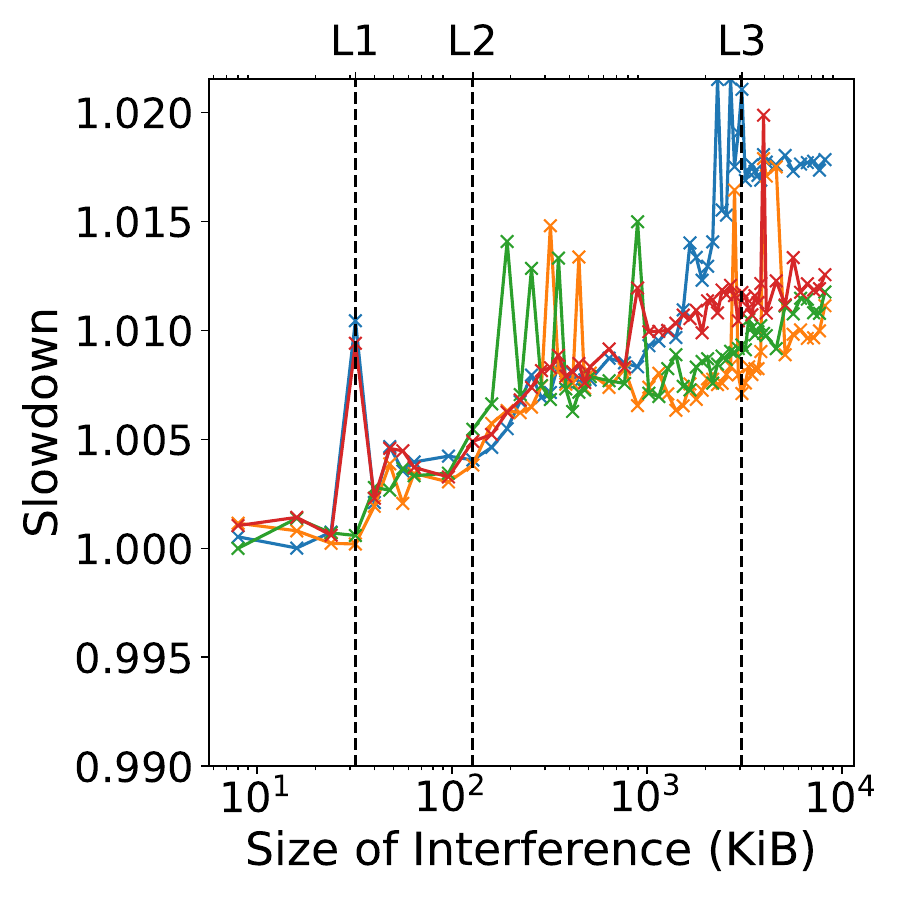}
            \captionsetup{justification=centering}
            \caption{rk3588: Way / Read}
            \label{fig:all-rk3588-way-sift-read-vga}
        \end{subfigure}
        \hfill
        \begin{subfigure}{0.24\textwidth}
            \centering
            \includegraphics[width=\textwidth]{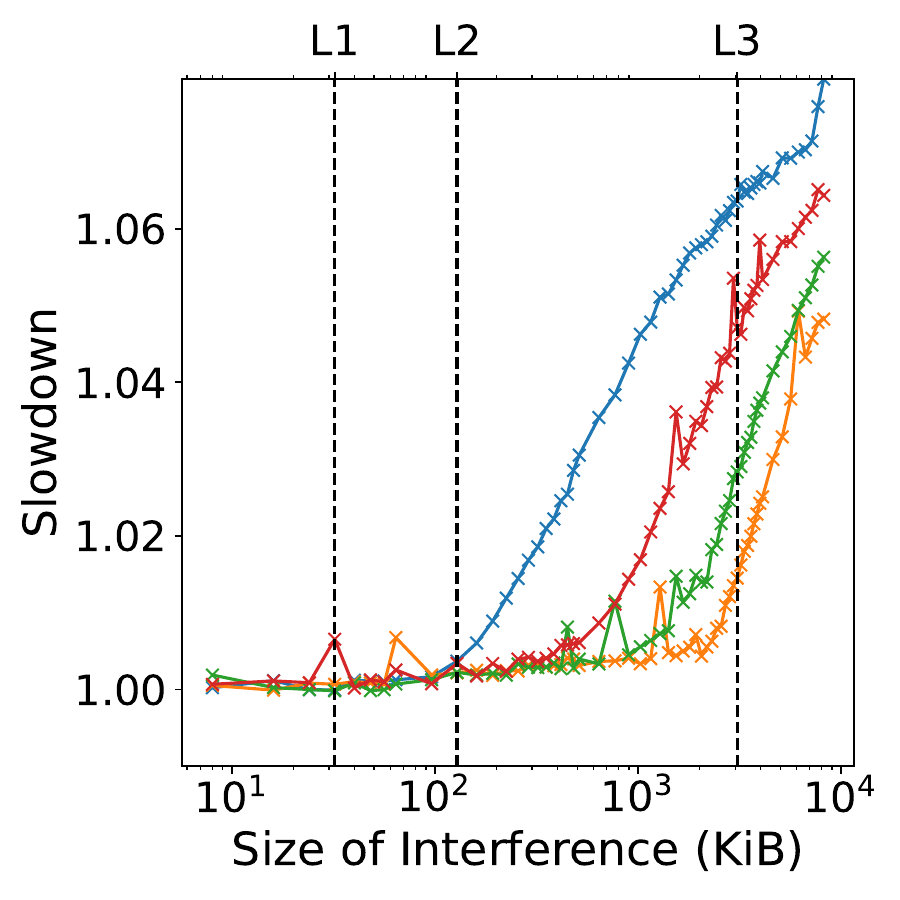}
            \captionsetup{justification=centering}
            \caption{rk3588: Way / Write}
            \label{fig:all-rk3588-way-sift-write-vga}
        \end{subfigure}
        \hfill
        \begin{subfigure}{0.24\textwidth}
            \centering
            \includegraphics[width=\textwidth]{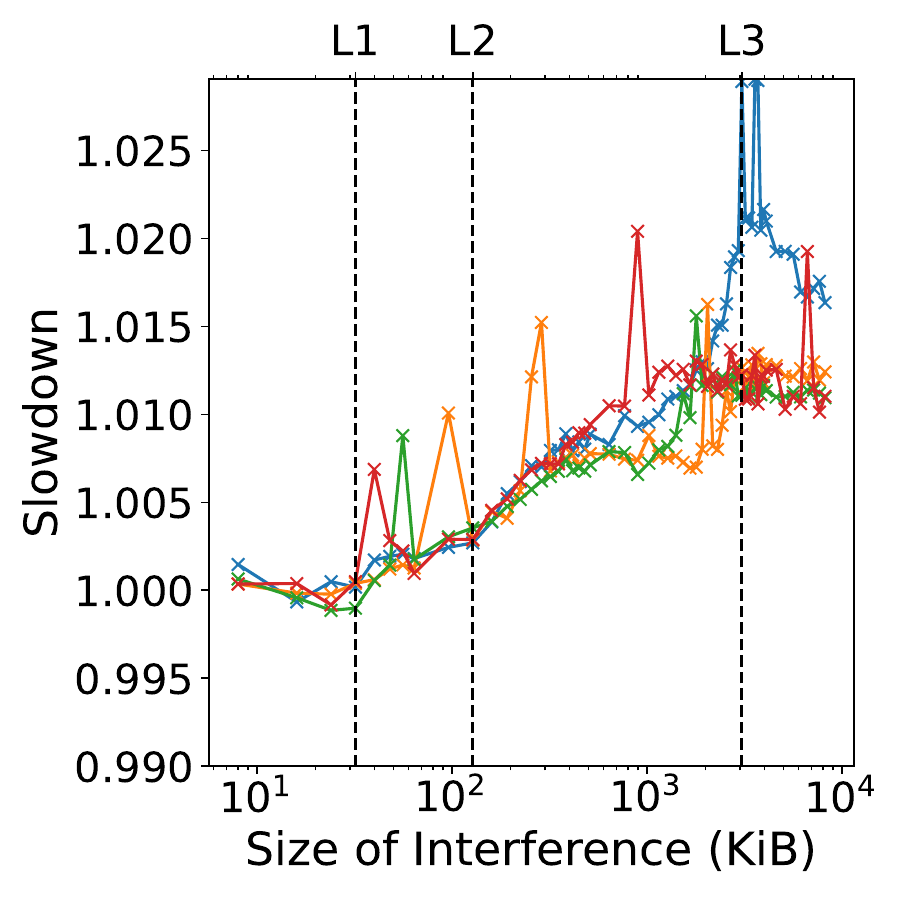}
            \captionsetup{justification=centering}
            \caption{rk3588: Way / Modify}
            \label{fig:all-rk3588-way-sift-modify-vga}
        \end{subfigure}
        \hfill
        \begin{subfigure}{0.24\textwidth}
            \centering
            \includegraphics[width=\textwidth]{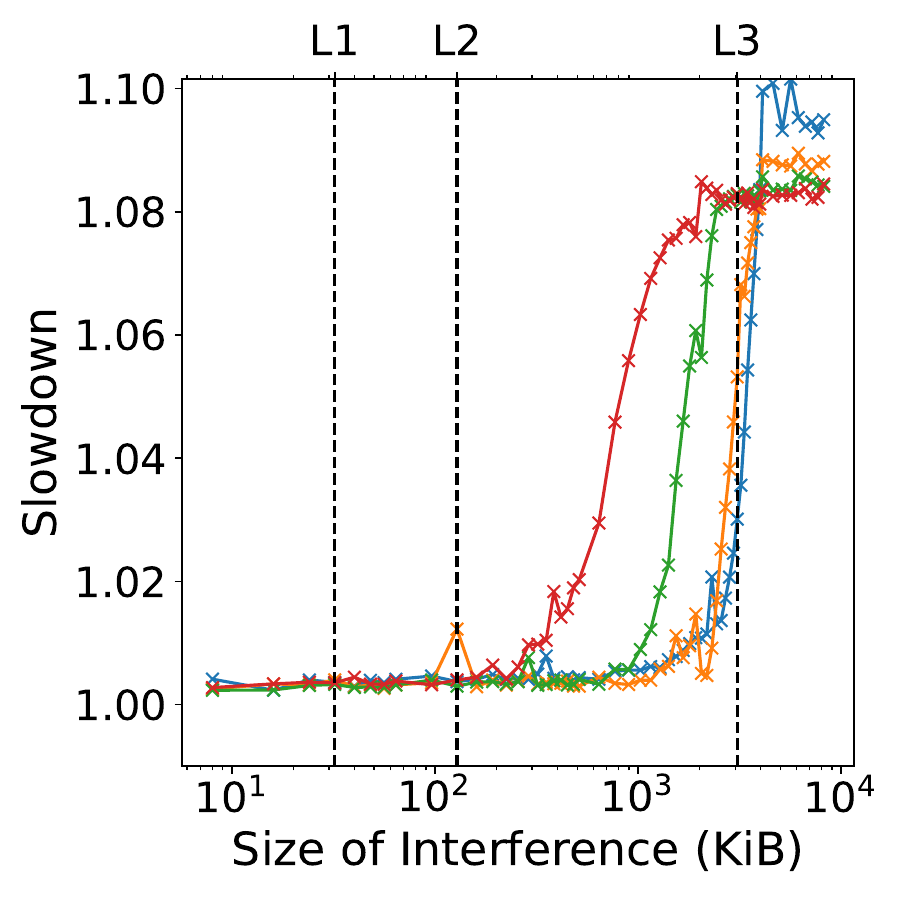}
            \captionsetup{justification=centering}
            \caption{rk3588: Way / Prefetch}
            \label{fig:all-rk3588-way-sift-prefetch-vga}
        \end{subfigure}
        \hfill
        
        \caption{Execution Slowdown on \textit{'Sift'} benchmark for \textit{'VGA'} dataset on \textit{'RK3588'} with Interferences and cache partitioning.}
        \label{fig:rk3588-sift-vga}
    \end{figure}

        \clearpage

        \subsection{ORIN}

    \begin{figure}[H]
        \begin{subfigure}{\textwidth}
            \centering
            \includegraphics[width=0.5\textwidth]{figures/set_subplot/legend.pdf}
        \end{subfigure}
        \centering
        
        \begin{subfigure}{0.24\textwidth}
            \centering
            \includegraphics[width=\textwidth]{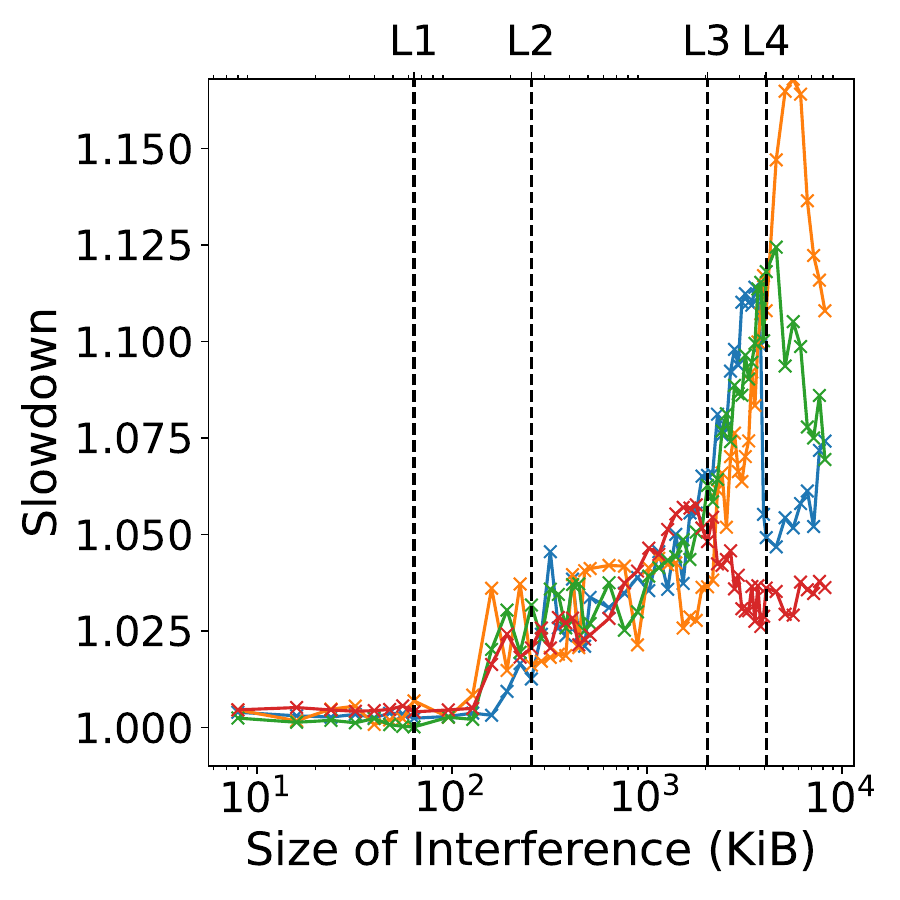}
            \captionsetup{justification=centering}
            \caption{orin: Way / Read}
            \label{fig:all-orin-way-disparity-read-cif}
        \end{subfigure}
        \hfill
        \begin{subfigure}{0.24\textwidth}
            \centering
            \includegraphics[width=\textwidth]{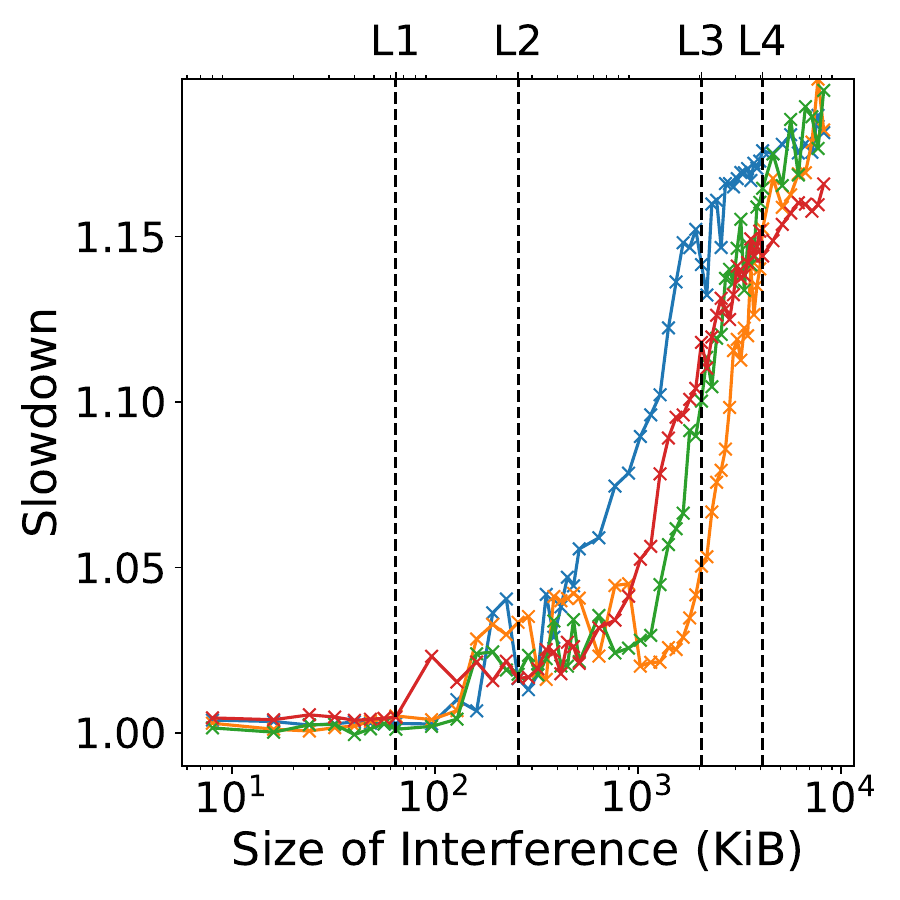}
            \captionsetup{justification=centering}
            \caption{orin: Way / Write}
            \label{fig:all-orin-way-disparity-write-cif}
        \end{subfigure}
        \hfill
        \begin{subfigure}{0.24\textwidth}
            \centering
            \includegraphics[width=\textwidth]{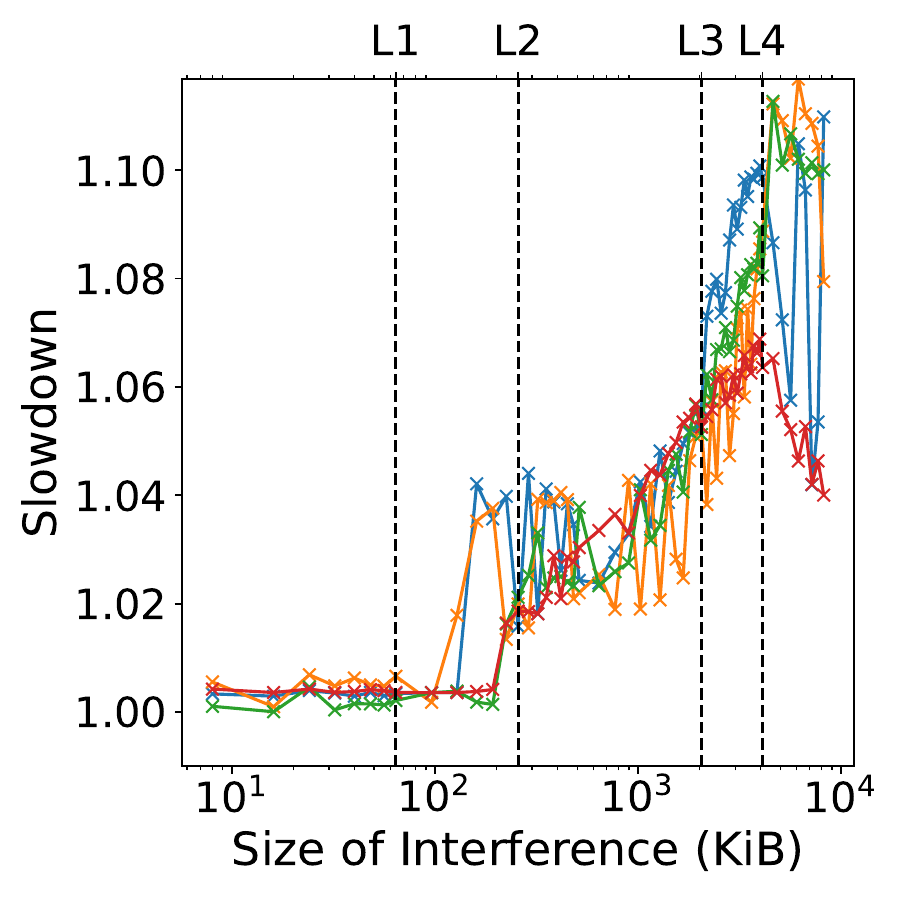}
            \captionsetup{justification=centering}
            \caption{orin: Way / Modify}
            \label{fig:all-orin-way-disparity-modify-cif}
        \end{subfigure}
        \hfill
        \begin{subfigure}{0.24\textwidth}
            \centering
            \includegraphics[width=\textwidth]{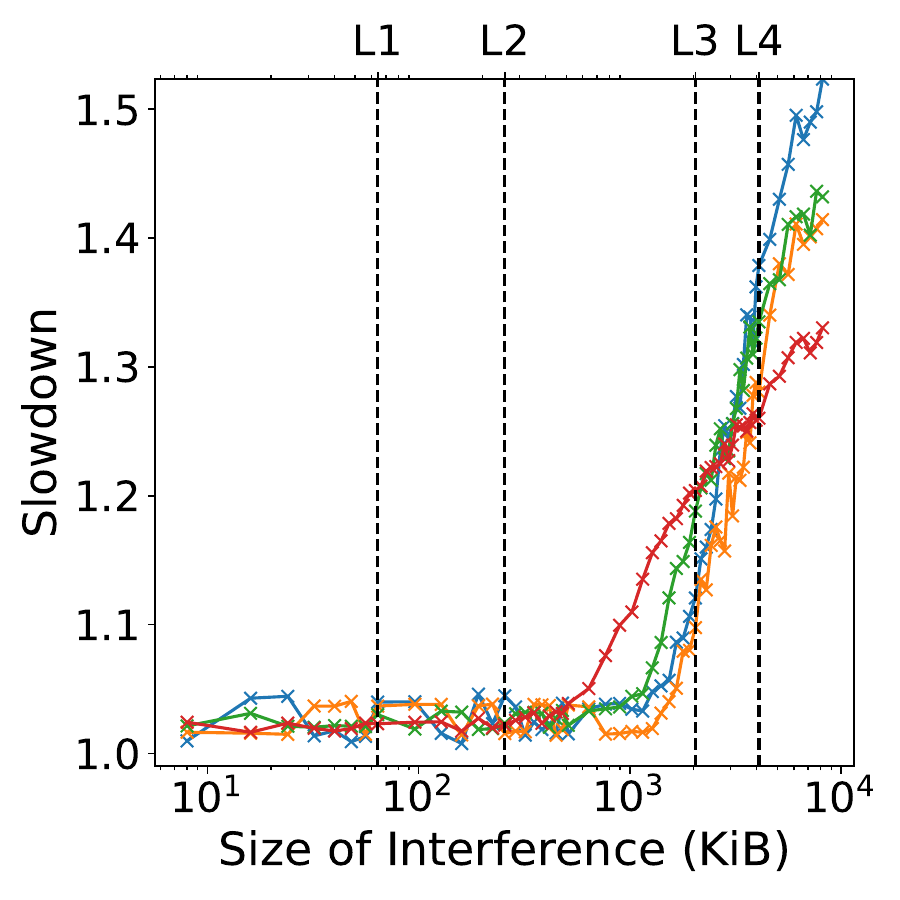}
            \captionsetup{justification=centering}
            \caption{orin: Way / Prefetch}
            \label{fig:all-orin-way-disparity-prefetch-cif}
        \end{subfigure}
        \hfill
        
        \caption{Execution Slowdown on \textit{'Disparity'} benchmark for \textit{'CIF'} dataset on \textit{'ORIN'} with Interferences and cache partitioning.}
        \label{fig:orin-disparity-cif}
    \end{figure}

    \begin{figure}[H]
        \begin{subfigure}{\textwidth}
            \centering
            \includegraphics[width=0.5\textwidth]{figures/set_subplot/legend.pdf}
        \end{subfigure}
        \centering
        
        \begin{subfigure}{0.24\textwidth}
            \centering
            \includegraphics[width=\textwidth]{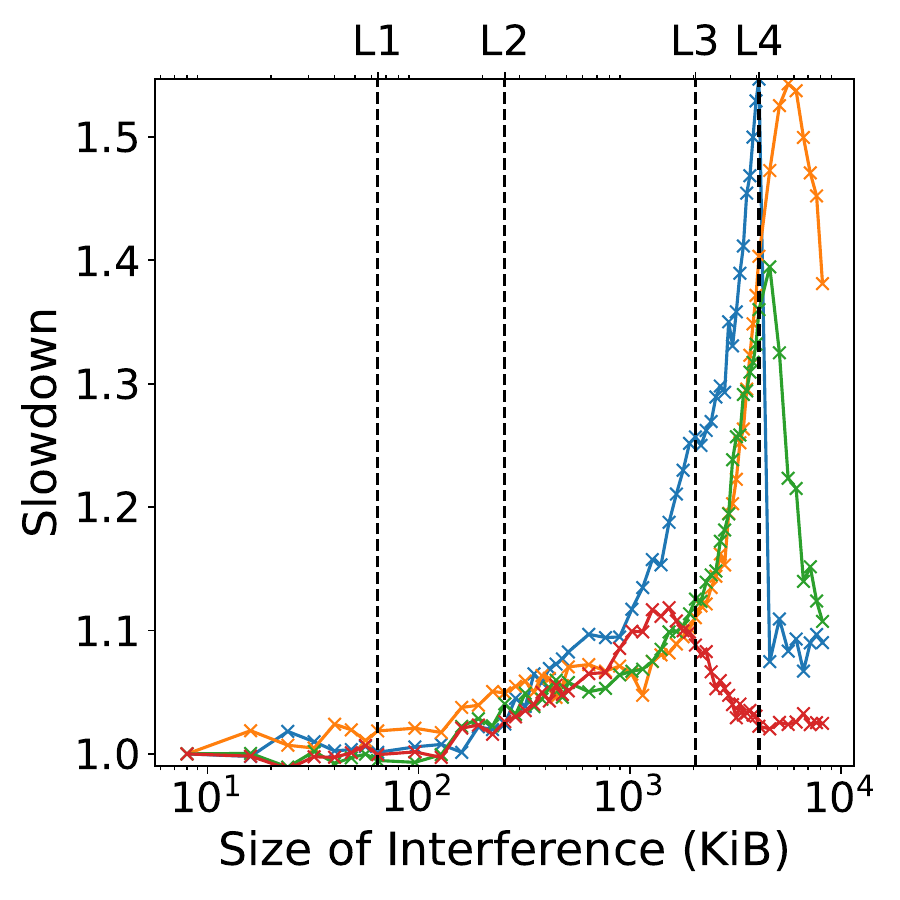}
            \captionsetup{justification=centering}
            \caption{orin: Way / Read}
            \label{fig:all-orin-way-mser-read-cif}
        \end{subfigure}
        \hfill
        \begin{subfigure}{0.24\textwidth}
            \centering
            \includegraphics[width=\textwidth]{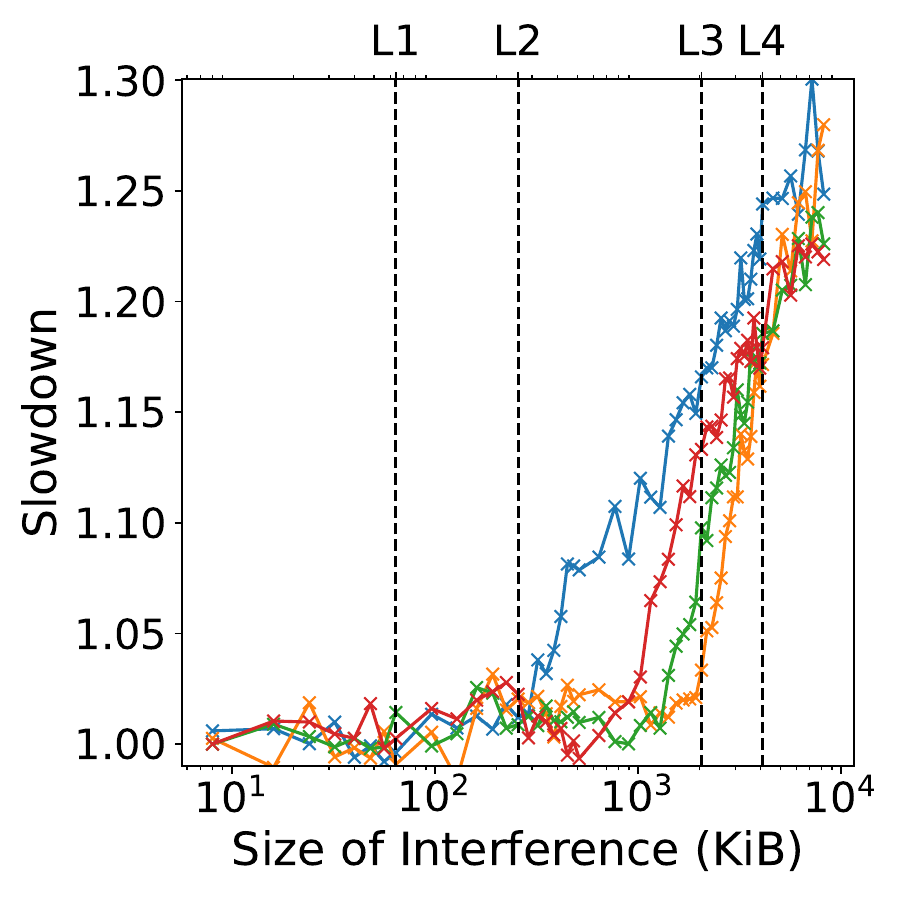}
            \captionsetup{justification=centering}
            \caption{orin: Way / Write}
            \label{fig:all-orin-way-mser-write-cif}
        \end{subfigure}
        \hfill
        \begin{subfigure}{0.24\textwidth}
            \centering
            \includegraphics[width=\textwidth]{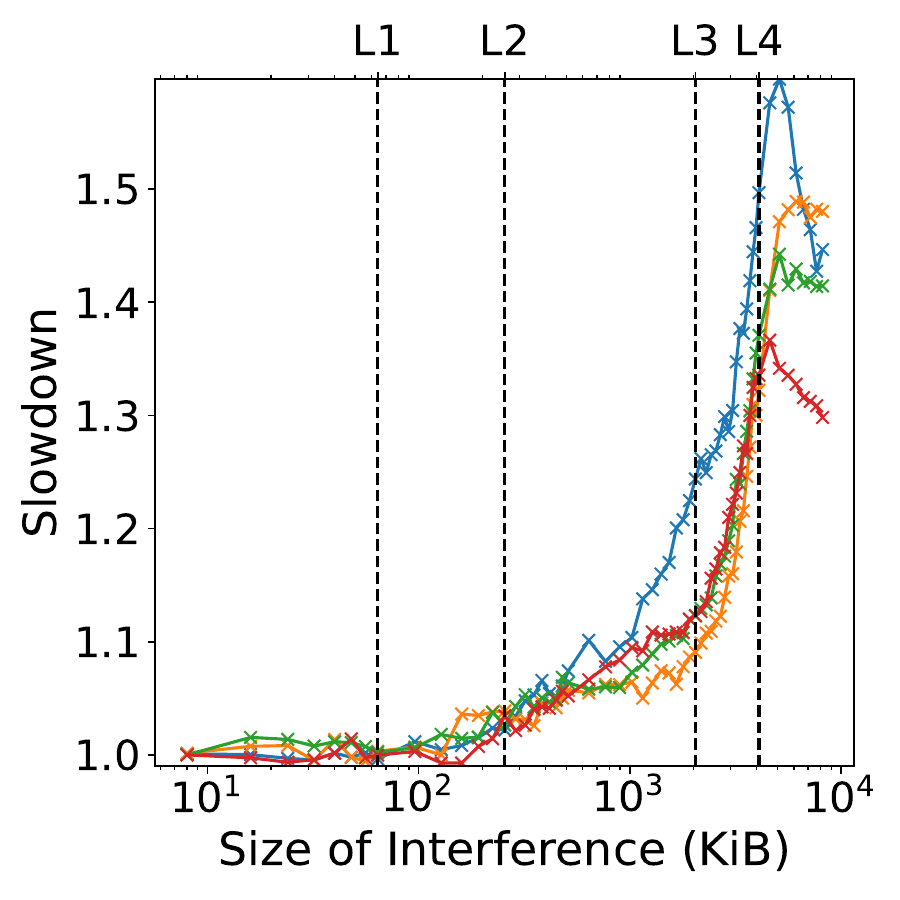}
            \captionsetup{justification=centering}
            \caption{orin: Way / Modify}
            \label{fig:all-orin-way-mser-modify-cif}
        \end{subfigure}
        \hfill
        \begin{subfigure}{0.24\textwidth}
            \centering
            \includegraphics[width=\textwidth]{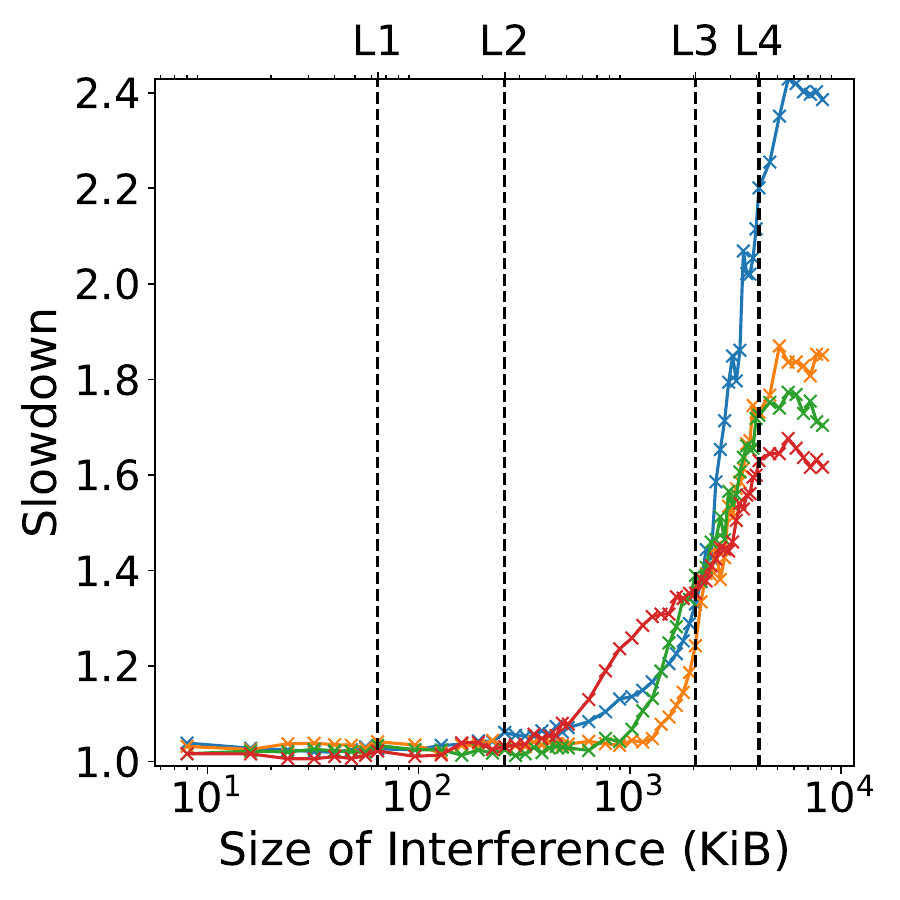}
            \captionsetup{justification=centering}
            \caption{orin: Way / Prefetch}
            \label{fig:all-orin-way-mser-prefetch-cif}
        \end{subfigure}
        \hfill
        
        \caption{Execution Slowdown on \textit{'Mser'} benchmark for \textit{'CIF'} dataset on \textit{'ORIN'} with Interferences and cache partitioning.}
        \label{fig:orin-mser-cif}
    \end{figure}

    \begin{figure}[H]
        \begin{subfigure}{\textwidth}
            \centering
            \includegraphics[width=0.5\textwidth]{figures/set_subplot/legend.pdf}
        \end{subfigure}
        \centering
        
        \begin{subfigure}{0.24\textwidth}
            \centering
            \includegraphics[width=\textwidth]{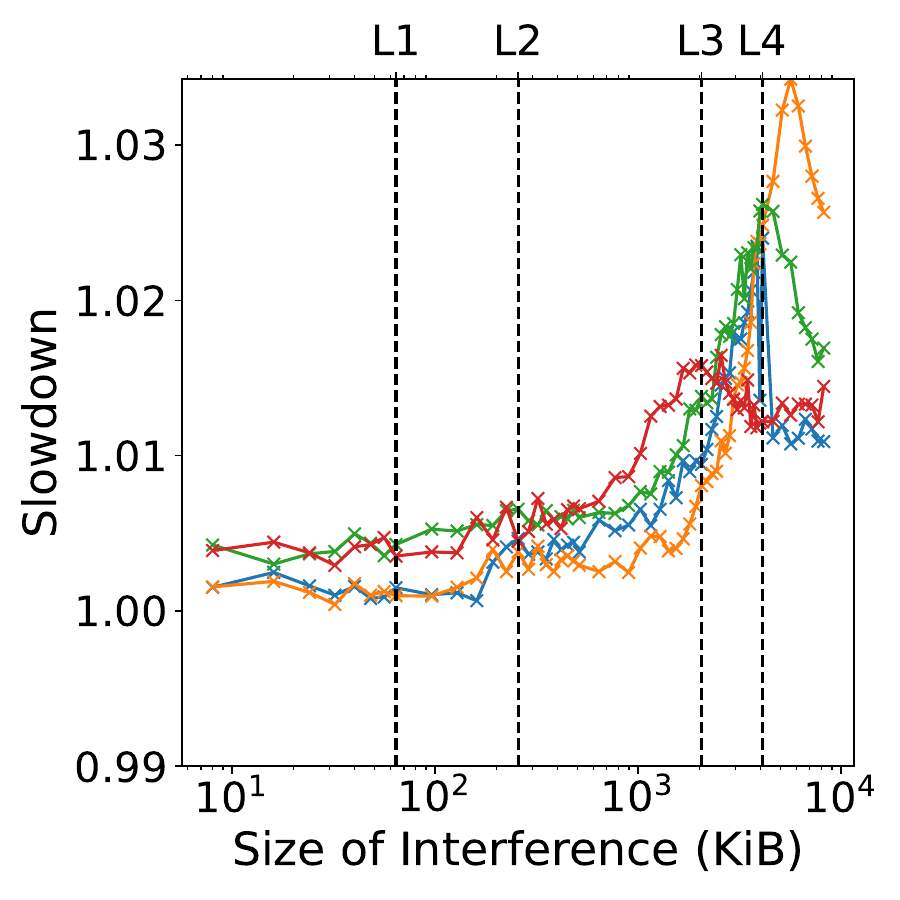}
            \captionsetup{justification=centering}
            \caption{orin: Way / Read}
            \label{fig:all-orin-way-tracking-read-cif}
        \end{subfigure}
        \hfill
        \begin{subfigure}{0.24\textwidth}
            \centering
            \includegraphics[width=\textwidth]{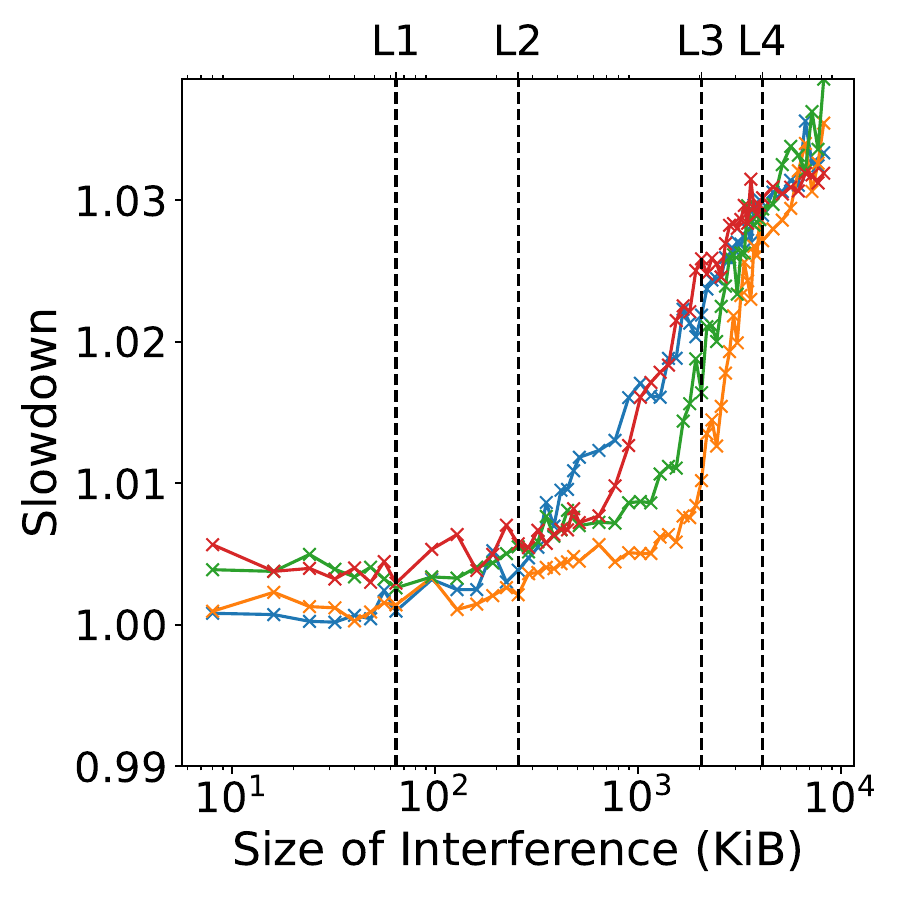}
            \captionsetup{justification=centering}
            \caption{orin: Way / Write}
            \label{fig:all-orin-way-tracking-write-cif}
        \end{subfigure}
        \hfill
        \begin{subfigure}{0.24\textwidth}
            \centering
            \includegraphics[width=\textwidth]{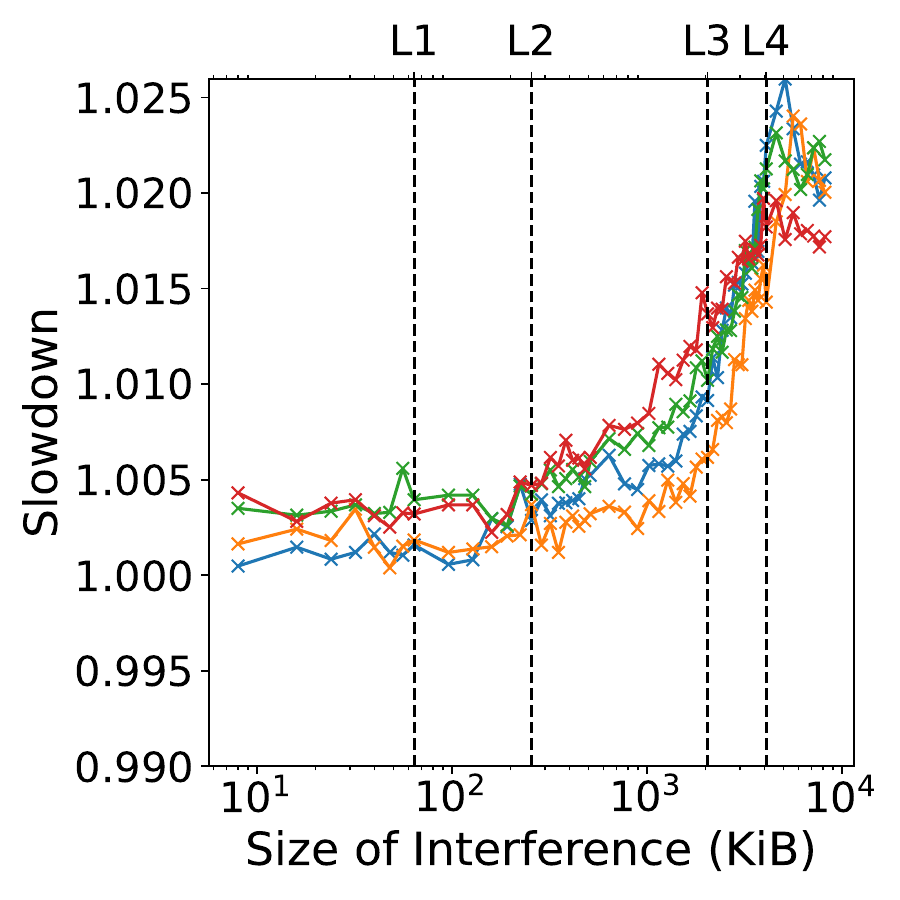}
            \captionsetup{justification=centering}
            \caption{orin: Way / Modify}
            \label{fig:all-orin-way-tracking-modify-cif}
        \end{subfigure}
        \hfill
        \begin{subfigure}{0.24\textwidth}
            \centering
            \includegraphics[width=\textwidth]{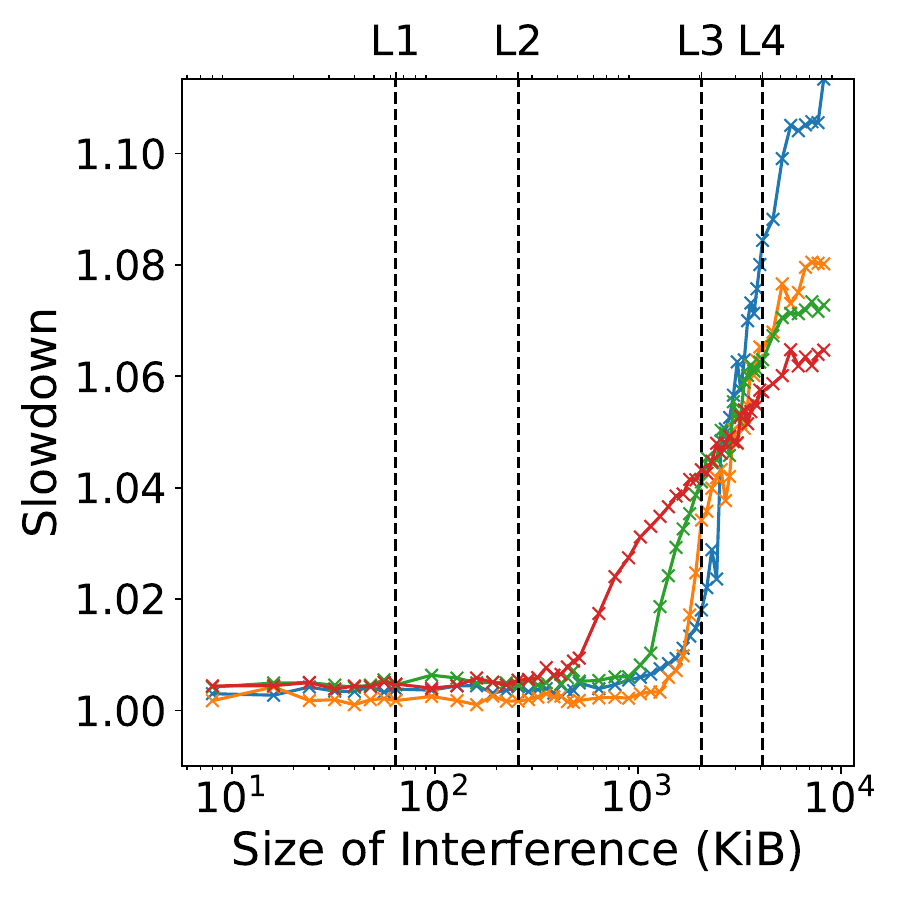}
            \captionsetup{justification=centering}
            \caption{orin: Way / Prefetch}
            \label{fig:all-orin-way-tracking-prefetch-cif}
        \end{subfigure}
        \hfill
        
        \caption{Execution Slowdown on \textit{'Tracking'} benchmark for \textit{'CIF'} dataset on \textit{'ORIN'} with Interferences and cache partitioning.}
        \label{fig:orin-tracking-cif}
    \end{figure}

    \begin{figure}[H]
        \begin{subfigure}{\textwidth}
            \centering
            \includegraphics[width=0.5\textwidth]{figures/set_subplot/legend.pdf}
        \end{subfigure}
        \centering
        
        \begin{subfigure}{0.24\textwidth}
            \centering
            \includegraphics[width=\textwidth]{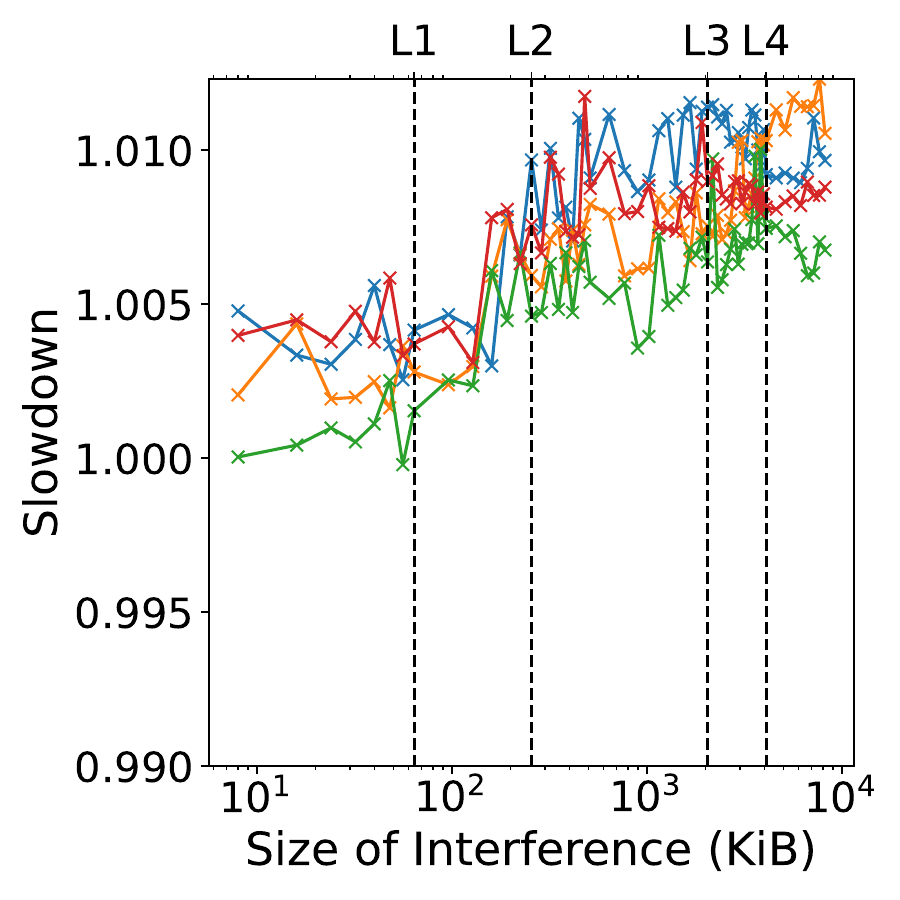}
            \captionsetup{justification=centering}
            \caption{orin: Way / Read}
            \label{fig:all-orin-way-sift-read-cif}
        \end{subfigure}
        \hfill
        \begin{subfigure}{0.24\textwidth}
            \centering
            \includegraphics[width=\textwidth]{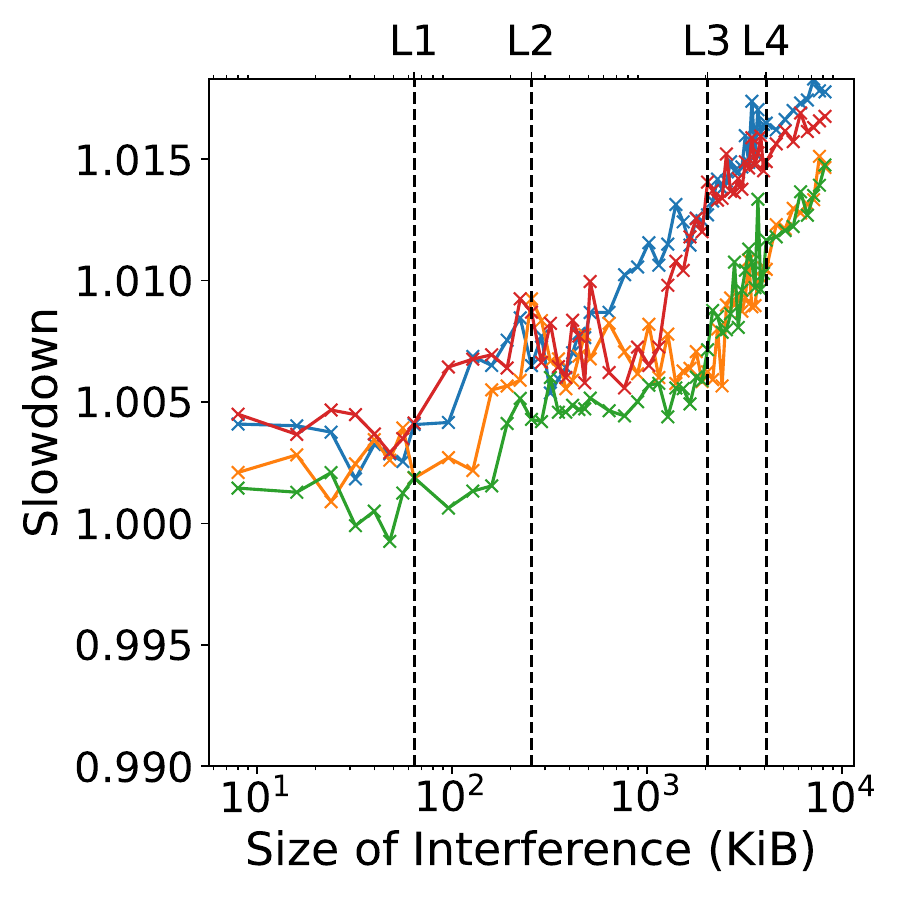}
            \captionsetup{justification=centering}
            \caption{orin: Way / Write}
            \label{fig:all-orin-way-sift-write-cif}
        \end{subfigure}
        \hfill
        \begin{subfigure}{0.24\textwidth}
            \centering
            \includegraphics[width=\textwidth]{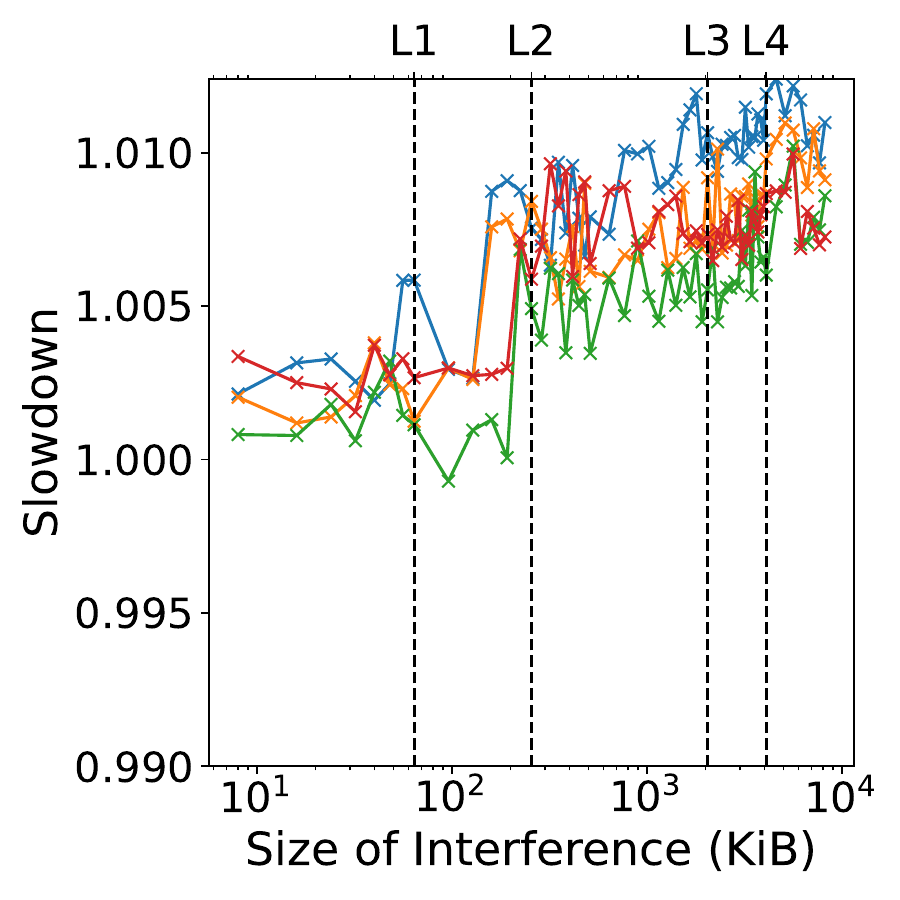}
            \captionsetup{justification=centering}
            \caption{orin: Way / Modify}
            \label{fig:all-orin-way-sift-modify-cif}
        \end{subfigure}
        \hfill
        \begin{subfigure}{0.24\textwidth}
            \centering
            \includegraphics[width=\textwidth]{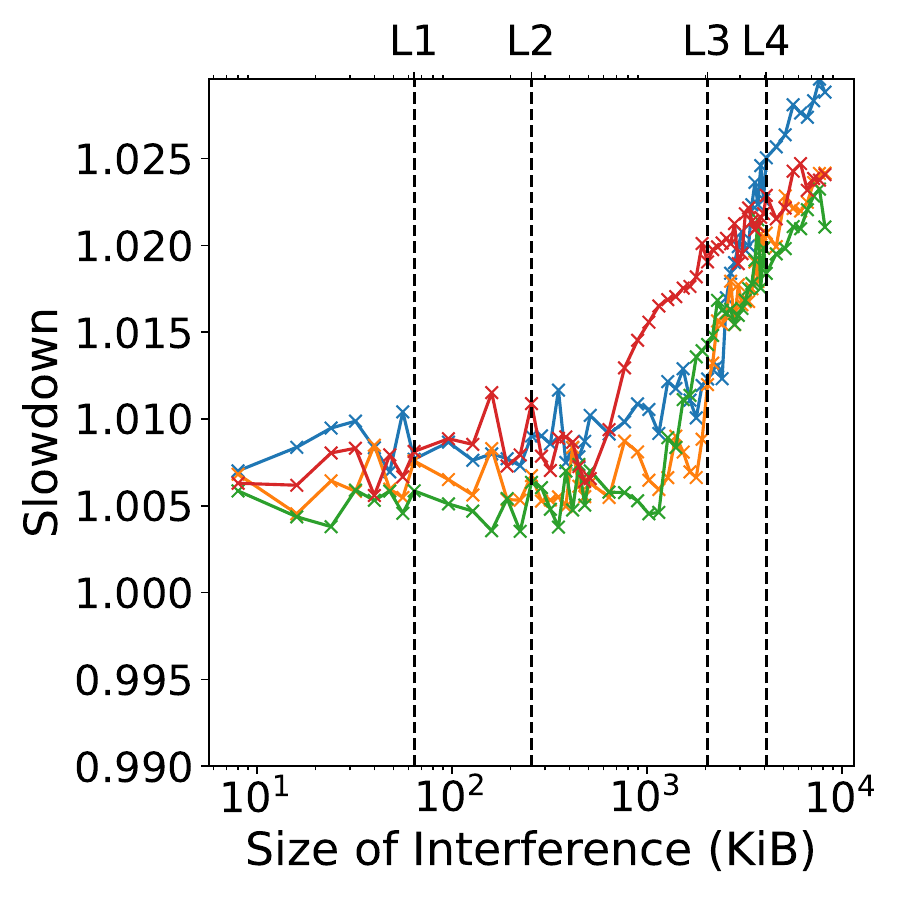}
            \captionsetup{justification=centering}
            \caption{orin: Way / Prefetch}
            \label{fig:all-orin-way-sift-prefetch-cif}
        \end{subfigure}
        \hfill
        
        \caption{Execution Slowdown on \textit{'Sift'} benchmark for \textit{'CIF'} dataset on \textit{'ORIN'} with Interferences and cache partitioning.}
        \label{fig:orin-sift-cif}
    \end{figure}

    \begin{figure}[H]
        \begin{subfigure}{\textwidth}
            \centering
            \includegraphics[width=0.5\textwidth]{figures/set_subplot/legend.pdf}
        \end{subfigure}
        \centering
        
        \begin{subfigure}{0.24\textwidth}
            \centering
            \includegraphics[width=\textwidth]{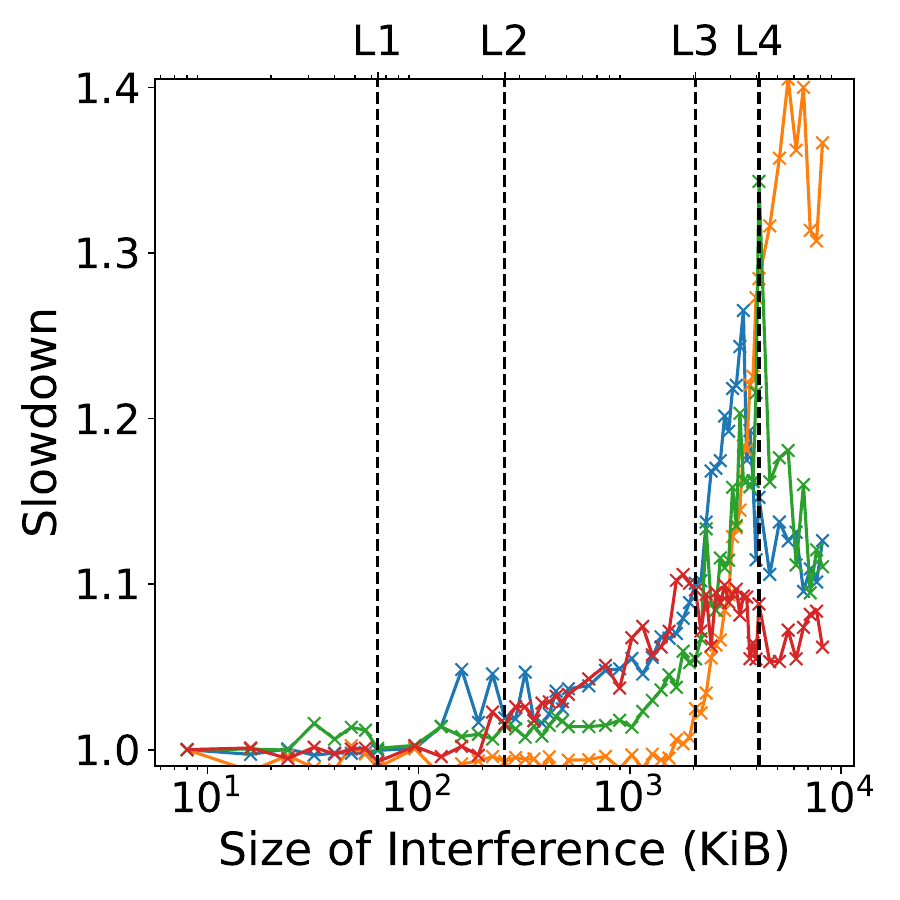}
            \captionsetup{justification=centering}
            \caption{orin: Way / Read}
            \label{fig:all-orin-way-disparity-read-vga}
        \end{subfigure}
        \hfill
        \begin{subfigure}{0.24\textwidth}
            \centering
            \includegraphics[width=\textwidth]{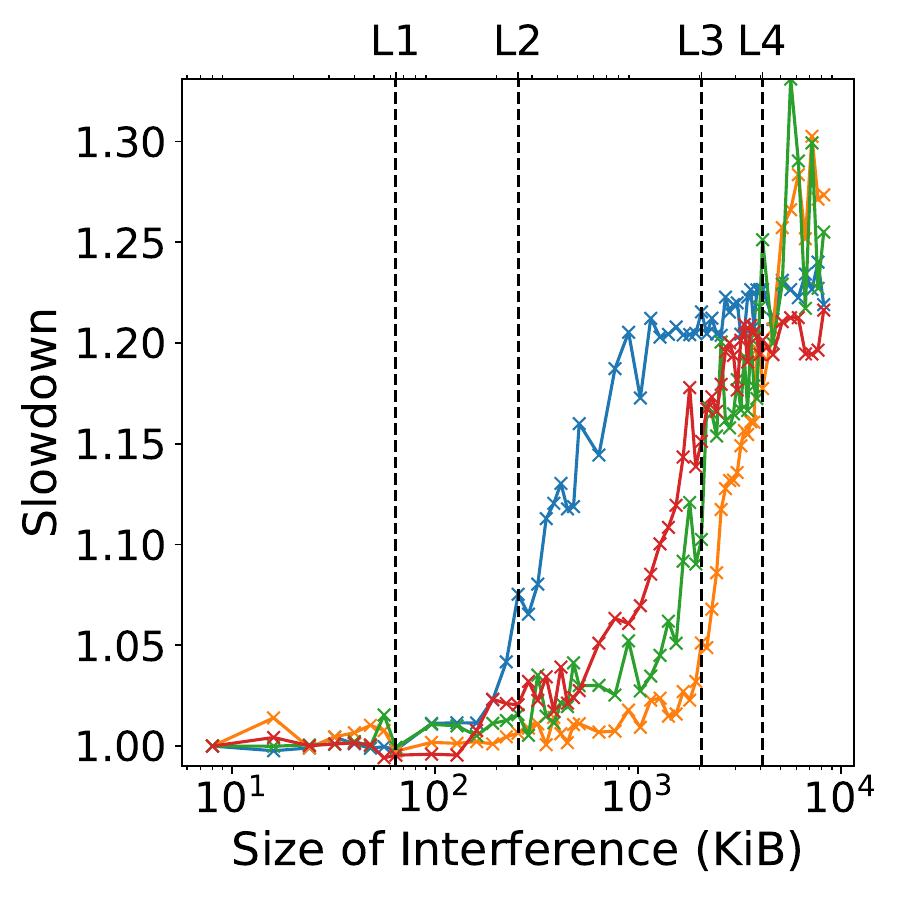}
            \captionsetup{justification=centering}
            \caption{orin: Way / Write}
            \label{fig:all-orin-way-disparity-write-vga}
        \end{subfigure}
        \hfill
        \begin{subfigure}{0.24\textwidth}
            \centering
            \includegraphics[width=\textwidth]{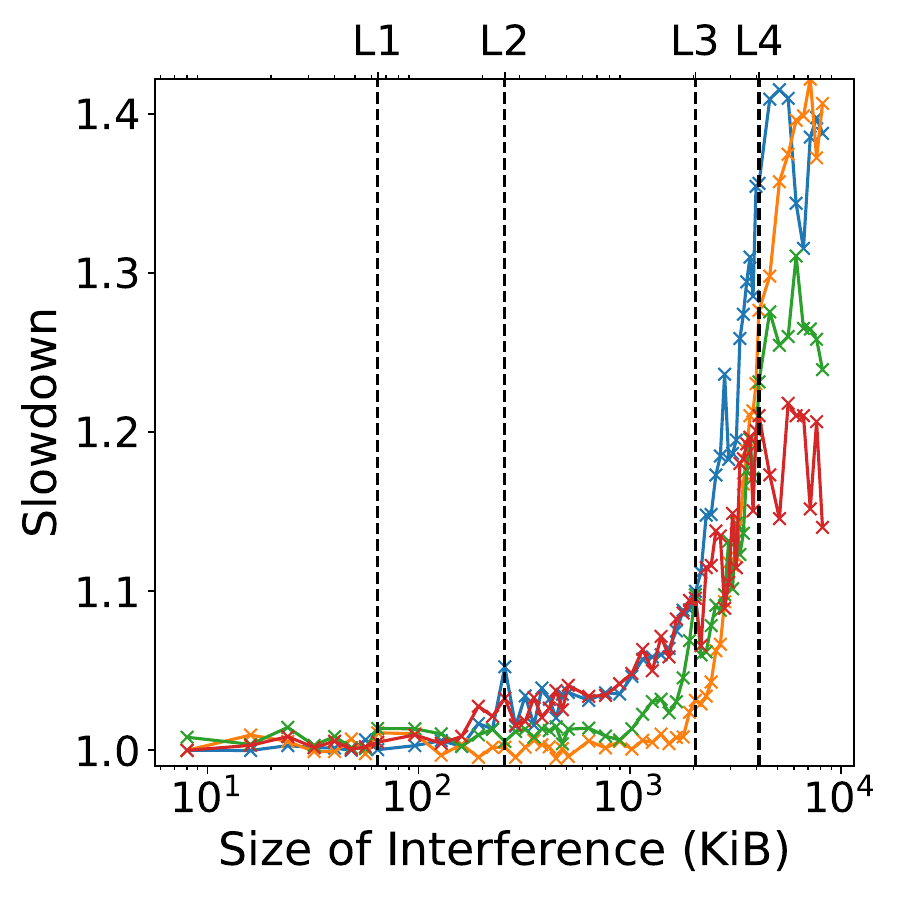}
            \captionsetup{justification=centering}
            \caption{orin: Way / Modify}
            \label{fig:all-orin-way-disparity-modify-vga}
        \end{subfigure}
        \hfill
        \begin{subfigure}{0.24\textwidth}
            \centering
            \includegraphics[width=\textwidth]{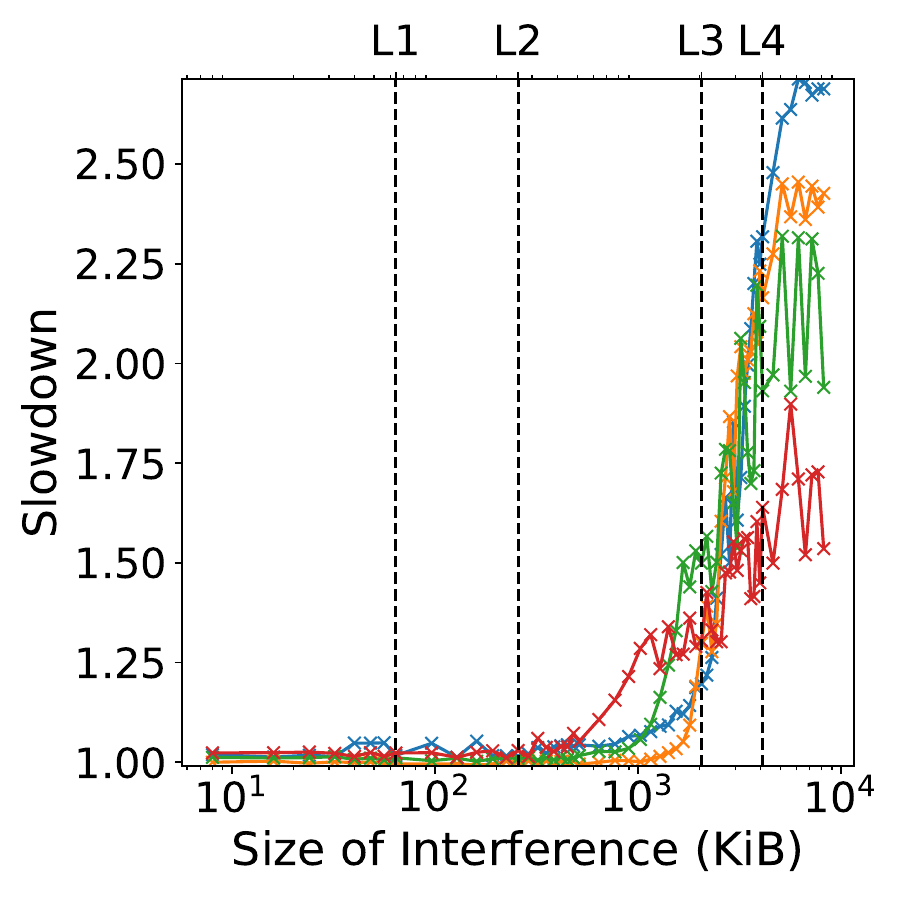}
            \captionsetup{justification=centering}
            \caption{orin: Way / Prefetch}
            \label{fig:all-orin-way-disparity-prefetch-vga}
        \end{subfigure}
        \hfill
        
        \caption{Execution Slowdown on \textit{'Disparity'} benchmark for \textit{'VGA'} dataset on \textit{'ORIN'} with Interferences and cache partitioning.}
        \label{fig:orin-disparity-vga}
    \end{figure}

    \begin{figure}[H]
        \begin{subfigure}{\textwidth}
            \centering
            \includegraphics[width=0.5\textwidth]{figures/set_subplot/legend.pdf}
        \end{subfigure}
        \centering
        
        \begin{subfigure}{0.24\textwidth}
            \centering
            \includegraphics[width=\textwidth]{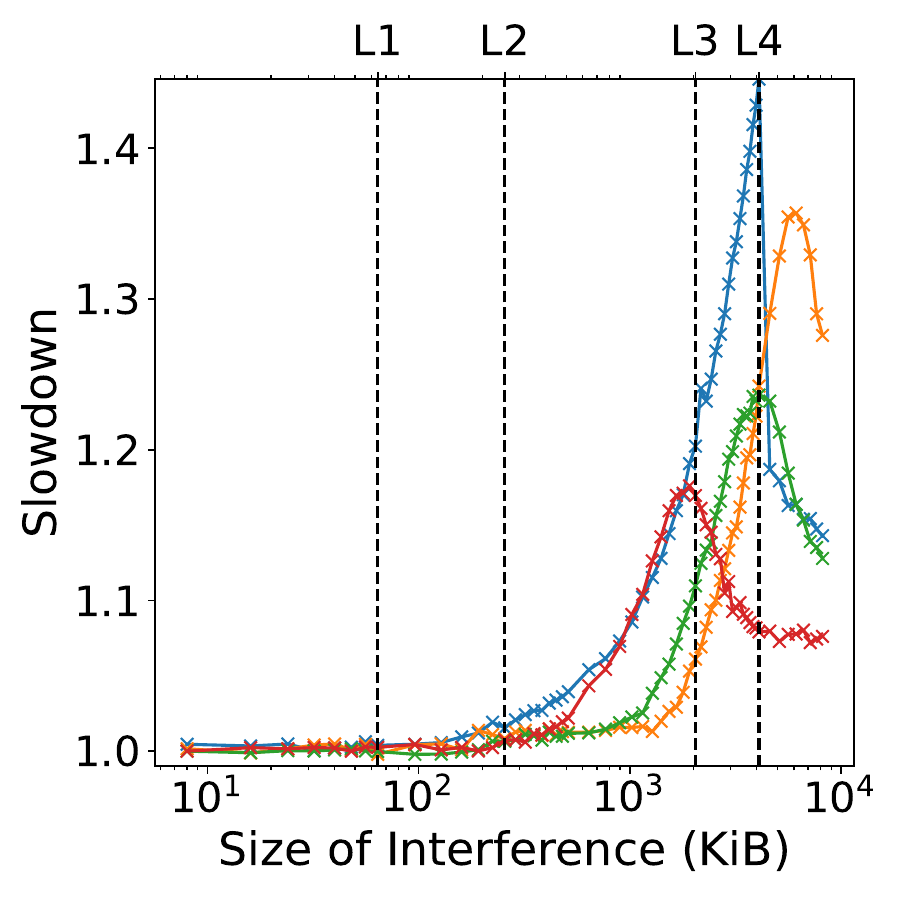}
            \captionsetup{justification=centering}
            \caption{orin: Way / Read}
            \label{fig:all-orin-way-mser-read-vga}
        \end{subfigure}
        \hfill
        \begin{subfigure}{0.24\textwidth}
            \centering
            \includegraphics[width=\textwidth]{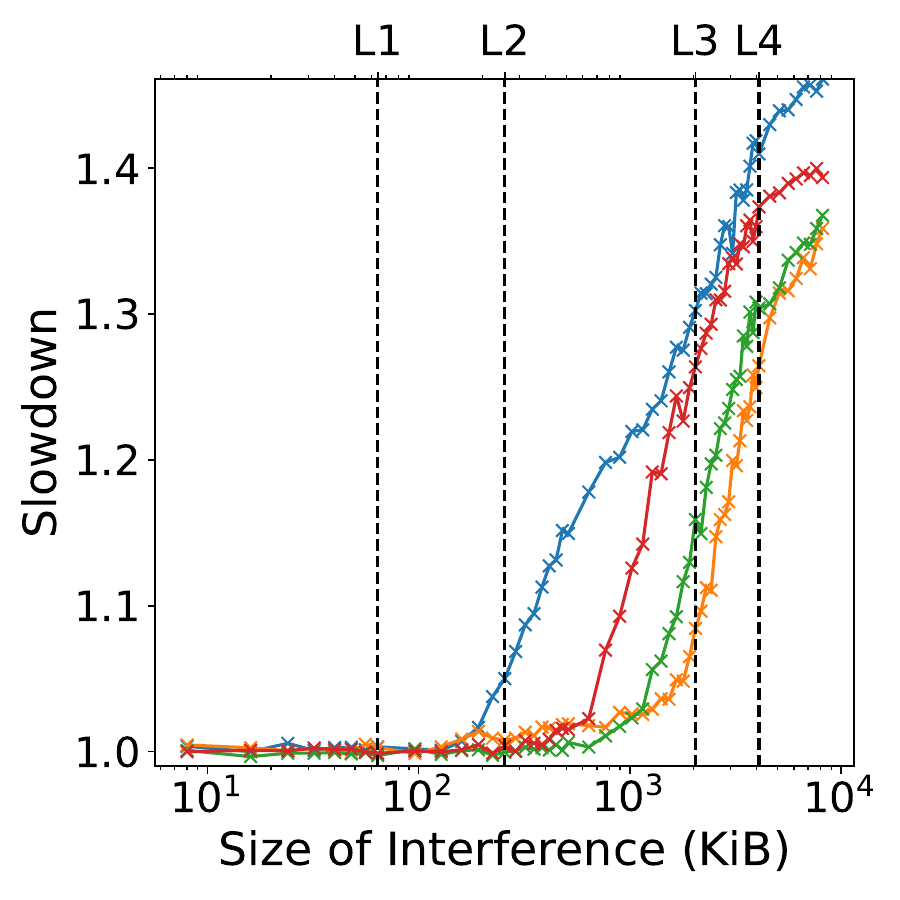}
            \captionsetup{justification=centering}
            \caption{orin: Way / Write}
            \label{fig:all-orin-way-mser-write-vga}
        \end{subfigure}
        \hfill
        \begin{subfigure}{0.24\textwidth}
            \centering
            \includegraphics[width=\textwidth]{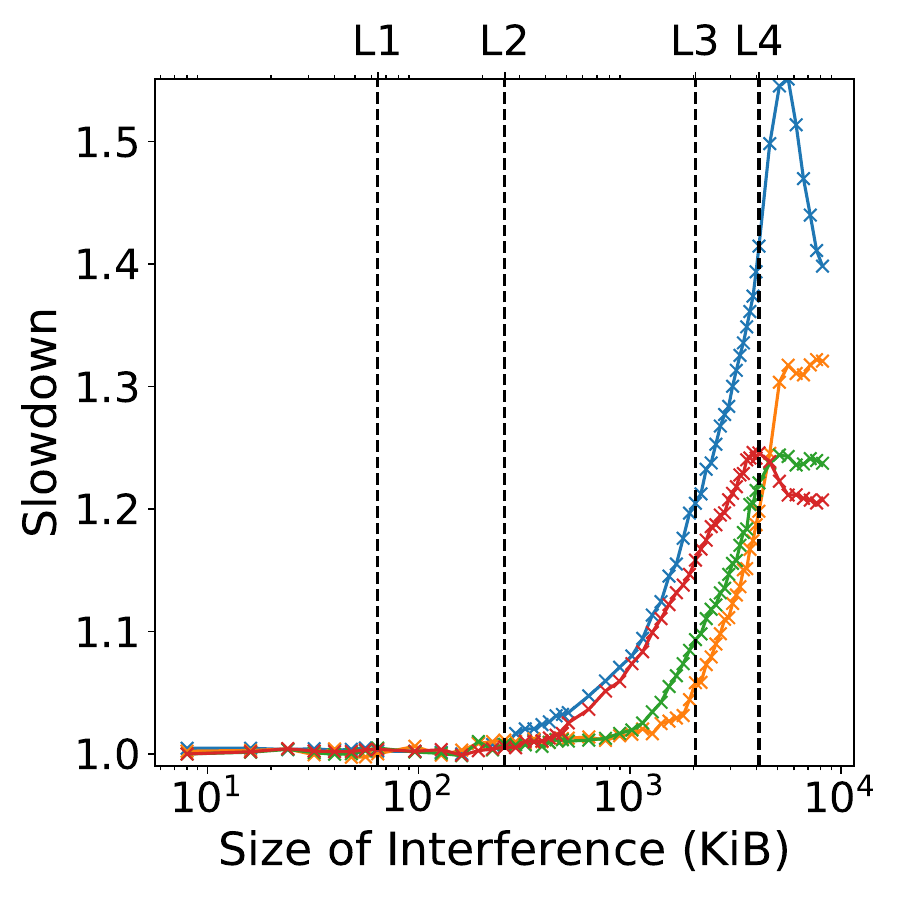}
            \captionsetup{justification=centering}
            \caption{orin: Way / Modify}
            \label{fig:all-orin-way-mser-modify-vga}
        \end{subfigure}
        \hfill
        \begin{subfigure}{0.24\textwidth}
            \centering
            \includegraphics[width=\textwidth]{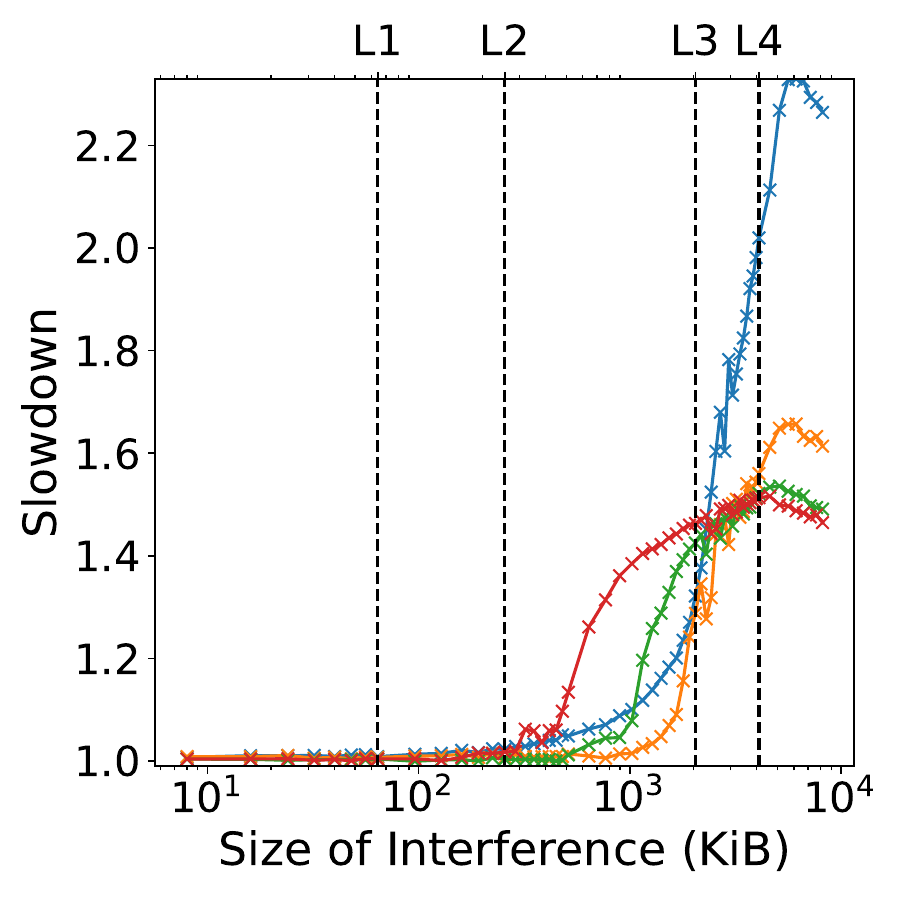}
            \captionsetup{justification=centering}
            \caption{orin: Way / Prefetch}
            \label{fig:all-orin-way-mser-prefetch-vga}
        \end{subfigure}
        \hfill
        
        \caption{Execution Slowdown on \textit{'Mser'} benchmark for \textit{'VGA'} dataset on \textit{'ORIN'} with Interferences and cache partitioning.}
        \label{fig:orin-mser-vga}
    \end{figure}

    \begin{figure}[H]
        \begin{subfigure}{\textwidth}
            \centering
            \includegraphics[width=0.5\textwidth]{figures/set_subplot/legend.pdf}
        \end{subfigure}
        \centering
        
        \begin{subfigure}{0.24\textwidth}
            \centering
            \includegraphics[width=\textwidth]{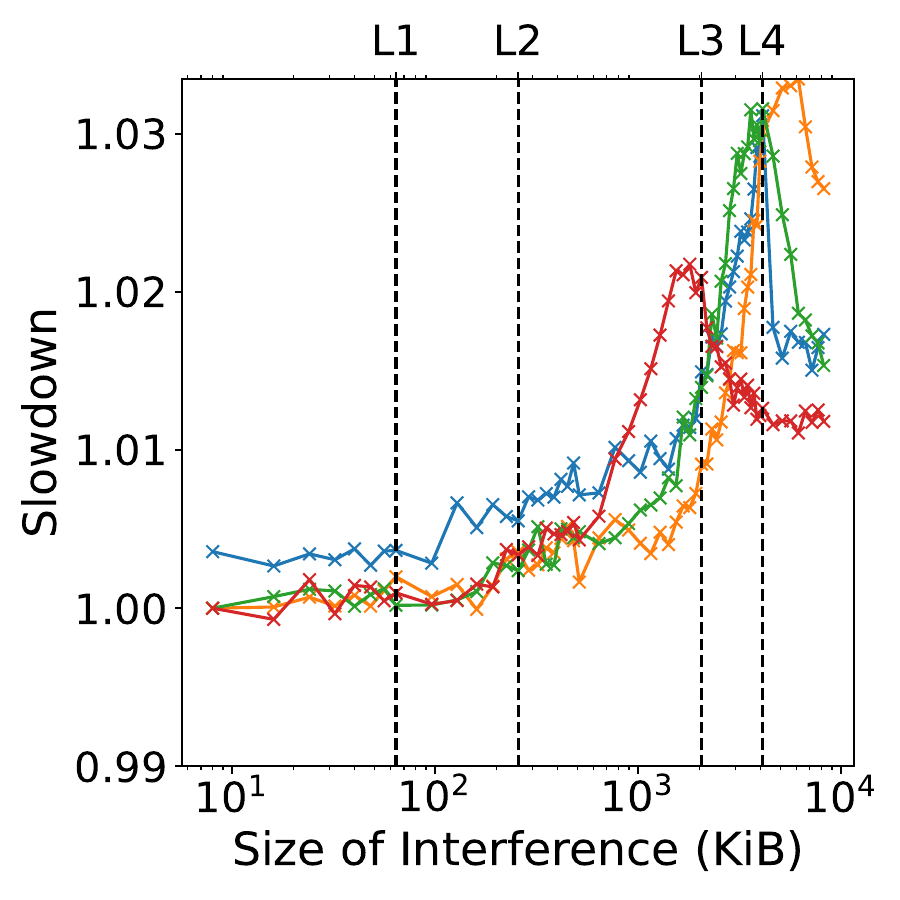}
            \captionsetup{justification=centering}
            \caption{orin: Way / Read}
            \label{fig:all-orin-way-tracking-read-vga}
        \end{subfigure}
        \hfill
        \begin{subfigure}{0.24\textwidth}
            \centering
            \includegraphics[width=\textwidth]{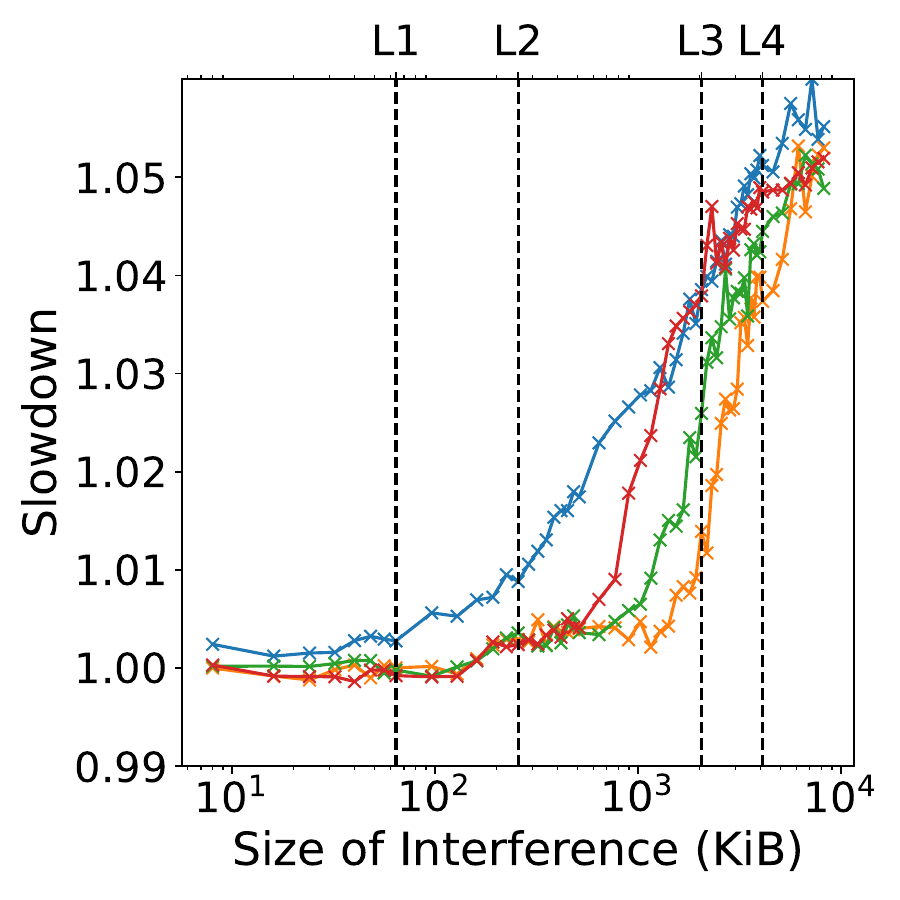}
            \captionsetup{justification=centering}
            \caption{orin: Way / Write}
            \label{fig:all-orin-way-tracking-write-vga}
        \end{subfigure}
        \hfill
        \begin{subfigure}{0.24\textwidth}
            \centering
            \includegraphics[width=\textwidth]{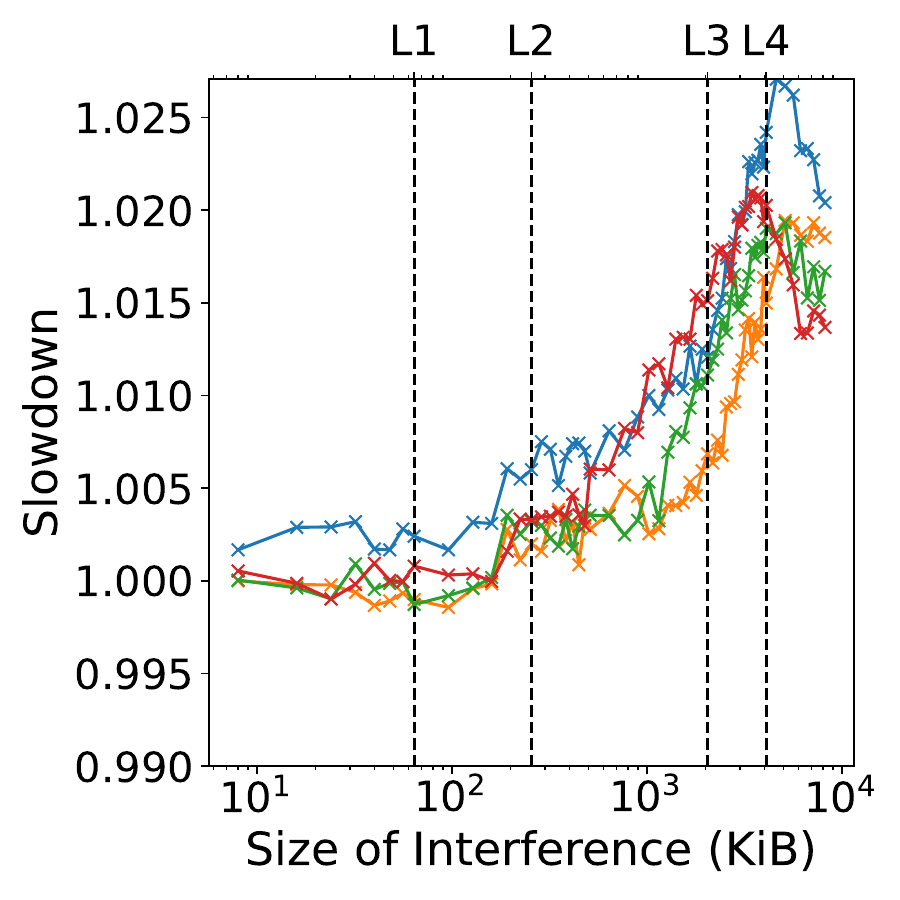}
            \captionsetup{justification=centering}
            \caption{orin: Way / Modify}
            \label{fig:all-orin-way-tracking-modify-vga}
        \end{subfigure}
        \hfill
        \begin{subfigure}{0.24\textwidth}
            \centering
            \includegraphics[width=\textwidth]{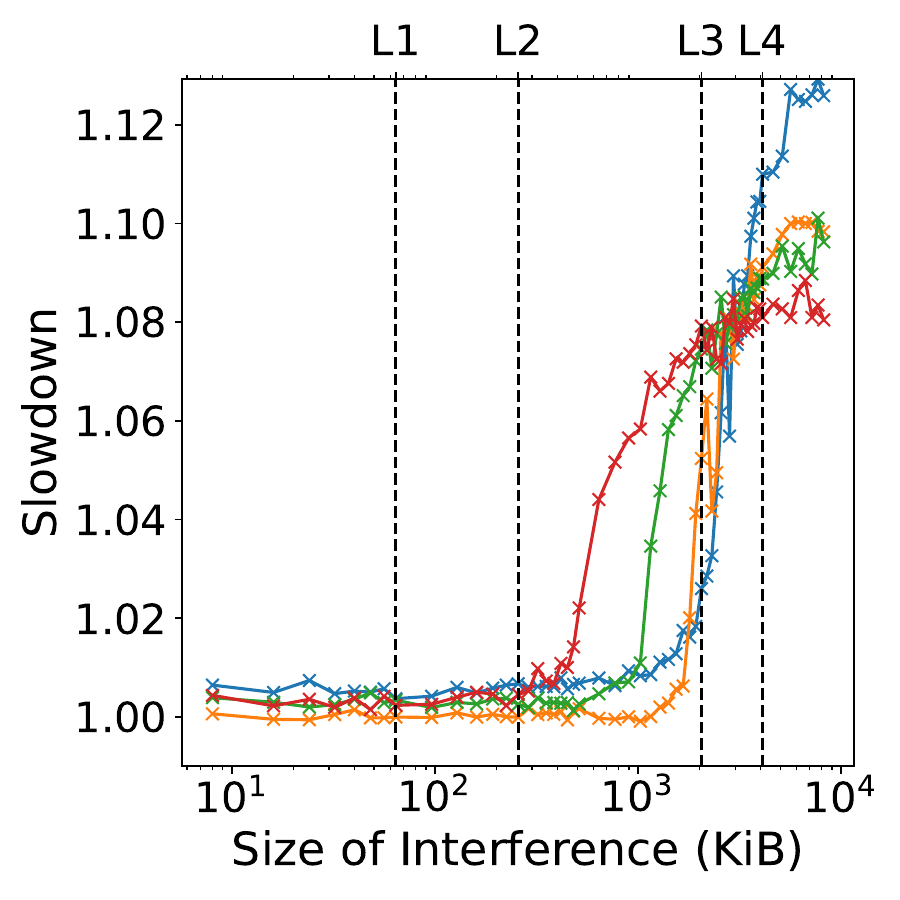}
            \captionsetup{justification=centering}
            \caption{orin: Way / Prefetch}
            \label{fig:all-orin-way-tracking-prefetch-vga}
        \end{subfigure}
        \hfill
        
        \caption{Execution Slowdown on \textit{'Tracking'} benchmark for \textit{'VGA'} dataset on \textit{'ORIN'} with Interferences and cache partitioning.}
        \label{fig:orin-tracking-vga}
    \end{figure}

    \begin{figure}[H]
        \begin{subfigure}{\textwidth}
            \centering
            \includegraphics[width=0.5\textwidth]{figures/set_subplot/legend.pdf}
        \end{subfigure}
        \centering
        
        \begin{subfigure}{0.24\textwidth}
            \centering
            \includegraphics[width=\textwidth]{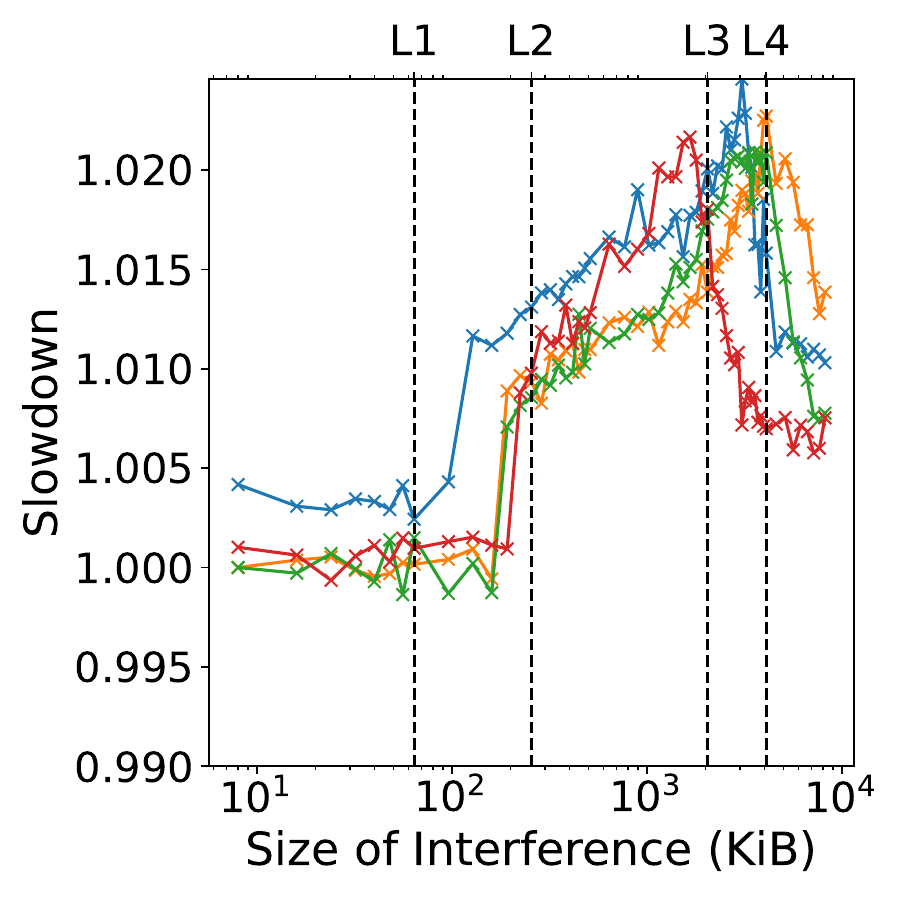}
            \captionsetup{justification=centering}
            \caption{orin: Way / Read}
            \label{fig:all-orin-way-sift-read-vga}
        \end{subfigure}
        \hfill
        \begin{subfigure}{0.24\textwidth}
            \centering
            \includegraphics[width=\textwidth]{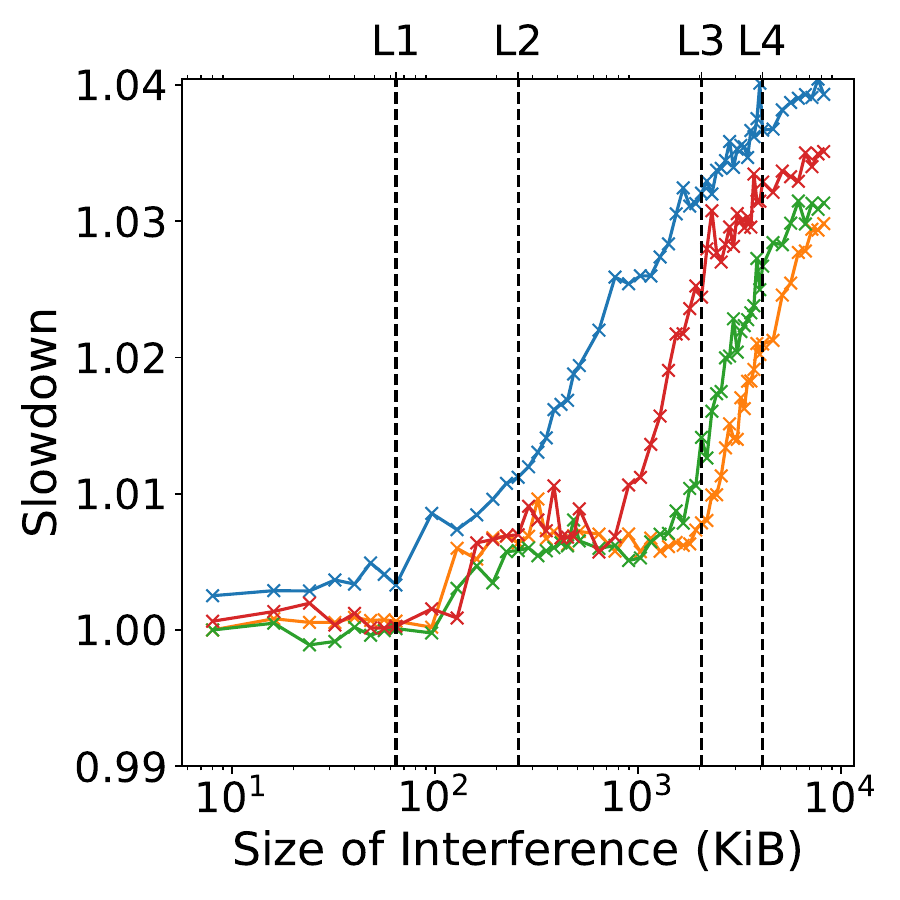}
            \captionsetup{justification=centering}
            \caption{orin: Way / Write}
            \label{fig:all-orin-way-sift-write-vga}
        \end{subfigure}
        \hfill
        \begin{subfigure}{0.24\textwidth}
            \centering
            \includegraphics[width=\textwidth]{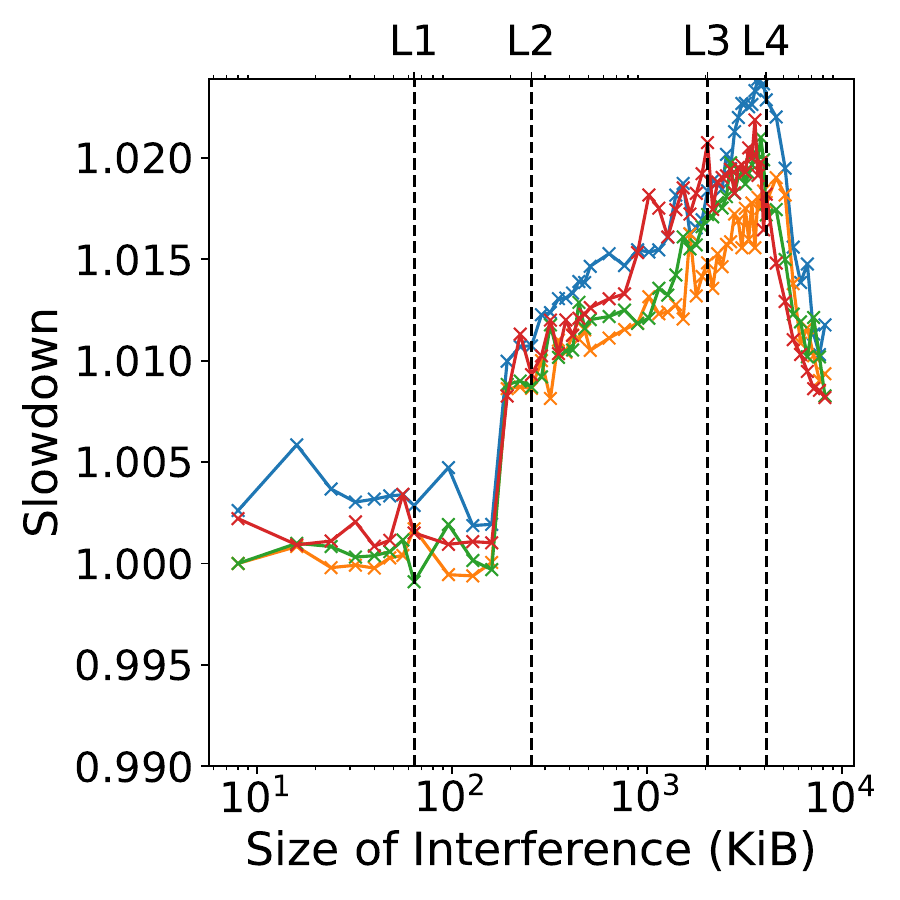}
            \captionsetup{justification=centering}
            \caption{orin: Way / Modify}
            \label{fig:all-orin-way-sift-modify-vga}
        \end{subfigure}
        \hfill
        \begin{subfigure}{0.24\textwidth}
            \centering
            \includegraphics[width=\textwidth]{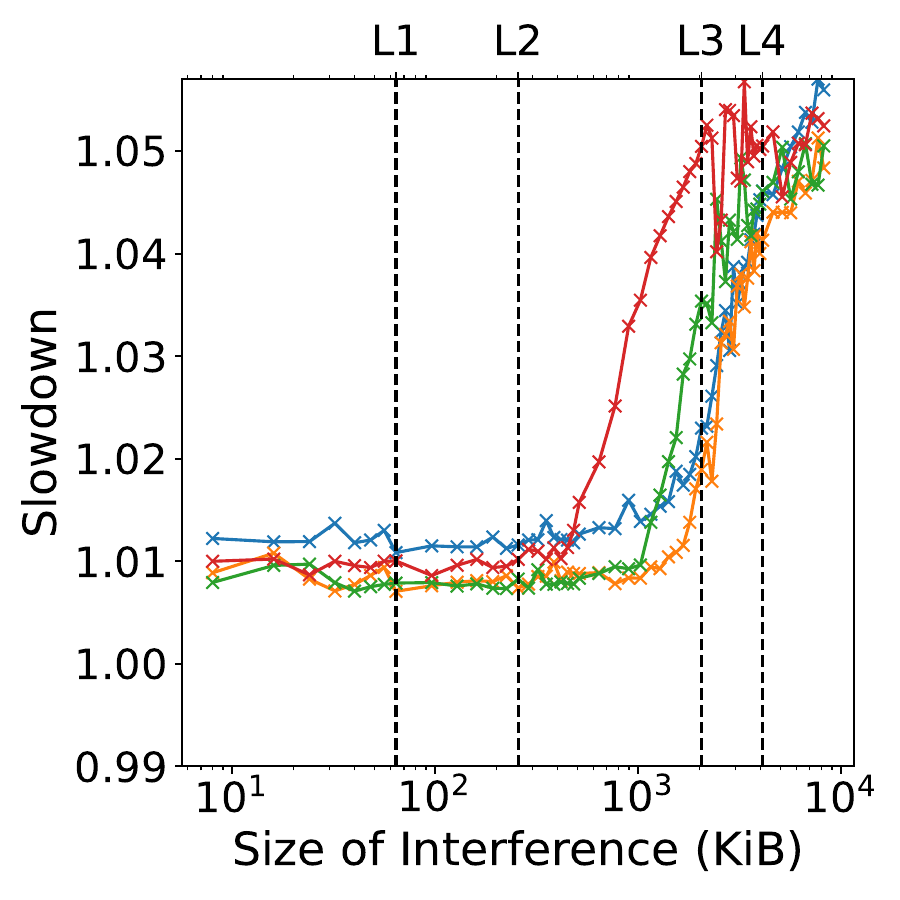}
            \captionsetup{justification=centering}
            \caption{orin: Way / Prefetch}
            \label{fig:all-orin-way-sift-prefetch-vga}
        \end{subfigure}
        \hfill
        
        \caption{Execution Slowdown on \textit{'Sift'} benchmark for \textit{'VGA'} dataset on \textit{'ORIN'} with Interferences and cache partitioning.}
        \label{fig:orin-sift-vga}
    \end{figure}

        \clearpage

        \subsection{ZCU102}

    \begin{figure}[H]
        \begin{subfigure}{\textwidth}
            \centering
            \includegraphics[width=0.5\textwidth]{figures/set_subplot/legend.pdf}
        \end{subfigure}
        \centering
        
        \begin{subfigure}{0.24\textwidth}
            \centering
            \includegraphics[width=\textwidth]{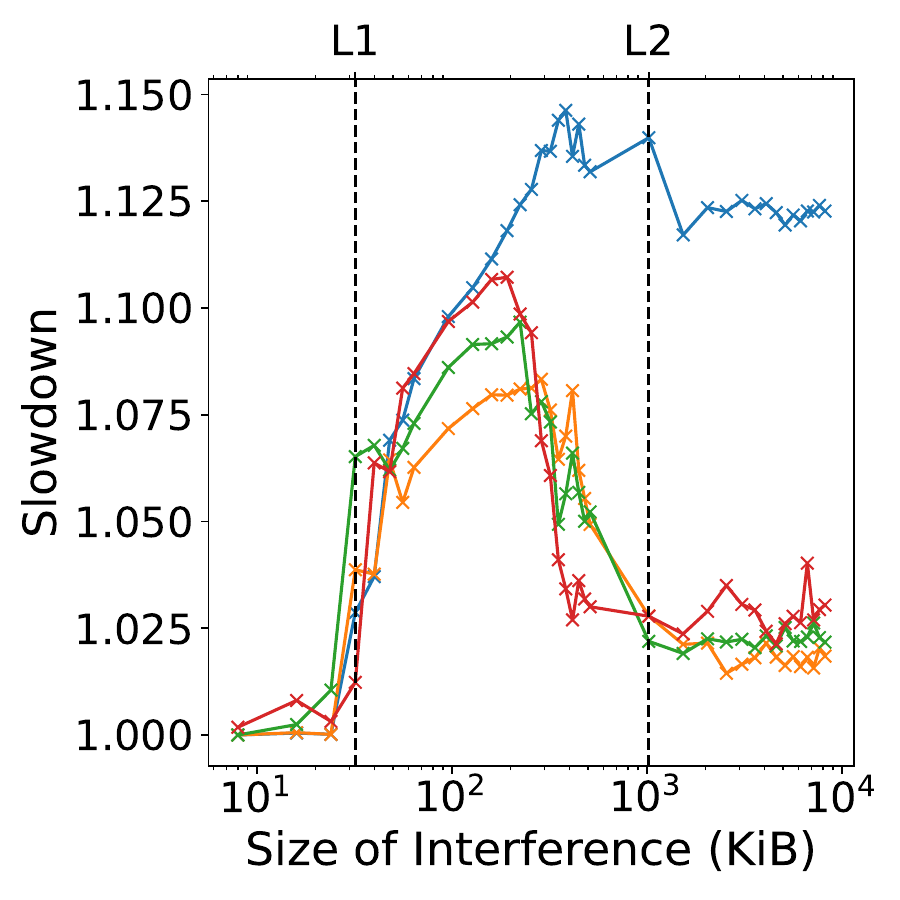}
            \captionsetup{justification=centering}
            \caption{zcu102: Set / Read}
            \label{fig:all-zcu102-set-disparity-read-cif}
        \end{subfigure}
        \hfill
        \begin{subfigure}{0.24\textwidth}
            \centering
            \includegraphics[width=\textwidth]{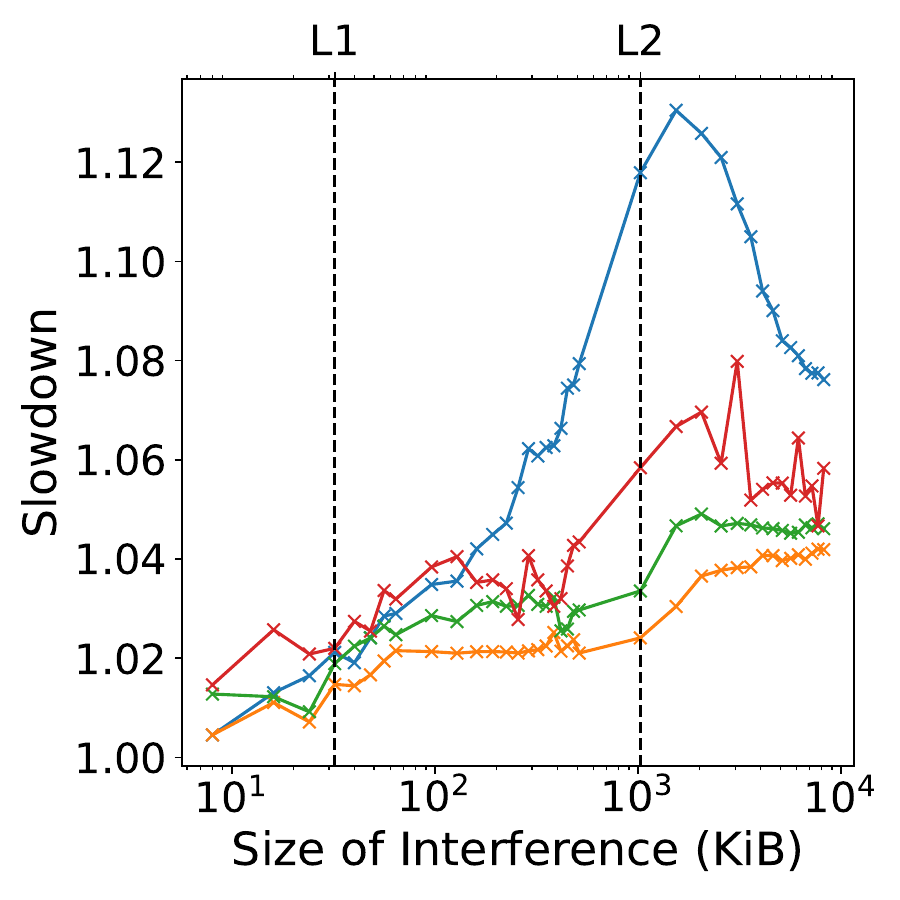}
            \captionsetup{justification=centering}
            \caption{zcu102: Set / Write}
            \label{fig:all-zcu102-set-disparity-write-cif}
        \end{subfigure}
        \hfill
        \begin{subfigure}{0.24\textwidth}
            \centering
            \includegraphics[width=\textwidth]{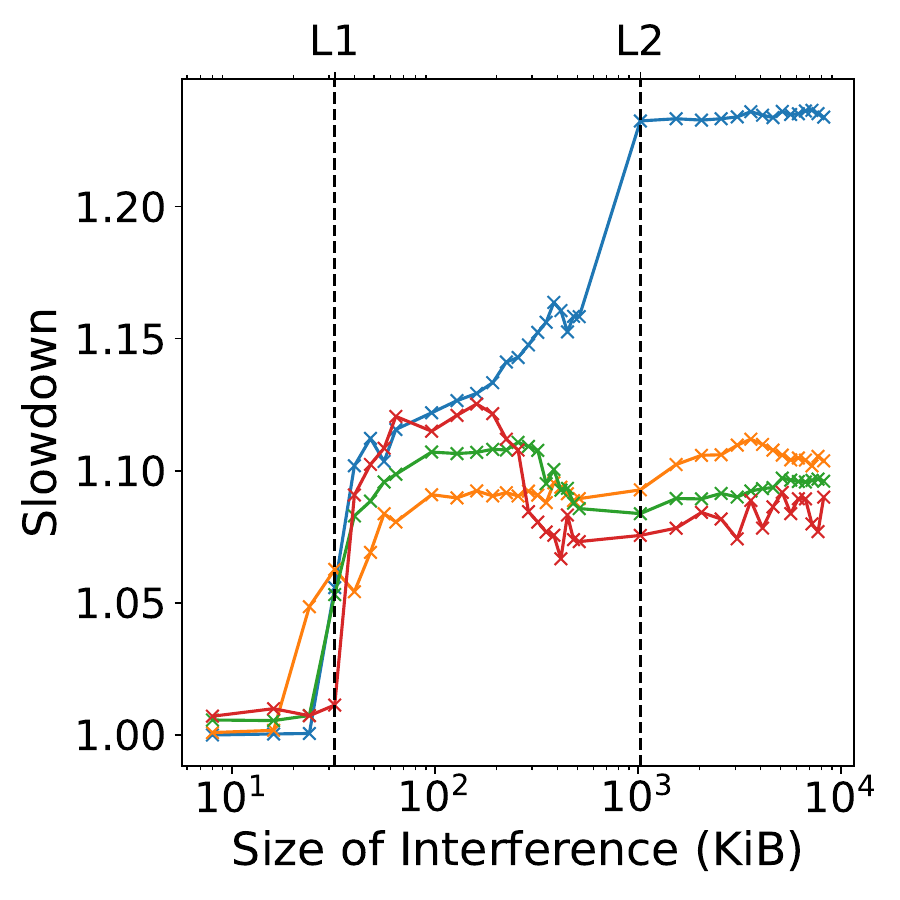}
            \captionsetup{justification=centering}
            \caption{zcu102: Set / Modify}
            \label{fig:all-zcu102-set-disparity-modify-cif}
        \end{subfigure}
        \hfill
        
        \caption{Execution Slowdown on \textit{'Disparity'} benchmark for \textit{'CIF'} dataset on \textit{'ZCU102'} with Interferences and cache partitioning.}
        \label{fig:zcu102-disparity-cif}
    \end{figure}

    \begin{figure}[H]
        \begin{subfigure}{\textwidth}
            \centering
            \includegraphics[width=0.5\textwidth]{figures/set_subplot/legend.pdf}
        \end{subfigure}
        \centering
        
        \begin{subfigure}{0.24\textwidth}
            \centering
            \includegraphics[width=\textwidth]{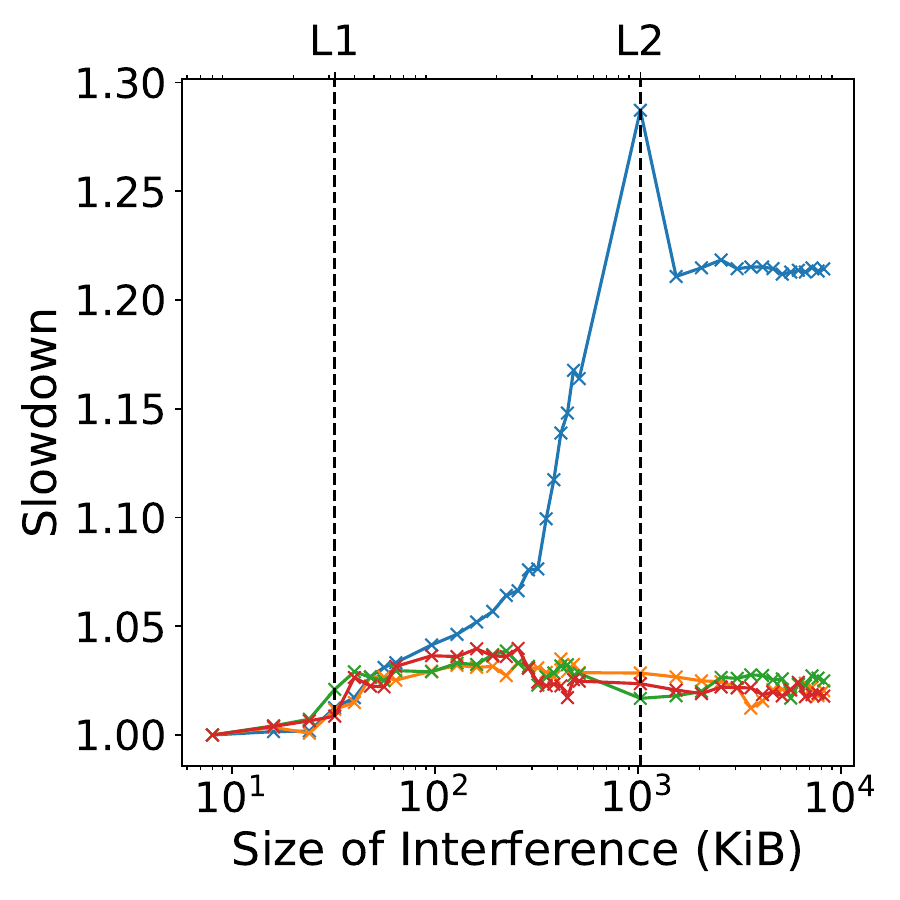}
            \captionsetup{justification=centering}
            \caption{zcu102: Set / Read}
            \label{fig:all-zcu102-set-mser-read-cif}
        \end{subfigure}
        \hfill
        \begin{subfigure}{0.24\textwidth}
            \centering
            \includegraphics[width=\textwidth]{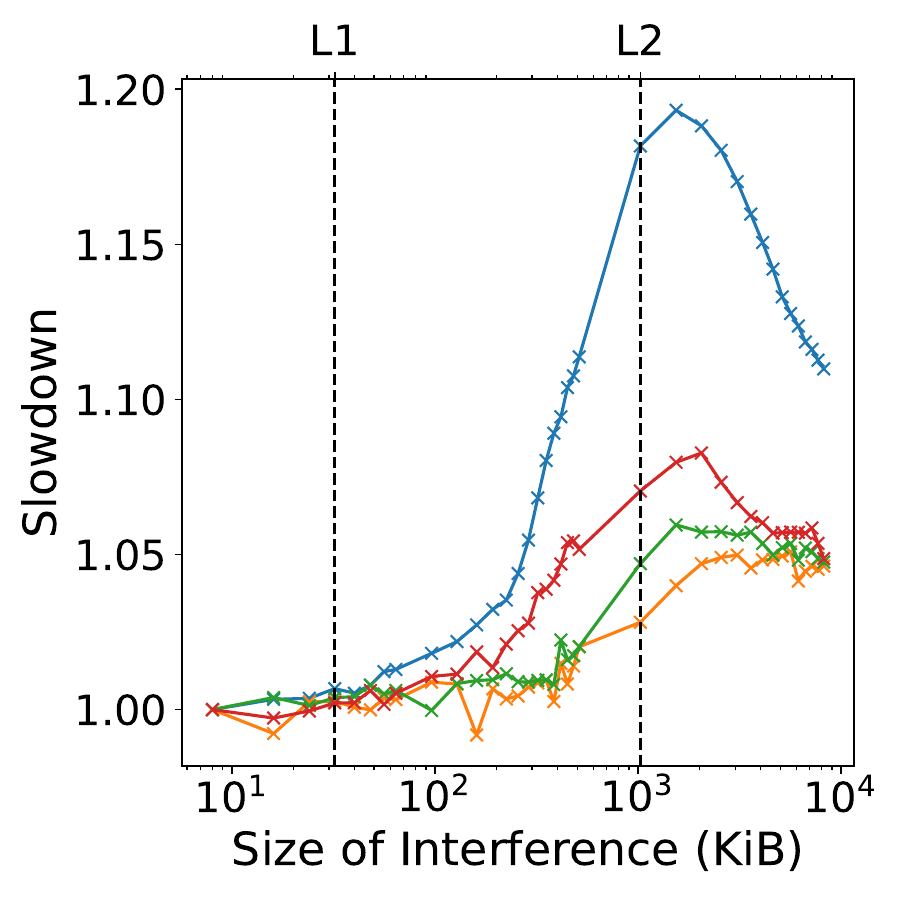}
            \captionsetup{justification=centering}
            \caption{zcu102: Set / Write}
            \label{fig:all-zcu102-set-mser-write-cif}
        \end{subfigure}
        \hfill
        \begin{subfigure}{0.24\textwidth}
            \centering
            \includegraphics[width=\textwidth]{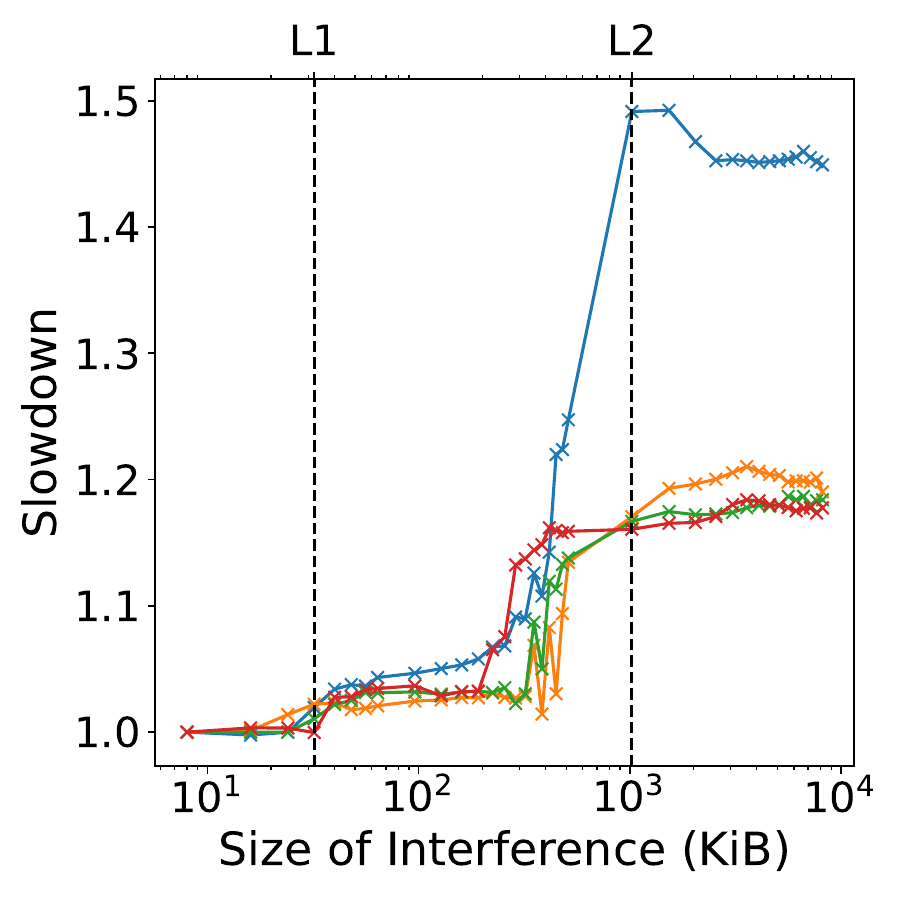}
            \captionsetup{justification=centering}
            \caption{zcu102: Set / Modify}
            \label{fig:all-zcu102-set-mser-modify-cif}
        \end{subfigure}
        \hfill
        
        \caption{Execution Slowdown on \textit{'Mser'} benchmark for \textit{'CIF'} dataset on \textit{'ZCU102'} with Interferences and cache partitioning.}
        \label{fig:zcu102-mser-cif}
    \end{figure}

    \begin{figure}[H]
        \begin{subfigure}{\textwidth}
            \centering
            \includegraphics[width=0.5\textwidth]{figures/set_subplot/legend.pdf}
        \end{subfigure}
        \centering
        
        \begin{subfigure}{0.24\textwidth}
            \centering
            \includegraphics[width=\textwidth]{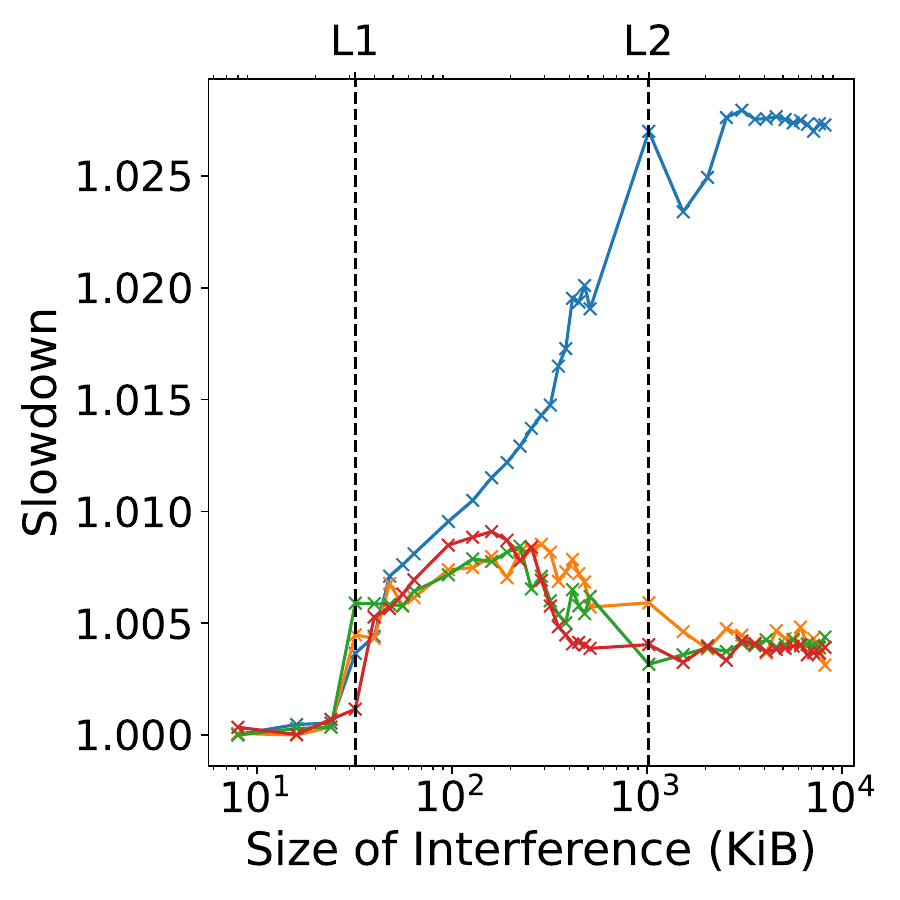}
            \captionsetup{justification=centering}
            \caption{zcu102: Set / Read}
            \label{fig:all-zcu102-set-tracking-read-cif}
        \end{subfigure}
        \hfill
        \begin{subfigure}{0.24\textwidth}
            \centering
            \includegraphics[width=\textwidth]{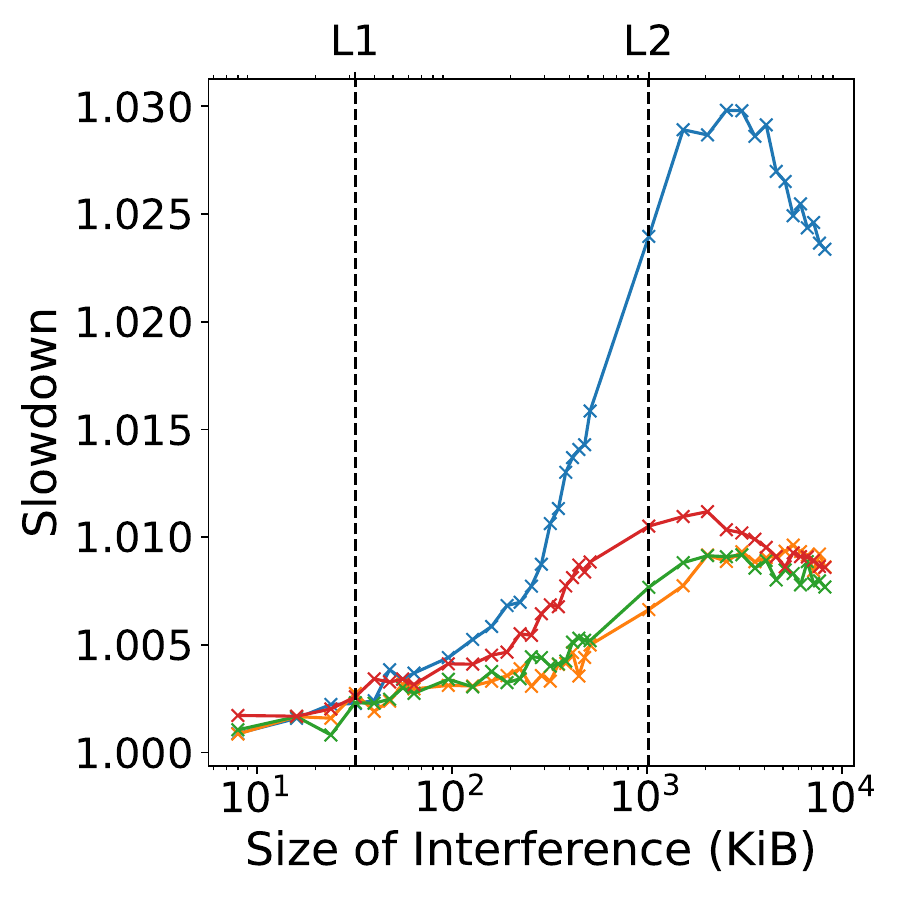}
            \captionsetup{justification=centering}
            \caption{zcu102: Set / Write}
            \label{fig:all-zcu102-set-tracking-write-cif}
        \end{subfigure}
        \hfill
        \begin{subfigure}{0.24\textwidth}
            \centering
            \includegraphics[width=\textwidth]{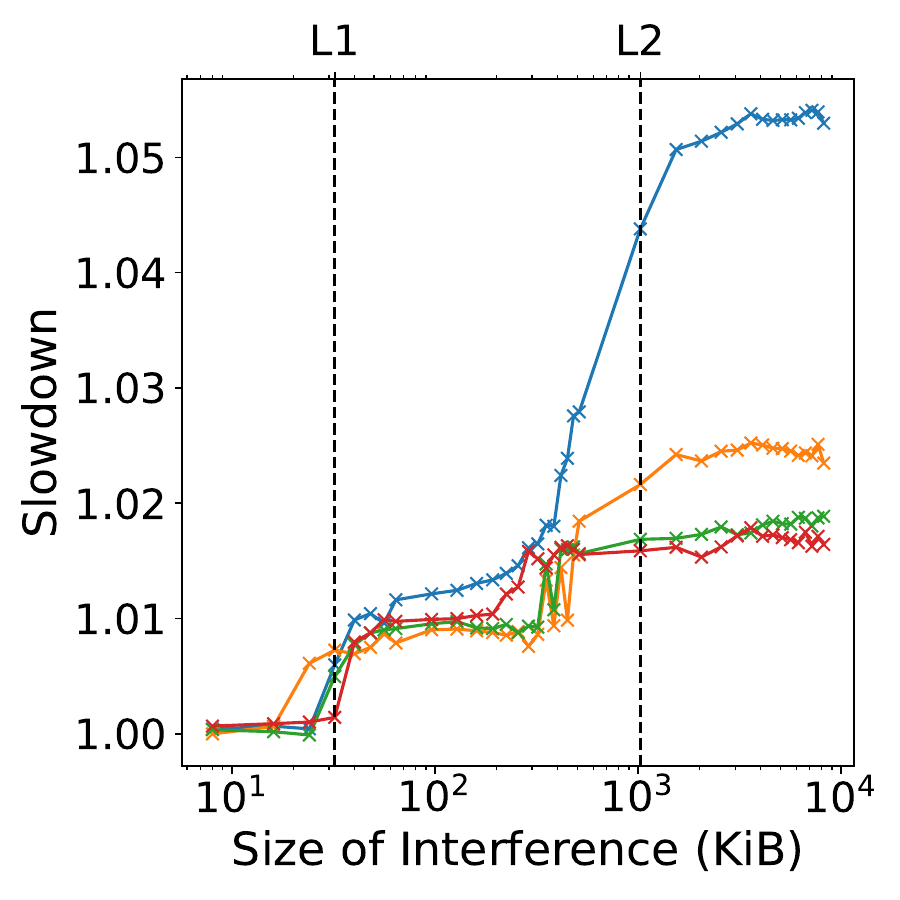}
            \captionsetup{justification=centering}
            \caption{zcu102: Set / Modify}
            \label{fig:all-zcu102-set-tracking-modify-cif}
        \end{subfigure}
        \hfill
        
        \caption{Execution Slowdown on \textit{'Tracking'} benchmark for \textit{'CIF'} dataset on \textit{'ZCU102'} with Interferences and cache partitioning.}
        \label{fig:zcu102-tracking-cif}
    \end{figure}

    \begin{figure}[H]
        \begin{subfigure}{\textwidth}
            \centering
            \includegraphics[width=0.5\textwidth]{figures/set_subplot/legend.pdf}
        \end{subfigure}
        \centering
        
        \begin{subfigure}{0.24\textwidth}
            \centering
            \includegraphics[width=\textwidth]{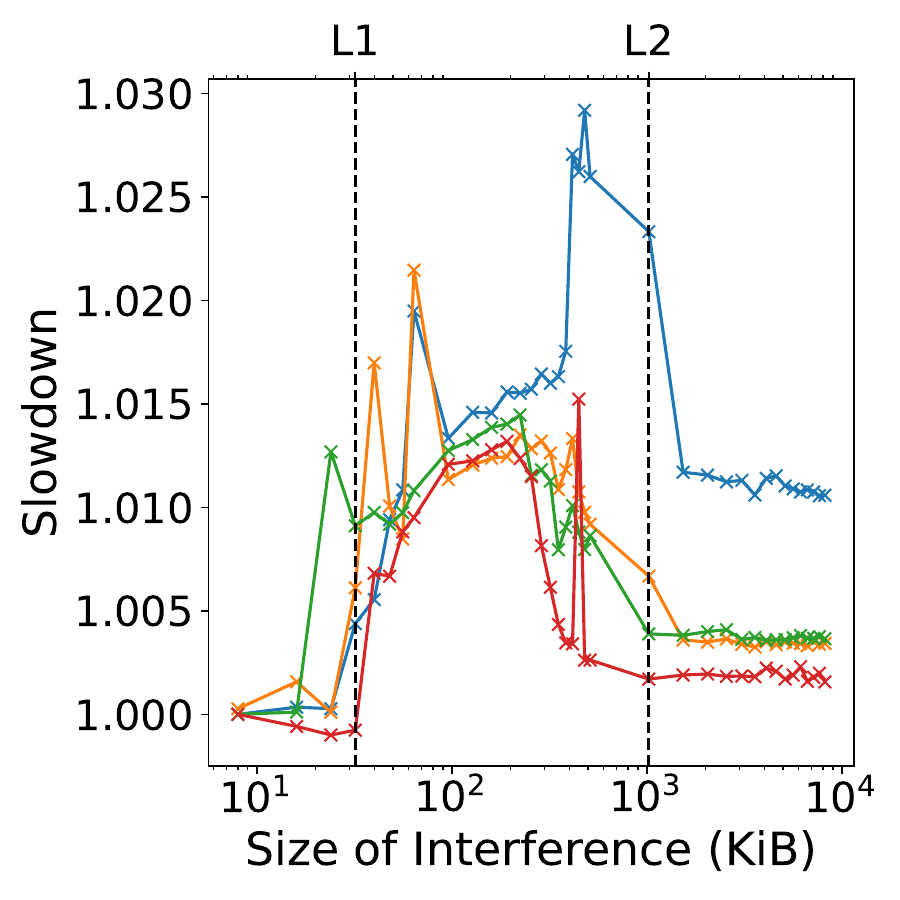}
            \captionsetup{justification=centering}
            \caption{zcu102: Set / Read}
            \label{fig:all-zcu102-set-sift-read-cif}
        \end{subfigure}
        \hfill
        \begin{subfigure}{0.24\textwidth}
            \centering
            \includegraphics[width=\textwidth]{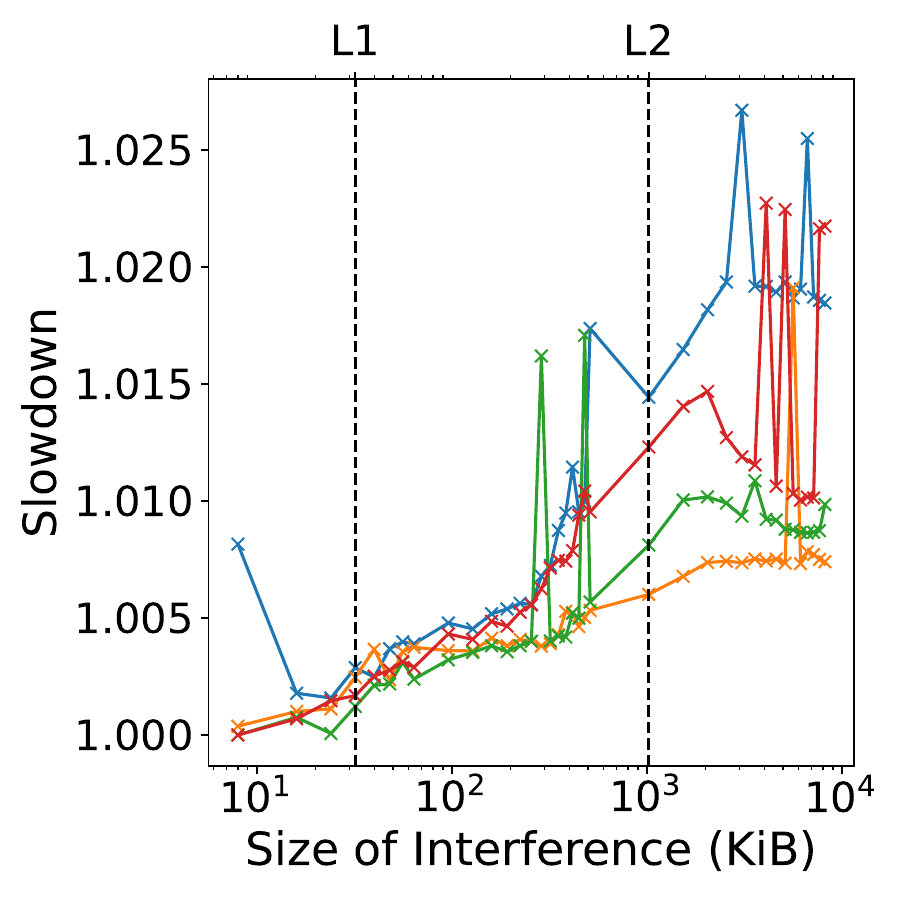}
            \captionsetup{justification=centering}
            \caption{zcu102: Set / Write}
            \label{fig:all-zcu102-set-sift-write-cif}
        \end{subfigure}
        \hfill
        \begin{subfigure}{0.24\textwidth}
            \centering
            \includegraphics[width=\textwidth]{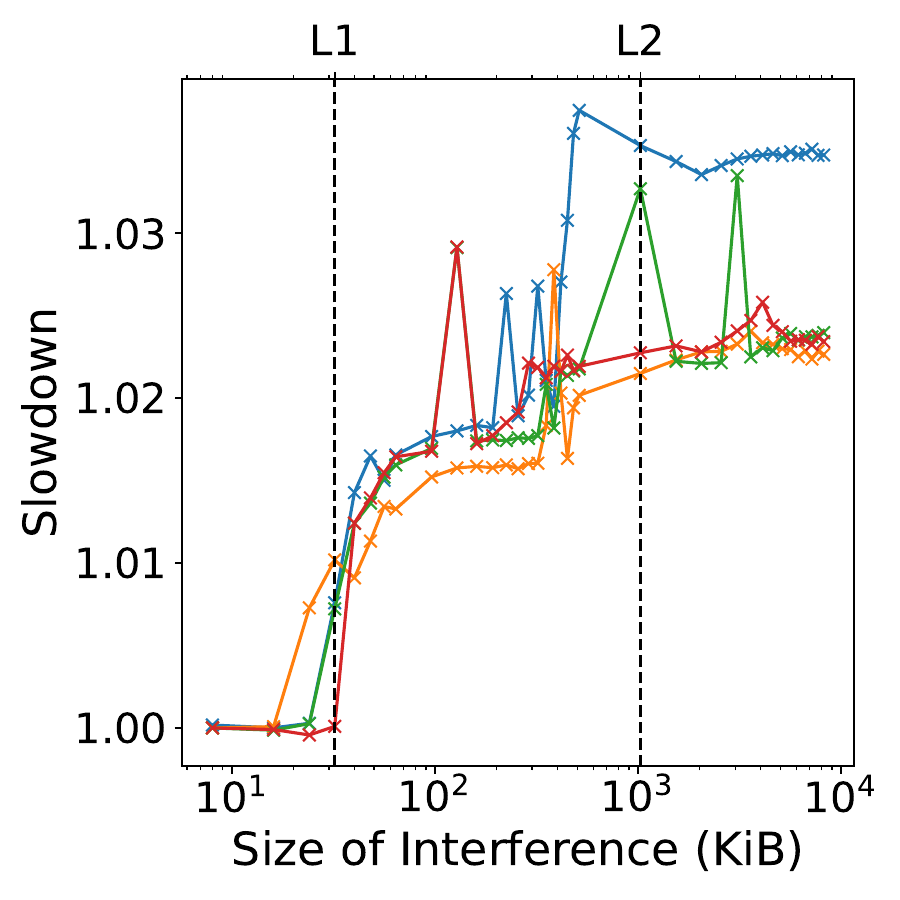}
            \captionsetup{justification=centering}
            \caption{zcu102: Set / Modify}
            \label{fig:all-zcu102-set-sift-modify-cif}
        \end{subfigure}
        \hfill
        
        \caption{Execution Slowdown on \textit{'Sift'} benchmark for \textit{'CIF'} dataset on \textit{'ZCU102'} with Interferences and cache partitioning.}
        \label{fig:zcu102-sift-cif}
    \end{figure}

    \begin{figure}[H]
        \begin{subfigure}{\textwidth}
            \centering
            \includegraphics[width=0.5\textwidth]{figures/set_subplot/legend.pdf}
        \end{subfigure}
        \centering
        
        \begin{subfigure}{0.24\textwidth}
            \centering
            \includegraphics[width=\textwidth]{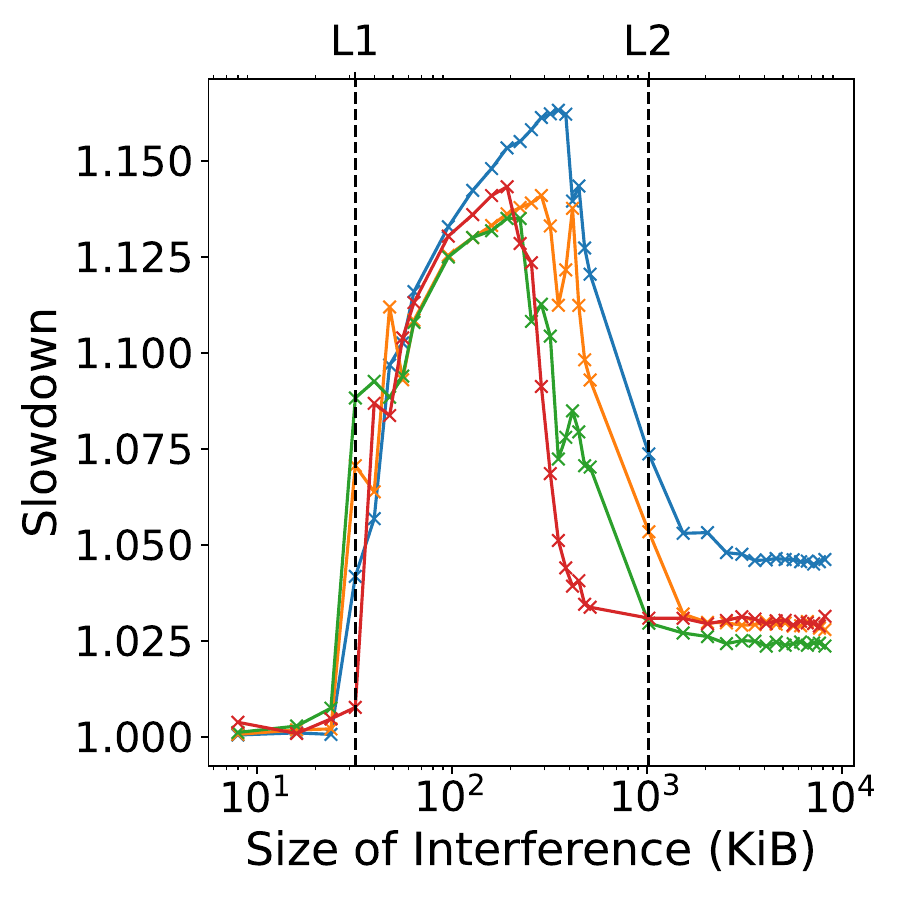}
            \captionsetup{justification=centering}
            \caption{zcu102: Set / Read}
            \label{fig:all-zcu102-set-disparity-read-vga}
        \end{subfigure}
        \hfill
        \begin{subfigure}{0.24\textwidth}
            \centering
            \includegraphics[width=\textwidth]{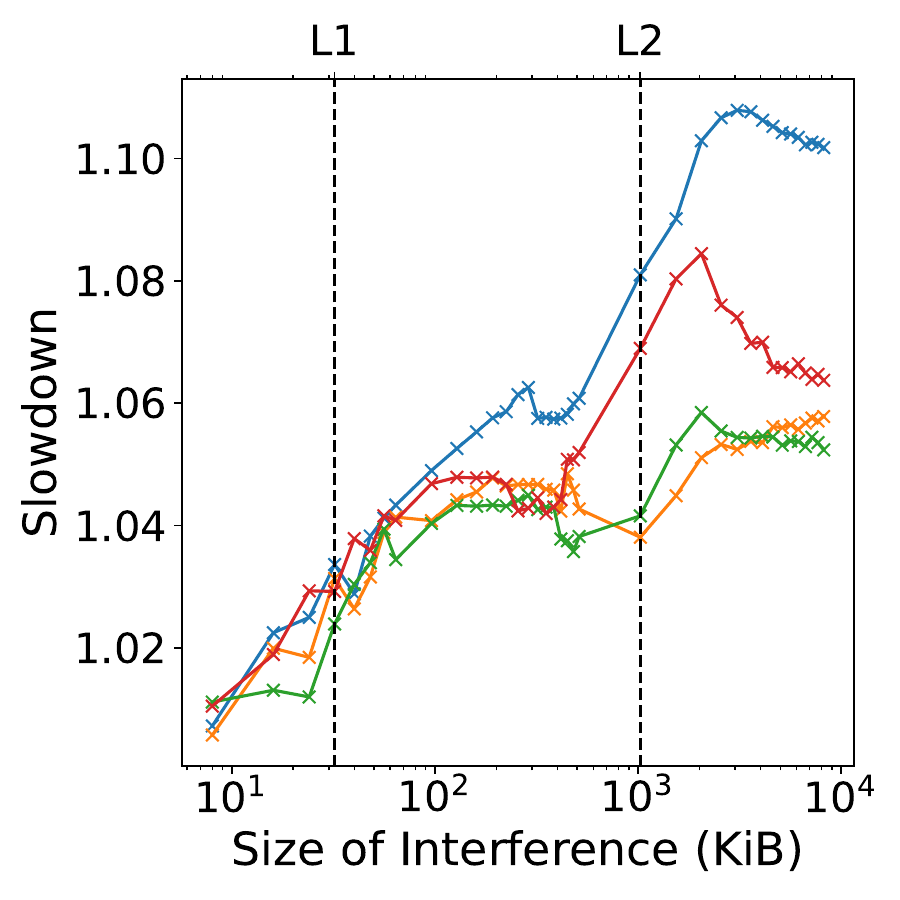}
            \captionsetup{justification=centering}
            \caption{zcu102: Set / Write}
            \label{fig:all-zcu102-set-disparity-write-vga}
        \end{subfigure}
        \hfill
        \begin{subfigure}{0.24\textwidth}
            \centering
            \includegraphics[width=\textwidth]{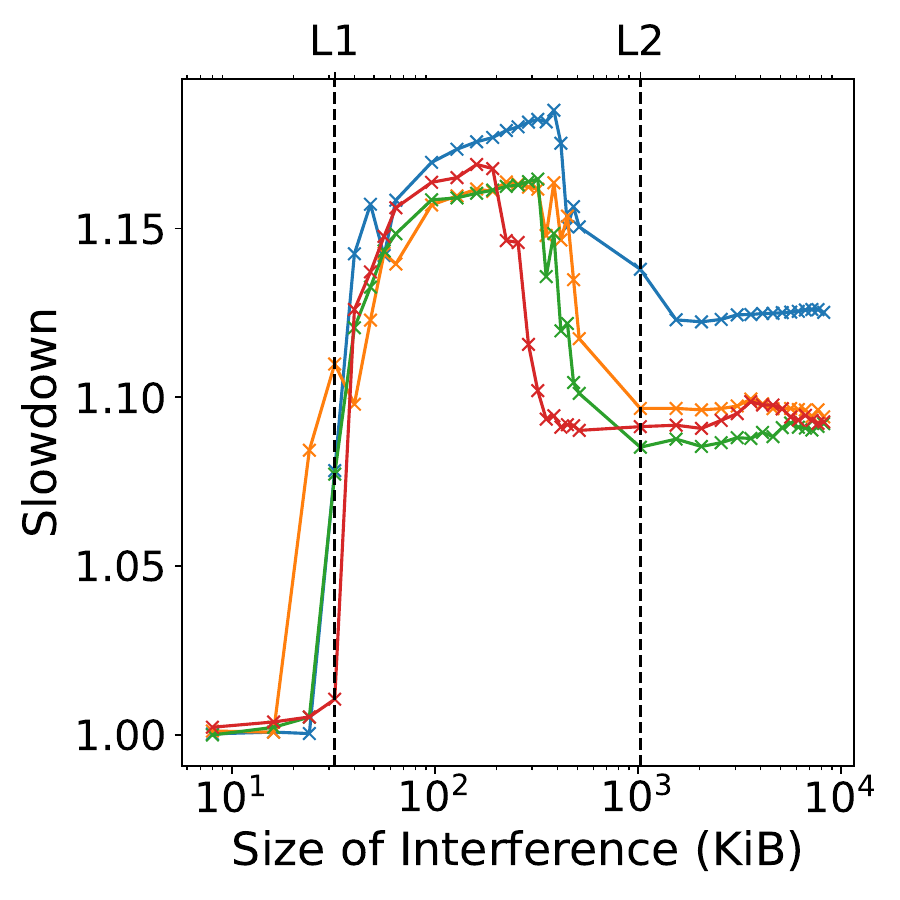}
            \captionsetup{justification=centering}
            \caption{zcu102: Set / Modify}
            \label{fig:all-zcu102-set-disparity-modify-vga}
        \end{subfigure}
        \hfill
        
        \caption{Execution Slowdown on \textit{'Disparity'} benchmark for \textit{'VGA'} dataset on \textit{'ZCU102'} with Interferences and cache partitioning.}
        \label{fig:zcu102-disparity-vga}
    \end{figure}

    \begin{figure}[H]
        \begin{subfigure}{\textwidth}
            \centering
            \includegraphics[width=0.5\textwidth]{figures/set_subplot/legend.pdf}
        \end{subfigure}
        \centering
        
        \begin{subfigure}{0.24\textwidth}
            \centering
            \includegraphics[width=\textwidth]{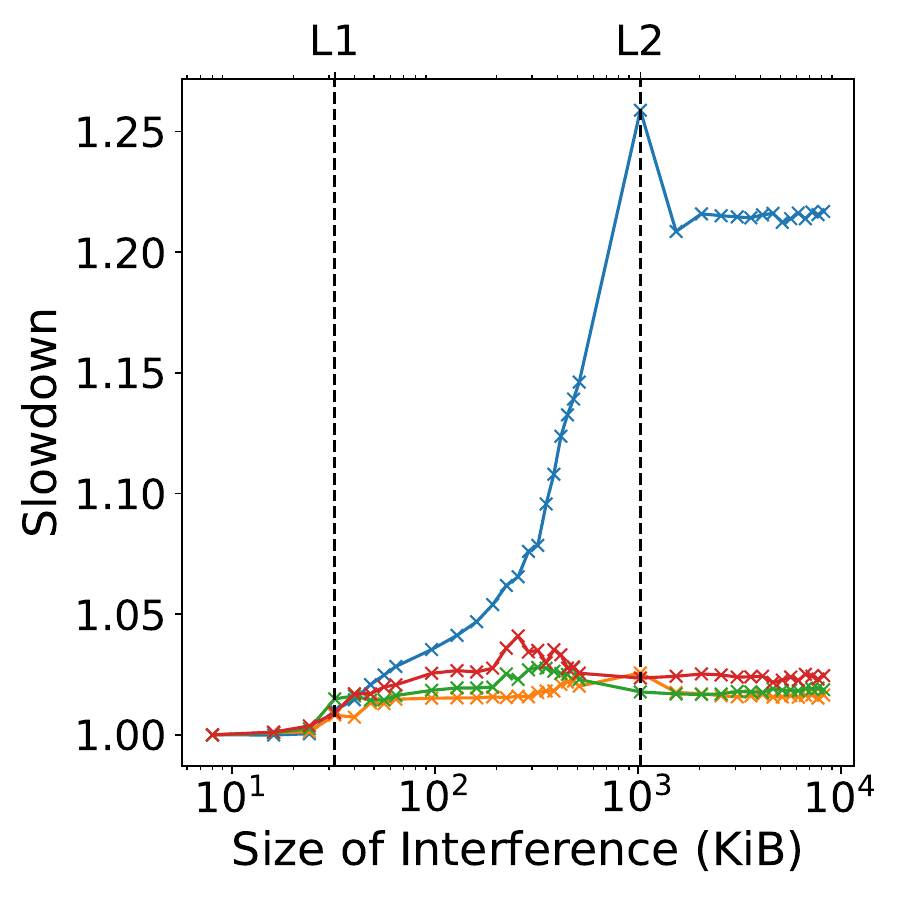}
            \captionsetup{justification=centering}
            \caption{zcu102: Set / Read}
            \label{fig:all-zcu102-set-mser-read-vga}
        \end{subfigure}
        \hfill
        \begin{subfigure}{0.24\textwidth}
            \centering
            \includegraphics[width=\textwidth]{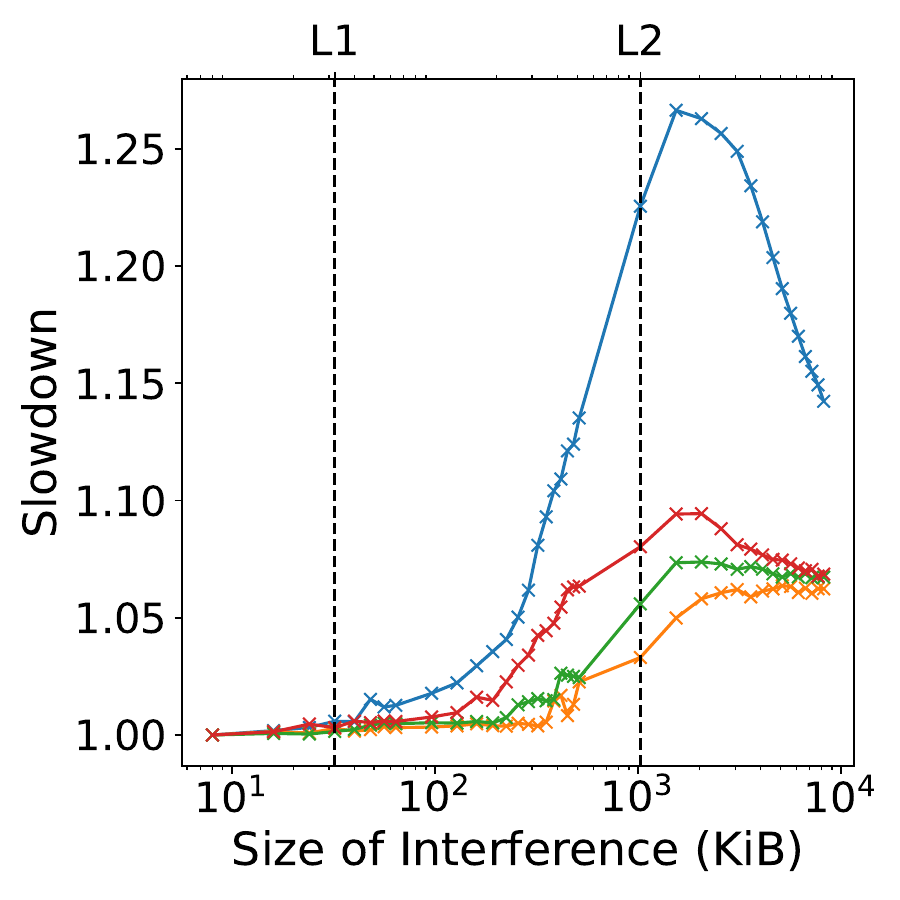}
            \captionsetup{justification=centering}
            \caption{zcu102: Set / Write}
            \label{fig:all-zcu102-set-mser-write-vga}
        \end{subfigure}
        \hfill
        \begin{subfigure}{0.24\textwidth}
            \centering
            \includegraphics[width=\textwidth]{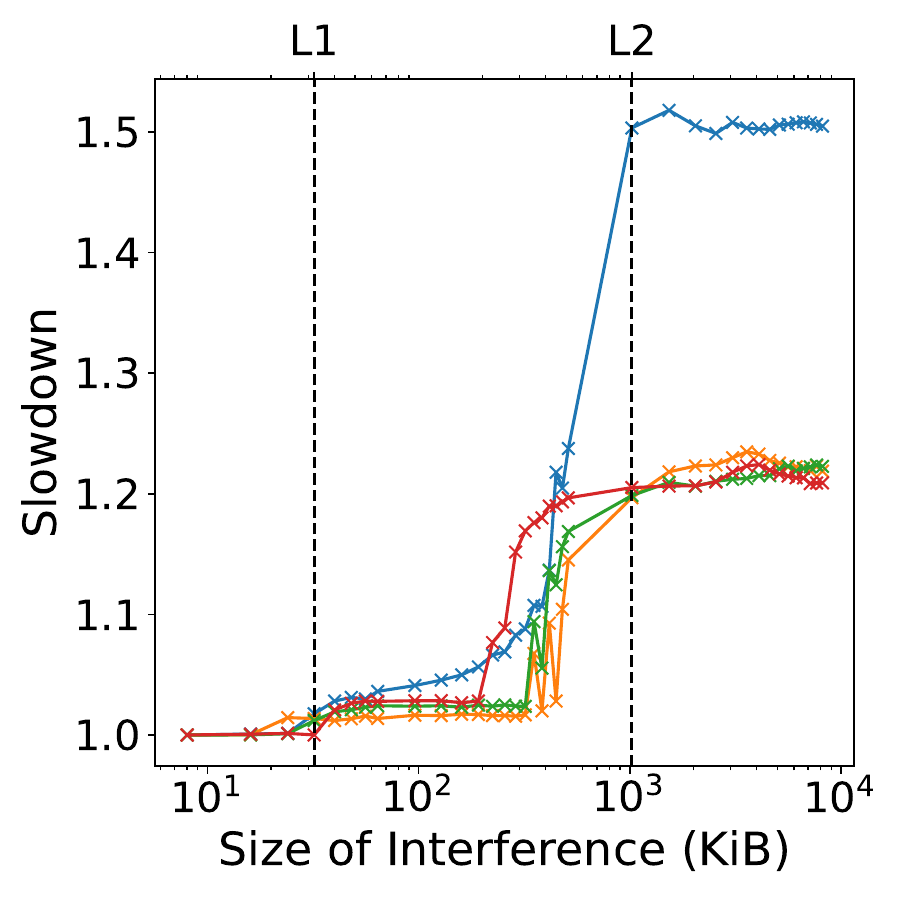}
            \captionsetup{justification=centering}
            \caption{zcu102: Set / Modify}
            \label{fig:all-zcu102-set-mser-modify-vga}
        \end{subfigure}
        \hfill
        
        \caption{Execution Slowdown on \textit{'Mser'} benchmark for \textit{'VGA'} dataset on \textit{'ZCU102'} with Interferences and cache partitioning.}
        \label{fig:zcu102-mser-vga}
    \end{figure}

    \begin{figure}[H]
        \begin{subfigure}{\textwidth}
            \centering
            \includegraphics[width=0.5\textwidth]{figures/set_subplot/legend.pdf}
        \end{subfigure}
        \centering
        
        \begin{subfigure}{0.24\textwidth}
            \centering
            \includegraphics[width=\textwidth]{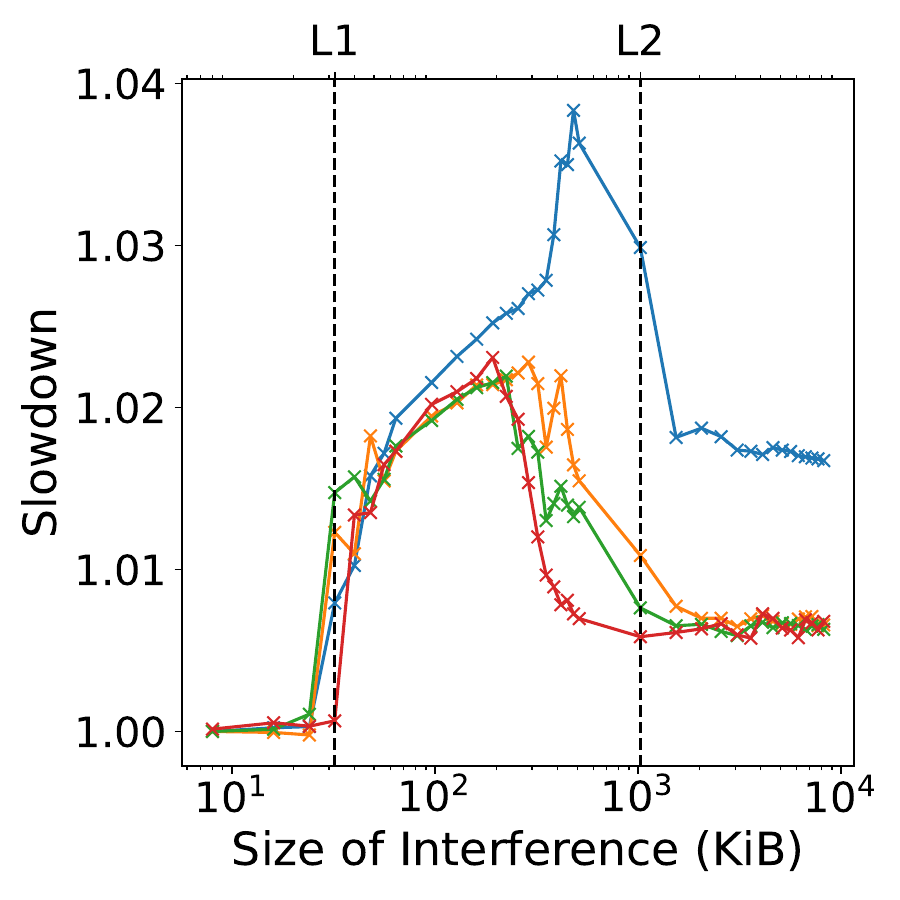}
            \captionsetup{justification=centering}
            \caption{zcu102: Set / Read}
            \label{fig:all-zcu102-set-tracking-read-vga}
        \end{subfigure}
        \hfill
        \begin{subfigure}{0.24\textwidth}
            \centering
            \includegraphics[width=\textwidth]{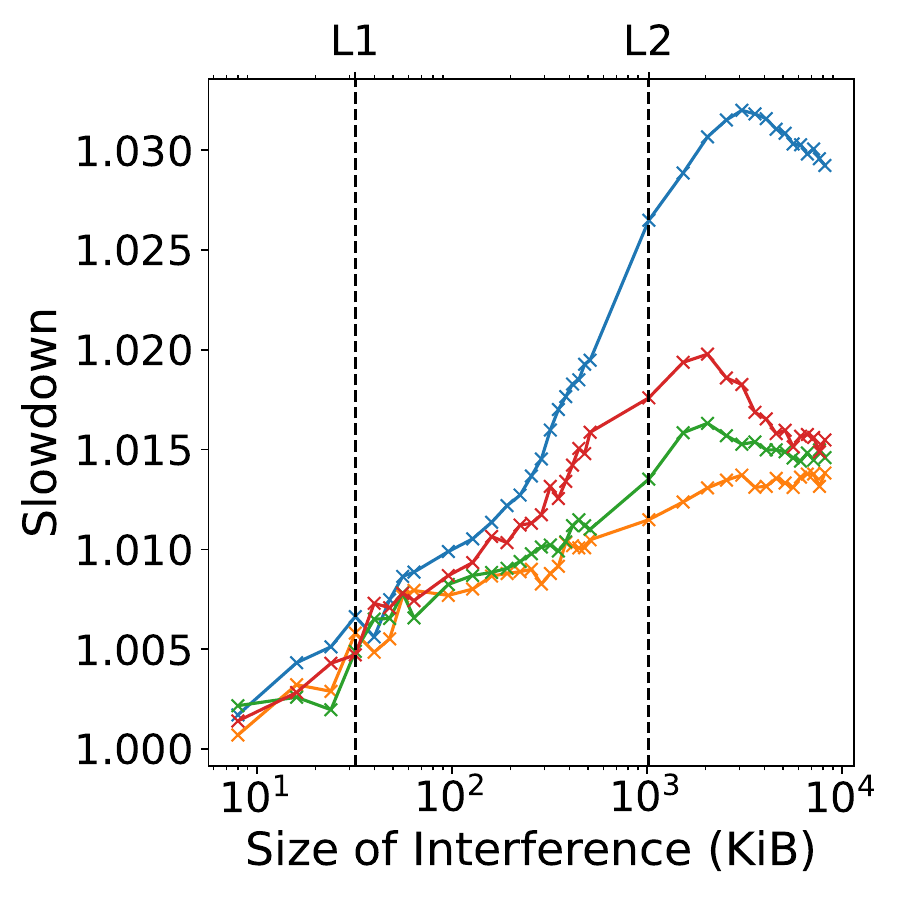}
            \captionsetup{justification=centering}
            \caption{zcu102: Set / Write}
            \label{fig:all-zcu102-set-tracking-write-vga}
        \end{subfigure}
        \hfill
        \begin{subfigure}{0.24\textwidth}
            \centering
            \includegraphics[width=\textwidth]{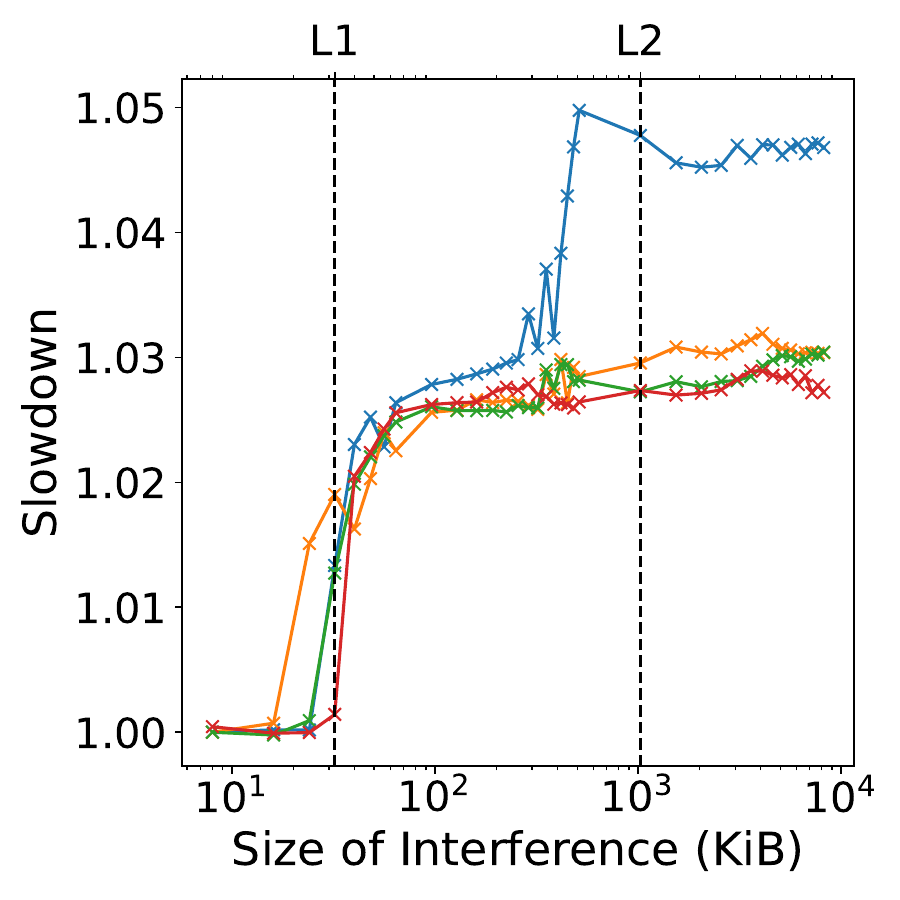}
            \captionsetup{justification=centering}
            \caption{zcu102: Set / Modify}
            \label{fig:all-zcu102-set-tracking-modify-vga}
        \end{subfigure}
        \hfill
        
        \caption{Execution Slowdown on \textit{'Tracking'} benchmark for \textit{'VGA'} dataset on \textit{'ZCU102'} with Interferences and cache partitioning.}
        \label{fig:zcu102-tracking-vga}
    \end{figure}

    \begin{figure}[H]
        \begin{subfigure}{\textwidth}
            \centering
            \includegraphics[width=0.5\textwidth]{figures/set_subplot/legend.pdf}
        \end{subfigure}
        \centering
        
        \begin{subfigure}{0.24\textwidth}
            \centering
            \includegraphics[width=\textwidth]{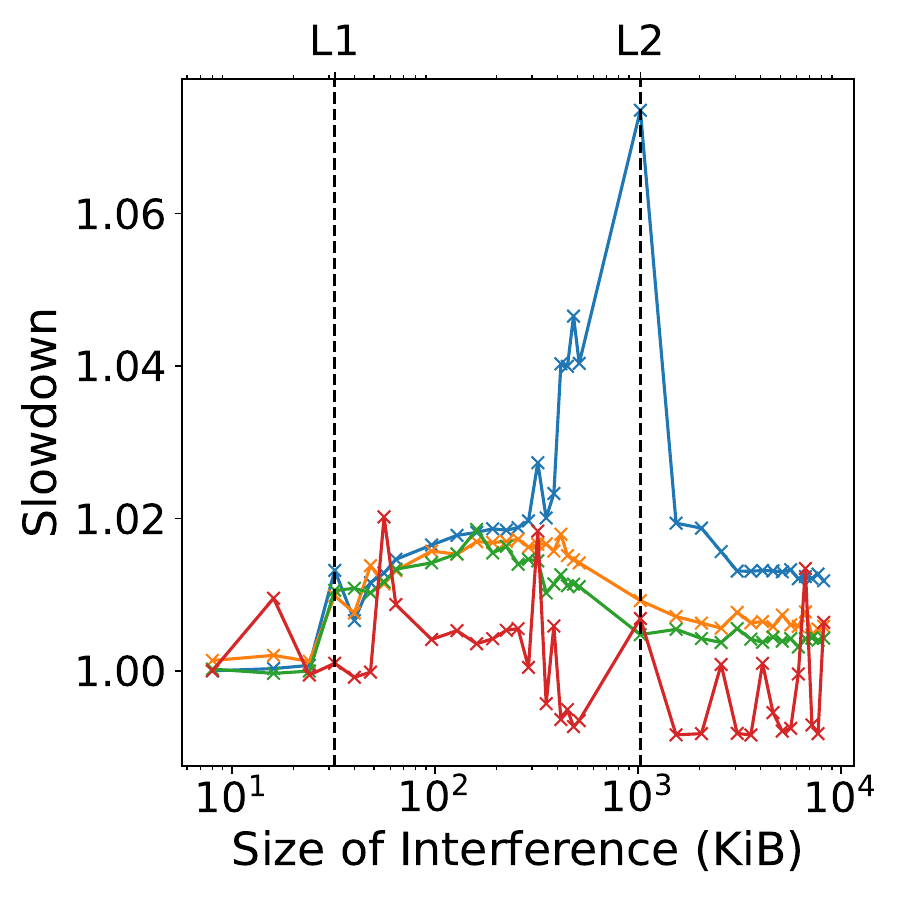}
            \captionsetup{justification=centering}
            \caption{zcu102: Set / Read}
            \label{fig:all-zcu102-set-sift-read-vga}
        \end{subfigure}
        \hfill
        \begin{subfigure}{0.24\textwidth}
            \centering
            \includegraphics[width=\textwidth]{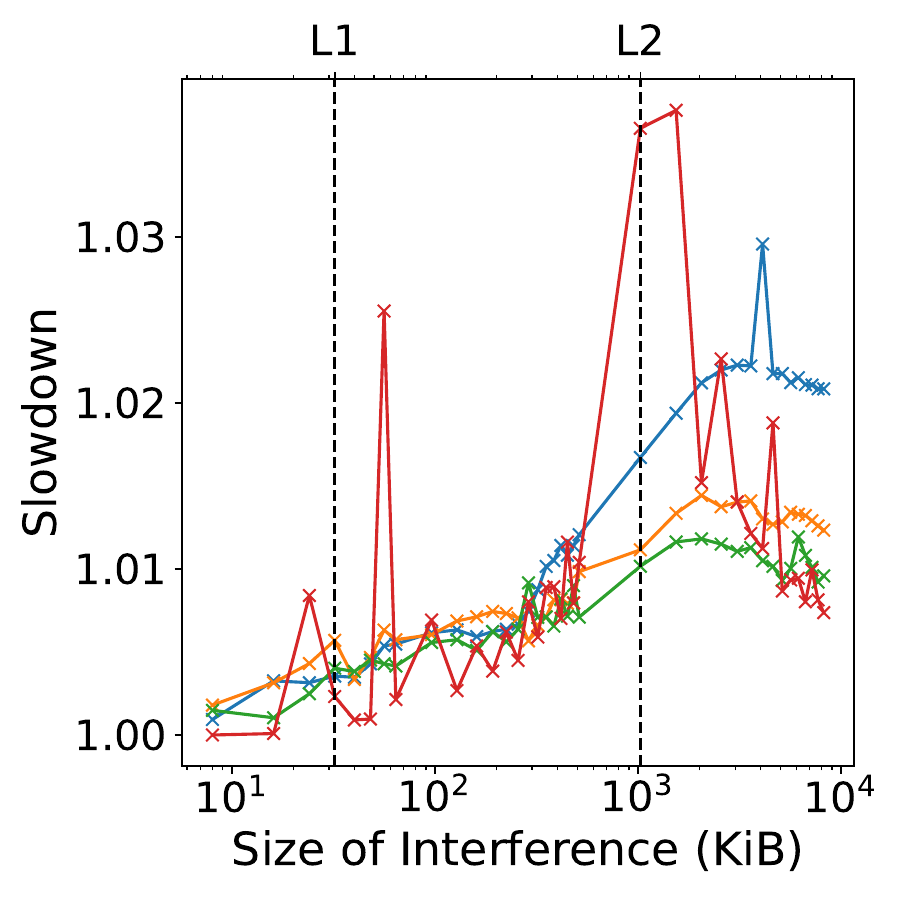}
            \captionsetup{justification=centering}
            \caption{zcu102: Set / Write}
            \label{fig:all-zcu102-set-sift-write-vga}
        \end{subfigure}
        \hfill
        \begin{subfigure}{0.24\textwidth}
            \centering
            \includegraphics[width=\textwidth]{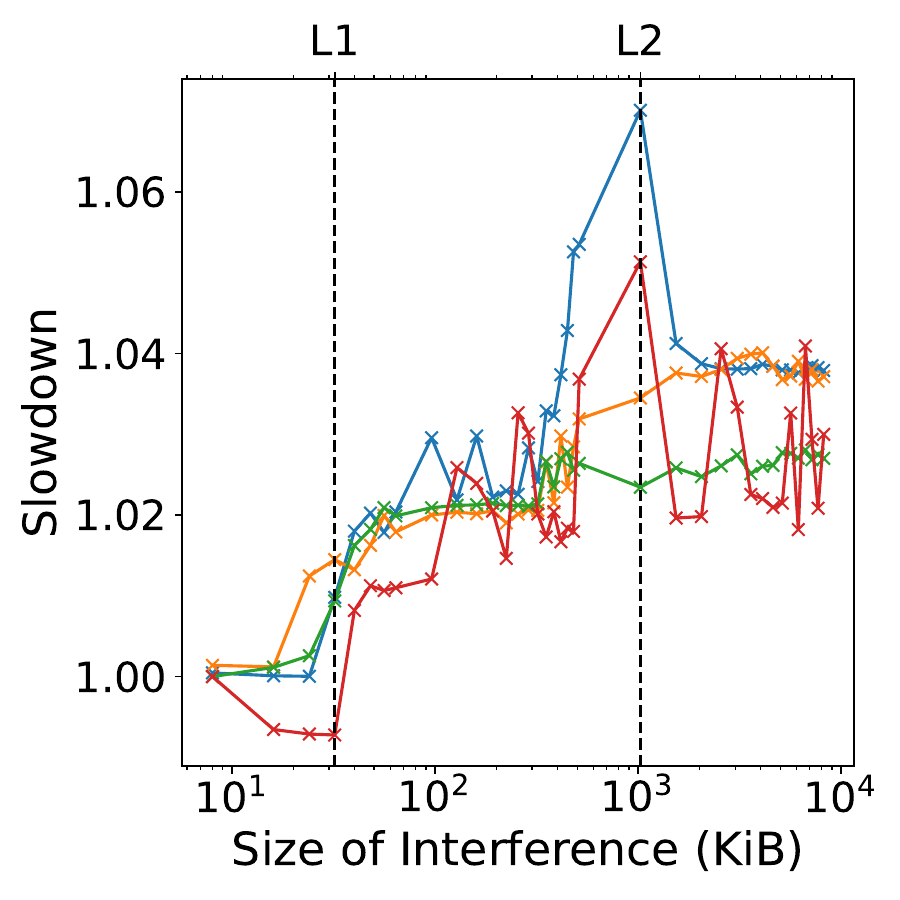}
            \captionsetup{justification=centering}
            \caption{zcu102: Set / Modify}
            \label{fig:all-zcu102-set-sift-modify-vga}
        \end{subfigure}
        \hfill
        
        \caption{Execution Slowdown on \textit{'Sift'} benchmark for \textit{'VGA'} dataset on \textit{'ZCU102'} with Interferences and cache partitioning.}
        \label{fig:zcu102-sift-vga}
    \end{figure}

        \clearpage